\newif\ifsubmission
\newif\ifanon
\newif\ifextendedabstract
\newif\ifstocproceedings

 \submissiontrue

\makeatletter
\@ifundefined{ifdraft}{
    \newif\iffastcompile

    \fastcompilefalse
}{}
\makeatother

\documentclass[11pt]{article}
\usepackage{headers}

\title{
Quantum Lazy Sampling and Path Recording for Any Group
}
\date{}

\ifanon
    \author{}
\else
    \newauthor{Ben Foxman}
    {
        Yale University, New Haven, CT, USA
    }
    \newauthor{Alex Lombardi}
    {
        Princeton University, Princeton, NJ, USA
    }
    \newauthor{Fermi Ma}
    {
        New York University, New York, NY, USA
    }
    \newauthor{Barak Nehoran}
    {
        Columbia University, New York, NY, USA
    }
    \newauthor{John Wright}
    {
        University of California Berkeley, Berkeley, CA, USA
    }
    \finishauthors{}
\fi

\begin{document}

\maketitle
\thispagestyle{empty}

\begin{abstract}
A central challenge in quantum algorithm analysis and cryptography is reasoning about algorithms with oracle access to a random group element (e.g. a random function, a random permutation, a random unitary). Can we efficiently simulate such algorithms? Can we determine what they know after $t$ queries? Classically, an important tool for this is \emph{lazy sampling}, where the oracle does not commit to the full group element at the beginning, but rather samples partial information about it on the fly. We study a quantum analog of lazy sampling: \emph{compressed oracles} (or \emph{recording oracles}), which are quantum data structures that allow such on-the-fly simulation for quantum queries.
Compressed oracles were originally introduced by Zhandry (CRYPTO '19) for random functions, were generalized to random unitaries by Ma-Huang (STOC '25) and to permutations by Carolan (STOC '26), and have been employed to great effect in security proofs and query complexity lower bounds due to their \emph{interpretability}. 

In this work, we define and analyze a \emph{general-purpose} and \emph{interpretable} path-recording oracle, derived from first principles, that perfectly simulates random elements of any closed subgroup of $U(N)$. 
Our path-recording oracle stores superpositions of $t$ input-output pairs $\ket{(x_1, y_1), \dots, (x_t, y_t)}$, which encode a Feynman path explored by the algorithm and thus transparently records the information that the algorithm may have learned from its queries. Our compressed oracle builds on a recent work of Grinko and Yoshida (QIP '26), who proposed a different kind of general-purpose compressed oracle without clear interpretability. Crucially for applications, we derive an operationally useful mathematical description of our update procedure in terms of the commutant of the group's tensor power representation.

One powerful feature of our path-recording oracle is that it enables direct \emph{comparisons} between compressed oracles for different groups, which gives a new technique for proving pseudorandomness results. For our main application, we formally relate the $S_N$ and $U(N)$ compressed oracles, yielding what is arguably the simplest construction to date of pseudorandom unitaries: the product $PC$ of a pseudorandom permutation and a random Clifford. This improves on the prior ``$PFC$'' construction of (Metger-Poremba-Sinha-Yuen, FOCS '24; Ma-Huang, STOC '25).
\end{abstract}
\thispagestyle{empty}

\newpage
\thispagestyle{empty}
\tableofcontents
\thispagestyle{empty}

\newpage

\iffastcompile 
\else
    \section{Introduction}

In algorithms, complexity, cryptography, and quantum information science, it is common to encounter an algorithm or experiment that has access to some form of random transformation. In the context of quantum computing, the algorithm is typically quantum, and the transformation itself may be either classical (a function $x\mapsto f(x)$) or quantum (an $n$-qubit unitary transformation or a quantum channel). Although these transformations, such as random Boolean functions or Haar-random unitaries, are not computationally efficient, they serve several important purposes:

\begin{itemize}
    \item They serve as an ideal model for justifying the correctness of algorithms. In practice, they can be replaced with an efficient transformation that is either provably pseudorandom or \textquote{scrambling} enough that the ideal model is accepted as a heuristic \cite{CCS:BelRog93,QIP:SMLBH25}. This captures a variety of applications from the classical use of hashing in data structures \cite{hashing-textbook} to theoretical justifications of quantum algorithms such as random circuit sampling \cite{Nature:google-quantum-supremacy}.
    \item They can be used to prove \emph{lower bounds} ruling out efficient black-box algorithms for computational problems (as \textquote{random instances} of black-box problems are often the hardest).
    For example, this can be used to rule out black-box quantum algorithms for NP that are faster than Grover search~\cite{SIAM:BBBV97,C:Zhandry19}.
    \item They can be used for analyzing the security of \emph{cryptographic constructions} such as the Luby-Rackoff construction of pseudorandom permutations/block ciphers \cite{SIAM:LubRac88,STOC:Carolan26} and the Metger-Poremba-Sinha-Yuen construction (and Ma-Huang analysis) of pseudorandom unitaries \cite{FOCS:MPSY24,STOC:MaHua25}.
    ~Indeed, most such security proofs proceed by a hybrid argument: switching a pseudorandom object (such as a PRF) to a fully random one (a random function) and then arguing that an adversary cannot break the idealized variant of the construction.
\end{itemize}
In this work, we seek to understand how to analyze the behavior of such algorithms. Classically, the basic tool/data structure for such analysis is called \textquote{lazy sampling}: instead of thinking of the exponential-size random object as being sampled \textquote{at the beginning}, only sample the pieces of the object that are needed for each step of the algorithm. For example,

\begin{itemize}
    \item Instead of sampling a random Boolean function $f: \{0,1\}^n \rightarrow \{0,1\}^n$, one can lazily sample its truth table. Whenever the algorithm wants to query $f(x)$, it (1) checks whether an input-output pair $(x,y)$ has been recorded in the existing partial truth table (and uses $y$ if so), and (2) otherwise samples a new random string $y$ and appends $(x,y)$ to the data structure.
    \item If instead the algorithm wants to query a random permutation $p: \{0,1\}^n \rightarrow \{0,1\}^n$, one can do exactly the same thing, except that each lazily sampled output $y$ should be sampled uniformly from all output strings that do not already appear in the data structure.
\end{itemize}
In general, this kind of lazy sampling can be done for any distribution $\mathcal D$ of functions such that the conditional distribution
\[p(y\mid x, (x_1, y_1), \hdots, (x_t, y_t)) = \underset{f\gets \mathcal D}{\mathrm{Pr}}\Big[f(x) = y  \mid f(x_1) = y_1, \hdots, f(x_t) = y_t\Big]
\]
is efficiently sampleable. For this efficient simulation, we think of $f$ as undetermined outside of the partial truth table $(x_1, y_1), \hdots, (x_t, y_t)$.  

\paragraph{Compressed Oracles and Recording Oracles} For quantum algorithms, the tool of choice for this problem has been \emph{compressed oracles} \cite{C:Zhandry19,STOC:MaHua25} (also known as \emph{recording oracles}). Compressed oracles are \emph{quantum data structures} that adapt the classical notion of lazy-sampling to allow quantum interactions. That is, they enable efficient simulation of quantum algorithms that interact (in a black-box manner) with a randomly chosen operation. 
In this work, this random operation will be a unitary 
sampled from some distribution $\calD$. Specifically, we will consider the case in which $\calD$ is chosen from 
the uniform measure over a finite group or, more generally, from the Haar measure over any compact Lie group.%
\footnote{
    Compact Lie groups are the groups that can be realized as closed subgroups of the unitary group in finite dimension, and include all finite groups.
}

Informally, a compressed oracle records in superposition the partial information that the algorithm \emph{would have discovered} about the group element $g$ if it were interacting with the real oracle. Rather than fully sampling $g$, the compressed oracle is stateful, maintaining this partial information on-the-fly. But unlike classical lazy-sampling, the \textquote{random choices} are made coherently, in superposition. As the algorithm queries the oracle, its internal state and the data structure become entangled.

The functionality of compressed oracles is that a quantum query algorithm should not be able to distinguish whether it is interacting with the real oracle or with the stateful compressed oracle. This allows us to use the compressed oracle to analyze the behavior of the algorithm if it were to query the real oracle.

\crefformat{desideratum}{(#2#1#3)}
\Crefformat{desideratum}{(#2#1#3)}
\crefrangeformat{desideratum}{(#3#1#4)--(#5#2#6)}
\Crefrangeformat{desideratum}{(#3#1#4)--(#5#2#6)}

\crefmultiformat{desideratum}
  {(#2#1#3)}
  { and (#2#1#3)}
  {, (#2#1#3)}
  {, and (#2#1#3)}

\Crefmultiformat{desideratum}
  {(#2#1#3)}
  { and (#2#1#3)}
  {, (#2#1#3)}
  {, and (#2#1#3)}

\begin{remark}[Desiderata for compressed oracles]
We list some key desirable properties of a compressed oracle.
\begin{enumerate}
    \item 
    \label[desideratum]{item:desideratum-compressed}
    \textbf{Polynomial-space encoding of the data structure.} This is the bare minimum requirement to consider the oracle \textquote{compressed}.
    \item 
    \label[desideratum]{item:desideratum-efficient}
    \textbf{Polynomial-time update algorithm.} This enables efficient simulation of quantum query algorithms accessing the oracle.
    \item 
    \label[desideratum]{item:desideratum-accurate}
    \textbf{Perfect simulation, or at least small simulation error.} That is, the algorithm should not be able to distinguish it from the real oracle. Ideally, this simulation is \emph{perfect}, but if not, then we would like it to be valid for \emph{as many queries} as possible.
    \item 
    \label[desideratum]{item:desideratum-interpretable}
    \textbf{Interpretability.} This property is somewhat subjective. Given a quantum query algorithm $A^{(\cdot)}$ that we are simulating using a compressed oracle, we would like the data structure to encode, in some understandable way, useful information about the algorithm's computation. This is often crucial for proving any sort of lower bound (including cryptographic security proofs) against quantum query algorithms. 
\end{enumerate}
\end{remark}
In the classical setting, lazy sampling algorithms indeed satisfy all of these properties (with property (2) requiring efficient conditional sampling). For example, after $t$ classical queries to a random function $f$, lazy-sampling the function has the effect of recording exactly the information of a $t$-size partial truth table of $f$. 
Unlike classical algorithms, quantum algorithms can learn information \emph{in superposition}, and even \emph{forget} the information previously learned.
Despite much recent effort, building and analyzing compressed oracles in the quantum setting has thus proved to be far more subtle. In this paper, we ask the following question:
\begin{center}
    \emph{
        Is there a general-purpose, \emph{interpretable} compressed oracle that allows one to \\ implement such 
        quantum 
        lazy-sampling 
        for every compact group?
    }
\end{center}

\subsection{Prior Work}
We briefly discuss some prominent examples of existing compressed oracles:

\paragraph{Random Boolean functions}
A seminal work of Zhandry \cite{C:Zhandry19} built compressed oracles for simulating quantum queries to random Boolean functions. Zhandry considered two related query models:
\begin{itemize}
    \item For a random function $f: \{0,1\}^n \rightarrow \{0,1\}$, queries to the $n$-qubit unitary $\ket{x}\mapsto (-1)^{f(x)} \ket{x}$.
    \item For a random function $f: \{0,1\}^n \rightarrow \{0,1\}^m$, queries to the $(n+m)$-qubit unitary $\ket{x}\ket{y} \mapsto \ket{x} \ket{y\oplus f(x)}$. 
\end{itemize}
\cite{C:Zhandry19} gives compressed oracles for both of these models. The compressed \emph{phase} oracle (and its generalization to a controlled phase oracle) unambiguously satisfies all four of the above desiderata. Regarding \Cref{item:desideratum-interpretable}, the compressed phase oracle simply records, for each input $x$ that has been queried by the algorithm, the parity of the number of times that it has been queried. The compressed \textquote{standard} oracle is written to satisfy properties \Cref{item:desideratum-compressed,item:desideratum-efficient,item:desideratum-accurate} and \emph{mostly} satisfies \Cref{item:desideratum-interpretable} up to some small error terms (see, e.g., \cite[Section 3]{AC:Unruh23}). 

These compressed oracles have been used to great effect in proving quantum query complexity lower bounds \cite{C:Zhandry19,EC:LiuZha19,AC:PenCaoXue25,QIP:TanWriZha25,JefZur25} as well as in (post-)quantum cryptographic constructions and security analyses \cite{C:Zhandry19,C:LiuZha19,EC:CFHL21,EC:HamLiuSin24,STOC:Carolan26}.

\paragraph{Haar-random unitaries}
A recent work of Ma and Huang \cite{STOC:MaHua25} constructs a compressed oracle that simulates Haar-random unitaries. Their construction achieves simulation error $O(t^2/N)$ for algorithms that make $t$ queries to an $N$-dimensional unitary, and has an extremely natural interpretation: the compressed oracle records superpositions of \emph{Feynman path information} $\ket{x_1}\mapsto \ket{y_1}, \hdots, \ket{x_t}\mapsto \ket{y_t}$ of the algorithm's queries, but \emph{forgets the order} of the queries (so only the set $\{(x_1, y_1), \hdots, (x_t,y_t)\}$ is recorded). Thus their compressed oracle satisfies properties \Cref{item:desideratum-compressed,item:desideratum-efficient,item:desideratum-accurate,item:desideratum-interpretable} except that it fails to achieve perfect simulation for \Cref{item:desideratum-accurate}. 

Moreover, their compressed oracle has already proved quite useful in quantum cryptographic constructions \cite{STOC:MaHua25,EC:ABGL25,C:ABGL25,TCC:HhaYam25,TCC:AnaGullin25,QIP:SMLBH25,BarGol26}, starting with the first provable construction of pseudorandom unitaries using post-quantum cryptography (analyzing the \textit{\textquote{$PFC$}} construction proposed by \cite{FOCS:MPSY24}).

\paragraph{Random permutations}
More recently, impressive progress has been made towards understanding compressed
\emph{permutations} \cite{AC:Unruh23,STOC:MajMalWal25,STOC:Carolan26}.%
\footnote{See also \cite{Rosmanis21} for a related work on quantum query lower bounds for permutations. We discuss this related work in more detail in \cref{sec:intro-tableau}.} 
Thus far, the model of choice has been a \textquote{standard} XOR oracle
\[ \ket{x}\ket{y}\ket{b}\mapsto \ket{x}\ket*{y\oplus \pi^{(-1)^b}(x)}\ket{b}
\]
for a random permutation $\pi: \{0,1\}^n \rightarrow \{0,1\}^n$. 
The state-of-the-art result here is a compressed permutation oracle of Carolan \cite{STOC:Carolan26} (as well as a compressed oracle of Unruh \cite{AC:Unruh23} that is not an isometry) that provably simulates random permutations for up to $O(N^{1/12})$ queries \cite{STOC:Carolan26}. Intriguingly, \cite{STOC:Carolan26} proves quantum query lower bounds for several problems that are tight relative to his compressed permutation oracle, which would imply tight lower bounds relative to random permutations if the simulation error could be improved.

We emphasize that each of the above compressed oracles was arrived at via a (very clever) ad-hoc argument tailored to the specific simulation problem at hand. Indeed, it has proved especially challenging to identify the right kind of construction/argument for compressed permutations, let alone anything more sophisticated that might come along in the future. 

\paragraph{Compressed Group Representations} A recent work of Grinko and Yoshida \cite{GriYos25} proposed what we call a \emph{tableau-recording oracle}: a general-purpose compressed oracle for simulating query access to a Haar-random element of a compact Lie group $G$ that records the partial information in the Fourier basis of $G$. We describe the relevant formalism in more detail in \cref{sec:results}, as it is essential to our work. Along the way to our main results, we also give a new proof of the main result of \cite{GriYos25}.

We emphasize that unlike prior compressed oracles, the tableau-recording oracle is not clearly interpretable (see \cref{sec:tech-overview-tableau} for more details). This presents a major obstruction to using it in applications such as query complexity lower bounds and cryptographic security proofs. 

\subsection{Our Results}
\label{sec:results}
In this work, we develop a general paradigm for designing and analyzing compressed oracles. In particular, we define an interpretable compressed oracle for a Haar-random element of an arbitrary compact Lie group $G$%
\footnote{
    We note, again, that this includes all finite groups as a special case.
}
that acts on $n$-qubit states via a representation $\rho: G \rightarrow U(N)$.%
\footnote{
    A \emph{unitary representation} of a group $G$ is any homomorphism from the group to the unitaries. That is, it requires that $\rho(g) \, \rho(h) = \rho(gh)$ for all group elements $g,h \in G$.
} 
This setting easily captures the three prominent examples described above, as random functions, random permutations,\footnote{To capture permutations, we instead consider the \emph{in-place} permutation oracle $\ket{x}\mapsto \ket{\pi(x)}$, which (along with its inverse) is algorithmically equivalent to the XOR oracle.} and Haar-random unitaries can all be described as Haar-random group representations.\footnote{We discuss later how to handle inverse queries, which are an \emph{anti}-representation of the group.} Our framework also captures several other natural examples of interest, including:
\begin{itemize}
    \item Haar-random \emph{orthogonal} unitaries $O \gets O(N)$ (unitaries with real entries).
    \item Random elements of the \emph{hyperoctohedral} group of products $PF$, where $P$ is a permutation unitary and $F$ is a binary phase unitary. This group arises naturally in \cite{STOC:MaHua25}.
    \item Similarly, the \emph{colored permutation} groups of products $PF$ where $F$ can be a $q$-ary or even continuous complex phase unitary.
    \item The Haar-random unitary cipher $\ket{k} \otimes \ket{x} \mapsto \ket{k} \otimes U_k\ket{x}$ for a collection of i.i.d.\ Haar-random unitaries $\{U_k\}$; this model represents an idealized pseudorandom unitary or encryption scheme. 
    \item Random elements of \emph{subgroups} of the symmetric group such as the alternating group $A_N$. 
\end{itemize}
Since our compressed oracle can be seen as a generalization of the Ma-Huang oracle,
we refer to our compressed oracle as the \textbf{path-recording oracle}. 

\begin{figure}[H]
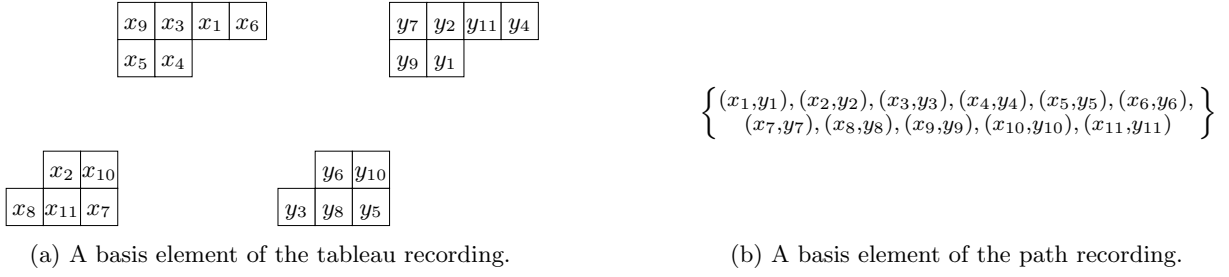

    \centering
    \begin{subfigure}{0.45\textwidth}
        \centering
        \Yboxdim{14pt}
        \footnotesize
        \Ytableau{
            {x_9}{x_3}{x_1}{x_6},
            {x_5}{x_4},
            ,
        }
        {
                 {x_2}{x_{10}},
            {x_8}{x_{11}}{x_7}
        }
        \Ytableau{
            {y_7}{y_2}{y_{11}}{y_4},
            {y_9}{y_1},
            ,
        }
        {
              {y_6}{y_{10}},
            {y_3}{y_8}{y_5}
        }
        \caption{A basis element of the tableau recording.}
        \label{fig:tableau_and_path_recording_examples_a}
    \end{subfigure}
    \hfill
    \begin{subfigure}{0.45\textwidth}
        \centering
        \vspace{1.4em}
        \begin{align*}
            \left\{
                \substack{
                    (x_1, y_1),\,
                    (x_2, y_2),\,
                    (x_3, y_3),\,
                    (x_4, y_4),\,
                    (x_5, y_5),\,
                    (x_6, y_6),\,
                    \\
                    (x_7, y_7),\,
                    (x_8, y_8),\,
                    (x_9, y_9),\,
                    (x_{10}, y_{10}),\,
                    (x_{11}, y_{11})
                }
            \right\}
        \end{align*}
        \vspace{1.4em}
        \caption{A basis element of the path recording.}
        \label{fig:tableau_and_path_recording_examples_b}
    \end{subfigure}
    \caption{In the two types of compressed oracles, the resulting data structures take different forms. The examples shown above correspond to the case in which the group is the unitary group.}
    \label{fig:tableau_and_path_recording_examples_ab}
\end{figure}

\needspace{2\baselineskip}
In fact, we derive both the tableau-recording oracle and the path-recording oracle together from first principles, establishing the following:
\begin{itemize}
\item For every unitary representation $\rho$ of a compact Lie group $G$, the tableau- and path-recording oracles perfectly simulate query access to a Haar-random $\rho(g)$ (and therefore, simulate each other). We identify simple Uhlmann transformations relating these compressed oracles to each other and to a na\"ive \textquote{uncompressed oracle} (see \cref{sec:tech-overview-tableau}).
\item In the special case $G = U(N)$, our framework easily recovers the Ma-Huang approximate path-recording oracle while avoiding their $O(t^2/N)$ simulation error.
\item Other settings of our path-recording oracle automatically recover Zhandry's compressed oracle as well as other natural ideal models. 
\end{itemize}

Finally, as our main application, we use our path-recording oracles to prove the standard security of a new, extremely simple construction of a pseudorandom unitary: the product $PC$ of a random permutation $P$ and a circuit $C$ sampled from a unitary $2$-design (such as a random Clifford circuit). 

\subsubsection{Tableau Recording}\label{sec:intro-tableau}
As a starting point, in \cref{sec:tableau-recording}, we give a new proof of the following theorem. 

\begin{theorem}
    [\cite{Har05,GriYos25}]
    \label{thm:main-tableau}
    Let $\rho$ be an $N$-dimensional unitary representation of a compact Lie group $G$. Let $A^{\rho(g)}$ denote a quantum query algorithm with oracle access to $\rho(g)$ for a Haar-random $g\gets G$ (for finite groups, the uniform distribution).

    Then, there is an isometry $Q_\rho^{\mathrm{Tab}}$ acting on the algorithm's query register $\reg A$ and a recording register $\reg R$ with the following properties:

    \begin{enumerate}
        \item After $t$ queries to $Q_\rho^{\mathrm{Tab}}$, the dimension of $\reg R$ is at most $N^{2t}$. 
        \item $Q_\rho^{\mathrm{Tab}}$ has an exact efficient implementation relative to compatible%
        \footnote{
            See \Cref{lemma:vilenkin-klimyk} for details on what it means to be compatible.
        }
        \emph{Clebsch-Gordan transforms}~\cite{BCH05,Har05} and \emph{dual Clebsch-Gordan transforms}~\cite{nguyen2023mixed, grinko2023gelfand} for tensoring with the representation $\rho$ and its dual $\bar{\rho}$, respectively.
        \item The mixed state obtained by running $A^{\rho(g)}$ for a Haar-random $g$ is \emph{identical} to that of running $A^{Q_\rho^{\mathrm{Tab}}}$ and tracing out the $\reg R$ register.
    \end{enumerate}
    Moreover, there is a variant of this isometry that also supports queries to the \emph{conjugate} representation $\rho(g)^*$, the inverse map $\rho(g)^{\dagger}$, the transpose map $\rho(g)^T$, and controlled versions of these unitaries. 
\end{theorem}
Thus, the Tableau-recording oracle \emph{always} satisfies desiderata \Cref{item:desideratum-compressed,item:desideratum-accurate}, and reduces property \Cref{item:desideratum-efficient} (efficient updates) to a previously studied problem in algorithmic representation theory. In particular, applying this framework to the defining representation of the unitary group $U(N)$, this gives an \emph{exact} stateful simulator for Haar-random unitaries, with an efficient implementation to arbitrary accuracy.

\paragraph{Related work to \cref{thm:main-tableau}} 
The tableau recording isometry $Q_\rho^{\mathrm{Tab}}$ was originally studied in \cite[Section 8.1.3]{Har05} in the different context of \textquote{generalized phase estimation}.%
\footnote{Harrow was interested in an algorithm --- called generalized phase estimation (GPE) --- that allowed for measuring (and performing operations controlled on) the hidden irrep label associated with a state $\ket{\psi} \in V \simeq \bigoplus_{\lambda} V_\lambda \otimes M_\lambda$ belonging to a completely \emph{reducible} representation of $G$, \emph{without} having to explicitly decompose $V$. Rather than maintaining a recording register $\reg R$ for simulating queries to a Haar-random $\rho(g)$, Harrow used $\reg R$ as a temporary workspace for GPE, which only requires one query to what we call $Q_\rho^{\mathrm{Tab}}$. In Section 8.1.3, Harrow observes the form of $Q_\rho^{\mathrm{Tab}}$ as part of a discussion on the relationship between the Clebsch-Gordan transform and the Fourier transform.}
Its application to compressed oracles and simulation, including the first proof of \cref{thm:main-tableau} and the first efficient $U(N)$ implementation, is due to the recent work~\cite{GriYos25}. Our proof uses a different formalism/approach compared to \cite{GriYos25}, proceeding from first principles by purifying the group element in the (countably infinite-dimensional) Hilbert space of square integrable functions on the group, $L^2(G)$, and then viewing it in the Fourier basis to recover the tableau-recording oracle (see \Cref{sec:tech-overview-tableau} for more details). This gives an explicit Uhlmann transformation (the Fourier transform) relating the compressed and na\"ive uncompressed representation oracles. 

Additionally, a recent work of Rosmanis \cite{Rosmanis21} considers the special case of random permutations and proves a quantum query lower bound for permutation inversion. While \cite{Rosmanis21} uses many of the techniques typical of the compressed oracle literature, the work only considers an uncompressed oracle, with the purification register consisting of a structured superposition of permutations that agree with a partial assignment $\{x_i \to y_i\}_{i \in [t]}$. Such states are not orthogonal, and so it does not define or study a compressed oracle. It does take a representation-theoretic view of the projectors onto low and high success probability, which correspond to projectors onto certain natural subspaces in the Fourier/tableau basis.

\paragraph{Other Query Types}
We briefly mention two improvements that we make to \cref{thm:main-tableau}. First, we give simple black-box reductions between the four query types $\rho(g)$, $\rho(g)^*$, $\rho(g)^{\dagger}$, and $\rho(g)^T$.

\begin{theorem}
    Given black-box access%
    \footnote{
        For complex conjugate queries and transpose queries, we technically need the complex conjugate, $\overline{Q_\rho^{\mathrm{Tab}}}$, of $Q_\rho^{\mathrm{Tab}}$. For all the cases we consider, it is possible to write $Q_\rho^{\mathrm{Tab}}$ using only real coefficients, which makes these the same. But even for general groups, this provides a useful reduction from conjugate and transpose queries to $\overline{Q_\rho^{\mathrm{Tab}}}$, which has a generic efficient implementation given any implementation of $Q_\rho^{\mathrm{Tab}}$.
    }
    to $Q_\rho^{\mathrm{Tab}}$,
    we can implement 
    $Q_{\rho^*}^{\mathrm{Tab}}$, 
    $Q_{\rho^T}^{\mathrm{Tab}}$, and
    $Q_{\rho^{\dagger}}^{\mathrm{Tab}}$
    each with a single application of $Q_\rho^{\mathrm{Tab}}$,
    and simple pre- and post-processing.
\end{theorem}

In the case of transpose queries, $Q_{\rho^T}^{\mathrm{Tab}}$, this pre- and post-processing only involves exchanging two registers, which is, of course, simple and efficient.
In the case of conjugate and inverse queries, this involves performing a bijective mapping between the labels and bases of two corresponding irreducible representations\footnote{A representation $\rho_\lambda$ is \emph{irreducible} if it has no non-trivial $G$-invariant subspaces. It turns out that for all compact Lie groups $G$, there is a countable list of all irreducible representations (or irreps) of $G$ up to isomorphism, and for particular groups of interest enumerating irreps and their bases often has a simple combinatorial interpretation.} (between an irrep and its \emph{dual irrep}), and its efficiency depends on the group and how its irreps are encoded.

\paragraph{The Unitary Group}
For the unitary case in particular, we show in \Cref{sec:tableau-efficient-unitary-group} that this mapping is a simple signed flip on the tableau data structure (which consists of simply rotating the data structure, with an attached $\pm 1$ phase depending on its parity; see \Cref{def:signed-flip-tableau}). This in particular allows us to show that the dual Clebsch-Gordan transform~\cite{nguyen2023mixed, grinko2023gelfand} for the unitary group can be very simply reduced to the (standard) Clebsch-Gordan transform.
\begin{lemma}
    For the unitary group, the dual Clebsch-Gordan transform, in the same basis needed for \Cref{thm:main-tableau}, is given by a \emph{signed flip} (the $\mathsf{Flip}$ operation in \Cref{def:signed-flip-tableau}), followed by a Clebsch-Gordan transform, followed by another signed flip. 
    That is,
    $
        \mathrm{dCG}
        =
        \mathsf{Flip}
        \cdot
        \mathrm{CG}
        \cdot
        \mathsf{Flip}
    $.
\end{lemma}
\noindent
Thus we can express the update of \Cref{thm:main-tableau} purely in terms of the Clebsch-Gordan transform without requiring a separate dual Clebsch-Gordan transform.
Moreover, the different query types are recovered by pre- and post-processing with a $\mathsf{Swap}$, a $\mathsf{Flip}$, or both.%
\footnote{
    Here, we are writing $\mathsf{Swap}$ to swap between two sets of registers in the recording, and $\mathsf{Flip}$ to apply in parallel to both sets of registers. Note that $\mathsf{Flip}$ itself is simply rearranging the entries on each side, plus a sign that is trivial to compute based on the parity of all the entries.
}

\begin{table}[H]
    \centering
    \begingroup
    \renewcommand{\arraystretch}{1.4}
    \begin{tabular}{p{3cm}p{5cm}}
        \toprule
        \textbf{Query type}
        & 
        \textbf{Update}
        \\
        \midrule
        Standard
        $\rho(g)$
        & 
        $Q_\rho^{\mathrm{Tab}}$ 
        \\
        Transpose 
        $\rho(g)^T$
        & 
        $
            \mathsf{Swap} 
            \cdot 
            Q_\rho^{\mathrm{Tab}}
            \cdot 
            \mathsf{Swap} 
        $ 
        \\
        Conjugate 
        $\rho(g)^*$
        & 
        $
            \mathsf{Flip} 
            \cdot 
            Q_\rho^{\mathrm{Tab}}
            \cdot 
            \mathsf{Flip} 
        $ 
        \\
        Inverse 
        $\rho(g)^{\dagger}$
        & 
        $
            \mathsf{Swap} 
            \cdot 
            \mathsf{Flip} 
            \cdot 
            Q_\rho^{\mathrm{Tab}}
            \cdot 
            \mathsf{Flip} 
            \cdot 
            \mathsf{Swap} 
        $ 
        \\
        \bottomrule
    \end{tabular}
    \endgroup
    \caption{Summary of the four different query types for the unitary group. In each case, one must only pre- and post-process with one or both of two simple operations.}
    \label{tab:intro-tableau-query-types}
\end{table}

\subsubsection{Path Recording}
The primary drawback of $Q_\rho^{\mathrm{Tab}}$ is \emph{interpretability}; given the contents of the recording register $\reg R$, it is hard to intuit and argue what information about $\rho(g)$ has actually been recorded. 

For our first main contribution, we give a \emph{second} compressed oracle $Q_\rho^{\mathrm{Path}}$ that provides better interpretability. In order to motivate $Q_\rho^{\mathrm{Path}}$, we briefly recall the \cite{STOC:MaHua25} approximate path-recording oracle for the defining representation of $U(N)$. In the \cite{STOC:MaHua25} oracle,
\begin{itemize}
    \item The compressed oracle register $\reg R$ is spanned by \emph{set states} $\ket{\{(x_1, y_1), \hdots, (x_t, y_t)\}}$, on which the restriction is imposed that the $y_i$ are all distinct.
    \item When a new query is made on an input $\ket{x}$, the recording register $\ket{S}$ is updated to a uniform superposition of $\ket{S \cup \{(x,y)\}}$ over all $y\notin S$, while the state $\ket{y}$ (entangled with the recording register $\ket{S\cup \{(x,y)\}}$) is returned as the output.
\end{itemize}
The rough intuition for this oracle is that the compressed purification $\reg R$ is supposed to record Feynman path information $x_1 \mapsto y_1, \hdots, x_t \mapsto y_t$ about the oracle queries; however, only the \emph{set} of pairs is recorded, which means that the \emph{order} of the queries has been forgotten by the data structure.

We build a path-recording oracle that, unlike the one in~\cite{STOC:MaHua25}, is \emph{exact} (no simulation error), and moreover fully generalizes to work for any compact Lie group. While for $G = U(N)$ (and $\rho$ the defining representation%
\footnote{
    The \emph{defining} or \emph{fundamental representation} of $U(N)$ is the one that represents each unitary as its own matrix $\rho(U) = U$ for every group element $U \in U(N)$.
}%
), the path-recording oracle records Feynman paths up to a \emph{permutation} symmetry (that is, a symmetry in the group algebra $\C[S_t]$), our general path-recording oracle records Feynman paths up to symmetries imposed by the \emph{commutant algebra} of $\rho^{\otimes t}$. 
This is the algebra $\calA_t$ of all operators that commute with the tensor power of $\rho(g)$ for all group elements
(see \Cref{tab:schur-weyl} for the commutant algebras of some important groups):
\[ 
    \mathcal A_t 
    = 
    \{A \in \mathrm{End}((\mathbb C^N)^{\otimes t}): A\cdot \rho(g)^{\otimes t} = \rho(g)^{\otimes t}\cdot A \text{ for all }g\in G\}.
    =: 
    \mathrm{End}_G((\mathbb C^N)^{\otimes t}) 
\]

\begin{table}[H]
\centering
\setlength{\tabcolsep}{6pt}
\renewcommand{\arraystretch}{1.2}
\scalebox{0.94}{
\begin{tabular}{
  >{\raggedright\arraybackslash}m{2.2cm}
  >{\raggedright\arraybackslash}m{1.0cm}
  >{\centering\arraybackslash}m{2.8cm}
  >{\centering\arraybackslash}m{3.7cm}
  >{\raggedright\arraybackslash}m{3.8cm}
  >{\raggedright\arraybackslash}m{1.4cm}
}
\hline
\multicolumn{2}{c}{
    \textbf{Group}
}
& 
\textbf{Tensor space} 
& 
\textbf{Representation} 
& 
\multicolumn{2}{c}{
    \textbf{Commutant algebra} 
}
\\
\hline

Unitary
& 
$U(N)$ 
& 
$V_N^{\otimes t}$ 
& 
$U^{\otimes t}$
& 
Symmetric Group Algebra
& 
$\mathbb{C}[S_t]$ 
\\
\noalign{\vspace{6pt}}

\phantom{\mbox{\tiny (Haar Cipher)}}
Unitary Product
\mbox{\tiny (Haar Cipher)}
& 
$U(N)^K$ 
& 
{$\Big(V_N^{\oplus K}\Big)^{\otimes t}$} 
& 
$\bigg(\sum\limits_{k \in [K]} \proj{k} \otimes U_k\bigg)^{\otimes t}$
& 
Colored Permutation Group Algebra
& 
{\small $\C[\Z_K \! \wr \! S_t]$} 
\\
\noalign{\vspace{6pt}}

Orthogonal
& 
$O(N)$ 
& 
$V_N^{\otimes t}$ 
&
\shortstack[c]{
\phantom{\tiny X}
\\
\phantom{\tiny X}
\\
$U_{g}^{\otimes t}$
\\
{\tiny $U_g$ is the standard embedding}
\\
{\tiny of $g \in O(N)$ as a unitary over $\mathbb{C}^t$}
}
& 
Brauer Algebra
& 
$B_t(N)$ 
\\
\noalign{\vspace{6pt}}

Colored \mbox{Permutations}
& 
$\Z_r \wr S_t$ 
& 
$V_{N}^{\otimes t}$ 
& 
\shortstack[c]{
\phantom{\tiny X}
\\
\phantom{\tiny X}
\\
\phantom{\tiny X}
\\
$(P_{\pi}F_{f})^{\otimes t}$
\\
{\tiny $P_{\pi}$ is the permutation matrix} 
\\
{\tiny of $\pi \in S_N$, and $F_{f}$ is}
\\
{\tiny diagonal with $r$'th roots of unity}
}
& 
Tanabe Algebra
& 
$T_t(N, r)$ 
\\
\noalign{\vspace{6pt}}

Symmetric
& 
$S_N$ 
& 
$(V_{N-1}\oplus V_1)^{\otimes t}$ 
& 
\shortstack[c]{
\phantom{\tiny X}
\\
\phantom{\tiny X}
\\
$P_{\sigma}^{\otimes t}$
\\
{\tiny $P_{\sigma}$ is the permutation matrix} 
\\
{\tiny of $\sigma \in S_N$}
}
& 
Partition Algebra
& 
$P_t(N)$ 
\\
\noalign{\vspace{6pt}}

Boolean Functions
& 
$\Z_2^N$ 
& 
{$\Big(V_1^{\oplus N}\Big)^{\otimes t}$} 
& 
$\bigg(\sum\limits_{x \in [N]} \scalebox{.8}{(-1)}^{f(x)} \proj{x}\bigg)^{\otimes t}$
& 
\multicolumn{2}{l}{Colored Even-Partition Algebra}
\\
\noalign{\vspace{6pt}}

\hline
\end{tabular}
}
\caption{Example instances of commutant relationships (Schur-Weyl dualities) between group actions on tensor spaces and their commutant algebras. The path-recording oracle for any compact Lie group can be described in terms of its corresponding commutant algebra.}
\label{tab:schur-weyl}
\end{table}

Unlike the \cite{STOC:MaHua25} oracle, the updates for our generalized path-recording also require reweighting different subspaces in an important representation-theoretic basis called the Schur basis (see \cref{sec:tech-overview-path} for details). For now, we informally state the form of our path-recording oracle to give a taste of the result.

\begin{theorem}[informal, see \cref{def:path-recording,thm:path-recording}]\label{thm:main-path}
    Let $\rho$ be an $N$-dimensional unitary representation of a compact Lie group $G$. Let $A^{\rho(g)}$ denote a quantum query algorithm with oracle access to $\rho(g)$ for a Haar-random $g\gets G$.

    Then, there is an isometry $Q_\rho^{\mathrm{Path}}$ acting on the algorithm's query register $\reg A$ and a recording register $\reg R$ with the following properties:

    \begin{enumerate}
        \item After $t$ queries to $Q_\rho^{\mathrm{Path}}$, the content of $\reg R$ is ordered Feynman path information $(x_1, \hdots, x_t, y_1, \hdots, y_t)$ for $x_i, y_i \in [N]$.
        \item The $t^{\mathrm{th}}$ application of $Q_\rho^{\mathrm{Path}}$ can be expressed in the following form:
        \begin{equation}\label{eq:path-update-intro}
        Q_{\rho,t}^{\mathrm{Path}}
        =
        \underbrace{
            \vphantom{
                \left(
                    \prod_{i=1}^t
                    \left(
                        \Omega_{\calA_i}
                    \right)
                \right)
            }
        \Big(
            \Lambda_{\calA_t}
        \Big)
        _{\reg{R}}
        }_{\operatorname{ReweightEnd}}
        \cdot
        \underbrace{
            \vphantom{
                \left(
                    \prod_{i=1}^t
                    \left(
                        \Omega_{\calA_i}
                    \right)_{\reg{R}_{\le i}}
                \right)
            }
        \Big(
            \Omega_{\calA_t}
        \Big)
        _{\reg{R}}
        }_{\operatorname{Symmetrize}}
        \cdot
        \underbrace{
            \vphantom{
                \left(
                    \prod_{i=1}^t
                    \left(
                        \Omega_{\calA_i}
                    \right)_{\reg{R}_{\le i}}
                \right)
            }
            \App
        }_{\operatorname{Append}}
        \cdot
        \underbrace{
            \vphantom{
                \left(
                    \prod_{i=1}^t
                    \left(
                        \Omega_{\calA_i}
                    \right)_{\reg{R}_{\le i}}
                \right)
            }
        \Big(
            \Lambda_{\calA_{t-1}}^+
        \Big)
        _{\reg{R}_{\leq t-1}}
        }_{\operatorname{ReweightStart}},
    \end{equation}
    where:
    \begin{itemize}
        \item $\mathcal A_t = \mathrm{End}_G((\mathbb C^N)^{\otimes t})$ denotes the commutant algebra of $\rho^{\otimes t}$.
        \item $\Lambda_{\calA_t}, \Lambda_{\calA_{t-1}}^+$ denote certain \emph{subspace reweighting} operators,
        \item $\mathrm{App} = \sum_{x_t, y_t} \ketbra{y_t}{x_t}_{\reg A} \otimes \ket{(x_t, y_t)}_{\reg R_t}$ denotes the (un-normalized) initialization of an EPR pair shared between the algorithm and the recording.
        \item $\Omega_{\calA_t}$ denotes a \emph{symmetrization} projection that can be described in terms of $\calA_t$ and represents \emph{forgetting} partial Feynman path information.
    \end{itemize}
        \item The mixed state obtained by running $A^{\rho(g)}$ for a Haar-random $g$ is \emph{identical} to that of running $A^{Q_\rho^{\mathrm{Path}}}$ and tracing out the $\reg R$ register. 
    \end{enumerate}
    Moreover, essentially the same update also supports queries to $\rho(g)^T$, while another variant additionally supports queries to $\rho(g)^*$ and $\rho(g)^{\dagger}$. 
\end{theorem}

We give more details in \cref{sec:tech-overview-path,sec:path}, but we emphasize that the symmetrization operation $\Omega_{\calA_t}$ has an extremely natural interpretation directly in the \textquote{path basis} $\ket{(x_1, y_1), \hdots, (x_t, y_t)}$. (\cref{thm:commutant-epr-proj-as-symmetrization}): 
Choose any basis for the commutant algebra (for instance, for group algebras, simply take the basis of the group elements), and apply every such basis element in superposition.%
\footnote{
    Technically, we apply each such basis element to the all the $x_i$'s and its \emph{dual transpose} to all the $y_i$'s (see \cref{thm:commutant-epr-proj-as-symmetrization} for details). For group algebras, this amounts to applying the same group element across both the $x$ and $y$ entries.
}
For instance, for the unitary group, $\Omega_{S_t}$ amounts to applying all possible permutations, in superposition, among the $t$ different $(x_i, y_i)$ pairs.
Thus, 
modulo understanding the subspace reweighting $\Lambda_{\calA_t}$, 
the update rule is highly interpretable.

\paragraph{Equivalence of Tableau and Path Recording}
We emphasize that as written, $Q_\rho^{\mathrm{Path}}$ is a mathematical description and is not necessarily algorithmic. On the other hand, we show that $Q_\rho^{\mathrm{Path}}$ is isometric to the algorithmic $Q_\rho^{\mathrm{Tab}}$ via two applications of the generalized Schur transform, special cases of which have been studied in \cite{BCH05,krovi2019efficient,fei2023efficientquantumalgorithmportbased, nguyen2023mixed,grinko2023gelfand,burchardt2025highdimensionalquantumschurtransforms}, and which can be built from Clebsch-Gordan transforms. We thus have that the updates to our path recording oracle can likewise be made efficient with an efficient Clebsch-Gordan transform.

\paragraph{Concurrent Work on Path Recording}
Concurrently to this work, we were made aware that the authors of \cite{GriYos25} will update \cite{GriYos25} to include a version of the path-recording oracle. Their path-recording oracle, like ours, is derived by use of a generalized Schur transform on the tableau-recording oracle. Their path recording update is defined by transporting the tableau update through this isometry. In the special case of $G = U(N)$, they also give a mathematical formula (different from ours) for the path basis update matrix coefficients. 
In this work, beyond defining the path-recording oracle for general compact Lie groups, we focus on providing an interpretable mathematical description of the path basis update rule (\Cref{thm:main-path}) and showing how to use it for security proofs and applications. 

\subsubsection{Approximations and Pseudorandomness of the $PC$ Ensemble}

Our second main contribution is analyzing $Q_\rho^{\mathrm{Path}}$ in special cases of interest by introducing \emph{approximations}. Our thesis is that for each group representation, knowing what the commutant algebra $\mathcal A_t$ looks like is the key to understanding what the exact updates to the Feynman path recording are effectively doing to the Feynman paths (when they record, when they forget, etc.), and how a more elementary approximate update can be derived.

As a proof of concept, in \cref{sec:approximation-specialization}, we show how to derive the \cite{STOC:MaHua25} path-recording oracle for Haar-random unitaries from our path-recording oracle (when specialized for $U(N)$) via a sequence of two approximations:
\begin{itemize}
    \item restricting our oracle to appending \emph{distinct} outputs $y_1, \hdots, y_t$ in the path basis, and
    \item replacing the subspace reweighting operators $\Lambda_{\calA_t}, \Lambda^+_{\calA_{t-1}}$ by \emph{scalars} (i.e., no reweighting),
\end{itemize}
which results in the Ma-Huang oracle, rewritten as
\[ Q^{\mathrm{Path}}_{MH} 
=  \Big( \sqrt t \cdot \Omega_{\mathbb C[S_{t}]}\Big)_{\reg R} \cdot \frac 1 {\sqrt{N-t+1}} \Big(\Pi_{\mathsf{Dist}}\Big)_{\reg R_Y} \mathsf{App}_{\reg A \reg R}.
\]
Indeed, one can essentially read off the algorithmic Ma-Huang update from this expression: the key idea is identifying the Ma-Huang set states $\{(x_1, y_1), \hdots, (x_t, y_t)\}$ as $S_t$-symmetrized path states $\frac 1 {\sqrt {t!}}\sum_{\pi \in S_t} \pi^{\otimes 2} \ket{x_1, \hdots, x_t, y_1, \hdots, y_t}$. 

Expanding on this, we show in \cref{sec:approximation-specialization} how the path-recording oracle can be used to \emph{automatically} derive three other compressed oracles that have been either implicitly or explicitly studied before: Zhandry's original compressed (phase) oracle, the ideal Haar cipher $\ket{k,x}\mapsto \ket{k} \otimes U_k\ket{x}$, and Haar-random diagonal unitaries $\ket{x} \mapsto u_x \ket{x}$. 

Finally, for our primary application, in \cref{sec:prus} we analyze the path-recording oracle for the in-place representation of the permutation group $S_N$:
\[ \rho(\pi) \cdot \ket{x} = \ket{\pi(x)}.
\]
This compressed oracle is particularly well-behaved---it turns out to approximately match the compressed unitary oracle!---on recording states restricted to a dense \textquote{distinct, nonplussed} subspace $\Pi_{\mathsf{DNP}}^{X,Y}$. Through a careful analysis, we are able to leverage this to prove the standard security of a simple new construction of pseudorandom unitaries:
\begin{theorem}[see \cref{thm:PRU-random-P}]\label{thm:main-PC-intro}
    For all adversaries $\mathsf{Adv}^{(\cdot)}$ making $t$ standard queries, the oracle distribution $P\cdot C$ for a uniformly random in-place permutation $P$ and $C$ sampled from any unitary $2$-design is $O(t^2/N)$-indistinguishable from a Haar-random unitary.
\end{theorem}
This improves on the $PFC$ construction of pseudorandom unitaries (PRUs) of Metger, Poremba, Sinha, and Yuen~\cite{FOCS:MPSY24} (proved secure by Ma and Huang~\cite{STOC:MaHua25}): we prove that the random phase $F$ is actually unnecessary, and $PC$ is already a PRU! By using a pseudorandom permutation for $P$, this yields, in our opinion, the simplest PRU construction that has been analyzed to date. 

Accordingly, the same proof gives a new, simple construction of unitary $t$-designs using $2t$-wise independent permutations:
\begin{theorem}[See \cref{cor:PRU-design}]\label{thm:PC-design-intro}
    Suppose that $P$ is sampled from a family of non-adaptive $(t, \epsilon)$ post-quantum pseudorandom permutations (respectively, $(t,\epsilon)$-adaptively secure post-quantum pseudorandom permutations) with \emph{statistical security} and that $C$ is sampled from an exact $2$-design. Then, $P\cdot C$ is an $\epsilon + O(t^2/N)$ approximate $t$-design (respectively, an $\epsilon + O(t^2/N)$ approximate adaptive $t$-design). 

    In particular (see \cite{FOCS:Zhandry12,FOCS:MPSY24}), the conclusions hold if $P$ is sampled from a family of $2t$-wise independent permutations, or $\frac \epsilon 2 \cdot N^{-t}$-almost $2t$-wise independent permutations. 
\end{theorem}

More broadly, because of its wide applicability, as well as its interpretability, we believe that our path-recording oracle will facilitate a variety of additional pseudorandomness results as well as new quantum query complexity lower bounds.

\subsection{AI Disclosure}
The majority of the work that went into this paper was done by the human authors in the Summer and Fall of 2025, before the current age of AI mathematics. The proof of the PC construction of pseudorandom unitaries came more recently in the Spring of 2026; the full proof structure and most of the technicalities were produced by the authors. In the later stages of this work, we used generative AI, including various versions of ChatGPT and Claude, to help with a few of the remaining technical pieces and proof simplifications. Every AI suggestion was properly vetted and verified by the authors, and the authors take full responsibility for the content and correctness of this paper.

    \section{Technical Overview}
In this section, we overview the main ideas behind our results and their proofs. 

\subsection{Tableau Recording}\label{sec:tech-overview-tableau}
We begin with the tableau recording oracle. Again, we remark that the tableau recording oracle for finite groups can be seen implicitly in \cite[Section 8.1.3]{Har05} and was also defined and analyzed in \cite{GriYos25}. We describe the construction for expository purposes, for some technical reasons (a different proof and some slight variations/improvements), and because it is connected to and will help explain our path-recording oracle.

First, given a representation $\rho: G\rightarrow U(N)$, why should we expect there to be a compressed oracle at all? The reason is a direct generalization of Zhandry's compressed oracle. Recall that in the case of the binary phase representation $\rho(f) \ket{x}= (-1)^{f(x)} \ket{x}$ of functions $f : \bit^n \to \bit$,%
\footnote{
    We will later see that we should treat such functions as group elements of the group $\Z_2^{N}$, for $N = 2^n$.
}
Zhandry's compressed oracle is constructed in two steps:

\begin{enumerate}
    \item First, \emph{purify} the random function $f$ to a function register $\mathsf F \simeq \mathbb C^{\{0,1\}^{2^n}}$; here, $\mathsf F$ is initialized to the uniform superposition $\ket{+}$ and a function query is represented by the unitary
    \[ Q_\rho \ket{x}\ket{f} = (-1)^{f(x)} \ket{x} \ket{f}.
    \]
    \item Second, apply the quantum Fourier transform (in this case, simply the Hadamard transform) to $\mathsf F$: the initialization is then $\ket{\emptyset}$ rather than $\ket{+}$, while a function query is now written as
    \[ Q_\rho^{\mathsf{Fourier}} \ket{x} \ket{D} = \ket{x} \ket{D \oplus \{x\}},
    \]
    where $D$ is interpreted as a subset of $\{0,1\}^n$, or \textquote{database}, and $D \oplus \{x\}$ adds $x$ to the database if $x \notin D$ and otherwise removes it.
\end{enumerate}
The magic is that in the Fourier basis, after $t$ queries, the database state $\ket{D}$ is always $t$-sparse. It can thus be stored in sparse form using $tn = t\log N$ space. (In addition, the update is extremely simple and can be implemented efficiently.)

How does this generalize to other group representations? The natural idea is to consider the same two-step process:

\begin{enumerate}
    \item First, purify the random group element $g$ to a \textquote{uniform superposition} on a register $\reg{G}$. In the case of finite groups, we write $\reg{G} = \mathbb C[G]$ to be a Hilbert space with basis $\{\ket{g}, g\in G\}$. In this language, the initial state is indeed the uniform superposition $\ket{+}$, and the purified query can be written as
    \[ Q_\rho \ket{x}\ket{g} = \rho(g) \ket{x} \otimes \ket{g}.
    \]
    In the case of infinite groups such as $U(N)$, this is not the right purification (it would have uncountable dimension); instead, we use the well-studied countable-dimension Hilbert space of square-integrable functions 
    \[ L^2(G) = \Big\{ f: G\rightarrow \mathbb C \text{ such that } \int_G |f(g)|^2 dg < \infty\Big\},
    \]
    where $dg$ denotes the Haar measure on $G$. This Hilbert space comes equipped with commuting left- and right- actions of $G$ (pre- and post- composition) that together define what is called the \textbf{regular representation} of $G$. This register is naturally initialized to the purified Haar measure ($f = \mathbf 1$) and the purified query is analogous to above:
    \[ Q_\rho \Big(\ket{x} \otimes \int_G f(g) \ket{g} dg\Big)= \int_G \rho(g) \ket{x} \otimes f(g) \ket{g} dg.
    \]
    \item Second, apply the quantum Fourier transform to $\reg{G}$. When generalized from finite cyclic groups to arbitrary compact Lie groups, the Fourier transform is an isomorphism
    \begin{align}
        \label{eq:fourier-transform}
        F: L^2(G) \overset{\sim}\longrightarrow \bigoplus_{\lambda \in \widehat G} V_\lambda \otimes V_\lambda^*,
    \end{align}
    where the index $\lambda \in \widehat G$ ranges over \emph{all irreducible representations} (or \emph{irreps}) of $G$ (up to isomorphism)%
    \footnote{
        For a compact Lie group, there are only countably many irreps.
    }
    and $V_\lambda$ denotes the vector space underlying the irreducible representation $\rho_\lambda$.%
    \footnote{
        Note that we use $\rho(g)$ to refer to the queried representation, which may or may not be reducible. On the other hand, $\rho_{\lambda}(g)$ is the irreducible representation corresponding to $\lambda \in \wh{G}$.
    }
    In other words, there exists an orthonormal Fourier basis spanned by states
    \begin{align}
        \label{eq:techover-fourier-basis}
        F^{-1} 
        \ket{\lambda, X, Y} 
        = 
        \sqrt{\dim(V_\lambda)} 
        \int_G 
        \bra{Y} 
        \rho_\lambda(g) 
        \ket{X} 
        \ket{g} 
        dg 
        \in 
        L^2(G)
    \end{align}
    for $\lambda \in \wh{G}$, and $\ket{X}, \ket{Y}$ ranging over a basis of $V_\lambda$, such that the Fourier transform decomposes $L^2(G)$ in this basis. 
\end{enumerate}
We call this basis $\bigoplus_\lambda V_\lambda \otimes V^*_\lambda$ the \textbf{tableau basis} by analogy to the case $G = U(N)$: the irreducible representations of $U(N)$ are indexed by Young diagrams $\lambda$, and the basis elements $\ket{X}$ of $V_\lambda$ correspond to semi-standard Young tableaux of shape $\lambda$ (see \Cref{sec:prelims_for_specific_algebras} for details on Young diagrams and tableaux). 

While this is a natural proposal, the big questions are:
\begin{itemize}
    \item Is this Fourier basis representation compressible?
    \item What is the form of the query $Q_\rho^{\mathrm{Tab}}$ in this basis?
\end{itemize}
The first question is implicitly answered by the resolution to the second, but let us answer it separately anyway since it will be relevant for the path-recording oracle. One way to see why the basis is compressible is to observe that all recording states obtained by making $t$ queries to $Q_\rho$ are in the span of the following \textquote{$t$-power matrix element states} of $L^2(G)$: 
\[ 
    \int_G 
    \bra{y_1, \hdots, y_t} 
    \rho(g)^{\otimes t} 
    \ket{x_1, \hdots, x_t} 
    \ket{g} 
    dg
    \,.
\]
This already places a limit on the \emph{dimension} of accessible states on the recording register, but it does not say how they appear in the Fourier basis. To understand this, we make use of the \emph{Schur transform} \cite{BCH05}
\begin{align}
    \label{eq:thechover-shur-transform}
    \schur: (\mathbb C^N)^{\otimes t} \overset{\sim}{\longrightarrow} \bigoplus_{\lambda \in \widehat G_t} V_\lambda \otimes V_\lambda^{\mathcal A_t},
\end{align}
where $\lambda \in \widehat G_t$ ranges over a finite subset of the possible irreps%
\footnote{For example, in the case $G = U(N)$ and $\rho$ is the defining representation, $\widehat G_t$ is the set of all Young diagrams with exactly $t$ boxes.} 
and $V_\lambda^{\mathcal A_t}$ is some multiplicity space (we will say more about $V_\lambda^{\mathcal A_t}$ later). 
In particular, this means that there is a Schur basis consisting of states $\schur^\dagger \ket{\lambda, X, S} \in (\mathbb C^N)^{\otimes t}$ (such that the tensor power $\rho(g)^{\otimes t}$ acts as the $\lambda$ irrep, mapping them to $ \schur^\dagger \Big(\ket{\lambda} \otimes \rho_\lambda(g) \ket{X} \otimes \ket{S}\Big)$), and such states together span $(\mathbb C^N)^{\otimes t}$. 

We can thus 
insert the identity $\schur^\dagger \schur$ and
write the $t$-power matrix element state as 
\begin{align*}
    \int_G 
    \bra{y_1, \hdots, y_t} 
    \rho(g)^{\otimes t} 
    \ket{x_1, \hdots, x_t} 
    \ket{g} 
    dg
    &
    =
    \int_G 
    \bra{y_1, \hdots, y_t}
    \;
    \schur^\dagger 
    \,
    \schur
    \;
    \rho(g)^{\otimes t} 
    \;
    \schur^\dagger 
    \,
    \schur
    \;
    \ket{x_1, \hdots, x_t} 
    \ket{g} 
    dg
    \,,
\end{align*}
where
\(
    \schur
    \;
    \rho(g)^{\otimes t} 
    \;
    \schur^\dagger 
    =
    \sum_{\lambda \in \wh{G}_t}
    \proj{\lambda}
    \otimes
    \rho_{\lambda}(G)
    \otimes
    \Id
\).
This implies that each $t$-power matrix element state is in the span of a restricted set of Fourier basis states (\Cref{eq:techover-fourier-basis}),
\begin{align*}
    \int_G 
    \bra{Y, T} 
    \Big(
        \rho_\lambda(g)
        \otimes 
        \Id 
    \Big)
    \ket{X,S} 
    \ket{g} 
    dg
    \;
    = 
    \;
    \delta_{S,T}
    \int_G 
    \bra{Y} 
    \rho_\lambda(g) 
    \ket{X} 
    \ket{g} 
    dg 
    \;
    \propto 
    \;
    \delta_{S,T} 
    \cdot 
    F^{-1} 
    \ket{\lambda, X, Y}
    \,,
\end{align*}
such that $\lambda$ comes from the restricted set of irreps $\widehat G_t$. This exactly says that its Fourier transform is supported on
\[ \bigoplus_{\lambda\in \widehat G_t} V_\lambda \otimes V^*_\lambda,
\]
a small finite subset of the Fourier basis,
of dimension at most $\dim((\mathbb C^N)^{\otimes 2t}) = N^{2t}$. So the purification is in fact sparse in the Fourier (tableau) basis! Moreover, it can be stored using space at most $O(t\log N)$.

Finally, we consider the update rule $Q_\rho^{\mathrm{Tab}}$. For the $(t+1)$th query, this must be an isometry which, for each $\lambda \in \widehat G_t$, maps
\[ 
    \ket{x} 
    \otimes 
    \ket{\lambda, X, Y} 
    \in 
    \mathbb C^N 
    \otimes 
    V_\lambda 
    \otimes 
    V_\lambda^* 
    \;\;
    \longrightarrow 
    \;\;
    \mathbb C^N 
    \otimes 
    \bigoplus_{\lambda^+ \in \widehat G_{t+1}} 
    V_{\lambda^+} 
    \otimes 
    V^*_{\lambda^+}
    \,, 
\]
whose basis elements look like
\[ \ket{y} \otimes \ket{\lambda^+, X^+, Y^+}.  
\]
Intuitively, upon the algorithm's query, we want to grow $X, Y$ \textquote{by one element each} to record an additional piece of information. It turns out that this update can be viewed as a composition of two instances of a well-studied map from representation theory: The Clebsch-Gordan transform \cite{BCH05,Har05} is an isomorphism describing the irrep decomposition%
\footnote{
    Typically, the Clebsch-Gordan transform is defined in the setting where both tensor factors are irreducible.
    Here, transforms for $\rho$ allow the second tensor factor to be \emph{reducible}, as discussed in \Cref{sec:CG-defn}. 
    This will be useful, for example, for $S_N$, whose natural representation on $\mathbb{C}^N$ is reducible.
    In the formal derivation, we sometimes express $\rho$ in an irrep basis for convenience, but this is not required algorithmically.
}
\[ \mathbb C^N \otimes V_\lambda \overset{\sim}{\longrightarrow} \bigoplus_{\lambda^+\in \widehat G_{t+1}} V_{\lambda^+}\otimes M_{\lambda, \lambda^+}
\]
for some multiplicity spaces $M_{\lambda, \lambda^+}$. For example, in the case $G = U(N)$ (where the $M_{\lambda, \lambda^+}$ all have dimension $1$), this describes the procedure for \textquote{inserting a new box} into a Young diagram/semi-standard Young tableau (as in \Cref{fig:Unitary-CG-update}). 

\begin{figure}[H]
    \centering
    \begin{tikzpicture}[baseline=(AX.base)]
        \node (AX) at (-.1,0) [label=above:{}] {
            \ytableaushort{
                {*(green)\boldsymbol{x}}
            }
        };
        \node (Acomma) at (.5,0) {$\otimes$};
        \node (AY) at (2,0) [label=above:{$X$}] {
            \ytableaushort{
                1123,
                25,
                7
            }
        };
        \node (Balpha1) at (7,0) {$\alpha_1$};
        \node (BX) at (8.5,0) [label=above:{$X^+_1$}] {
            \ytableaushort{
                11{\boldsymbol{x}}3,
                22{*(green)5},
                7
            } 
        };
        \node (Bcomma) at (10.25,0) {$+ \;\; \alpha_2$};
        \node (BY) at (12,0) [label=above:{$X^+_2$}] {
            \ytableaushort{
                1123,
                2{\boldsymbol{x}},
                5{*(green)7}
            } 
        };
        \node (Bdots) at (14,0) {$+ \;\; \cdots$};
        \draw[->] (AY.east) to node[midway,above] {Clebsch-Gordan} node[midway,below] {\textquote{add a box}} (Balpha1.west);
    \end{tikzpicture}
    \caption{
        Example Clebsch-Gordan update for the unitary group,
        with the
        newly added box highlighted. 
        Roughly speaking, each shape $\lambda$ 
        (Young 
        diagram) that 
        contains one extra box not in 
        the original, and each valid filling 
        (tableau) 
        with the contents 
        of $X$ plus the new $x$ value, receive some weight. 
    }
\label{fig:Unitary-CG-update}
\end{figure}
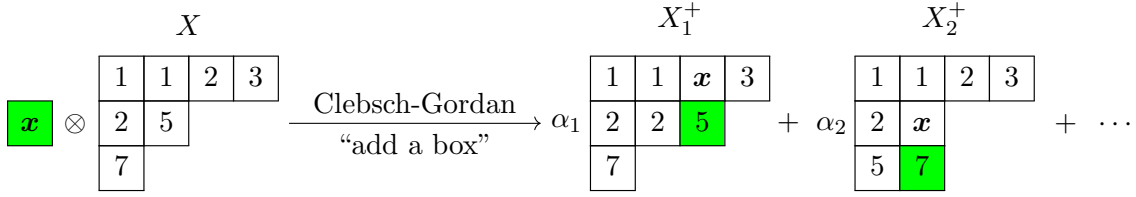

In general, applying this transformation to $\ket{x, \lambda, X}$ will produce a superposition of elements of the form\barak{Should we change the multiplicity here and in the figure to be $r$ instead of $m$ so that it matches the notation used in the body of the paper?}
\[ \ket{\lambda, \lambda^+, X^+, m}
\]
with $\ket{X^+} \in V_{\lambda^+}$ and $\ket{m} \in M_{\lambda, \lambda^+}$. This (along with $\ket{Y}$), of course, does not match the type of vector that the update rule is supposed to output.

However, we can arrive at a valid element of $\bigoplus_{\lambda^+} V_{\lambda^+}\otimes V^*_{\lambda^+}$ by \emph{inverting} another irrep decomposition
\[ \bigoplus_\lambda V_\lambda^* \otimes M_{\lambda, \lambda^+} \rightarrow  \mathbb C^N\otimes V_{\lambda^+}^*.
\]
We refer to this map as an inverse \emph{dual} Clebsch-Gordan transform.\footnote{In the body, we technically define the dual Clebsch-Gordan transform to decompose $V_{\lambda^+}\otimes (\mathbb C^N)^*$ into (primal) irreps. Prior work on the dual Clebsch-Gordan transform assumes the representation of $G$ on $\mathbb C^N$ is irreducible, implicitly identifies the dual irrep on $(\mathbb C^N)^*$ with a pre-defined irrep space $V_{\mu^*}$, and decomposes $V_{\lambda^+}\otimes V_{\mu^*}$, which roughly forces a different choice of basis on the dual space. To align with prior work, we translate from our dual Clebsch-Gordan transform to the previously defined one at the level of matrix coefficients.} Applying this map to $\ket{\lambda^+, \lambda, Y, m}$ yields an element of $\mathbb C^N \otimes \bigoplus_{\lambda^+ \in \widehat G_t} V_{\lambda^+}\otimes V^*_{\lambda^+}$. It turns out that this composition (see \cref{fig:temp1}) is equal to $Q_\rho^{\mathrm{Tab}}$!

\begin{figure}[hbt]
    \centering
    \begin{quantikz}[wire types={q,q,q,n,n,q,q},classical
    gap=0.08cm]
    \lstick{$\ket{\lambda}$} &  &  & \ctrl{2} &  & \permute{3,5,1,4,2}\push{\phantom{-----}} 
    &  & \ctrl{2} 
    &  &  & \rstick{$\ket{\lambda^{+}}$} \\
    \lstick{$\ket{X}$} &  & \permute{2, 1} &  &  & \push{\phantom{--------}} 
    &  &  
    &  &  & \rstick{$\ket{X^{+}}$} \\
    \lstick{$\ket{Y}$} &  &  & \gate[3]{\substack{\phantom{\mathrm{dC}^{\dagger}}\\C\\\phantom{\mathrm{dC}^{\dagger}}}} & \qw{\phantom{X}\lambda^{+}} & \push{\phantom{--------}} 
    &  & \gate[3]{\substack{\phantom{\mathrm{dC}^{\dagger}}\\\mathrm{dC}^{\dagger}\\\phantom{\mathrm{dC}^{\dagger}}}} 
    &  &  & \rstick{$\ket{Y^{+}}$} \\
    &  &  &  & \qw{\phantom{r}m} & \setwiretype{q} &  &  & \setwiretype{n} &  & \\
    &  & \push{\phantom{--}} & \qw & \qw{\phantom{X}X^{+}} & \setwiretype{q}\push{\phantom{--------}} 
    &  &  
    & \permute{3} \setwiretype{n} &  &  \\
    \lstick{$\ket{\mu}$} &  &  & \ctrl{-1} &  &  
    &  & \ctrl{-1} 
    &  &  & \rstick{$\ket{\mu}$} \\
    \lstick{$\ket{x}$} &  & \permute{-1} \setwiretype{n} &  &  &  
    &  &  
    & \push{\phantom{--}} & \setwiretype{q} & \rstick{$\ket{y}$} \\
    \end{quantikz}
    \caption{
        A single query can be decomposed into a Clebsch--Gordan transformation
        followed by the inverse dual-Clebsch-Gordan operation, 
        though not on the same registers.
        The top three registers are those of the recording oracle's data structure, while the bottom two are the algorithm's registers.
        The algorithm's multiplicity register (in cases where it exists) is not shown here since it is not touched by the update.
        For unitaries, $\mu$ is the single-box fundamental irrep that acts as $\rho_{\Box}(U) = U$.
    }
    \label{fig:temp1}
\end{figure}
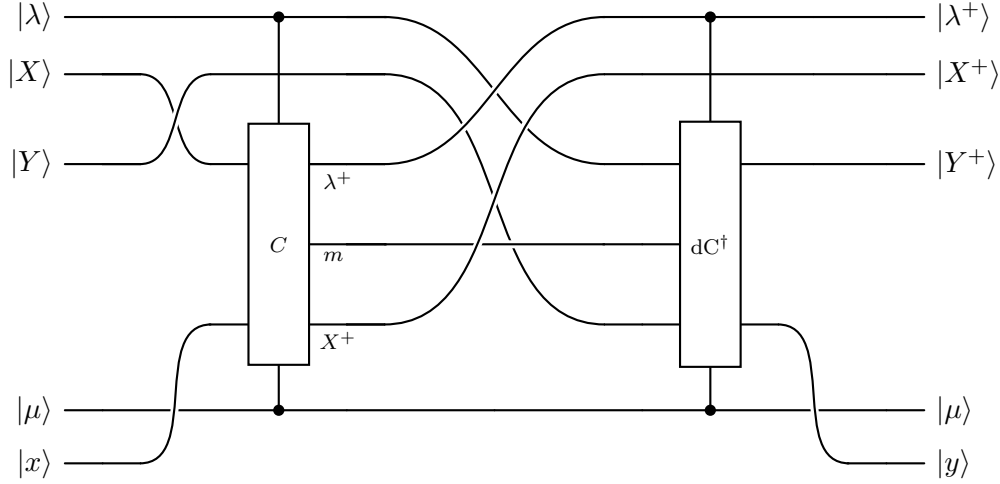

In \cite{Har05}, Harrow gives an abstract algebraic proof that these maps are equivalent (in the case of finite groups). In \cref{sec:tableau-recording}, we give a proof via a direct calculation in the Fourier basis. This is somewhat necessary in order to reason about algorithmic efficiency, because the Clebsch-Gordan transform is only defined up to choices of basis for the irrep and multiplicity spaces.\footnote{Harrow implicitly forces a particular choice of basis for the multiplicity space, but algorithmically we are free to choose another basis. } We prove that the composition correctly computes $Q_\rho^{\mathrm{Tab}}$ whenever the Clebsch-Gordan and dual Clebsch-Gordan transforms are compatibly chosen to satisfy a certain identity on their matrix coefficients (see \cref{lemma:vilenkin-klimyk}). Since, in the case of $G = U(N)$, a standard choice of Clebsch-Gordan transform on the Gelfand-Tsetlin basis of $V_\lambda$ satisfies this matrix coefficient identity \emph{and} has an efficient algorithmic implementation \cite{BCH05,burchardt2025highdimensionalquantumschurtransforms}, we conclude that $Q_{U(N)}^{\mathrm{Tab}}$ has an efficient algorithm. 

\begin{figure}[H]
    \centering
    \begin{tikzpicture}[baseline=(AX.base)]
        \node (AX) at (0,0) [label=above:{$X$}] {
            \ytableaushort{
                1123,
                25,
                7
            }
        };
        \node (Acomma) at (1.5,0) {$,$};
        \node (AY) at (3,0) [label=above:{$Y$}] {
            \ytableaushort{
                1224,
                34,
                5
            }
        };
        \node (BX) at (9,0) [label=above:{$X^+$}] {
            \ytableaushort{
                11{\boldsymbol{x}}3,
                22{*(green)5},
                7
            } 
        };
        \node (Bcomma) at (10.5,0) {$,$};
        \node (BY) at (12,0) [label=above:{$Y^+$}] {
            \ytableaushort{
                1224,
                {\boldsymbol{y}}3{*(green)4},
                5
            }
        };
        \draw[->,bend left] (AY.east) to node[midway,above] {query $x$} (BX.west);
    \end{tikzpicture}
    \caption{
        When updating the tableau recording, both tableaux grow in 
        the same way, such that the resulting shapes will always be identical. 
        For the unitary group, this (very roughly) means 
        that we add a box in the same place 
        (in superposition over valid locations) for both the $X$ and $Y$ tableaux.
    }
\label{fig:tableau_update}
\end{figure}
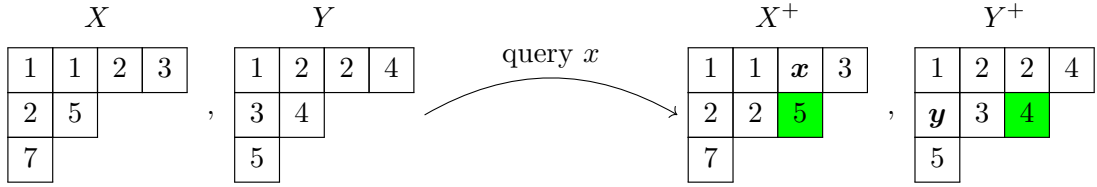

In fact, rather than using direct implementations of a dual Clebsch-Gordan transform \cite{nguyen2023mixed,grinko2023gelfand}, we observe a simple identity (\cref{claim:flip-gives-dual-CG}) that efficiently reduces the $U(N)$ dual Clebsch-Gordan transform to the ``primal'' one, in the Gelfand-Tsetlin (or Young tableau) basis. This gives an efficient unitary compressed oracle in terms of only an efficient Clebsch-Gordan transform. 

\subsubsection{Other Query Types} 
What if we want to be able to query the complex conjugate $\rho(g)^*$, the inverse $\rho(g)^\dagger$, or the transpose $\rho(g)^T$ of the representation? And what if we want to be able to perform controlled queries? 

\paragraph{Complex Conjugate Queries}
We present two distinct ways to handle conjugate queries. We defer the second approach to the ``transpose and inverse queries'' discussion.

For the first approach, complex conjugate queries $\rho(g)^*$ are themselves another representation of the group, which, for unitary representations, is also known as the \emph{dual representation} $\rho^*(g)$. 
So we can consider the same derivation, but for $\rho^*$ instead of $\rho$. 
This has us switch the roles of the Clebsch-Gordan and dual Clebsch-Gordan transforms. Thus each $\rho^*(g)$ query can be implemented as in \Cref{fig:temp1}, but with a dual Clebsch-Gordan followed by an inverse Clebsch-Gordan.

We also derive an alternative interpretation of this update for the special case $G = U(N)$: as referenced above, we show that a simple signed rotation (a $\mathsf{Flip}$ operation) of the Young tableaux is sufficient to implement a dual Clebsch-Gordan transform \emph{in terms of} the primal Clebsch-Gordan transform. This means that (1) we can reduce conjugate queries to standard queries by simply flipping the Young tableaux $\ket{X}$ and $\ket{Y}$ before and after the query, and (2) even standard queries can be implemented relative to \emph{just} a primal Clebsch-Gordan transform, without any explicit reference to a dual Clebsch-Gordan transform. 

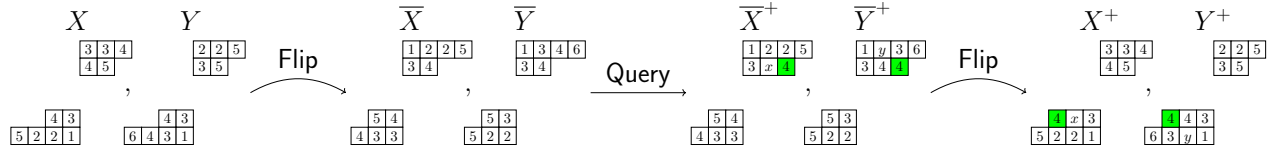
\begin{figure}[H]
    \centering
    \scalebox{.5}{
    \begin{tikzpicture}[baseline=(AX.base)]
        \node (CX) at (-9,0) [label=above:{\huge $\;\,X$}] {
            \Ytableau{
                334,
                45,
            }
            {
                    ,
                  43,
                5221,
            }
        };
        \node (Ccomma) at (-7.5,0) {\huge $,$};
        \node (CY) at (-6,0) [label=above:{\huge $\;\,Y$}] {
            \Ytableau{
                225,
                35,
            }
            {
                    ,
                  43,
                6431,
            }
        };
        \node (AX) at (0,0) [label=above:{\huge $\overline{X}$}] {
            \Ytableau{
                1225,
                34,
                ,
            }
            {
                 54,
                433
            }
        };
        \node (Acomma) at (1.5,0) {\huge $,$};
        \node (AY) at (3,0) [label=above:{\huge $\overline{Y}$}] {
            \Ytableau{
                1346,
                34,
                ,
            }
            {
                 53,
                522
            }
        };
        \node (BX) at (9,0) [label=above:{\huge $\phantom{}^{\phantom +}\overline{X}^+$}] {
            \Ytableau{
                1225,
                3{x},
                ,
            }
            {
                 54,
                433
            }
        };
        \node at (BX) {
            \gyoung(:<>::::::<>,:::::!<\Yfillcolour{green}>;4,,,,:<>)
        };
        \node (Bcomma) at (10.5,0) {\huge $,$};
        \node (BY) at (12,0) [label=above:{\huge $\phantom{}^{\phantom +}\overline{Y}^+$}] {
            \Ytableau{
                1y36,
                34,
                ,
            }
            {
                 53,
                522
            }
        };
        \node at (BY) {
            \gyoung(:<>::::::<>,:::::!<\Yfillcolour{green}>;{4},,,,:<>)
        };
        \node (DX) at (18,0) [label=above:{\huge $\phantom{}^{\phantom +}X^+$}] {
            \Ytableau{
                334,
                45,
            }
            {
                    ,
                {x}3,
                5221,
            }
        };
        \node at (DX) {
            \gyoung(:<>::::::<>,,,,:!<\Yfillcolour{green}>;{4},:<>)
        };
        \node (Dcomma) at (19.5,0) {\huge $,$};
        \node (DY) at (21,0) [label=above:{\huge $\phantom{}^{\phantom +}Y^+$}] {
            \Ytableau{
                225,
                35,
            }
            {
                      ,
                    43,
                63{y}1,
            }
        };
        \node at (DY) {
            \gyoung(:<>::::::<>,,,,:!<\Yfillcolour{green}>;{4},:<>)
        };
        \draw[->] (AY.east) to node[midway,above] {{\huge $\mathsf{Query}$}} (BX.west);
        \draw[->,bend left] (CY.east) to node[midway,above] {{\huge $\mathsf{Flip}$}} (AX.west);
        \draw[->,bend left] (BY.east) to node[midway,above] {{\huge $\mathsf{Flip}$}} (DX.west);
    \end{tikzpicture}
    }
    \caption{Example complex conjugate query for the unitary group.}
\label{fig:tableau_update_neg}
\end{figure}

Of course, we may want to make queries to both $\rho(g)$ and $\rho^*(g)$, or more generally to query both in superposition, as 
$
    \proj{0} 
    \otimes 
    \rho(g) 
    + 
    \proj{1} 
    \otimes 
    \rho^*(g) 
$.
This is merely another representation of the group, and our framework above also works for this representation. The (generalized) Schur transform that results is closely related to the mixed Schur transform of~\cite{fei2023efficientquantumalgorithmportbased, nguyen2023mixed,grinko2023gelfand} 
(more precisely, it is the direct sum of mixed Schur transforms for different sequences of $\rho(g)$ and $\rho^*(g)$). As before, each query is still a pair of Clebsch-Gordan transforms (the same Clebsch-Gordan transforms as for the standard query $\rho(g)$), depending on the control qubit. 

\paragraph{Controlled Queries}
Controlled queries can be handled similarly, taking the representation 
$
    \rho_{\mathsf{controlled}}(g)
    \coloneqq
    \proj{0} 
    \otimes 
    \Id
    + 
    \proj{1} 
    \otimes 
    \rho(g).
$
Since this is also a representation of the group, we can similarly apply the framework to this representation. This will, as expected,%
\footnote{
    Of course, since we already know what a controlled query \emph{should} do, rerunning this framework for the controlled representation $\rho_{\mathsf{controlled}}(g)$ just to recover this behavior may be overkill. However, we say this to emphasize that we do not need any specialized ad hoc modifications for each new query type.
} 
recover a tableau recording with an update that performs a normal $\rho(g)$ update on the $\ket{1}$ branch and does nothing on the $\ket{0}$ branch. 

\paragraph{Transpose and Inverse Queries}
Unlike complex conjugate queries, which are representations of the group, transpose and inverse queries are anti-representations,%
\footnote{
    Unlike representations, which have $\rho(g)\; \rho(h) = \rho(gh)$, anti-representations multiply in reverse order: $\rho^\dagger(g) \;\rho^\dagger(h) = \rho^\dagger(hg)$
}
and therefore do not fall directly into the framework above.
However, we make use of the following simple idea: implementing a transpose query can be thought of as conjugating a \emph{forward query} by the following (anti-linear) isometry on $L^2(G)$
\[ \mathsf{Transpose} \cdot \int_G f(g) \ket{g} dg = \int_G f(g^{-1})^* \ket{g} dg.
\]
In order to reason about this, we observe that $\mathsf{Transpose}$ implements the following simple involution on the Fourier basis (see \Cref{sec:fourier-unitary-identities}):
\[ \mathsf{Transpose} \cdot F^{-1}\ket{\lambda,X, Y} = F^{-1} \ket{\lambda, Y, X}.
\]
Thus, performing a transpose query in the Tableau basis simply requires swapping the roles of $\ket{X}$ and $\ket{Y}$!%
\footnote{
    Since $\mathsf{Transpose}$ is anti-linear (or conjugate-linear) and is not a physically implementable operation, it turns out that rather than performing the update rule $Q_\rho^{\mathrm{Tab}}$ on the swapped registers, one needs to perform the complex conjugate $\overline{Q_\rho^{\mathrm{Tab}}}$; but any circuit for $Q_\rho^{\mathrm{Tab}}$ can be generically converted into a circuit for $\overline{Q_\rho^{\mathrm{Tab}}}$; following \cite{C:Zhandry25}, we consider this to be a form of ``black-box access'' to the update. Moreover, in all the cases we consider, it is possible to write $Q_\rho^{\mathrm{Tab}}$ in a basis in which it is purely real, in which case $\overline{Q_\rho^{\mathrm{Tab}}} = Q_\rho^{\mathrm{Tab}}$ and this distinction becomes moot.
} 

\begin{figure}[H]
    \centering
    \scalebox{.5}{
    \begin{tikzpicture}[baseline=(AX.base)]
        \node (CX) at (-9,0) [label=above:{\huge $\;\,X$}] {
            \Ytableau{
                334,
                45,
            }
            {
                    ,
                  43,
                5221,
            }
        };
        \node (Ccomma) at (-7.5,0) {\huge $,$};
        \node (Clambda) at (-7.2,2.7) {\huge $(\lambda)$};
        \node (CY) at (-6,0) [label=above:{\huge $\;\,Y$}] {
            \Ytableau{
                225,
                35,
            }
            {
                    ,
                  43,
                6431,
            }
        };
        \node (AX) at (0,0) [label=above:{\huge $\;\,Y$}] {
            \Ytableau{
                225,
                35,
            }
            {
                    ,
                  43,
                6431,
            }
        };
        \node (Acomma) at (1.5,0) {\huge $,$};
        \node (Alambda) at (1.8,2.7) {\huge $(\lambda)$};
        \node (AY) at (3,0) [label=above:{\huge $\;\,X$}] {
            \Ytableau{
                334,
                45,
            }
            {
                    ,
                  43,
                5221,
            }
        };
        \node (BX) at (9,0) [label=above:{\huge $\;\,\phantom{}^{\phantom +}Y^+$}] {
            \gyoung(:<>:::::<>,::::::!<\Yfillcolour{green}>;{5},,,,<>:<>)
        };
        \node at (BX) {
            \Ytableau{
                225,
                3{x},
            }
            {
                    ,
                  43,
                6431,
            }
        };
        \node (Bcomma) at (10.5,0) {\huge $,$};
        \node (Blambda) at (10.8,2.7) {\huge $(\lambda)$};
        \node (BY) at (12,0) [label=above:{\huge $\;\,\phantom{}^{\phantom +}X^+$}] {
            \gyoung(:<>:::::<>,::::::!<\Yfillcolour{green}>;{5},,,,<>:<>)
        };
        \node at (BY) {
            \Ytableau{
                {y}33,
                44,
            }
            {
                    ,
                  43,
                5221,
            }
        };
        \node (DX) at (18,0) [label=above:{\huge $\;\,\phantom{}^{\phantom +}X^+$}] {
            \gyoung(:<>:::::<>,::::::!<\Yfillcolour{green}>;{5},,,,<>:<>)
        };
        \node at (DX) {
            \Ytableau{
                {y}33,
                44,
            }
            {
                    ,
                  43,
                5221,
            }
        };
        \node (Dcomma) at (19.5,0) {\huge $,$};
        \node (Dlambda) at (19.8,2.7) {\huge $(\lambda)$};
        \node (DY) at (21,0) [label=above:{\huge $\;\,\phantom{}^{\phantom +}Y^+$}] {
            \gyoung(:<>:::::<>,::::::!<\Yfillcolour{green}>;{5},,,,<>:<>)
        };
        \node at (DY) {
            \Ytableau{
                225,
                3{x},
            }
            {
                    ,
                  43,
                6431,
            }
        };
        \draw[->] (AY.east) to node[midway,above] {{\huge $\mathsf{Query}$}} (BX.west);
        \draw[->,bend left] (CY.east) to node[midway,above] {{\huge $\mathsf{Swap}$}} (AX.west);
        \draw[->,bend left] (BY.east) to node[midway,above] {{\huge $\mathsf{Swap}$}} (DX.west);
    \end{tikzpicture}
    }
    \caption{Example transpose query for the unitary group.}
\end{figure}
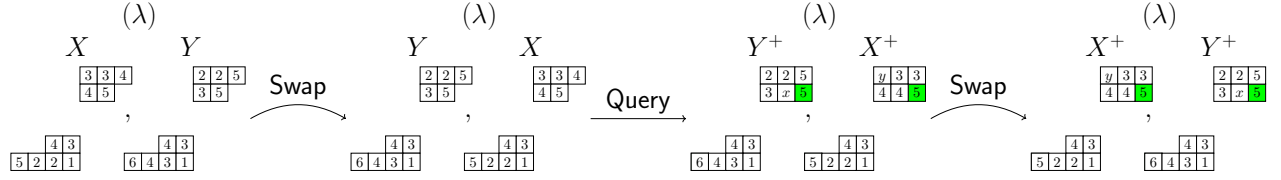

Similarly, we can perform an inverse query by implementing a conjugate query with $\ket{X}$ and $\ket{Y}$ reversed. In fact any anti-representation can be handled this way.

Finally, we remark that a similar idea can be used to derive a conjugate query in the Tableau basis, in terms of a (complex conjugated) standard query, using the anti-linear isometry
\[ \mathsf{Conj} \cdot \int_G f(g) \ket{g} dg = \int_G f(g)^* \ket{g} dg.
\]
We refer the reader to \cref{sec:tableau-other-queries} for more details. 

\subsection{Path Recording}\label{sec:tech-overview-path}
While the Tableau recording oracle is mathematically beautiful, satisfies perfect simulation, and can be algorithmically efficient, it seems quite difficult to prove operationally useful properties of this compressed oracle. In contrast, Zhandry's compressed oracle and the Ma-Huang path-recording oracle have easily interpretable \textquote{database states} on the recording register that can be used for proving quantum query lower bounds and pseudorandomness results. For example, in the Ma-Huang path-recording oracle,
\begin{itemize}
    \item The recording register contains superpositions of \textquote{set states} of the form $\ket{\{(x_1, y_1), \hdots, (x_t, y_t)\}}$.
    \item To make a query on an input $\ket{x}$, and database state $\ket{D}$, one does the following:
    \begin{itemize} 
    \item Create an EPR pair in superposition over $y\notin \{y_1, \hdots, y_t\}$\footnote{We remark that while $y\notin \{y_1, \hdots, y_t\}$ is enforced to make this update procedure an isometry, this is \emph{not} perfectly consistent with a Haar-random unitary (for example, if you query $U^{\otimes t}\ket{x_1, x_2, \hdots, x_t}$ you may observe the same output $y$ multiple times). As we will show, this is one out of two sources of simulation error in the Ma-Huang oracle.}
    \item Swap one of the $\ket{y}$'s into the algorithm's register (replacing $\ket{x}$) 
    \item Add the pair $(x,y)$ (using the second $\ket{y}$) to the database set $D$.
    \end{itemize}
\end{itemize}
One might naturally ask: why are they stored as \emph{sets} rather than ordered tuples? This is easy to understand in the case of \textbf{non-adaptive queries} applied to \textbf{distinct inputs}: if you query a (purified) Haar-random unitary on distinct inputs $\ket{x_1, \hdots, x_t}$, this results in the uncompressed recording%
\footnote{
    Here and throughout the paper, we use the notation $\vec x := (x_1, \dots, x_t)$ to denote a tuple of entries.
}
\begin{align} 
    \int_U 
    U\ket{x_1} 
    \otimes 
    \hdots 
    \otimes 
    U\ket{x_t} 
    \otimes 
    \ket{U} 
    dU 
    &
    = 
    \sum_{y_1, \hdots, y_t} 
    \int_U 
    \bra{\vec y} 
    U^{\otimes t} \ket{\vec x} 
    \cdot 
    \ket{\vec y} 
    \otimes 
    \ket{U} 
    dU 
    \\
    &
    = 
    \sum_{y_1, \hdots, y_t} 
    \ket{\vec y} 
    \otimes 
    \int_U 
    \bra{\vec y}
    U^{\otimes t} \ket{\vec x}
    \cdot
    \ket{U} 
    dU 
    \\
    &
    = 
    \sum_{y_1, \hdots, y_t} 
    \ket{\vec y} 
    \otimes 
    \int_U 
    \;
    \prod_{i = 1}^{t}
    \Big(
        \bra{y_i}
        U 
        \ket{x_i}
    \Big)
    \cdot
    \ket{U} 
    dU
    \,,
\end{align}
and indeed the recording register has \emph{forgotten} the ordering of $x_1, \hdots, x_t$ (but \emph{remembers} the matching between $x_i$ and $y_i$). Even more explicitly, let us try to compare the uncompressed $L^2(G)$ recording with a na\"ive path recording by defining the following map $\Theta_t: (\mathbb C^N)^{\otimes 2t} \rightarrow L^2(G)$.
\[ \Theta_t \ket{x_1, \hdots, x_t, y_1, \hdots, y_t} = \int_U \bra{\vec y} U^{\otimes t} \ket{\vec x} \cdot \ket{U} dU.
\]
Then, if we think of a permutation $\pi\in S_t$ as acting on $\ket{x_1, \hdots, x_t}$ by permuting the registers
\[ S(\pi) \ket{x_1, \hdots, x_t} = \ket{x_{\pi^{-1}(1)}, \hdots, x_{\pi^{-1}(t)}}, 
\]
we have that
\begin{align}
    \Theta_t \cdot \Big( S(\pi) \otimes S(\pi) \Big) \ket{\vec x, \vec y} &= \int_U \bra{\vec y} S(\pi)^{\dagger} U^{\otimes t} S(\pi) \ket{\vec x} \cdot \ket{U} dU \\
    &= \int_U \bra{\vec y}  U^{\otimes t} \ket{\vec x} \cdot \ket{U} dU  \\
    &= \Theta_t \ket{\vec x, \vec y}.
\end{align}
This holds because every permutation $S(\pi)$ on the registers \emph{commutes} with the tensor power operator $U^{\otimes t}$. In fact, let us write this identity slightly differently:
\begin{equation}
    \Theta_t \cdot \Big(S(\pi) \otimes \Id\Big) \ket{\vec x, \vec y} = \Theta_t \cdot \Big( \Id \otimes S(\pi)^T\Big) \ket{\vec x, \vec y},
\end{equation}
which can be seen to hold by renaming $S(\pi) \ket{\vec y} \rightarrow \ket{\vec y}$. This means that the state $\ket{\vec x, \vec y}$ contains redundant information and should not be used as-is in a recording oracle.

However, if we instead used the state
\[ \ket{\{(x_1, y_1), \hdots, (x_t, y_t)\}} \coloneqq \frac 1 {\sqrt{t!}}\sum_{\pi \in S_t} \Big(S(\pi) \otimes S(\pi) \Big) \ket{\vec x, \vec y},
\]
then because we have symmetrized with respect to the symmetric group $S_t$, we have that 
\[(S(\pi) \otimes S(\pi)) \ket{\{(x_1, y_1), \hdots, (x_t, y_t)\}} = \ket{\{(x_1, y_1), \hdots, (x_t, y_t)\}},\] so these states satisfy the same symmetries as the states $\Theta_t \ket{x,y}$. Indeed, the image of the projector%
\footnote{
    It is straightforward to check from the definition that this operator is idempotent
    (that is, that $(\Omega_{S_t})^2 = \Omega_{S_t}$) and Hermitian.
}
\begin{align}
    \label{eq:techover:omega-symmetric-algebra}
    \Omega_{S_t} 
    = 
    \frac{1}{t!} 
    \sum_{\pi\in S_t} 
    S(\pi) \otimes S(\pi)
    \,,
\end{align} 
is precisely the set of states $\ket{\psi}$ satisfying the identity
\begin{equation}
    (S(\pi) \otimes \Id) \ket{\psi} = (\Id \otimes S(\pi)^T)\ket{\psi}. \label{eq:EPR-symmetry}
\end{equation}
Although our argument so far has not proved it, it turns out that the image of $\Omega_{S_t}$ is \emph{precisely} the subspace of valid recording states for \emph{arbitrary} quantum query algorithms, even those that are adaptive and may query on non-distinct inputs. 

\oldparagraph{What is $\Omega_{S_t}$?} Let us pause and try to better understand the $\Omega_{S_t}$ operator. The right tool for this job is \textbf{Schur-Weyl duality}. Schur-Weyl duality for the unitary representation $U^{\otimes t}$ on $(\mathbb C^N)^{\otimes t}$ amounts to employing the following facts:
\begin{enumerate}
    \item The \emph{only} operators on $(\mathbb C^N)^{\otimes t}$ that commute with every $U^{\otimes t}$ are linear combinations of the permutation elements $S(\pi)$. In other words, the $U(N)$-commutant algebra of $(\mathbb C^N)^{\otimes t}$ is the group algebra $\mathbb C[S_t]$. 
    \item Conversely, the \emph{only} operators on $(\mathbb C^N)^{\otimes t}$ that commute with every $S(\pi)$ are linear combinations of the diagonal action operators $U^{\otimes t}$. 
\end{enumerate}
According to Schur-Weyl duality, this turns out to imply that the space $(\mathbb C^N)^{\otimes t}$ decomposes in the following way:
\begin{equation}
    (\mathbb C^N)^{\otimes t} \simeq \bigoplus_\lambda V_\lambda^{U(N)} \otimes V_\lambda^{S_t},
\end{equation}
where $V_\lambda^{U(N)}$ is the space of an irreducible representation of $U(N)$, $V_\lambda^{S_t}$ is a \emph{corresponding} irreducible representation space of $S_t$. The isometry implementing this mapping is the \emph{Schur transform} of \Cref{eq:thechover-shur-transform}. In this decomposition, $U^{\otimes t}$ acts as
\[ U^{\otimes t}\mid_{V_\lambda^{U(N)} \otimes V_\lambda^{S_t}} = \rho_\lambda(U) \otimes \Id.
\]
That is, $U^{\otimes t}$ acts properly as the irrep
transformation $\rho_\lambda(U)$ on $V_\lambda^{U(N)}$ and ignores the commutant register $V_\lambda^{S_t}$. On the other hand, the permutation elements $S(\pi)$ act on the other tensor factor:
\[ S(\pi)\mid_{V_\lambda^{U(N)} \otimes V_\lambda^{S_t}} = \Id \otimes \rho_\lambda(\pi).
\]
With this decomposition, we can understand the projection $\Omega_{S_t}$: since states $\ket{\psi} \in (\mathbb C^N)^{\otimes t} \otimes (\mathbb C^N)^{\otimes t}$ in the image of $\Omega_{S_t}$ satisfy \cref{eq:EPR-symmetry} and the $S(\pi)$ span the commutant algebra, we conclude that $(A\otimes \Id) \ket{\psi} = (\Id \otimes A^T)\ket{\psi}$ for every $A \in \mathbb C[S_t]$. These operators $A$ are precisely those that act only on the commutant irreps:
\[ A = \bigoplus_\lambda \Id_{V_\lambda^{U(N)}} \otimes A_\lambda, \qquad A_\lambda \in \mathrm{End}(V_\lambda^{S_t})
\]
Thus, for any state $\ket{\psi} \in (\mathbb C^N)^{\otimes t} \otimes (\mathbb C^N)^{\otimes t}$, which we can generically write (in the Schur-transformed basis) as 
\(
    \ket{\psi} 
    = 
    {
    \sum_{\lambda, \lambda'} 
    \alpha_{\lambda, \lambda'} 
    \ket{\lambda, \lambda'} 
    \otimes 
    \ket{\psi_{\lambda, \lambda'}}
    }
\),
for 
\(
    \ket{\psi_{\lambda, \lambda'}}
    \in
    V_\lambda^{U(N)}
    \otimes 
    V_{\lambda'}^{U(N)} 
    \otimes 
    V_\lambda^{S_t}
    \otimes 
    V_{\lambda'}^{S_t}
\),
we have that
\begin{align*}
    (
        A
        \otimes
        \Id
    )
    \ket{\psi} 
    &
    \;
    = 
    \;
    \sum_{\lambda, \lambda'} 
    \alpha_{\lambda, \lambda'} 
    \ket{\lambda, \lambda'} 
    \otimes
    (
        \Id_{V_\lambda^{U(N)}} 
        \otimes 
        \Id_{V_{\lambda'}^{U(N)} } 
        \otimes 
        A_\lambda
        \otimes 
        \Id_{V_{\lambda'}^{S_t}} 
    )
    \ket{\psi_{\lambda, \lambda'}}
    \\
    =
    \;
    (
        \Id
        \otimes
        A^T
    )
    \ket{\psi} 
    &
    \;
    = 
    \;
    \sum_{\lambda, \lambda'} 
    \alpha_{\lambda, \lambda'} 
    \ket{\lambda, \lambda'} 
    \otimes
    (
        \Id_{V_\lambda^{U(N)}} 
        \otimes 
        \Id_{V_{\lambda'}^{U(N)} } 
        \otimes 
        \Id_{V_{\lambda}^{S_t}} 
        \otimes 
        A_{\lambda'}^T
    )
    \ket{\psi_{\lambda, \lambda'}}
\end{align*}
We can see, therefore, that any state 
satisfying \cref{eq:EPR-symmetry} must be such that:
\begin{itemize}
    \item only the blocks $\lambda = \lambda'$ have nonzero weight (consider $A = \Pi_\lambda$ that projects onto a fixed $\lambda$-block), and
    \item each component $\ket{\psi_\lambda}$ is an EPR state on the commutant registers $V_\lambda^{S_t}\otimes V_\lambda^{S_t}$ in tensor product with an arbitrary state on the $U(N)$ registers (because the EPR state is the unique state, up to overall phase, satisfying $(\Id \otimes A_\lambda^T) \ket{\psi_\lambda}= (A_\lambda \otimes \Id) \ket{\psi_\lambda}$ for all linear maps $A_\lambda$).
\end{itemize}
Thus, we refer to $\Omega_{S_t}$ as the \textbf{commutant EPR projector}. 

\paragraph{Our Path-Recording Oracle} 
With this thought process in mind, we are ready to introduce our exact path-recording compressed oracle. As before, we assume that our goal is to simulate query-access to a single unitary representation of $G$
\[ Q_\rho (\ket{x} \otimes \int_G f(g) \ket{g} dg ) = \int_G \rho(g) \ket{x} \otimes f(g) \ket{g} dg.
\]
By our earlier argument, we know that by applying the Fourier transform $F$ on the $L^2(G)$ register, after $t$ queries the recording register will be supported on
\[ \bigoplus_{\lambda \in \widehat G_t} V_\lambda \otimes V_\lambda^*,
\]
where $\widehat G_t$ denotes exactly the set of irreps that appear in the Schur-Weyl decomposition of $(\mathbb C^N)^{\otimes t}$:
\[ (\mathbb C^N)^{\otimes t} \simeq \bigoplus_{\lambda \in \widehat G_t} V_\lambda^G \otimes V_\lambda^{\mathcal A_t}.
\]
Compared to the case of $G = U(N)$ (and $\rho$ the defining representation) discussed earlier, this Schur-Weyl decomposition is defined with respect to the commutant algebra
\[ \mathcal A_t = \{A \in \mathrm{End}((\mathbb C^N)^{\otimes t}): A\cdot \rho(g)^{\otimes t} = \rho(g)^{\otimes t}\cdot A \text{ for all }g\in G\}.
\]
For the defining representation of $U(N)$, we  have that $\calA_t = \C[S_t]$, while for other group representations, the commutant will be another semi-simple diagram algebra (see \Cref{tab:schur-weyl} for more examples).
For a general representation (just as in the special case before), one can observe that uncompressed recording states obey symmetries imposed by $\mathcal A_t$:
\[
    \Theta_t \cdot (A\otimes \Id)=\Theta_t \cdot (\Id\otimes A^T) \text{ for all } A \in \mathcal A_t.
\]
Therefore, we should aim for a path-recording oracle with database states in the subspace in which these two commutant actions have been identified, which is exactly the commutant EPR subspace that is the image of $\Omega_{\mathcal A_t}$ in the Schur basis.

We now derive a path-recording oracle from the tableau-recording oracle.
To begin, we will define an isometry
\[
    \Gamma_t: 
    \bigoplus_{\lambda \in \widehat G_t} 
    V_\lambda^G 
    \otimes 
    (V_\lambda^G)^* 
    \rightarrow 
    (\mathbb C^N)^{\otimes t} 
    \otimes 
    ((\mathbb C^N)^*)^{\otimes t}
\]
that maps from the tableau basis to the path basis.
Let $\ket{\lambda, X, Y}$ be a basis vector in the tableau basis, where $\lambda \in \widehat G_t$.
The isometry $\Gamma_t$ acts as follows:
\begin{enumerate}
    \item 
    It first creates an EPR state
    \[ \ket{\mathrm{EPR}_\lambda} = \frac 1 {\sqrt{\dim(V_\lambda^{\mathcal A_t})}}\sum_{S \in \mathcal B(V_\lambda^{\mathcal A_t})} \ket{S} \ket{S^*},
    \]
    where $\mathcal B(V_\lambda^{\mathcal A_t})$ is a basis for $V_\lambda^{\mathcal A_t}$, and we interpret $\ket{S^*}\in (V_\lambda^{\mathcal A_t})^*$ as a dual vector, and appends it to the tableau basis vector.
    This produces the state $\ket{\lambda, X, Y}\ket{\mathrm{EPR}_\lambda}$, which is an element of the space $V_\lambda^G \otimes (V_\lambda^G)^* \otimes V_{\lambda}^{\mathcal A_t} \otimes (V_{\lambda}^{\mathcal A_t})^*$.
    \item 
    Next, it applies the inverse Schur transform $\schur^\dagger$ to $V_\lambda^G \otimes V_\lambda^{\mathcal A_t}$
    and the inverse dual Schur transform $\schur^T$ to $(V_\lambda^G \otimes V_\lambda^{\mathcal A_t})^*$.
    The resulting state is an element of $(\mathbb C^N)^{\otimes t} \otimes ((\mathbb C^N)^*)^{\otimes t}$, and can therefore be written as a superposition over $t$-tuples of pairs $\ket{(x_1, y_1), \dots, (x_t, y_t)}$. 
\end{enumerate}

Having defined $\Gamma_t$, it is straightforward to state the path-recording oracle as%
\footnote{
    Here, again, $F$ is the Fourier transform on $L^2(G)$ as before (\Cref{eq:fourier-transform}).
}
\begin{equation*}
    Q_\rho^{\mathrm{Path}}
    \coloneqq 
    (\Id\otimes \Gamma_{t+1}) 
    \,
    Q_\rho^{\mathrm{Tab}} 
    \,
    (\Id \otimes  \Gamma_t^\dagger)
    = 
    (\Id\otimes (\Gamma_{t+1} \cdot F) )
    \,
    Q_\rho 
    \,
    (\Id \otimes (F^\dagger \cdot \Gamma_t^\dagger))
    \,.
\end{equation*}
In other words, if the purification register is in the path basis,  $Q_\rho^{\mathrm{Path}}$ simply applies $\Gamma_t^{\dagger}$ to map it to the tableau basis.
It then applies the tableau-basis oracle,
and then applies $\Gamma_{t+1}$ to convert the purification register back to the path basis.

\paragraph{Interpreting $Q_\rho^{\mathrm{Path}}$} Of course, the main reason for defining the path-recording oracle is to be able to interpret the recording states and the effect of an oracle query. To do this, we prove that $Q_\rho^{\mathrm{Path}}$ has the following form

\begin{theorem}[See \cref{def:path-recording,thm:path-recording}]\label{thm:tech-overview-path}
    The path-recording oracle $Q_\rho^{\mathrm{Path}}$ can be expressed as

    \[
    Q_{\rho,t+1}^{\mathrm{Path}}
    =
    \sum_{x,y\in [N]}
    \ketbra{y}{x}
    \otimes
    \Lambda_{t+1} \cdot 
    \Omega_{\calA_{t+1}} \cdot 
    \App^{(t)}_{x,y} \cdot \Lambda_t^{+} \cdot 
    \Omega_{\calA_t},
    \]
    where:
    \begin{itemize}
        \item $\mathcal A_{t}$ denotes the $G$-commutant algebra of $(\mathbb C^N)^{\otimes t}$. 
        \item $\Omega_{\calA_t}$ denotes the (controlled-$\lambda$) projection onto $\ket{\mathrm{EPR}_\lambda}$ in the Schur basis,
        \item $\App^{(t)}_{x,y}$ denotes the map appending $\ket{(x,y)}$ to the database state.%
        \footnote{
            Note the database here contains an \emph{ordered} list of $(x,y)$ pairs, and this operator merely appends to that list. As we will see, the operator $\Omega_{\calA_t}$ resymmetrizes the list, removing this ordering (and anything else that is not to be recorded).
        }
        \item $\Lambda_t = \bigoplus_{\lambda \in \widehat G_t} \sqrt{\frac{\dim(V_\lambda^{\mathcal A_t})}{\dim(V_\lambda^{G})}}\cdot (\Pi_\lambda\otimes \Id)$, where $\Pi_\lambda$ denotes projecting onto the $\lambda$-component in the Schur basis.\footnote{Since \(\Omega_{A_t}\) has already projected onto equal \(\lambda\)-labels on
the two Schur bases, it is equivalent here to project onto the \(\lambda\)
block on either half or both.}
        \item $\Lambda_t^+=\bigoplus_{\lambda \in \widehat G_t} \sqrt{\frac{\dim(V_\lambda^{G})}{\dim(V_\lambda^{\mathcal A_t})}}\cdot (\Pi_\lambda \otimes \Id)$ is (pseudo-)inverse to $\Lambda_t$
    \end{itemize}
\end{theorem}
In other words, assuming that the recording register was already in the image of $\Omega_{\mathcal A_t}$ (which will be true if it was obtained by running an algorithm that queries $Q_\rho^{\mathrm{Path}}$), the update is the following sequence of operations.
\begin{itemize}
    \item Reweight subspaces in the Schur basis.
    \item Initialize an EPR state $\sum_{y \in [N]} \ket{y} \ket{y}$,
    \item Append $(x,y)$ to the database using the query input and half of the EPR state. The other half of the EPR state will be returned as the response.
    \item Re-impose the symmetry $\Omega_{\mathcal A_{t+1}}$ on the recording register.\footnote{We remark that the recording register was previously in the image of $\Omega_{\mathcal A_t}$, so intuitively $\Omega_{\mathcal A_{t+1}}$ is only imposing additional symmetries involving the last register.}
    \item Reweight the (new) subspaces in the Schur basis again.
\end{itemize}
We refer the reader to \cref{sec:path} for the formal proof of \cref{thm:tech-overview-path}. The high-level idea is to relate the isometry $\Gamma_t \cdot F$ with the map $\Theta_t$ from earlier (which aimed to analogize the path state $\ket{x,y}$ with the recording state $\int_G \bra{y} \rho(g)^{\otimes t} \ket{x} \cdot \ket{g} dg$) by proving that
\[ 
    \Gamma_t 
    \cdot
    F 
    \cdot
    \Theta_t 
    = 
    \Lambda_t 
    \cdot
    \Omega_{\mathcal A_t}
    \,.
\]
At this point, there are two remaining questions to answer regarding interpretation of this formula for $Q_\rho^{\mathrm{Path}}$ for general groups:
\begin{enumerate}
    \item The commutant EPR projector $\Omega_{\calA_t}$ may appear somewhat mysterious in the path basis outside of the case of $G = U(N)$ and $\calA_t = \C[S_t]$ in the earlier example.
    \item The subspace reweighting operators $\Lambda_t$ are applied in the Schur basis, which has no simple/obvious interpretation in the path basis.
\end{enumerate}
For the first question, we show that $\Omega_{\calA_t}$ has a clear general-purpose form directly in the path basis, generalizing the formula $\Omega_{S_t} = \frac 1 {t!} \sum_{\pi \in S_t} S(\pi) \otimes S(\pi)$ for the case $G = U(N)$. This allows us to think of $\Omega_{S_t}$ as an operation in the standard (\textquote{path}) basis rather than one the Schur basis. 

This is accomplished as follows: let $\mathcal B(\mathcal A_t)$ denote a basis for the commutant algebra. For example, this basis can consist of permutations $\pi \in S_t$ for the case $G = U(N)$ and $\calA = \C[S_t]$.  
For the case $G = S_N$, the commutant algebra $\mathcal A_t = P_t(N)$ is the \emph{partition algebra} \cite{halverson2005partition}, 
which has a basis consisting of \emph{set-partition diagrams} (see \Cref{fig:symmetric-and-partition-bases_b}). 

\begin{figure}[H]
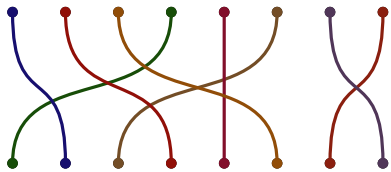
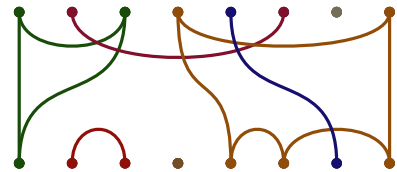

    \centering
    \begin{subfigure}{0.45\textwidth}
        \centering
        \begin{equation*}
        \begin{tikzinline}[scale=1,baseline]
            \makenodes{8}{1cm}{-1cm}{.7cm}

            \connectset[
                color=darkgreen!80!darkbrown
            ]
            {
                B1, 
                T4
            }
            \connectset[
                color=darkblue!80!darkbrown
            ]
            {
                B2, 
                T1
            }
            \connectset[
                color=darkbrown!80!darkbrown
            ]
            {
                B3, 
                T6
            }
            \connectset[
                color=darkred!80!darkbrown
            ]
            {
                B4, 
                T2
            }
            \connectset[
                color=darkpurple!80!darkbrown
            ]
            {
                B5, 
                T5
            }
            \connectset[
                color=darkorange!80!darkbrown
            ]
            {
                B6, 
                T3
            }
            \connectset[
                color=red!60!black!60!darkbrown
            ]
            {
                B7, 
                T8
            }
            \connectset[
                color=blue!80!black!30!darkbrown
            ]
            {
                B8, 
                T7
            }
        \end{tikzinline}
        \end{equation*}
        \caption{An example basis element of the symmetric group algebra, $\C[S_t]$, for $t = 8$.}
        \label{fig:symmetric-and-partition-bases_a}
    \end{subfigure}
    \hfill
    \begin{subfigure}{0.45\textwidth}
        \centering
        \begin{equation*}
        \begin{tikzinline}[scale=1,baseline]
            \makenodes{8}{1cm}{-1cm}{.7cm}

            \connectset[
                color=darkgreen!80!darkbrown
            ]
            {
                B1, 
                T1,
                T3
            }
            \connectset[
                color=darkgreen!80!darkbrown
            ]
            {
                B1, 
                T3
            }
            \connectset[
                height=0,
                color=darkbrown!80!darkbrown
            ]
            {
                B4, 
                B4 
            }
            \connectset[
                color=darkred!80!darkbrown
            ]
            {
                B2, 
                B3 
            }
            \connectset[
                height=.8,
                color=darkpurple!80!darkbrown
            ]
            {
                T6,
                T2
            }
            \connectset[
                color=darkorange!80!darkbrown
            ]
            {
                B5, 
                B6, 
                B8,
                T4,
                T8
            }
            \connectset[
                color=darkorange!80!darkbrown
            ]
            {
                B8,
                T8
            }
            \connectset[
                color=darkblue!80!darkbrown
            ]
            {
                B7, 
                T5 
            }
            \connectset[
                height=0,
                color=darkyellow!80!darkbrown
            ]
            {
                T7, 
                T7 
            }
        \end{tikzinline}
        \end{equation*}
        \caption{An example basis element of the partition algebra, $P_t(N)$, for $t = 8$.}
        \label{fig:symmetric-and-partition-bases_b}
    \end{subfigure}
    \caption{The symmetric group algebra, $\C[S_t]$, the partition algebra, $P_t(N)$, and other diagram algebras can be described in terms of a basis of partition diagrams that partition $t$ top vertices and $t$ bottom vertices into components. See \Cref{sec:diagram-algebras} for more details.}
    \label{fig:symmetric-and-partition-bases_ab}
\end{figure}
\noindent
Using a variant of the Schur orthogonality relations we show (see \cref{thm:commutant-epr-proj-as-symmetrization}) that in general,
\[ 
    \Omega_{\mathcal A_t} = \sum_{a \in \mathcal B(\mathcal A_t)} S(a) \otimes S(a^*)^T,
\]
where $a^*$ denotes a \emph{dual basis element} with respect to the trace pairing $\tr_{\mathcal A_t}(ab)$.%
\footnote{
    $\tr_{\mathcal A_t}(ab)$ is defined to be the trace of the \textquote{multiplication by $ab$ map} on $\mathcal A_t$. The dual basis element $a^*$ corresponding to a basis element $a \in \calB(\calA_t)$ is the unique element of $\calA_t$ such that $\tr_{\mathcal A_t}(a^* \; b) = \delta_{ab}$ where $b$ runs over the basis of the algebra, $\calB(\calA_t)$.
} 
As an example, for $G = U(N)$, the elements of $\mathcal B(\mathbb C[S_t])$ are orthogonal under this trace pairing, except that $\tr([\pi] \cdot [\pi^{-1}]) = t!$ . We thus have that $[\pi]^* = \frac{1}{t!} [\pi^{-1}]$, recovering the form of $\Omega_{S_t}$ in \Cref{eq:techover:omega-symmetric-algebra}. Of course, this operator $\Omega_{\mathcal A_t}$ depends on the group $G$ and its representation, but understanding it simply amounts to understanding the commutant algebra $\mathcal A_t$ (in other words, the algebraic symmetries of $\rho^{\otimes t}$) and does not require the Schur basis. For example, we study this operator closely for the case $G = S_N$ in \cref{sec:prus}, as well as some simpler cases in \cref{sec:approximation-specialization}.

Thus, what remains in order to understand the exact path-recording oracle is to understand the reweighting operator $\Lambda_{\mathcal A_t}$. As we discuss next, if one is willing to make \emph{approximations}, this challenge can be surmounted as well.

\paragraph{\cite{STOC:MaHua25} as an approximation of our oracle} How does our path-recording oracle (for the case $G = U(N)$) compare with the Ma-Huang oracle? Here is a direct comparison:
\begin{itemize}
    \item The \cite{STOC:MaHua25} oracle and our oracle both begin by creating an EPR state over outputs $\ket{y}$; however, \cite{STOC:MaHua25} restricts $y$ to not collide (in the standard basis) with the $y$ values in the existing recording state.
    \item The \cite{STOC:MaHua25} oracle then adds the pair $\ket{(x,y)}$ to an unordered \emph{set}, while our path-recording oracle first appends $\ket{(x,y)}$ to an \emph{ordered list} or \emph{$t$-tuple} and then symmetrizes with respect to $S_t$ (permuting the pairs into all possible positions in the list). On states with distinct $y_i$, these operations are the same up to a multiplicative scalar.
    \item Finally, our oracle contains two subspace reweightings (one at the beginning and one at the end) that the \cite{STOC:MaHua25} oracle does not have.
\end{itemize}
Thus, the fact that the \cite{STOC:MaHua25} oracle approximates ours (for $G = U(N)$) follows from two facts that we prove in \cref{sec:approximation-specialization}:
\begin{enumerate}
    \item For any $t$-query adversary $\mathcal A^{Q_\rho^{\mathrm{Path}}}$, the probability that the $y_i$ in the recording register are all distinct is at least $1-O(t^2/N)$.
    \item The rescaling operators $\Lambda_t$ are \emph{scalar multiples of the identity}, up to multiplicative error $(1 \pm O(t^2/N))$.
\end{enumerate}
Conversely, this gives a method of deriving the Ma-Huang approximate path-recording oracle from first principles:
\begin{enumerate}
    \item \cref{thm:tech-overview-path} describes a general exact path-recording oracle.
    \item For a simpler form with easier interpretability, approximate the $\Lambda_{S_t}$ reweightings with scalar multiples of the identity, resulting in an operator $\widetilde Q_\rho^{\mathrm{Path}}$.
    \item Without the subspace reweighting, $\widetilde Q_\rho^{\mathrm{Path}}$ is not an isometry; however, it \emph{is} an isometry when inserted $y$ values are restricted to be distinct. This gives the interpretable---and, upon inspection, efficiently implementable---approximate path-recording oracle
    \[ Q_\rho^{MH} = \frac 1 {\sqrt{N-t}}
    \sum_{x,y\in [N]}
    \ketbra{y}{x}
    \otimes
    \sqrt{t+1} \cdot \Omega_{\calA_{t+1}} \cdot \Pi_{\mathrm{Dist}}^Y \cdot 
    \App^{(t)}_{x,y},
    \]
    which is precisely the path-recording oracle of~\cite{STOC:MaHua25}.
\end{enumerate}
\noindent
See \Cref{sec:approximation-specialization}, for additional examples of how to derive path recording oracles for other groups in this way.

\subsection{Pseudorandomness of the $PC$ Ensemble}
Finally, we describe our approach for proving that our $PC$ ensemble is a pseudorandom unitary (\cref{thm:main-PC-intro}).
Standard security of pseudorandom unitaries says that an adversary cannot distinguish the ensemble from Haar random with any polynomial number of standard queries, even adaptive ones.%
\footnote{
    There is also a notion of strong security which allows other query types.
    As is, this PC construction, much like the PFC construction of~\cite{STOC:MaHua25} does not satisfy strong PRU security, and for the same reason.
    If we want strong PRU security, we conjecture that arguments along these lines can show that $C_1 \cdot P \cdot C_2$ satisfies the stronger security notions, much like the strong $C_1 \cdot P \cdot F \cdot C_2$ construction of~\cite{STOC:MaHua25}.
}
To gain some intuition, we can start by asking: what is the difference between querying a random permutation unitary%
\footnote{
    The permutation here is an element of $G = S_N$ (not to be confused with the commutant algebra $\C[S_t]$ that we considered earlier for the case where $G = U(N)$). For each such permutation, it acts on $\C^N$ by permuting the $N$ basis vectors in the standard basis.
}
$P$ and querying a Haar-random unitary $U$? There are two simple, detectable differences:
\begin{itemize}
    \item If you query any permutation unitary $P$ on a uniform superposition%
    \footnote{
        We will always use the plus state to denote the uniform positive superposition in the computational basis, $\ket{+} \coloneqq \frac{1}{\sqrt{N}} \sum_{z \in [N]} \ket{z}$. 
    }
    $\ket{+}\in \mathbb C^N$, the result is $\ket{+}$. However, for Haar-random unitary $U$, the state $U \ket{+}$ is a Haar-random state in $\mathbb C^N$.
    \item If you query any permutation $P^{\otimes 2}$ on two copies the same computational basis state $\ket{x} \ket{x}$, the resulting state $\ket{P(x)}\ket{P(x)}$ has the same computational basis state $y = P(x)$ appearing on both registers. However, for Haar-random $U$, the state $U^{\otimes 2} \ket{x} \ket{x}$ is two copies of a Haar-random state, which will almost never agree in the computational basis. 
\end{itemize}
However, conveniently, prepending $P$ with a random Clifford circuit (or any unitary $2$-design) $C$ thwarts both of these attacks:
\begin{itemize}
    \item If you query $P\cdot C$ on an arbitrary pure state $\ket{\psi}$ (independent of $P, C$), $C\ket{\psi}$ has almost no overlap with $\ket{+}$.
    \item If you query $(P\cdot C)^{\otimes 2}$ on an arbitrary state $\ket{\psi}\in (\mathbb C^{N})^{\otimes 2}$, the state $C^{\otimes 2}\ket{\psi}$ with high probability has no collisions in the computational basis. That is, it is almost completely contained within the \emph{distinct subspace} $\Pi_{\mathrm{Dist}} = \mathrm{Span}\{\ket{x_1, x_2}: x_1\neq x_2\}$. 
\end{itemize}
Of course, just because these two attacks are thwarted does not necessarily mean that \emph{all} possible attacks are thwarted.
Perhaps a more clever attack, which makes queries on something more sophisticated,
could distinguish a random permutation from a random unitary?

We show that this is not the case:
At a high level, we use our path-recording formalism, specialized to both groups $S_N$ and $U(N)$, to prove that these are in fact the \emph{only obstructions} to a random permutation $P$ looking like a random unitary, and so multiplying $P$ by $C$ does in fact result in a pseudorandom unitary. More formally, for any $t$, we define the \textquote{distinct, nonplussed} subspace $\mathsf{DNP}\subset (\mathbb C^N)^{\otimes t}$ to be the intersection of the distinct subspace with the \textquote{nonplussed subspace} $\mathsf{NoPlus} = \im( (\Id - \ketbra{+})^{\otimes t})$.
We must argue both that it suffices to consider only the distinct nonplussed subspace, and that permutations and unitaries are statistically indistinguishable on this subspace. We sketch the main points of the argument below (and see \Cref{sec:prus} for the proof in detail).

We know from \Cref{thm:main-path} that queries to a Haar random unitary $U \gets U(N)$ are perfectly indistinguishable from queries the the path-recording oracle $Q^{\mathrm{Path}}_{U(N)}$, and similarly, that queries to a random permutation $P \gets S_N$ are perfectly indistinguishable from those to $Q^{\mathrm{Path}}_{S_N}$.
To prove that $P \cdot C$ is indistinguishable from a Haar random unitary, we can thus take any $t$-query adversary $\mathsf{Adv}_t$ which may attempt to distinguish them, and 
\emph{directly compare} the path-recording purifications%
\footnote{
    The (right-) invariance of the Haar measure of $U(N)$ says that for a unitary $U$ drawn from the Haar measure and any other unitary $C$, $U \cdot C$ is identically distributed to $U$. Thus queries to $Q^{\mathrm{Path}}_{U(N)} \,\cdot\, C$ are perfectly indistinguishable from queries to $U \cdot C$ (by \Cref{thm:main-path}), and consequently from those to the Haar random unitary $U$.
}
\begin{align} 
    \label{eq:techover-adv-path-states-perm-vs-unitary}
    \ket{\mathsf{Adv}_t^{Q^{\mathrm{Path}}_{S_N} \,\cdot\, C}} 
    \qquad 
    \text{and}
    \qquad 
    \ket{\mathsf{Adv}_t^{Q^{\mathrm{Path}}_{U(N)} \,\cdot\, C}}
    \,.
\end{align}
Crucially, the two path-recording oracles are acting on the \emph{same} Hilbert space $((\mathbb C^N)^{\otimes t})^{\otimes 2}$ in their recording registers, which is what allows us to make the direct comparison. We proceed in three steps:

\begin{enumerate}
    \item 
    \textbf{Projecting onto the Distinct Nonplussed Subspace.} 
    First, we prove that for any possible adversary making $t$ queries to either of the two path-recording oracles, the recording register will only ever have at most $O(t^2/N)$ mass outside the distinct nonplussed subspace $\mathsf{DNP}^{\otimes 2} \subset ((\mathbb C^N)^{\otimes t})^{\otimes 2}$.

    The distinct nonplussed subspace $\mathsf{DNP}$, as the intersection of two simpler subspaces (the distinct subspace, $\mathsf{Dist}$, and the nonplussed subspace, $\mathsf{NoPlus}$) is somewhat tricky to reason about on its own. 
    But by bounding the \emph{Friedrichs angle} between the two subspaces, 
    we show that the projection $\Pi_{\mathsf{DNP}}^{X,Y}$ onto the distinct nonplussed subspace can be approximated by the product $\Pi_{\mathsf{Dist}}^{X,Y} \cdot \Pi_{\mathsf{NoPlus}}^{X,Y}$ of projections onto the two larger subspaces.

    We then show a \emph{ricochet property} of the path-recording oracles. This says, roughly, that any unitary $C$ that is applied before each path-recording oracle query can instead be thought of as if it were applied in parallel as $C^{\otimes t}$ to the $X$ part of the recording register. 
    We can then use the 2-design property of random Cliffords to argue that the $X$ part of the recording register will, with high probability, be in the distinct subspace, and separately in the nonplussed subspace. 
    A direct calculation also shows the same for the $Y$ part of the recording register.
    We then argue by a quantum union bound that the recording register must, with similarly high probability, be in the subspaces simultaneously.

    Thus, we may apply the projection $\Pi_{\mathsf{DNP}}^{X,Y}$ onto the distinct nonplussed subspace, while incurring error at most $O(t^2/N)$.
    \item 
    \textbf{The Symmetrization Operators $\Omega_{\calA_t}$ on the Distinct Nonplussed Subspace.} 
    Second, we prove that the commutant EPR projectors $\Omega_{S_t}$ and $\Omega_{P_t(N)}$ are in fact \emph{equal} on the subspace $\mathsf{DNP}^{\otimes 2}$, meaning
    \[ 
        \Omega_{S_t} \; \Pi_{\mathsf{DNP}}^{X,Y} = \Omega_{P_t(N)} \; \Pi_{\mathsf{DNP}}^{X,Y}
        \,.
    \]
    In our opinion, this is the most important step of the analysis. 
    Consider the operators $\Omega_{S_t}$ and $\Omega_{P_t(N)}$ written in the diagram bases of the two commutant algebras (see \Cref{fig:symmetric-and-partition-bases_ab}):
    \begin{align}
        \Omega_{S_t} 
        &
        =
        \frac{1}{t!} 
        \sum_{a \in S_t} 
        S(a) \otimes S(a)
        &
        \Omega_{P_t(N)}
        &
        =
        \sum_{a \in \mathcal B(P_t(N))} 
        S(a) \otimes S(a^*)^T
        \label{eq:techover-pc-omega-sym-vs-part}
    \end{align}
    Since the partition diagrams of $P_t(N)$ can be arbitrary partitions of the $2t$ vertices ($t$ top/output vertices and $t$ bottom/input vertices), they 
    range over a much larger set than the permutation diagrams of $\C[S_t]$ (which require a perfect matching between the $t$ input vertices and the $t$ output vertices in order to be a permutation in $S_t$). Thus $\Omega_{P_t(N)}$ ranges over a much larger basis than does $\Omega_{S_t}$. How could they be equal?

    But observe something important about how these set-partition diagrams act on $(\C^N)^{\otimes t}$: Suppose a diagram $a \in \mathcal B(P_t(N))$ has a component with at least two vertices on the bottom row (say, vertices numbered $i$ and $j$ when counting from the left). Then its representation $S(a)$ on the space $(\C^N)^{\otimes t}$ has the form $\sum_{z \in [N],\, \dots} \ketbra{\dots}{\dots, z, \dots, z, \dots}$ with the two $z$'s appearing (at least) on the $i$'th and $j$'th positions. But such an operator acts as exactly 0 unless the state it acts on has a collision (in the computational basis) on those two registers. So we know that on the distinct subspace, such diagrams must drop out of the sum on the right of \Cref{eq:techover-pc-omega-sym-vs-part}.
    Now suppose instead that a diagram $a \in \mathcal B(P_t(N))$ has some component consisting of only a single vertex on the bottom row (say, at vertex numbered $i$ from the left). Then its representation $S(a)$ on $(\C^N)^{\otimes t}$ has the form $\sum_{z \in [N],\, \dots} \ketbra{\dots}{\dots, z, \dots}$ with the $z$ appearing only at position $i$ in the bra of the operator. Such an operator acts as exactly 0 unless the state it acts on has a plus state $\ket{+}$ in position $i$. So we know that on the \emph{nonplussed subspace}, such diagrams must drop out of the right-hand sum of \Cref{eq:techover-pc-omega-sym-vs-part}. On the \emph{distinct nonplussed subspace}, \emph{both} kinds of set-partition diagrams are annihilated. It is not hard at this point to convince yourself that the only diagrams remaining are those in which each component has exactly one vertex on the bottom row, and one vertex on the top row of the diagram, in other words, \emph{permutation diagrams}.

    We are not quite done yet, however, since we are now left with 
    \[
        \sum_{
            \substack{
                a \in \mathcal B(P_t(N))
                \\
                a \text{ is a permutation diagram}
            }
        } 
        S(a) \otimes S(a^{*_{P_t(N)}})^T
    \]
    where the dual relationship $a^{*_{P_t(N)}}$ is still with respect to the trace pairing of the partition algebra $P_t(N)$ (and not of the symmetric group algebra $\C[S_t]$). 
    Nevertheless, we prove that for every permutation $\pi \in S_t$, the dual element $[\pi]^{*_{P_t(N)}}$ is a linear combination of $\frac{1}{t!}[\pi^{-1}]$ (matching the $S_t$ case) and \emph{non-permutation} elements of $\mathcal B(P_t(N))$ (which, as we have argued, act as zero on the distinct nonplussed subspace $\mathsf{DNP}$). 

    Thus, after our restriction to $\mathsf{DNP}^{\otimes 2}$, the path-recording updates are \emph{identical} up to the subspace reweighting operators.

    \item 
    \textbf{The Reweighting Operators $\Lambda_{\calA_t}$ on the Distinct Nonplussed Subspace.}
    At this point, we already know that $\Lambda_{S_t}$ is approximately a constant. We additionally show that $\Lambda_{P_t(N)}$ is approximately constant (indeed, the same constant) on the subspace $\mathsf{DNP}^{\otimes 2}$.

    This conclusion comes from viewing the distinct nonplussed subspace from the standpoint of the representation theory of $S_N$: it turns out that in the Schur basis of $S_N$ and $P_t(N)$, the projector $\Pi_{\mathsf{DNP}}$ projects $\bigoplus_\lambda V_\lambda^{S_N} \otimes V_\lambda^{P_t(N)}$ precisely onto the \textquote{full-box} irrep labels $\lambda$. In more detail, each irrep label $\lambda \in \widehat{P}_t(N)$ corresponds to a Young diagram with at most $t$ boxes below the top row.%
    \footnote{
        by the common convention of indexing irreps of $S_N$ with Young diagrams of exactly $N$ boxes. Often, we ignore the top row of boxes, and instead count only the boxes that are moved to the second row or further. 
    }
    We show that the distinct nonplussed projector $\Pi_{\mathsf{DNP}}$ projects onto the irreps $\lambda$ with \emph{exactly} $t$ boxes below the top row. Moreover, we can compute that the dimensions of such irreps are very close (up to a multiplicative $(1 \pm O(t^2/N))$) between $S_N$ and $U_N$ and between $P_t(N)$ and $\C[S_t]$, showing that $\Lambda_{S_t}$ and $\Lambda_{P_t(N)}$ are approximately the same, up to error $O(t^2 / N)$. 
\end{enumerate}

We can thus conclude that the two purified states in \Cref{eq:techover-adv-path-states-perm-vs-unitary}, when querying the path-recording oracles for either a random unitary or a random permutation (preceded with a random Clifford), have distance at most $O(t^2/N)$ when tracing out the purification, which is negligible when $t = \poly(n)$ and $N = 2^n$. Thus the $t$-query adversary cannot distinguish $P \cdot C$ from a Haar random unitary with anything better than negligible advantage. We finally conclude by replacing the random permutation $P$ with either a pseudorandom permutation to get a pseudorandom unitary or a $2t$-wise independent permutation to get a unitary $t$-design.

\fi 

\newcommand{\AdvApp}{\mathsf{Adv}_{t}^{\App}}
\newcommand{\Dist}{\mathsf{Dist}}
\newcommand{\dist}{\mathsf{dist}}
\newcommand{\Hist}{\mathsf{Hist}}
\newcommand{\DNP}{\mathsf{DNP}}
\newcommand{\NoPlus}{\mathsf{NoPlus}}
\newcommand{\End}{\mathrm{End}}

\section{Preliminaries}

Throughout this paper, we consider the following setting. Let $G$ be a compact Lie group. Without loss of generality, $G$ can be thought of as a closed subgroup of the group of unitary matrices on a finite-dimensional Hilbert space. Compact Lie groups admit a unique normalized Haar measure that allows us to define uniform averages over $G$. We use $\int_G dg$ to denote integration under the Haar measure on $G$. This setting captures the following special cases of interest to quantum computation:

\begin{itemize}
    \item $G = U(N)$ is the group of $N\times N$ unitary matrices.
    \item $G = O(N)$ is the group of $N\times N$ real orthogonal matrices.
    \item $G = (U(1))^N$ is the group of $N\times N$ diagonal unitary matrices. 
    \item Any finite group $G$, in which case $\int_G f(g)  dg = \frac 1 {|G|} \sum_g f(g)$. This includes the permutation group $S_N$, which can be embedded as $N \times N$ permutation matrices, as well as the abelian group $\{\pm 1\}^N$ embedded as diagonal sign matrices. 
\end{itemize}

We will consider a finite-dimensional representation $\rho: G\mapsto U(N)$ of the group $G$. Our goal is to perform efficient on-the-fly simulation of a quantum algorithm $A$ with ``oracle access to $\rho(g)$ for a random $g$'', where $g$ is sampled (once) from the Haar measure and used for all queries. By efficient on-the-fly simulation, we mean a stateful simulation such that the $t$-th query that $A$ makes can be simulated by a $\poly(t,\log(N))$-time algorithm operating on a $\poly(t,\log(N))$-qubit Hilbert space. 

\subsection{Notation for compact groups}

\begin{definition}[Haar measure] \label{def:haar-measure}
    Let $G$ be a compact Lie group.
    We say that a (regular Borel) measure $\mu$ is left-invariant if for any measureable subset of the group, $S \subseteq G$, and for every $g \in G$, $\mu(gS) = \mu(S)$. Similarly, we say that $\mu$ is right-invariant if for all $S \subseteq G$, and for every $g \in G$, $\mu(Sg) = \mu(S)$.
    The \emph{Haar measure} on $G$ is the unique such measure (up to a rescaling) which is both left-invariant and right-invariant.
\end{definition}

\begin{definition}[Group element kets]\label{def:group-element-kets}
Let $G$ be a compact Lie group with normalized Haar measure $dg$.
We write $\{\ket{g} : g \in G\}$ for kets labeled by group elements, with
$\braket{g}{g'}=\delta_{g,g'}$ (so $\braket{g}{g}=1$).
To make this notation consistent between the finite and infinite cases,
we write $dg$ to denote a ``half-measure''. Then for any (square-integrable) function $f: G \mapsto \mathbb{C}$, we define
\[
\ket{f}=\int_G f(g)\ket{g}\,dg
\]
with the convention that
\[
\braket{f_1}{f_2}=\int_G \overline{f_1(g)}f_2(g)\,dg.
\]
Thus, $\braket{f}{f}=1$ when $\int_G|f(g)|^2\,dg=1$. For finite $G$, this means that $\ket{f}=\frac{1}{|G|}\sum_{g\in G}f(g)\ket{g}$ and we identify unit-norm basis vectors $\ket{g} = \sqrt{|G|}\cdot \ket{f_g}$, where $f_g(g') = \delta_{g, g'}$. 
\end{definition}

\begin{definition}[$L^2$ function space]
$L^2(G)$ is defined to be the space of
square-integrable functions $f : G \to \mathbb{C}$. Each function is represented by $\ket{f}$ as in \Cref{def:group-element-kets}.
\end{definition}

\begin{definition}[Regular representation]
The \emph{left regular representation}
\[
L : G \to U(L^2(G))
\]
is defined by
\[
(L_g f)(x) = f(g^{-1}x), \qquad f \in L^2(G), \ g,x \in G.
\]
This gives $L^2(G)$ the structure of a unitary representation of $G$. We note that unlike all other representations considered in this work, the regular representation may be infinite-dimensional. 
\end{definition}

\subsection{Purified Oracle Algorithms (or Adversaries)}
We formalize the notion of an algorithm, or adversary, with oracle access to a unitary $\mathcal O_{\reg A \reg R}$ acting on the algorithm's query register $\reg A$ along with a purification register $\reg R$ that is hidden from the algorithm. This formalism enables all of our compressed oracle definitions and will be used explicitly in~\cref{subsec:update_rule_for_adversaries,subsec:adversary-updates-DNP,subsec:putting_together}. 

\begin{definition}[Purified Oracle Adversaries]
    \label{def:oracle_adversary}
    \normalfont 
    A $t$-query adversary, denoted $\mathsf{Adv}_t^{\mathcal{O}}$, is specified by a list of $n + m$ qubit unitaries $(A_1, \dots, A_t)$. $A_i$ acts on a $n$-qubit \textit{query register} $\reg{A}$, along with a $m$-qubit \textit{workspace register} $\reg{B}$. The purified oracle $\mathcal{O}$ acts on both the adversary's query register, as well as a \textit{database register} $\mathsf{R}$, which is inaccessible to the adversary. After $t$ queries to $\mathcal{O}$, the state of the adversary is%
    \footnote{
        We are using the conventions that $\tr[\ket{\psi}] := \tr[\proj{\psi}\,]$ and that $\prod_{i=1}^t X_i := X_t \, X_{t-1} \cdots X_2 \, X_1$.
    }
    \begin{equation}
        \tr_{\reg{R}}\left[\mathsf{Adv}_t^{\mathcal{O}} \left(\ket{0}_{\reg{AB}} \otimes \ket{\emptyset}_{\reg{R}}\right) \right] = \tr_{\reg{R}}\left[\left(\prod_{i=1}^t \mathcal{O}_{\reg{AR}} \cdot A_{i\; \reg{AB}}\right) \left(\ket{0}_{\reg{AB}} \otimes \ket{\emptyset}_{\reg{R}}\right) \right]
    \end{equation}
\end{definition}
In this paper, we study purified oracle adversaries $\mathsf{Adv}_t^{\mathcal O}$ that have access to an oracle $\mathcal O$ corresponding to a purified random group representation $\rho(g)$. 

\subsection{Peter-Weyl Theorem and the Fourier Transform}
The first step towards understanding how to simulate group representations is defining the Fourier transform, which is an algorithmic form of the Peter--Weyl theorem. 

\begin{theorem}[Peter--Weyl]
There is a unitary isomorphism of representations
\[
L^2(G) \;\simeq\; \bigoplus_{\lambda\in\widehat{G}} V_\lambda \otimes V_\lambda^*,
\]
where $L^2(G)$ denotes the left regular representation and $g$ acts as $\rho_\lambda(g) \otimes \Id_{V^*_\lambda}$ on the right hand side. 
Here, $\widehat{G}$ refers to the set of isomorphism classes of irreducible representations of $G$. 
\end{theorem}

\begin{definition}[Fourier Transform]

We define the Fourier transform as the Peter-Weyl decomposition isomorphism in a particular basis. For every irreducible representation $\lambda$, choose an orthonormal basis $\{\,|X\rangle : X\in \mathcal{I}_\lambda\,\}$ of $V_\lambda$,
with dual basis $\{\ket{Y^*} : Y\in \mathcal{I}_\lambda\,\}$ of $V_\lambda^*$, where $\ket{Y^*} \in V_\lambda^*$ acts as $\bra{Y}$ on $V_\lambda$. This leads to a basis $\ket{\lambda, X, Y}$ of the right hand side $\bigoplus_{\lambda\in\widehat{G}} V_\lambda \otimes V_\lambda^*$ of the form
\[ \ket{\lambda, X, Y} = \ket{X, Y^*} \in V_\lambda \otimes V_\lambda^*. 
\]
In this basis, it turns out that the isomorphism $F$ in the Peter-Weyl theorem can be written as 
\[F^{-1} |\lambda,X,Y\rangle = \sqrt{d_\lambda}\;
\int_G \!\bra{Y}\rho_\lambda(g)\ket{X}\;|g\rangle\, dg.
\]
\end{definition}
We define this isomorphism $F$ to be the Fourier transform. By the unitarity of the Fourier transform, the equation above implies the \textbf{Schur orthogonality relations}:

\begin{lemma}[Schur orthogonality for compact groups] For all $\lambda, \lambda', X, X', Y, Y'$, we have
\[ \sqrt{d_\lambda} \sqrt{d_{\lambda'}} \int_G  \overline{\bra{Y}\rho_\lambda(g)\ket{X}}\bra*{Y'}\rho_{\lambda'}(g)\ket*{X'} dg = \delta_{\lambda, \lambda'} \cdot \delta_{X, X'} \cdot \delta_{Y, Y'}. 
\]
\end{lemma}

In \cref{sec:schur-orthogonality}, we derive generalized Schur orthogonality relations for arbitrary semisimple algebras.

\subsubsection{Properties of the Fourier Transform}\label{sec:fourier-unitary-identities} In this section, we consider the following simple isometries defined on $L^2(G)$:
\begin{align} \mathsf{Conj}: \ket{f} &\mapsto \int_G f(g)^* \ket{g} dg \\ \mathsf{Inv}: 
\ket{f} &\mapsto \int_G f(g^{-1}) \ket{g} dg \\
\mathsf{Transpose}: \ket{f} &\mapsto \int_G f(g^{-1})^* \ket{g} dg
\end{align}
where the third map is the composition of the first two. Note that $\mathsf{Conj}$ and $\mathsf{Transpose}$ are anti-linear (or conjugate-linear) rather than linear maps. We now describe how these maps act on the Fourier basis. For the first, 
\[ \mathsf{Conj} \cdot F^{-1} \ket{\lambda, X, Y} = \sqrt{d_\lambda} \int_G \bra{Y}\rho_\lambda(g)\ket{X}^*\;|g\rangle\, dg = \sqrt{d_\lambda} \int_G \bra{Y}\rho_\lambda(g)^*\ket{X}\;|g\rangle\, dg = F^{-1} \ket{\lambda^*, X, Y},
\]
where $\lambda^*$ denotes a dual representation acting as $\rho_{\lambda^*}(g) = \rho_\lambda(g)^*$ on an appropriate dual basis, and we have identified $\ket{X}$ non-canonically as a basis element of $V_{\lambda^*}$. For the second, we have
\[\mathsf{Inv} \cdot F^{-1} \ket{\lambda, X, Y} = \sqrt{d_\lambda} \int_G \bra{Y}\rho_\lambda(g^{-1})\ket{X}\;|g\rangle\, dg = \sqrt{d_\lambda} \int_G \bra{Y}\rho_\lambda(g)^\dagger \ket{X}\;|g\rangle\, dg = F^{-1} \ket{\lambda^*, Y, X},
\]
via the identity $\bra{Y} U^\dagger\ket{X} = \bra{X} U^* \ket{Y}$, where $\ket{Y}$ and $\ket{X}$ are identified as elements of $V_\lambda^*$ and $(V_\lambda^*)^*$, respectively. Finally, this tells us that the third map is given by
\[ \mathsf{Transpose} \cdot F^{-1} \ket{\lambda, X, Y} = F^{-1} \ket{\lambda, Y, X}.
\]

\subsection{Complete Reducibility and Schur-Weyl Duality}

In order to simulate oracle access to $\rho$, we will reduce to the case where $\rho$ is an \emph{irreducible} representation of $G$. To do this, we make use of the following result. Let $\widehat{G}$ denote the set of all finite-dimensional irreducible representations of $G$. 

\begin{theorem}[Complete reducibility for compact Lie groups, {\cite[Section 4.10]{Lie-groups-Hall-book}}]\label{thm:complete-reducibility}
Let $G$ be a compact Lie group and let $\rho$ be a finite\mbox{-}dimensional unitary representation. Then $\rho$ decomposes as a finite direct sum of irreducible representations
\[
  \rho \;\simeq\; \bigoplus_{i=1}^k \rho_{\lambda_i} .
\]
for $\lambda_i \in \widehat{G}$ (not necessarily distinct). We write $\mathrm{Irr}(\rho)$ to be the subset of $\widehat{G}$ which appears in the decomposition of $\rho$ with non-zero multiplicity.
\end{theorem}
Finally, of particular interest to us is the decomposition of \emph{tensor power representations}

\[
\rho^{\otimes t}(g) = \rho(g)^{\otimes t}, \qquad g \in G
\]
given a fixed ``base'' representation $\rho$. It turns out that the decomposition of $\rho^{\otimes t}$ into irreducible representations can be understood in terms of the commutant algebra of the (tensor power) representation, defined below. 

\begin{definition}[Commutant algebra]
Let $\rho: G \to U(N)$ be a unitary representation of a compact Lie group $G$.
For any integer $t \ge 1$, consider the tensor power representation $\rho^{\otimes t}$ acting on $(\mathbb{C}^N)^{\otimes t}$.
The \emph{commutant algebra} $\mathcal{A}_t$ of $\rho^{\otimes t}$ is the set of all
$N^t \times N^t$ matrices that commute with every $\rho^{\otimes t}(g)$:
\[
\mathcal{A}_t
\;=\; \mathrm{End}_G((\mathbb C^N)^{\otimes t}) = 
\{\, X \in \mathbb{C}^{N^t \times N^t} :
X \rho^{\otimes t}(g) = \rho^{\otimes t}(g) X
\text{ for all } g \in G \,\}.
\]
\end{definition}

By definition, the commutant algebra $\mathcal A_t$ acts on $(\mathbb C^N)^{\otimes t}$ in a way that commutes with the action of $G$. Throughout the paper, we make use of Schur-Weyl duality, which explains the structure of the representation $\rho^{\otimes t}$ using the commutant algebra $\mathcal A_t$. 

\begin{theorem}[Schur--Weyl Duality]\label{thm:Schur-Weyl}
Let $G$ be a compact Lie group and $\rho$ be a finite-dimensional unitary representation $\rho: G \to U(V)$. Let $\mathcal A_t \simeq \mathrm{End}_G(V^{\otimes t})$ denote the commutant algebra.

Then, the space $V^{\otimes t}$ decomposes as
\[
V^{\otimes t}
\;\simeq\;
\bigoplus_{\lambda} V_{G}^{\lambda} \otimes V_{\calA_t}^{\lambda},
\]
where each $V_{G}^{\lambda}$ is an irreducible representation of $G$ and each $V_{\calA_t}^{\lambda}$ is an irreducible representation of $\mathcal{A}_t$.
Under this decomposition, $G$ and $\mathcal A_t$ act as
\[
\rho^{\otimes t}(g) = \bigoplus_{\lambda} \rho_\lambda(g) \otimes \Id_{V_{\calA_t}^{\lambda}},
\qquad
X = \bigoplus_{\lambda} \Id_{V_{G}^{\lambda}} \otimes \rho_\lambda(X),
\quad X \in \mathcal{A}_t.
\]
In particular, the $\mathcal A_t$-commutant of $V^{\otimes t}$ can be described as the following finite-dimensional algebra:
\[ \mathrm{End}_{\mathcal A_t}(V^{\otimes t}) = \mathrm{Span}\{ \rho(g)^{\otimes t}\}_{g\in G}. 
\]
\end{theorem}
This theorem is a standard consequence of complete reducibility along with the double centralizer theorem \cite[Section 4.1.5]{Goodman-Wallach-book}. 

\begin{example}
    \label{ex:classical_schur_weyl_duality}
    Take $G = U(N)$, and let $\rho$ be the fundamental representation, satisfying $\rho(U) = U$. Classical Schur-Weyl duality~\cite{schur1927rationalen, weyl1939classical} proves that the commutant algebra $\mathcal{A}_t$ of $\rho^{\otimes t}$ is the algebra spanned by all \textit{permutation matrices} on the $t$ registers:
    \begin{equation}
        \label{eq:permutation_representation}
        \mathcal{A}_t = \text{span}\{X_{\pi}: \pi \in S_t\},
        \quad 
        X_{\pi}\ket{i_1, i_2, \dots, i_t} = \ket{i_{\pi^{-1}(1)},  i_{\pi^{-1}(2)}, \dots, i_{\pi^{-1}(t)}}
    \end{equation}
    which is itself a representation of the symmetric group algebra $\C[S_t]$ known as the \textit{permutation representation}. 
\end{example}

\subsection{Diagram Algebras and the Schur Representation}
\label{sec:diagram-algebras}
Throughout the paper, we consider other instances of Schur-Weyl duality which generalize the classical $(U(N), \C[S_t])$-duality in \Cref{ex:classical_schur_weyl_duality}. In these settings, the relevant commutant algebras are obtained by extending the basis of $\C[S_t]$ (which consists of all permutations in $S_t$) to more general \textit{partition diagrams}:
\begin{definition}[Partition Diagram]
    A \textit{partition diagram} is a set partition of the $2n$ vertices $ \{1,\dots,n\}\ \cup\ \{1',\dots,n'\}$. Partition diagrams are typically visualized using a $n \times 2$ grid. We refer to the top row of the diagram as the \textit{output} row, and the bottom row of the diagram as the \textit{input} row. 
\end{definition}
\begin{figure}[H]
    \centering
    \begin{tikzpicture}[scale=1]
        \makenodes[iolabels]{9}{1cm}{0cm}{1cm}

        \connectset[color=darkred]{T1, B1, B5}
        \connectset[color=darkorange]{T2, T5}
        \connectset[color=darkgreen]{B2, T4}
        \connectset[color=darkblue]{T3, B4, B8}
        \connectset[color=darkbrown]{T6, B6}
        \connectset[color=darkpurple]{T7, B9, T9}
    \end{tikzpicture}
    \caption{Example of a partition diagram. \\
    Note that the connected components are important, but the specific graph connecting them is not.}
\label{fig:partition_alg_example_set}
\end{figure}

\begin{figure}[H]
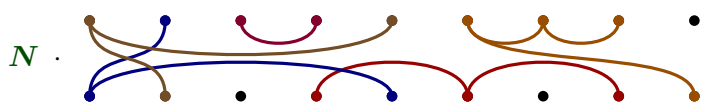

    \centering
    \begin{equation}
    \phantom{
        \textcolor{darkgreen}{\boldsymbol{N}} \; \cdot \; .
    }
    \begin{tikzinline}[scale=1,baseline]
        \makenodes{9}{2.5cm}{1.5cm}{1cm}

        \connectset[color=darkred]{B3, B6}
        \connectset[color=darkorange,height=.4]{T6, B7, T7, T8}
        \connectset[color=darkgreen]{B5, B8}
        \connectset[color=darkblue]{B1, T2}
        \connectset[color=darkbrown]{T1, T5, B4}
        \connectset[color=darkpurple,height=.4]{T3, T4}

        \makeverticalconnectors[dashed]{9}{1.5cm}{1cm}{1cm}

        \makenodes{9}{1cm}{0cm}{1cm}

        \connectset[color=darkgreen!80!darkbrown]{T2, T5}
        \connectset[color=darkblue!80!darkbrown]{B1, B5, T1}
        \connectset[color=darkbrown!80!darkbrown]{B2, T4}
        \connectset[color=darkorange!80!darkbrown]{T7, T9, B9}
        \connectset[color=darkred!80!darkbrown]{B4, B8, T3}
        \connectset[color=darkpurple!80!darkbrown]{T6, B6}
    \end{tikzinline}
    \phantom{
        \textcolor{darkgreen}{\boldsymbol{N}} \; \cdot \; .
    }
    \end{equation}
    \begin{equation*}
        =
    \end{equation*}
    \begin{equation}
    \phantom{
        \textcolor{darkgreen}{\boldsymbol{N}} \; \cdot \; .
    }
    \begin{tikzinline}[scale=1,baseline]
        \makenodes{9}{2cm}{1cm}{1cm}

        \connectset[color=darkred]{B3, B6}
        \connectset[color=darkorange,height=.4]{T6, B7, T7, T8}
        \connectset[color=darkgreen]{B5, B8}
        \connectset[color=darkblue]{B1, T2}
        \connectset[color=darkbrown]{T1, T5, B4}
        \connectset[color=darkpurple,height=.4]{T3, T4}

        \makenodes{9}{1cm}{0cm}{1cm}

        \connectset[color=darkgreen!80!darkbrown]{T2, T5}
        \connectset[color=darkblue!80!darkbrown]{B1, B5, T1}
        \connectset[color=darkbrown!80!darkbrown]{B2, T4}
        \connectset[color=darkorange!80!darkbrown]{T7, T9, B9}
        \connectset[color=darkred!80!darkbrown]{B4, B8, T3}
        \connectset[color=darkpurple!80!darkbrown]{T6, B6}
    \end{tikzinline}
    \phantom{
        \textcolor{darkgreen}{\boldsymbol{N}} \; \cdot \; .
    }
    \end{equation}
    \begin{equation*}
    =
    \end{equation*}
    \begin{equation}
    \textcolor{darkgreen}{\boldsymbol{N}} \; \cdot \; \phantom{.}
    \begin{tikzinline}[scale=1,baseline]
        \makenodes{9}{1cm}{0cm}{1cm}

        \connectset[color=darkred]{B4, B6, B8}
        \connectset[color=darkorange,height=.4]{T6, T7, T8, B9}
        \connectset[color=darkblue]{T2, B1, B5}
        \connectset[color=darkbrown]{T1, T5, B2}
        \connectset[color=darkpurple,height=.4]{T3, T4}
    \end{tikzinline}
    \phantom{
        \textcolor{darkgreen}{\boldsymbol{N}} \; \cdot \; .
    }
    \end{equation}
    \caption{Multiplication of partition diagrams. The green component is confined to the middle layer and drops out to become an extra factor of $N$.} 
\label{fig:partition_alg_example2}
\end{figure}

\begin{definition}[Multiplication of Diagrams]
    \label{def:mult_of_diagrams}
    Given two partition diagrams $D_1$ and $D_2$, the product diagram $D_1D_2$ is defined by identifying the vertices in the bottom row of $D_1$ with the vertices in the top row of $D_2$ (see~\cref{fig:partition_alg_example2}). The vertices in the identified middle rows are then removed—if $c$ connected components are removed this way, the product is multiplied by $N^c$. Here, $N$ is a free parameter, although we will typically identify $N$ with the dimension of the group in~\cref{tab:schur-weyl-partition}.  
\end{definition}

\noindent Note that when $D_1$ and $D_2$ are \textit{permutation diagrams}, where each component in the input row is connected to exactly one component in the output row, \Cref{def:mult_of_diagrams} coincides with the usual composition rule for permutations in $S_t$.  

\begin{definition}[Diagram Algebras~\cite{KS08}]
    \label{def:diagram_algebra}
     Fix an integer $t \ge 0$. Let $S$ be a subset of partition diagrams with $t$ columns. The \textit{diagram algebra generated by $S$} is the $\mathbb{C}$-algebra
    \begin{equation}
        \operatorname{Alg}_{\mathbb{C}}(S)
        =
        \operatorname{span}_{\mathbb{C}}
        \{D_1D_2\cdots D_k : k \ge 0,\; D_i \in S\},
    \end{equation}
    using the multiplication rule defined in~\cref{def:mult_of_diagrams}.
\end{definition}
For example, the symmetric group algebra $\C[S_t]$ is a diagram algebra with a basis given by the set of all \textit{permutation diagrams} with $t$ columns, i.e. diagrams whose connected components all contain exactly one vertex in the input and output rows. Other examples of diagram algebras include the \textit{Brauer algebra} $B_{t}(N)$ (\cite{Brauer1937AlgebrasSemisimpleContinuousGroups}), spanned by partition diagrams where every connected component has size 2, and the \textit{partition algebra} $P_t(N)$, (\cite{Martin1996StructurePartitionAlgebras}), which is spanned by the set of all partition diagrams. In \Cref{sec:unitary_haar_cipher}, we also consider a generalization of partition diagrams where each connected component is assigned a \textquote{color} $k \in [K]$. For more details, see \Cref{sec:unitary_haar_cipher}. 

Just as diagram algebras generalize the symmetric group algebra $\C[S_t]$, the \textit{Schur representation} generalizes the permutation representation (\cref{ex:classical_schur_weyl_duality}):

\needspace{2\baselineskip}
\begin{definition}[The Schur Representation]
    \label{def:schur_rep}
    Let $A$ be a diagram algebra. The \textit{Schur representation}%
    \footnote{
        This representation is sometimes also called the \textquote{permutation representation} in the literature, since it generalizes the permutation representation of the symmetric group. For more general diagram algebras, however, it goes beyond permuting registers. Furthermore, even for the symmetric group, there is a chance for confusion with the representation as \emph{permutation matrices} (in-place permutations of the standard basis vectors). Other sources call this the \textquote{natural representation}, but of course, which representation is natural depends on the algebra and its presentation. To disambiguate, we therefore follow~\cite{foxman2026efficient} in calling it the Schur representation.
    }
    $S(\cdot)$ is an $N^t$-dimensional representation, defined on a diagram $D$ as follows:
    \begin{equation}
        \label{eq:schur_rep}
    S(D)
    =
    \sum_{\vec{x},\vec{y}\in [N]^t}
    \left(
        \prod_{\substack{u,v\in \{1,\dots,n,1',\dots,n'\}\\ \text{$u$, $v$ in the same block in $D$}}}
        \delta_{z_u,z_v}
    \right)
    \ket{\vec{x}}\bra{\vec{y}},
\end{equation}
with 
\begin{equation}
    z_r
    =
    \begin{cases}
        x_r 
        & 
        r \in  \{1,\dots,n\}
        \\
        y_r
        & 
        r \in  \{1',\dots,n'\}
    \end{cases}
\end{equation}
In other words, $S(D)_{\vec{x},\vec{y}} = 1$ if the labeling of vertices given by $\vec{x},\vec{y}$ is constant on every connected component of $D$, and $0$ otherwise.  
\end{definition}

For example, 

\begin{figure}[H]
    \centering
    \[
    S\!\left(
    \vcenter{\hbox{
    \scalebox{0.85}{
    \begin{tikzpicture}[scale=1]
        \makenodes[iolabels]{6}{1cm}{0cm}{1cm}

        \connectset[color=darkred]{T1, B1, B5}
        \connectset[color=darkorange]{T2, T5}
        \connectset[color=darkgreen]{B2, T4}
        \connectset[color=darkblue]{T3, B4}
    \end{tikzpicture}
    }}}
    \right)
     = \sum_{\vec{x} \in [N]^7} \ketbra{x_1, x_2, x_3, x_4, x_2, x_6}{x_1, x_4, x_5, x_3, x_1, x_7}\] 
    \label{fig:schur_rep_example}
\end{figure}

\begin{table}[h]
\centering
\setlength{\tabcolsep}{6pt}
\renewcommand{\arraystretch}{1.2}
\scalebox{0.94}{
\begin{tabular}{
  >{\raggedright\arraybackslash}m{2.2cm}
  >{\raggedright\arraybackslash}m{1.0cm}
  >{\centering\arraybackslash}m{2.8cm}
  >{\centering\arraybackslash}m{3.7cm}
  >{\raggedright\arraybackslash}m{3.8cm}
  >{\raggedright\arraybackslash}m{1.4cm}
}
\hline
\multicolumn{2}{c}{
    \textbf{Group}
}
& 
\textbf{Tensor space} 
& 
\textbf{Representation} 
& 
\multicolumn{2}{c}{
    \textbf{Commutant algebra} 
}
\\
\hline

Unitary
& 
$U(N)$ 
& 
$V_N^{\otimes t}$ 
& 
$U^{\otimes t}$
& 
Symmetric Group Algebra
& 
$\mathbb{C}[S_t]$ 
\\
\noalign{\vspace{6pt}}

\phantom{\mbox{\tiny (Haar Cipher)}}
Unitary Product
\mbox{\tiny (Haar Cipher)}
& 
$U(N)^K$ 
& 
{$\Big(V_N^{\oplus K}\Big)^{\otimes t}$} 
& 
$\bigg(\sum\limits_{k \in [K]} \proj{k} \otimes U_k\bigg)^{\otimes t}$
& 
Colored Permutation Group Algebra
& 
{\small $\C[\Z_K \! \wr \! S_t]$} 
\\
\noalign{\vspace{6pt}}

Orthogonal
& 
$O(N)$ 
& 
$V_N^{\otimes t}$ 
&
\shortstack[c]{
\phantom{\tiny X}
\\
\phantom{\tiny X}
\\
$U_{g}^{\otimes t}$
\\
{\tiny $U_g$ is the standard embedding}
\\
{\tiny of $g \in O(N)$ as a unitary over $\mathbb{C}^t$}
}
& 
Brauer Algebra
& 
$B_t(N)$ 
\\
\noalign{\vspace{6pt}}

Colored \mbox{Permutations}
& 
$\Z_r \wr S_t$ 
& 
$V_{N}^{\otimes t}$ 
& 
\shortstack[c]{
\phantom{\tiny X}
\\
\phantom{\tiny X}
\\
\phantom{\tiny X}
\\
$(P_{\pi}F_{f})^{\otimes t}$
\\
{\tiny $P_{\pi}$ is the permutation matrix} 
\\
{\tiny of $\pi \in S_N$, and $F_{f}$ is}
\\
{\tiny diagonal with $r$'th roots of unity}
}
& 
Tanabe Algebra
& 
$T_t(N, r)$ 
\\
\noalign{\vspace{6pt}}

Symmetric
& 
$S_N$ 
& 
$(V_{N-1}\oplus V_1)^{\otimes t}$ 
& 
\shortstack[c]{
\phantom{\tiny X}
\\
\phantom{\tiny X}
\\
$P_{\sigma}^{\otimes t}$
\\
{\tiny $P_{\sigma}$ is the permutation matrix} 
\\
{\tiny of $\sigma \in S_N$}
}
& 
Partition Algebra
& 
$P_t(N)$ 
\\
\noalign{\vspace{6pt}}

Boolean Functions
& 
$\Z_2^N$ 
& 
{$\Big(V_1^{\oplus N}\Big)^{\otimes t}$} 
& 
$\bigg(\sum\limits_{x \in [N]} \scalebox{.8}{(-1)}^{f(x)} \proj{x}\bigg)^{\otimes t}$
& 
\multicolumn{2}{l}{Colored Even-Partition Algebra}
\\
\noalign{\vspace{6pt}}

\hline
\end{tabular}
}
\caption{Schur-Weyl dualities between different compact Lie groups and their commutant algebras. In all cases, the commutant algebra is the Schur representation (\cref{def:schur_rep}) of the given diagram algebra. See \Cref{sec:unitary_haar_cipher} for the definition of the Schur representation of $\C[\Z_K \wr S_t]$. In many of these cases, the given commutant algebra relation only holds when $t$ is sufficiently small relative to $N$. For larger $t$, the commutant instead becomes a quotient algebra of the one listed; see \Cref{sec:prelims_for_specific_algebras} for more details and \Cref{sec:diagonal-unitaries} for an example in which this becomes important.}
\label{tab:schur-weyl-partition}
\end{table}

\subsection{Trace forms and dual bases for semisimple algebras}

These preliminaries are adapted from \cite[Section 5]{halverson2005partition}. 

In this section, we work over a finite-dimensional semisimple algebra $A$ over the complex numbers; this means that $A$ is isomorphic to a direct product of finite-dimensional complex matrix algebras.

Let $A$ be any such algebra, and let $a\in A$. We define the \emph{regular trace} of $a$ by
\[
    \tr_A(a) := \tr(L_a),
\]
where $\tr$ denotes the trace of a linear operator and $L_a$ denotes left multiplication by $a$.  This gives a
symmetric bilinear pairing
\[
    \langle a,b\rangle_A := \tr_A(ab).
\]
The pairing is nondegenerate because $A$ is semisimple.

\begin{definition}[Dual basis]
    \label{def:dual_basis}
    Given a basis $\mathcal B(A)$ of $A$, we write $a^*$ for the dual basis element corresponding to $a\in B(A)$; that is,
\[
    \tr_A(a b^*)=\delta_{a,b}
    \qquad
    \text{for all }a,b\in \mathcal B(A).
\]
\end{definition}

Throughout the paper, dual bases are taken with respect to the regular trace $\tr_A$. We will later use this dual basis to interpret the commutant EPR projector $\Omega_{A_t}$. 

\begin{example}[Group algebras]\label{ex:group-algebra-dual}
Let $A=\mathbb C[G]$ be the group algebra of a finite group $G$, with basis
$\{[g]:g\in G\}$.  Left multiplication by $[g]$ permutes this basis.  Hence
\begin{equation}
    \label{eq:reg_trace_for_group}
     \tr_{\mathbb C[G]}([g])
    =
    \begin{cases}
        |G|, & g=e,\\
        0, & g\ne e.
    \end{cases}
\end{equation}
Therefore, for $a=\sum_{g\in G} a_g[g]$,
\[
    \tr_{\mathbb C[G]}(a)=|G|\,a_e.
\]
The dual basis is then given by
\[
    [g]^*=\frac{1}{|G|}[g^{-1}].
\]
\end{example}

\subsection{The (dual) Clebsch--Gordan transform and Clebsch--Gordan coefficients}
\label{sec:CG-defn}

Let $(\rho_\lambda,V_{G}^{\lambda})$ and $(\rho_\mu,V_{G}^{\mu})$ be irreducible unitary representations of $G$. By complete reducibility (\cref{thm:complete-reducibility}), we have a decomposition
\[
  V_{G}^{\lambda} \otimes V_{G}^{\mu} \simeq  \bigoplus_{\lambda^+ \in \widehat{G}}\!\bigl(M_{\lambda,\mu}^{\lambda^+}\otimes V_{G}^{\lambda^+}\bigr),
\]
where $G$ acts as $\Id_{M_{\lambda, \mu}^{\lambda^+}} \otimes \rho_{\lambda^+}$ on the right hand side. In this setting, we let $m_{\lambda,\mu}^{\lambda^+} = \dim(M_{\lambda, \mu}^{\lambda^+})$ and define a standard basis $\{\ket{r}\}$ for $M_{\lambda,\mu}^{\lambda^+}$.

\begin{definition}[Clebsch--Gordan transform]\label{def:clebsch-gordan}

A \emph{Clebsch--Gordan transform} is any family of unitary, $G$-respecting isomorphisms
\[
  C_{\lambda,\mu}:\; V_{G}^{\lambda} \otimes V_{G}^{\mu} \xrightarrow{\ \simeq\ } 
  \bigoplus_{\lambda^+}\bigl(M_{\lambda,\mu}^{\lambda^+}\otimes V_{G}^{\lambda^+}\bigr),
\]
meaning that for all $g\in G$, 
\[ C_{\lambda, \mu}
    \Big(  
    \rho_{\lambda}(g)
    \otimes 
    \rho_{\mu}(g)
    \Big)
    C^\dagger_{\lambda, \mu} = \bigoplus_{\lambda^+ \in \widehat{G}}\bigl(\Id_{M_{\lambda, \mu}^{\lambda^+}} \otimes \rho_{\lambda^+}(g)\bigr)
\]
\end{definition}
The Clebsch--Gordan transform is unique up to independent unitary changes of bases on each multiplicity space $M_{\lambda,\mu}^{\lambda^+}$. We will use $C = \sum_{\lambda, \mu} \ketbra{\lambda} \otimes \ketbra{\mu} \otimes C_{\lambda, \mu}$ 
to denote the Clebsch--Gordan transform defined on all $V_{G}^{\lambda}\otimes V_{G}^{\mu}$ simultaneously. 

\paragraph{Clebsch--Gordan Coefficients} Next, we give notation for the matrix elements of $C$ in fixed bases for all of the underlying spaces. Under our conventions from earlier, we use  
\[
  \{\ket*{X}\}_X,\quad
  \{\ket*{x}\}_{x},\quad
  \{\ket*{r,X^+}\}_{r,X^+}
\]
as bases for $V_{G}^{\lambda}$, $V_{G}^{\mu}$, and $M_{\lambda,\mu}^{\lambda^+}\!\otimes V_{G}^{\lambda^+}$. We can then write $C$ as an explicit matrix:

\[
  C = \sum_{\substack{\lambda, \mu, \lambda^+\\ X, x \\ r, X^+}}
  \cgcoeffs \cdot \cgketbras,
\]
where we call the matrix elements $\cgcoeffs$ the \textbf{Clebsch--Gordan coefficients} for $G$. 

In the event that $m_{\lambda, \mu}^{\lambda^+} = 1$ for all $\lambda^+$, we may drop the $\ket{r}$ from our notation. 

\paragraph{Dual Clebsch--Gordan Transform and Coefficients}

Finally, we define a related \emph{dual} Clebsch--Gordan transform and state an identity involving Clebsch--Gordan and dual Clebsch--Gordan coefficients that we make use of to analyze our compressed representations. Departing slightly from prior work \cite{nguyen2023mixed,grinko2023gelfand} but more in line with \cite{Har05}, we first define the dual Clebsch-Gordan transform using dual vector spaces. Then, we fix an irreducible representation labeling convention so that for every $\lambda$, there is a ``dual label'' $\lambda^*$ such that $V_{G}^{\lambda^*} \simeq \left(V_{G}^{\lambda}\right)^*$ and the action of $G$ on $\left(V_{G}^{\lambda}\right)^*$ is the dual representation $(\rho_{\lambda^*}(g)\cdot v^*)(w) =  v^*( \rho_\lambda(g)^\dagger \cdot w) = (\rho_\lambda(g) v)^*(w)$. Thus, in an explicit dual basis, $\rho_{\lambda^*}(g)$ acts as $\rho_\lambda(g)^*$. 

Under these conventions, we then define ``dual Clebsch-Gordan coefficients'' on the underlying irrep spaces (with no explicit dual spaces).

\begin{definition}[Dual Clebsch--Gordan transform]\label{def:dualclebsch-gordan}

A \emph{Dual Clebsch--Gordan transform} is any family of unitary, $G$-respecting isomorphisms
\[
  dC_{\lambda,\mu}:\; V_{G}^{\lambda} \otimes (V_{G}^{\mu})^* \xrightarrow{\ \simeq\ } 
  \bigoplus_{\lambda^-}\bigl(M_{\lambda,\mu,*}^{\lambda^-}\otimes V_{G}^{\lambda^-}\bigr),
\]
meaning that for all $g\in G$, 
\[ dC_{\lambda, \mu}
    \Big(  
    \rho_{\lambda}(g)
    \otimes 
    \rho^*_{\mu}(g)
    \Big)
    dC^\dagger_{\lambda, \mu} = \bigoplus_{\lambda^- \in \widehat{G}}\bigl(\Id_{M_{\lambda, \mu,*}^{\lambda^-}} \otimes \rho_{\lambda^-}(g)\bigr)
\]
\end{definition}

To be compatible with prior work, given a collection of isomorphisms $(V_G^\mu)^*\simeq V_G^{\mu^*}$ for irrep label pairs $(\mu, \mu^*)$ (suppressed for simplicity), we abuse notation and identify this dual Clebsch--Gordan transform with a map

\[ dC_{\lambda, \mu}: V_{G}^{\lambda}\otimes V_{G}^{\mu^*} \rightarrow \bigoplus_{\lambda^- \in \widehat{G}} (M_{\lambda, \mu^*}^{\lambda^-} \otimes V_{G}^{\lambda^-}).
\]
In other words, we have

\[
  dC = \sum_{\substack{\lambda, \mu, \lambda^-\\ Y, y \\ r, Y^-}}
  \dcgcoeffs \cdot \dcgketbras,
\]
with
\[ \dcgcoeffs = \cgcoeff{\lambda}{\mu^*}{\lambda^-}{Y}{y}{r}{Y^-}
\]
and
\[ dC_{\lambda, \mu}
    \Big(  
    \rho_{\lambda}(g)
    \otimes 
    \rho_{\mu^*}(g)
    \Big)
    dC^\dagger_{\lambda, \mu} = \bigoplus_{\lambda^- \in \widehat{G}}\bigl( \Id_{M_{\lambda, \mu^*}^{\lambda^-}} \otimes \rho_{\lambda^-}(g)\bigr).
\]
Up to an identification of bases of $\left(V_{G}^{\mu}\right)^*$ and $V_{G}^{\mu^*}$, this leaves us with the same degrees of freedom in choosing a dual Clebsch--Gordan transform as we have for choosing a Clebsch--Gordan transform.

\paragraph{An additional property} For our compressed representations, we work with choices of Clebsch--Gordan and dual Clebsch--Gordan whose coefficients satisfy an additional identity.

\begin{lemma}
\label{lemma:vilenkin-klimyk}
    There exist choices of $C$, $dC$ satisfying the identities 

    \begin{align}
        \bra*{\lambda^+, \mu, X^+, x
        }
        dC^\dagger 
        \ket*{\lambda^+, \mu, \lambda, r, X 
        } 
        = 
        \sqrt{
            \frac {
                \dim
                \left(
                    V_{G}^{\lambda}
                \right)
            }
            {
                \dim
                \left(
                    V_{G}^{\lambda^+}
                \right)
            }
        }
        \cgcoeffs^*
    \end{align}

\end{lemma}
\noindent In fact, \cite{VK92} give the identity

\begin{align*}
    \bra*{\lambda^+, \mu, X^+, x
    }
    dC^\dagger 
    \ket*{\lambda^+, \mu, \lambda, r, X 
    } 
    = 
    \sqrt{
        \frac {
            \dim
            \left(
                V_{G}^{\lambda}
            \right)
        }
        {
            \dim
            \left(
                V_{G}^{\lambda^+}
            \right)
        }
    }
    \cgcoeffs
\end{align*}
for choices of $C, dC$ with real coefficients. We state this relaxed condition in case this allows for additional efficient instantiations. 

\paragraph{Generalized Clebsch--Gordan}
We also use generalized Clebsch--Gordan and dual Clebsch--Gordan transforms in which the second tensor factor is an arbitrary finite-dimensional representation $\rho$, rather than an irrep. These decompose $V_G^\lambda \otimes V_\rho$ and $V_G^\lambda \otimes V_\rho^*$, respectively. The compatibility identity of \Cref{lemma:vilenkin-klimyk} has the same form with $\mu$ replaced by $\rho$. In our algorithmic statements, the transforms act in the specified query basis and coherently over the irrep label $\lambda$. For a reducible $\rho$, any change of basis needed is assumed to be included in the transforms.

\subsection{The Generalized Schur Transform}
We will also make use of a generalization of the Schur transform, defined below.

\begin{definition}[Schur transform]\label{def:schur-transform}
Let $(\rho, V)$ be a finite-dimensional (not necessarily irreducible) unitary representation of $G$. 
A \emph{Schur transform} is any family of unitary, $G$-respecting isomorphisms
\[
    \schur_{\rho,t}:
    \; 
    V^{\otimes t} 
    \xrightarrow{\ \simeq\ }
    \bigoplus_{\lambda} 
    \bigl(
        V_{G}^{\lambda} 
        \otimes 
        V_{\calA_t}^{\lambda} 
    \bigr),
\]
meaning that for all $g\in G$, 
\[ 
    \schur_{\rho,t}
    \left(  
        \rho(g)^{\otimes t}
    \right)
    \schur_{\rho,t}^\dagger
    = 
    \bigoplus_{\lambda \in \widehat{G}}
    \bigl(
        \rho_{\lambda}(g)
        \otimes 
        \Id_{V_{\calA_t}^{\lambda}}
    \bigr) 
\]
\end{definition}
We will use $\schur_{\rho} = \sum_{t \ge 0} \ketbra{t} \otimes \schur_{\rho,t}$ to denote the Schur transform defined on all $t$ simultaneously, extending $\schur_{\rho,t}$ for smaller $t$ to act as identity for all but the first $t$ registers. By convention, we take $\schur_{\rho,0}$ to be the isometry that takes no input and produces the label for the trivial 1-dimensional irrep of $G$, with no multiplicity.

\paragraph{Schur Transform for a Locally Block-Diagonal Representation}
We will often consider cases in which $\rho$ is already block-diagonal, with $V= \bigoplus_{\mu \in \mathrm{Irr}(\rho)} V_{G}^{\mu} \otimes M^{\mu}$ and $\rho(g) = \sum_{\mu \in \mathrm{Irr}(\rho)} \proj{\mu} \otimes \rho_{\mu}(g)\otimes \Id_{M^{\mu}(g)}$.
This Schur transform is equivalent to the one described above up to the local transformation that block-diagonalizes each copy of $V$ (which will often be efficient, but not always).

The local multiplicity registers $\{M^{\mu}_{i}\}_{i \in [t]}$ will end up being collected as part of the total multiplicity $V_{\calA_t}^{\lambda}$. We will sometimes ignore the local multiplicity registers for notational simplicity, since at least in principle, a Schur transform that considers them can be built from one that does not by tacking them onto the $V_{\calA_t}^{\lambda}$ register at the end (though this may not yield the desired basis for $V_{\calA_t}^{\lambda}$).

The representation $\rho$ determines which subset of irreps will appear (or equivalently, the set of available irreps will determine the representation). 
Furthermore, in this basis, we can view the Schur transform as not explicitly depending on the representation $\rho$, since we can view it as accepting any irrep in $\widehat{G}$, not just those that appear in $\rho$ (though of course, if $\widehat{G}$ has infinite cardinality, any actual implementation must choose some subset of $\widehat{G}$ to allow).
We will then just write $\schur_{G, t}$ and $\schur_{G} = \sum_{t} \ketbra{t} \otimes \schur_{G, t}$, or just $\schur$ when the group is clear from context.

The Schur transform can then be written as a map from sequences of the form
\begin{align}
    (\vec{\mu}, \vec{x})
    :=
    (
        (\mu_1, x_1),
        (\mu_2, x_2),
        \dots,
        (\mu_t, x_t)
    )
\end{align}
to superpositions of states of the form $\ket{\lambda, T_{\calA_t}, T_{G}}$, where $\lambda \in \widehat{G}$, $\ket{T_{\calA_t}} \in V_{\calA_t}^{\lambda}$, and $\ket{T_{G}} \in V_{G}^{\lambda}$, with coefficients 
$
    \bra{
        \lambda, 
        T_{\calA_t}, 
        T_{G}
    }
    \schur
    \ket{
        \vec{\mu}, 
        \vec{x}
    }
$.

\subsection{Young Tableaux and Bratteli Paths}

\label{sec:prelims_for_specific_algebras}

The irreducible representations of the groups we study (and their corresponding commutant algebras) are indexed by integer partitions, typically visualized with \textit{Young diagrams}. For example, the following Young diagram corresponds to the partition $(5, 2, 2, 1)$ of $N = 10$: 
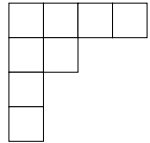
\begin{center}
    \Ydiagram{4, 2, 1, 1}
\end{center}
We use $|\lambda|$ to denote the total number of boxes in $\lambda$, $\lambda_i$ to denote the number of boxes in the $i$th row, and $\lambda'_i$ to denote the number of boxes in the $i$th column. 

\paragraph{The Symmetric Group}
\begin{definition}[Standard Young Tableau]
    Let $\lambda$ be a Young diagram. A \textit{standard Young tableau} (SYT) corresponding to $\lambda$ is a filling of the Young diagram with the numbers $\{1, 2, \dots |\lambda|\}$, such that all rows of $\lambda$ are filled in increasing order from left to right, and all columns of $\lambda$ are filled in increasing order from top to bottom. 
\end{definition}
For example, the following is one valid standard Young tableau corresponding to the Young diagram above: 
\begin{center}
    \ytableaushort{
     1356,
     27,
     4,
     8
    }
\end{center}
An equivalent characterization of an SYT $T$ is as a chain of Young diagrams
\begin{equation}
    \emptyset, A_1, A_2, \dots, A_{|\lambda|} = \lambda
\end{equation}
where $A_i$ is the shape formed by the boxes of $T$ whose labels are at most $i$. For the SYT above, the corresponding chain is given by
\begin{equation}
    \emptyset,\;\; \Ydiagram{1},\;\; \Ydiagram{1, 1},\;\; \Ydiagram{2, 1},\;\; \Ydiagram{2, 1, 1},\;\; \Ydiagram{3, 1, 1},\;\; \Ydiagram{4, 1, 1},\;\; \Ydiagram{4, 2, 1},\;\; \Ydiagram{4, 2, 1, 1}
\end{equation}

\begin{lemma}[Hook Length Formula]
    \label{def:hook_length_formula}
    Given a Young diagram $\lambda$, the \textit{hook length formula} $f^{\lambda}$ is 
    \begin{equation}
        f^\lambda = \frac{|\lambda|!}{\prod_{b \in \lambda} h(b)}
    \end{equation}
    where $h(b)$ is the \textit{hook length} of the box $b$. The hook length is equal to the number of boxes in the same row strictly to the right of $b$, plus the number of boxes in the same column strictly below $b$, plus one.
\end{lemma}

\begin{lemma}[\cite{Sag01}]
    \label{lem:dim_symmetric_group}
    For any $\lambda \in \wh{\C[S_t]}$, $f^\lambda$ counts the number of standard Young tableaux corresponding to $\lambda$. Equivalently, 
    $
        \dim
        \left(
            V_{S_{|\lambda|}}^{\lambda}
        \right)
        = 
        f^\lambda
    $.
\end{lemma}

\paragraph{The Unitary Group}
\begin{definition}[Semi-Standard Young Tableau]
    \label{def:ssyt}
    Let $\lambda = (\lambda_1 \ge \lambda_2 \ge \cdots \ge \lambda_N)$ be a weakly decreasing sequence of integers, not necessarily nonnegative. Write $\lambda_{(+)} = (\lambda_1, \dots, \lambda_p)$ for the positive entries of $\lambda$ (so $\lambda_p > 0 \ge \lambda_{p+1}$),
    and $\lambda_{(-)} = (\lambda_{N+1-q}, \dots, \lambda_{N})$ for the negative entries of $\lambda$ (so $\lambda_{N-q} \ge 0 > \lambda_{N+1-q}$),
    and let $\mu = (\mu_1 \ge \cdots \ge \mu_q)$, where $\mu_j := -\lambda_{N+1-j}$, denote the partition formed from the negative entries of $\lambda$, reversed and negated.

    A \emph{semi-standard Young tableau} (SSYT)%
    \footnote{
        Semi-standard Young tableaux are often defined only with positive rows and boxes, while tableaux with negative rows are called \emph{rational} semistandard tableaux~\cite{king1970generalized,STEMBRIDGE198779,KWON2008713}.
        We generally allow negative rows, and specify explicitly when negative rows are not present. We will thus often drop the word \textquote{rational} and refer to both as simply semistandard Young tableau.
    }
    of shape $\lambda$ with entries in $[N] := \{1, \dots, N\}$ consists of:
    \begin{itemize}
        \item a filling of the Young diagram of $\lambda_{(+)}$ (the \emph{positive tableau}) with entries in $[N]$, weakly increasing along each row (left to right) and strictly increasing down each column (top to bottom); and
        \item a filling of the Young diagram of $\mu$ (the \emph{negative tableau}) with entries in $[N]$, weakly decreasing along each row (left to right) and strictly decreasing down each column (top to bottom),
    \end{itemize}
    subject to the following compatibility condition. For $x \in [N]$, let $r^+(x)$ denote the largest row index $i \in \{1,\dots,p\}$ (rows of $\lambda_{(+)}$ numbered from the top) containing the entry $x$ in the positive tableau, or $r^+(x) := 0$ if $x$ does not appear there. Define $r^-(x)$ analogously, as the largest row index $j \in \{1, \dots, q\}$ (rows of $\mu$ numbered from the top, or equivalently, rows of $\lambda_{(-)}$) containing $x$ in the negative tableau, or $0$ if $x$ does not appear there. We require
    \begin{align}
        r^+(x) + r^-(x) \;\le\; x \qquad \text{for all } x \in [N].
        \label{eq:rational-ssyt-compatibility1}
    \end{align}
\end{definition}

\begin{figure}[H]
    \centering
    \begin{align*}
        \Ytableau{
            2566,
            37,
            5,
            ,
        }{
             65,
            331
        }
    \end{align*}
    \caption{Example of a semistandard tableau with shape $(4, 2, 1, 0, 0, -2, -3)$, where $N = 7$.}
    \label{fig:rational-semistandard-young-tableau}
\end{figure}

Let the \emph{complement} of a negative column be a positive column containing all the entries of $[N]$ besides those in the original column (arranged in order). Then the condition in \Cref{eq:rational-ssyt-compatibility1} is equivalent to saying that shifting the tableau by one column (by converting the first negative column to its complement) produces a valid ordering in the resulting positive tableau.

\begin{figure}[H]
    \centering
    \begin{tikzpicture}[baseline=(AX.base)]
        \node (CY) at (-6,0) [label=above:{\huge }] {
        \Ytableau{
            2566,
            37,
            5,
            ,
        }{
             65,
            331
        }
        };
        \node (AX) at (0,0) [label=above:{\huge }] {
        \Ytableau{
            22566,
            337,
            45,
            6,
            7
        }{
             6,
            33
        }
        };
        \draw[->] (CY.east) to node[midway,above] {{ $\mathsf{Shift}$}} (AX.west);
    \end{tikzpicture}
    \caption{Shifting the semistandard Young tableau by a column maintains a valid order in the positive tableau.}
    \label{fig:shifting-rational-ssyt}
\end{figure}
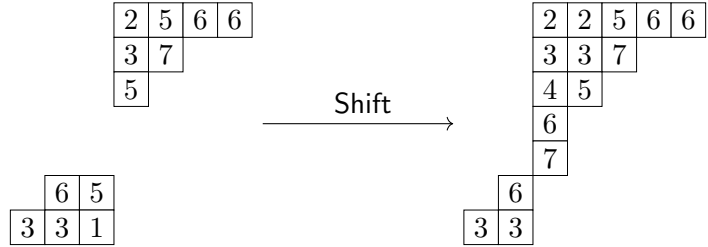

The Weyl Dimension Formula~\cite{weyl1939classical} counts the number of semi-standard Young tableaux of a certain shape $\lambda$ and thus gives the dimension of the corresponding irreducible representation.

\begin{lemma} \label{lemma:weyl-dim-form}
    Let $\lambda$ be a Young diagram. The number of semi-standard Young tableaux of shape $\lambda$ is
    \begin{align*}
        \dim
        \left(
            V_{U(N)}^{\lambda}
        \right)
        =
        \prod_{i = 1}^N
        \prod_{j = i+1}^N
        \frac
        {
            (\lambda_i - i)
            -
            (\lambda_j - j)
        }
        {
            j - i
        }
    \end{align*}
\end{lemma}

\paragraph{The Partition Algebra}
The partition algebra $P_t(N)$ is semisimple whenever $N > 2t - 2$ \cite{halverson2020set}. In this regime, the irreps of $P_t(N)$ are indexed by $N$-box Young diagrams such that all but at most $t$ boxes are in the first row [\cite{halverson2005partition}, Theorem 2.24b], i.e.
\begin{equation}
    \widehat{P_t(N)} = \{\lambda: |\lambda| = N, \;\lambda_0 \ge N - t\}
\end{equation}
We will adopt the convention that when $\lambda$ refers to an irrep of the partition algebra, the first row is ignored. With this convention, the irreps of the partition algebra are instead indexed by all Young diagrams with at most $t$ boxes, i.e. 
\begin{equation}
    \widehat{P_t(N)} = \{\lambda: |\lambda| \le t\}
\end{equation}
With this convention, the dimension of an irrep $\lambda \in \widehat{P_t(N)}$ is \cite{halverson2020set} given as follows:
\begin{lemma}
    \label{lem:dim_formula_partition_algebra}
    \begin{equation}
        \dim\left(V^{\lambda}_{P_t(N)}\right) = f^{\lambda} \sum_{i = |\lambda|}^{t} \stirling{t}{i}\binom{i}{|\lambda|}
    \end{equation}
    where $f^{\lambda}$ is the hook length formula in~\cref{def:hook_length_formula}, and $\stirling{t}{i}$ is a Stirling number of the second kind. Note that if $|\lambda| = t$, then $\dim\left(V^{\lambda}_{P_t(N)}\right) = f^{\lambda}$.
\end{lemma}
\noindent We will sometimes say an irrep $\lambda$ is a \textit{full box} irrep of $P_t(N)$ if $|\lambda| =t $. 

\subsection{Clebsch-Gordan Coefficients in the Gelfand-Tsetlin Basis of Unitaries}

\begin{lemma}[\cite{VK92}]
    \label{lemma:cg-in-gt-basis}
    Let $G$ be the unitary group $U(N)$. Let $X$ be a semi-standard Young tableau of shape $\lambda$ which may have positive and/or negative rows, and let $\boxed{x}$ be a Young tableau of shape $\mu = \boxed{\phantom{x}}$, the standard irrep of $U(N)$.
    We can write the Young tableau as a Gelfand-Tsetlin pattern, that is, a sequence of $N$ shapes $(X_{}^{\le \ell})_{\ell \in [N]}$, where each $X_{}^{\le \ell}$ is the shape of the $X$ tableau when it is restricted to entries at most $\ell$ (and removing $N - \ell$ empty rows such that it has $\ell$ rows). For each shape $X_{}^{\le \ell}$, let $X_{k}^{\le \ell}$ be the number of boxes on the $k$\ith row (which may be positive or negative).

    In the Gelfand-Tsetlin basis, the Clebsch-Gordan coefficient $\cgcoeff{\lambda}{\mu}{\lambda^+}{X}{x}{r}{X^+}$ for tensoring a single box Young tableau $\boxed{x}$ has the following properties:
    \begin{enumerate}
        \item The transition from shape $\lambda$ to $\lambda^+$ has no multiplicity, so $r$ is redundant. 
        \item It has non-zero value only when $\lambda^+$ differs from $\lambda$ by adding a single box (or removing a negative box) 
        \item It has non-zero value only when $X^+$ has the form 
            \begin{align*}
                X^+ 
                = 
                (
                    X_{}^{\le 1}, 
                    X_{}^{\le 2}, 
                    \dots, 
                    X_{}^{\le x - 1}, 
                    X_{}^{\le x}, 
                    X_{}^{\le x + 1},
                    \dots, 
                    X_{}^{\le N}
                )
                +
                (
                    0, 
                    0, 
                    \dots, 
                    0, 
                    \Box_{\Delta_{x}}, 
                    \Box_{\Delta_{x + 1}},
                    \dots, 
                    \Box_{\Delta_{N}}
                )
                \,,
            \end{align*}
            where 
            ``$+ \Box_{\Delta_{z}}$''
            means ``add a box to the $\Delta_{z}$\ith row'' (or remove a negative box), for some $(\Delta_{x} \in [x], \dots, \Delta_{N} \in [N])$.
            That is, every shape starting with the $x$\ith one will have a net increase of one box.
        \item In this case, the Clebsch-Gordan coefficients can be written as a product of reduced Wigner coefficients, or scalar factors. Namely,
            \begin{align*}
                &
                \hspace*{-2cm}
                \cgcoeff{\lambda}{\mu}{\lambda^+}{X}{x}{r}{X^+}
                = 
                \left\lvert
                    \frac{
                        \displaystyle\prod_{k = 1}^{x - 1}
                        \left(
                            X_{k}^{\le x - 1}
                            -
                            X_{\Delta_{x}}^{\le x}
                            +
                            \Delta_{x}
                            -
                            k
                            -
                            1
                        \right)
                    }{
                        \displaystyle\prod_{
                            \substack{
                                k = 1 
                                \\
                                k \ne \Delta_{x}
                            }
                        }^{x}
                        \left(
                            X_{k}^{\le x}
                            -
                            X_{\Delta_{x}}^{\le x}
                            +
                            \Delta_{x}
                            -
                            k
                        \right)
                    }
                \right\rvert
                ^{\frac{1}{2}}
                \\
                &
                \hspace*{-1cm}
                \phantom{=}
                \cdot
                \;
                \prod_{z > x}
                \left\lvert
                    \frac{
                        \displaystyle\prod_{
                            \substack{
                                k = 1 
                                \\
                                k \ne \Delta_{z - 1}
                            }
                        }^{z - 1}
                        \left(
                            X_{k}^{\le z - 1}
                            -
                            X_{\Delta_{z}}^{\le z}
                            +
                            \Delta_{z}
                            -
                            k
                            -
                            1
                        \right)
                    }{
                        \displaystyle\prod_{
                            \substack{
                                k = 1 
                                \\
                                k \ne \Delta_{z}
                            }
                        }^{z}
                        \left(
                            X_{k}^{\le z}
                            -
                            X_{\Delta_{z}}^{\le z}
                            +
                            \Delta_{z}
                            -
                            k
                        \right)
                    }
                \right\rvert
                ^{\frac{1}{2}}
                \left\lvert
                    \frac{
                        \displaystyle\prod_{
                            \substack{
                                k = 1 
                                \\
                                k \ne \Delta_{z}
                            }
                        }^{z}
                        \left(
                            X_{k}^{\le z}
                            -
                            X_{\Delta_{z - 1}}^{\le z - 1}
                            +
                            \Delta_{z - 1}
                            -
                            k
                        \right)
                    }{
                        \displaystyle\prod_{
                            \substack{
                                k = 1 
                                \\
                                k \ne \Delta_{z - 1}
                            }
                        }^{z - 1}
                        \left(
                            X_{k}^{\le z - 1}
                            -
                            X_{\Delta_{z - 1}}^{\le z - 1}
                            +
                            \Delta_{z - 1}
                            -
                            k
                            -
                            1
                        \right)
                    }
                \right\rvert
                ^{\frac{1}{2}}
                (-1)^{\Delta_{z} > \Delta_{z - 1}}
            \end{align*}
    \end{enumerate}

    Similarly, in the Gelfand-Tsetlin basis, the dual Clebsch-Gordan coefficients $$\dcgcoeff{\lambda}{\mu}{\lambda^-}{X}{x}{r}{X^-} = \cgcoeff{\lambda}{\bar{\mu}}{\lambda^-}{X}{x}{r}{X^-}$$ (where $\bar{\mu}$ is the dual of the standard representation) have the properties:
    \begin{enumerate}
        \item The transition from shape $\lambda$ to $\lambda^-$ has no multiplicity. 
        \item It has non-zero value only when $\lambda^-$ differs from $\lambda$ by removing a single box (or adding a negative box) 
        \item It has non-zero value only when $X^-$ has the form 
            \begin{align*}
                X^- 
                = 
                (
                    X_{}^{\le 1}, 
                    X_{}^{\le 2}, 
                    \dots, 
                    X_{}^{\le x - 1}, 
                    X_{}^{\le x}, 
                    X_{}^{\le x + 1},
                    \dots, 
                    X_{}^{\le N}
                )
                -
                (
                    0, 
                    0, 
                    \dots, 
                    0, 
                    \Box_{\Delta_{x}}, 
                    \Box_{\Delta_{x + 1}},
                    \dots, 
                    \Box_{\Delta_{N}}
                )
                \,,
            \end{align*}
            where 
            ``$- \Box_{\Delta_{z}}$''
            means ``remove a box from the $\Delta_{z}$\ith row'' (or add a negative box), for some $(\Delta_{x} \in [x], \dots, \Delta_{N} \in [N])$.
            That is, every shape starting with the $x$\ith one will have a net decrease of one box.
        \item In this case, as before, the Clebsch-Gordan coefficients can be written as a product of reduced Wigner coefficients, or scalar factors. Namely,
            \begin{align*}
                &
                \hspace*{-2cm}
                \dcgcoeff{\lambda}{\mu}{\lambda^-}{X}{x}{r}{X^-}
                = 
                \left\lvert
                    \frac{
                        \displaystyle\prod_{k = 1}^{x - 1}
                        \left(
                            X_{k}^{\le x - 1}
                            -
                            X_{\Delta_{x}}^{\le x}
                            +
                            \Delta_{x}
                            -
                            k
                        \right)
                    }{
                        \displaystyle\prod_{
                            \substack{
                                k = 1 
                                \\
                                k \ne \Delta_{x}
                            }
                        }^{x}
                        \left(
                            X_{k}^{\le x}
                            -
                            X_{\Delta_{x}}^{\le x}
                            +
                            \Delta_{x}
                            -
                            k
                        \right)
                    }
                \right\rvert
                ^{\frac{1}{2}}
                \\
                &
                \hspace*{-1cm}
                \phantom{=}
                \cdot
                \;
                \prod_{z > x}
                \left\lvert
                    \frac{
                        \displaystyle\prod_{
                            \substack{
                                k = 1 
                                \\
                                k \ne \Delta_{z - 1}
                            }
                        }^{z - 1}
                        \left(
                            X_{k}^{\le z - 1}
                            -
                            X_{\Delta_{z}}^{\le z}
                            +
                            \Delta_{z}
                            -
                            k
                        \right)
                    }{
                        \displaystyle\prod_{
                            \substack{
                                k = 1 
                                \\
                                k \ne \Delta_{z}
                            }
                        }^{z}
                        \left(
                            X_{k}^{\le z}
                            -
                            X_{\Delta_{z}}^{\le z}
                            +
                            \Delta_{z}
                            -
                            k
                        \right)
                    }
                \right\rvert
                ^{\frac{1}{2}}
                \left\lvert
                    \frac{
                        \displaystyle\prod_{
                            \substack{
                                k = 1 
                                \\
                                k \ne \Delta_{z}
                            }
                        }^{z}
                        \left(
                            X_{k}^{\le z}
                            -
                            X_{\Delta_{z - 1}}^{\le z - 1}
                            +
                            \Delta_{z - 1}
                            -
                            k
                            +
                            1
                        \right)
                    }{
                        \displaystyle\prod_{
                            \substack{
                                k = 1 
                                \\
                                k \ne \Delta_{z - 1}
                            }
                        }^{z - 1}
                        \left(
                            X_{k}^{\le z - 1}
                            -
                            X_{\Delta_{z - 1}}^{\le z - 1}
                            +
                            \Delta_{z - 1}
                            -
                            k
                            +
                            1
                        \right)
                    }
                \right\rvert
                ^{\frac{1}{2}}
                (-1)^{\Delta_{z} > \Delta_{z - 1}}
            \end{align*}
    \end{enumerate}
\end{lemma}

\subsection{The Fourier Basis of Semisimple Algebras}

\begin{definition}
    [The Fourier Basis]
    Let $\calA $ be a finite-dimensional semisimple algebra, i.e.
    \begin{equation}
        \label{eq:fourier_basis}
        \calA \overset{f}{\simeq} \bigoplus_{\lambda \in \wh{\calA}} \mathrm{End}(V^{\lambda})
    \end{equation}
    A \textit{Fourier basis} $ \left\{
        E_{i,j}^{\lambda}
        \right\}_{\lambda, i, j}$ for $\calA$ is the preimage of the natural matrix unit basis $\{\ketbra{i}{j}_{\lambda}\}_{\lambda, i, j}$ under the isomorphism $f$. 
\end{definition}
In general, the isomorphism in~\cref{eq:fourier_basis} is not unique, and so the Fourier basis is only defined relative to $f$. Moreover, the matrix unit basis $\{\ketbra{i}{j}_{\lambda}\}_{\lambda, i, j}$ implicitly assumes a choice of basis for each irrep space $V^\lambda$.\footnote{In algorithmic representation theory, a standard choice of basis for each irrep space is a \textit{subalgebra adapted basis}~\cite{DiaconisRockmore1990, moore2003genericquantumfouriertransforms, BCH05, maslen2016efficientcomputationfouriertransforms, foxman2026efficient}, which is constructed according to the branching rules when restricting an irrep of $\calA$ to an irrep of a subalgebra $\calB$. We will not require this assumption.} 

\begin{lemma}[\cite{foxman2026efficient}]
    \label{lem:properties-of-Fourier-basis}
    Let 
    $\calA_k$
    be a finite-dimensional semisimple algebra, and let
    $
        \left\{
        E_{i,j}^{k,\lambda}
        \right\}
        _{
            \begin{subarray}{l}
                \lambda \in \wh{\calA}_k
                \\ 
                i, j \in \calB(V_{\calA_k}^{\lambda})
            \end{subarray}
        }
    $
    be a Fourier basis.
    Then, we have that
    \begin{enumerate}
        \item\cite[Lemma 2.13b]{foxman2026efficient}: 
        \label{item:Fourier-basis-self-dual}
        $
        \left(
            E_{i,j}^{k,\lambda}
        \right)^*
        =
        \frac{
            1
        }{
            \dim(
                V_{\calA_k}^{\lambda}
            )
        }
        E_{j,i}^{k,\lambda}
        \allowdisplaybreaks
        $
        , where $*$ denotes the dual basis element (\cref{def:dual_basis}). 
        \item 
        \cite[Section 4.5.2]{foxman2026efficient}:
        $
        S
        \left(
            E_{i,j}^{k,\lambda}
        \right)^T
        =
        S
        \left(
        E_{j,i}^{k,\lambda}
        \right)
        \allowdisplaybreaks
        $
        , where $S(\cdot)$ denotes the Schur representation (\cref{def:schur_rep}).
        \item 
        \cite[Lemma 2.13a]{foxman2026efficient}:
        $
        E_{i,j}^{k,\lambda}
        E_{\ell,m}^{k,\mu}
        =
        \delta_{\lambda \mu}
        \,
        \delta_{j\ell}
        \,
        E_{i,m}^{k,\lambda}
        \label{eq:algebra-fourier-basis-product}
        \allowdisplaybreaks
        $.
    \end{enumerate}
\end{lemma}

\subsection{Ratios of Irrep Dimensions for the Partition and Symmetric Group Algebras}
\begin{lemma}
    \label{lem:irrep-ratios}
    Consider the $(S_N, P_t(N))$-Schur duality, where $t^2 \leq N/4$ (in this range, $P_t(N)$ is always semisimple). 
    For any Young diagram $\lambda$ with a full number of boxes $|\lambda| = t$,  
    \begin{align}
        \frac
        {
            \dim
            \left(
                V^{\lambda}_{P_t(N)}
            \right)
        }
        {
            \dim
            \left(
                V^{(N - t, \lambda)}_{S_N}
            \right)
        }
        =
        \frac{t!}{N^t}
        \left(
            1
            \pm
            O
            \left(
                \frac{t^2}{N}
            \right)
        \right)
        \,,
    \end{align}
    where $(N - t, \lambda)$ is the Young diagram with an extra first row of $N-t$ boxes.

    Similarly, for the $(U(N), \C[S_t])$-Schur duality, we have that for all irreps,%
    \footnote{
        Note that in this case, \emph{all} irreps correspond to Young diagrams with a full number of boxes ($|\lambda| = t \;\; \forall \lambda$).
    }
    \begin{align}
        \frac
        {
            \dim
            \left(
                V^{\lambda}_{\C[S_t]}
            \right)
        }
        {
            \dim
            \left(
                V^{\lambda}_{U(N)}
            \right)
        }
        =
        \frac{t!}{N^t}
        \left(
            1
            \pm
            O
            \left(
                \frac{t^2}{N}
            \right)
        \right)
    \end{align}
\end{lemma}
\begin{proof}
    When $|\lambda| = t$, we have that
    \begin{align}
        \dim
        \left(
            V^{(N - t, \lambda)}_{S_N}
        \right)
        &
        =
        \frac{N!}{
            \prod_{(i,j) \in \lambda} h_{\lambda}(i,j)
            \cdot 
            \prod_{i=1}^{N - t} 
            \left(
            N - t - i + 1 + \lambda_i'
            \right)
        }
        \,, \tag{\cref{lem:dim_symmetric_group}}
        \\
        \dim
        \left(
            V^{\lambda}_{P_t(N)}
        \right)
        &
        =
        \dim
        \left(
            V^{\lambda}_{S_t}
        \right)
        = 
        f^{\lambda}
        =
        \frac{t!}{\prod_{(i,j) \in \lambda} h_{\lambda}(i,j)} 
        \,,
        \tag{\cref{lem:dim_formula_partition_algebra}, \cref{lem:dim_symmetric_group}}
        \\
        \dim
        \left(
            V^{\lambda}_{U(N)}
        \right)
        &
        =
        \prod_{(i,j) \in \lambda}
        \frac{N + j - i}{h_{\lambda}(i,j)}
        \, \tag{\cref{lemma:weyl-dim-form}}.
    \end{align}
    where $\lambda'_i$ is the $i$'th column length of $\lambda$.
    \noindent Therefore, the irrep dimension ratio for the full box irreps of the $(S_N, P_t(N))$ duality is
    \begin{align}
        \frac
        {
            \dim
            \left(
                V^{\lambda}_{P_t(N)}
            \right)
        }
        {
            \dim
            \left(
                V^{(N - t, \lambda)}_{S_N}
            \right)
        }
        &
        =
        \frac
        {
            t!
            \cdot 
             \prod_{i=1}^{N - t} 
            \left(
            N - t - i + 1 + \lambda_i'
            \right)
        }
        {N!}
        \\
        &
        =
        \frac
        {
            t! \cdot (N - t)!
            \cdot 
             \prod_{i=1}^{N - t} 
            \left(
            N - t - i + 1 + \lambda_i'
            \right)
        }
        {N!  \cdot 
             \prod_{i=1}^{N - t} 
            \left(
            N - t - i + 1
            \right)}
        \\
        &
        =
        \frac{1}{\binom{N}{t}}
        \cdot 
        \prod_{i=1}^{N - t} 
        \left( \frac
        {
            N - t - i + 1 + \lambda_i'
        }
        {
            N - t - i + 1
        }
        \right)
        \\
        &
        =
        \frac{1}{\binom{N}{t}}
        \cdot 
        \prod_{i=1}^{t} 
        \left( \frac
        {
            N - t - i + 1 + \lambda_i'
        }
        {
            N - t - i + 1
        }
        \right)
        \tag{$\lambda_i' =0$ for $i > t$}
        \\
        &
        =
         \frac{1}{\binom{N}{t}}
        \cdot 
        \left(
            1 \pm 
                O\left(\frac{t}{N}\right)
        \right)^t
        \tag{$\lambda_i' \le |\lambda|$}
        \\
        &
        =
         \frac{1}{\binom{N}{t}}
        \cdot 
        \left(
            1 \pm 
                O\left(\frac{t^2}{N}\right)
        \right)
        \tag{$t^2 = O(N)$}
        \\
        & 
        = 
        \frac{t!}{N^t} \cdot \left(
            1 \pm 
                O\left(\frac{t^2}{N}\right)
        \right)
    \end{align}
    For the $(U(N), \C[S_t])$ duality, 
    \begin{align}
        \frac
        {
            \dim
            \left(
                V^{\lambda}_{S_t}
            \right)
        }
        {
            \dim
            \left(
                V^{\lambda}_{U(N)}
            \right)
        }
        &
        =
        \frac
        {t!}
        {\prod_{(i,j) \in \lambda}
        (N + j - i)}
        \\
        &
        =
        \frac{t!}{N^t}
        \cdot
        \left(
            1 
            \pm 
            O
            \left(
                \frac{t^2}{N}
            \right)
        \right)
        \tag{$|j - i| < |\lambda|$}
    \end{align}
\end{proof}

\section{The Tableau Recording Oracle}\label{sec:tableau-recording}

Let $G$ be a compact Lie group and let $\rho: G \rightarrow U(N)$ be a finite-dimensional unitary representation of $G$. We wish to give an efficient simulation of query access to $\rho(g)$ for $g$ sampled from the Haar measure of $G$ (where $g$ is re-used for all queries). In fact, we wish to simulate (controlled) access to all of $\rho(g), \rho(g)^\dagger, \rho(g)^*$, and $\rho(g)^\top$. 

\subsection{The Uncompressed Representation Oracle}\label{sec:uncompressed-oracle}
 We first consider the following purification of the above process:

\begin{itemize}
    \item The adversary's register is $V= \mathbb C^N$, containing the input on which they wish to evaluate $\rho(g)$.
    \item The recording register is the 
    Hilbert space $L^2(G)$ 
    (which may be countably infinite-dimensional), 
    initialized to the Haar measure $dg = \ket{\bf 1} \in L^2(G)$ using the constant $f(g) = 1$ function.
    \item A query is represented by the unitary 
    \begin{align}
        Q_\rho \Big(\ket{x}\otimes \ket{f}\Big)= \int_G f(g) \cdot \rho(g) \ket{x} \otimes \ket{g} dg.
    \end{align}
    \noindent Since each $\rho(g)$ is a unitary, this is in fact a norm-preserving transformation. We also use the following notation to denote $Q_\rho$:
    \begin{align}
        Q_\rho = \int_G \rho(g) \otimes \ketbra{g} dg.
    \end{align}

\end{itemize}

\noindent In fact, using $Q_\rho$, we can also easily describe a process that has query access to $\rho(g), \rho(g)^\dagger, \rho(g)^*$, and $\rho(g)^\top$:%
\footnote{
    See \Cref{sec:fourier-unitary-identities} for the definitions of $\mathsf{Inv}$, $\mathsf{Conj}$, and $\mathsf{Transpose}$.
}

\begin{itemize}
    \item Controlled access to $\rho(g)^\dagger$ is simulated by (controlled) $\Big(\Id_V \otimes \mathsf{Inv}\Big) Q_\rho \Big(\Id_V \otimes \mathsf{Inv}\Big)$,
    \item Controlled access to $\rho(g)^*$ is simulated by (controlled) $\Big(\Id_V \otimes \mathsf{Conj}\Big) Q_\rho \Big(\Id_V \otimes \mathsf{Conj}\Big)$,
    \item Controlled access to $\rho(g)^\top$ is simulated by (controlled) $\Big(\Id_V \otimes \mathsf{Transpose}\Big) Q_\rho \Big(\Id_V \otimes \mathsf{Transpose}\Big)$.
\end{itemize}

In this section, our goal is to give an equivalent simulation of this oracle using the Clebsch-Gordan transform. For the derivation, we express $\rho$ in an irreducible decomposition,%
\footnote{
    This choice is for the derivation only. When generalized transforms for $\rho$ are supplied directly in the query basis, no separate efficient implementation of this decomposition is assumed.
}
i.e., $V \simeq \bigoplus_{\mu} M^{\mu} \otimes V_G^{\mu}$, where each $\mu \in \widehat{G}$ is an irreducible representation that appears in $V$ with multiplicity $\dim(M^{\mu})$. We represent vectors $\ket{\psi} \in V$ using the basis $\ket{\mu} \otimes \ket{m} \otimes \ket{x}$ for $\ket{m} \in M^{\mu}$ and $\ket{x} \in V_G^{\mu}$.

In this basis, we can re-write $Q_\rho$ as
\begin{align}
    Q_\rho = 
    \sum_\mu 
    \Big( 
        \proj{\mu} 
        \otimes 
        \Id_{M_\mu}
        \otimes 
        \int_G 
        \rho_\mu(g) 
        \otimes 
        \proj{g} 
        dg 
    \Big) 
    = 
    \sum_\mu 
    \proj{\mu} 
    \otimes
    \Id
    \otimes 
    Q_{\rho_\mu}. 
\end{align}

    \subsection{The Tableau Recording Oracle}
The uncompressed recording oracle uses an extremely large, potentially even infinite-dimensional, Hilbert space in its purification, and is thus certainly not efficient in general. However, it turns out that in the Fourier basis (that is, considering the image of $L^2(G)$ under the Fourier transform $F$) we get a recording oracle that is efficient whenever the Clebsch--Gordan transform $C$ is efficient!%
\footnote{
    Technically, we need compatible Clebsch–Gordan and dual Clebsch–Gordan transforms, as discussed below.
}
Indeed, when zero queries have been made, the recording register is in the trivial irrep state $F \cdot \ket{\bf 1} = \ket{\lambda = \mathsf{triv}, X=1, Y=1}$, because $\ket{\bf 1} \in L^2(G)$ is the unique normalized state that is invariant under $G$-translation. 

Formally, we define the tableau recording oracle $Q_\rho^{\mathrm{Tab}}$ \cite{GriYos25} as follows.

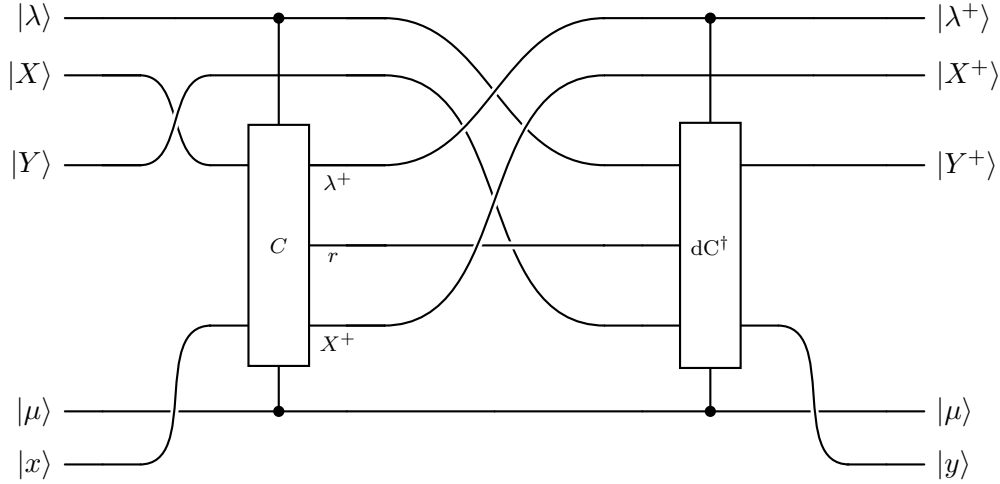
\begin{figure}[H]
    \centering
    \begin{quantikz}[wire types={q,q,q,n,n,q,q},classical
    gap=0.08cm]
    \lstick{$\ket{\lambda}$} &  &  & \ctrl{2} &  & \permute{3,5,1,4,2}\push{\phantom{-----}} 
    &  & \ctrl{2} 
    &  &  & \rstick{$\ket{\lambda^{+}}$} \\
    \lstick{$\ket{X}$} &  & \permute{2, 1} &  &  & \push{\phantom{--------}} 
    &  &  
    &  &  & \rstick{$\ket{X^{+}}$} \\
    \lstick{$\ket{Y}$} &  &  & \gate[3]{\substack{\phantom{\mathrm{dC}^{\dagger}}\\C\\\phantom{\mathrm{dC}^{\dagger}}}} & \qw{\phantom{X}\lambda^{+}} & \push{\phantom{--------}} 
    &  & \gate[3]{\substack{\phantom{\mathrm{dC}^{\dagger}}\\\mathrm{dC}^{\dagger}\\\phantom{\mathrm{dC}^{\dagger}}}} 
    &  &  & \rstick{$\ket{Y^{+}}$} \\
    &  &  &  & \qw{\phantom{r}r} & \setwiretype{q} &  &  & \setwiretype{n} &  & \\
    &  & \push{\phantom{--}} & \qw & \qw{\phantom{X}X^{+}} & \setwiretype{q}\push{\phantom{--------}} 
    &  &  
    & \permute{3} \setwiretype{n} &  &  \\
    \lstick{$\ket{\mu}$} &  &  & \ctrl{-1} &  &  
    &  & \ctrl{-1} 
    &  &  & \rstick{$\ket{\mu}$} \\
    \lstick{$\ket{x}$} &  & \permute{-1} \setwiretype{n} &  &  &  
    &  &  
    & \push{\phantom{--}} & \setwiretype{q} & \rstick{$\ket{y}$} \\
    \end{quantikz}
    \caption{
        A single query can be decomposed into a Clebsch--Gordan transformation
        followed by an inverse dual-Clebsch--Gordan operation, 
        though not on the same registers. The adversary's multiplicity register is not shown since it is not touched by the update.
    }
    \label{fig:CG-update}
\end{figure}

\begin{itemize}
    \item As before, the adversary's register contains linear combinations of $\ket{\mu} \otimes \ket{m} \otimes \ket{x}$ for $\ket{m} \in M^{\mu}$ and $\ket{x} \in V_{G}^{\mu}$. 
    \item The recording register now contains the following sub-registers:
    \begin{itemize}
        \item A register containing a label $\lambda \in \widehat{G}$ of an irreducible representation $\rho_\lambda$. While there are infinitely many such $\lambda$, it turns out that after $t$ queries to the tableau recording oracle, at most $N^t$ such $\lambda$ are ``reachable,'' resulting in a finite-dimensional Hilbert space after any finite number of  queries, which can be encoded in at most $t\log(N)$ qubits.
        \item Two registers $X$ and $Y$ containing elements of $V_{\lambda}$ and $V_\lambda^{*}$ controlled on $\ket{\lambda}$. We also non-canonically identify $V_\lambda^*$ with $V_\lambda$ by choosing a fixed basis of $V_\lambda$ with corresponding dual basis of $V_\lambda^*$. 

        $V_{\lambda}$ has dimension at most $N^t$ for any $\lambda$ reachable after $t$ queries, and can also therefore be encoded in at most $t \log(N)$ qubits.
    \end{itemize}
    \item The update rule is given by the composition of the following two maps (see \Cref{fig:CG-update}):
    \begin{itemize}
        \item First, given $\ket{\mu, m, x}\otimes \ket{\lambda, X, Y}$, apply $C \ket{\lambda, \mu, X, x}$. This results in a superposition of states of the form $\ket{\lambda, \mu, \lambda^+, r, X^+}$ (see \cref{def:clebsch-gordan}). 
        \item Then, on the registers containing $\lambda^+, \mu, \lambda, r, Y$, apply $dC^\dagger$, 
        the inverse of the dual Clebsch--Gordan transform.
        This results in a superposition of states of the form $\ket{\lambda^+, \mu, Y^+, y}$.
        \item The registers containing $\ket{\mu, m, y}$ are returned to the adversary, while the registers containing $\ket{\lambda^+, X^+, Y^+}$ are the updated state of the recording oracle. 
    \end{itemize}
\end{itemize}

\subsubsection{Equivalence of the uncompressed and tableau recording oracles}
Having defined our two recording oracles, we are ready to state the main result of this section. 
\begin{theorem}[Clebsch--Gordan Updates for Tableau Recordings]
\label{thm:CG_updates}
    Let $F: L^2(G) \rightarrow \bigoplus_\lambda V_\lambda \otimes V_\lambda^*$ denote the Fourier transform. Then, the uncompressed representation oracle is isometric to the tableau recording oracle via the map $F$: in particular, $(\Id \otimes F) \cdot Q_\rho \cdot (\Id\otimes F)^{-1}$ is equal to the tableau recording oracle update rule of a Clebsch--Gordan transform followed by an inverse dual Clebsch--Gordan transform.
\end{theorem}
This theorem is similar to the main result of \cite{GriYos25}, but additionally proves that the Fourier transform is an explicit Uhlmann transformation relating the tableau oracle to the uncompressed oracle. 

\begin{proof}
To show this, we expand out
\begin{align}
    \bra{\mu,m,y,\lambda^+,X^+,Y^+} (\Id \otimes F) \cdot Q_\rho \cdot (\Id \otimes F)^\dagger \ket{\mu,m,x, \lambda, X,Y} \label{eq:goal}
\end{align}
using the following expressions:
\begin{align}
    (\Id \otimes F)^\dagger \ket{\mu,x, \lambda, X,Y}
    &= \ket{\mu,x} \otimes \Bigg(\sqrt{\dim(V_{G}^{\lambda})} \int   \bra{Y} \rho_{\lambda}(g) \ket{X} \cdot \ket{g} \Bigg)
\end{align}

\begin{align}
    (\Id \otimes F)^\dagger \ket{\mu,y,\lambda^+,X^+,Y^+} &= \ket{\mu,y} \otimes \Bigg(\sqrt{\dim(V_{G}^{\lambda^+})} \int   \bra{Y^+} \rho_{\lambda^+}(g) \ket{X^+} \cdot \ket{g} \Bigg).
\end{align}
Plugging in the definition of $Q$, these imply that
\begin{align}
    (\ref{eq:goal}) = &\int_G \bra{\mu,m,y} \Big( 
    \ketbra{\mu, m} \otimes \rho_{\mu}(g)
    \Big) \ket{\mu,m,x} \\
    &
    \cdot
    \Bigg(
    \sqrt{
    \text{d}(V_{G}^{\lambda}) \cdot \text{d}(V_{G}^{\lambda^+})
    } \bra{Y^+} \overline{\rho_{\lambda^+}(g)}
    \ket{X^+} 
    \bra{Y} \rho_{\lambda}(g) 
    \ket{X}
    \Bigg) dg \\
    =& \sqrt{
    \text{d}(V_{G}^{\lambda}) \cdot \text{d}(V_{G}^{\lambda^+})
    }\int_G 
    \bra{Y^+} \overline{\rho_{\lambda^+}(g)}
    \ket{X^+}
    \underbrace{\bra{y} \rho_{\mu}(g) \ket{x}  
    \bra{Y} \rho_{\lambda}(g) 
    \ket{X}}_{(*)} dg \label{eq:update-matrix-entry}
\end{align}
Let us expand out
\begin{align}
    (*) &= \bra{\lambda,Y,\mu,y}\Big( 
    \ketbra{\lambda} 
    \otimes 
    \rho_{\lambda}(g)
    \otimes 
    \ketbra{\mu}
    \otimes
    \rho_{\mu}(g)
    \Big)
    \ket{\lambda,X,\mu,x}\\
    &= \bra{\lambda,\mu,Y,y}
    C^\dagger \cdot 
    \Bigg(
    C
    \Big( 
    \ketbra{\lambda} 
    \otimes 
    \ketbra{\mu}
    \otimes 
    \rho_{\lambda}(g)
    \otimes
    \rho_{\mu}(g)
    \Big)
    C^\dagger 
    \Bigg)
    \cdot C
    \ket{\lambda,\mu,X,x}\\
    &= \cgbra
    C^\dagger \cdot 
    \Bigg(  \ketbra{\lambda} \otimes \ketbra{\mu} \otimes
    \sum_{\lambda^{\#}, r \in [m_{\lambda, \mu}^{\lambda^{\#}}]} \ketbra{r} \otimes \ketbra*{\lambda^{\#}} \otimes  
    \rho_{\lambda^{\#}}(g)
    \Bigg)
    \cdot C
    \cgket \\
    &= \cgbra
    C^\dagger \cdot 
    \Bigg(  \ketbra{\lambda} \otimes \ketbra{\mu} \otimes
    \sum_{\lambda^{\#}, r \in [m_{\lambda, \mu}^{\lambda^{\#}}]} \ketbra{r} \otimes \ketbra*{\lambda^{\#}} 
    \\
    & \hspace{3.3cm}
    \otimes  
    \sum_{X^{\#}, Y^{\#}}
    \ketbra*{Y^{\#}}
    \rho_{\lambda^{\#}}(g)
    \ketbra*{X^{\#}}
    \Bigg)
    \cdot C
    \cgket\\
    &= \sum_{
    \substack{
    \lambda^{\#}, X^{\#},Y^{\#},\\
    r \in [m_{\lambda, \mu}^{\lambda^{\#}}]
    }
    }
    \cgbra 
    C^\dagger \ket*{\lambda,\mu,\lambda^{\#},r,Y^{\#}} 
    \cdot \cgcoeff{\lambda}{\mu}{\lambda^{\#}}{X}{x}{r}{X^{\#}}
    \cdot 
    \bra*{Y^{\#}}
    \rho_{\lambda^{\#}}(g)
    \ket*{X^{\#}}
\end{align}
Plugging this back into (\ref{eq:update-matrix-entry}), we have
\begin{align}
    (\ref{eq:goal}) = & \sqrt{\frac{
    \text{d}(V_{G}^{\lambda})}{\text{d}(V_{G}^{\lambda^+})}
    } \sum_{
    \substack{
    \lambda^{\#}, X^{\#},Y^{\#},\\
    r \in [m_{\lambda, \mu}^{\lambda^{\#}}]
    }
    } 
    \cgbra 
    C^\dagger \ket*{\lambda,\mu,\lambda^{\#},r,Y^{\#}}  \cgcoeff{\lambda}{\mu}{\lambda^{\#}}{X}{x}{r}{X^{\#}} \\
    & 
    \hspace{3cm} \cdot \underbrace{\Bigg(\text{d}(V_{G}^{\lambda^+})
    \cdot \int_G 
    \bra{Y^+} \overline{\rho_{\lambda^+}(g)}
    \ket{X^+}
    \bra*{Y^{\#}}
    \rho_{\lambda^{\#}}(g)
    \ket*{X^{\#}} dg\Bigg)}_{
     1 \ \text{if} \
    (\lambda^+,X^+,Y^+) = (\lambda^{\#},X^{\#},Y^{\#}), 
    \ \text{else} \
    0 
    }\\
    =& 
    \sqrt{\frac{
    \text{d}(V_{G}^{\lambda})}{\text{d}(V_{G}^{\lambda^+})}
    } \sum_{
    r \in [m_{\lambda, \mu}^{\lambda^+}]
    } 
    \cgbra 
    C^\dagger \ket*{\lambda,\mu,\lambda^+,r,Y^+}  \cgcoeffs.
\end{align}
Now, we invoke the dagger of \Cref{lemma:vilenkin-klimyk} to obtain 

\begin{align}
    (\ref{eq:goal}) = &
    \sum_{
    r \in [m_{\lambda^{+},\lambda, \mu}]
    } 
    \bra*{\lambda^+,\mu,Y^+,y}
    dC^\dagger \ket*{\lambda^+,\mu,\lambda,r,Y} \cgcoeffs.
\end{align}

Finally, we observe that these matrix entries correspond to the composition of the Clebsch--Gordan and inverse dual Clebsch--Gordan unitaries applied to the appropriate sub-registers indicated below:

\begin{align}
    C =& \sum_{\lambda, \lambda^+, \mu, r, X, X^+, x} \cgcoeffs \cdot \cgketbras
\end{align}

\begin{align}
    dC^\dagger =& \sum_{\lambda, \lambda^+, \mu, r, Y, Y^+, y} \bra*{\lambda^+,\mu,Y^+,y} dC^\dagger
    \ket{\lambda^+,\mu,\lambda,r,Y} \cdot \ketbra{\lambda^+, \mu, Y^+, y}{\lambda^+, \mu, \lambda, r, Y}.
\end{align}
This completes the derivation of the Clebsch--Gordan update rule. 
The same calculation applies directly to generalized transforms for $\rho$, replacing $\rho_\mu$ by $\rho$ and omitting the local irrep and multiplicity labels $\mu,m$.
\end{proof}

\subsubsection{Other query types}\label{sec:tableau-other-queries} The derived update rule implies that we can implement compressed conjugate, transpose, and inverse queries with the (linear%
\footnote{
    as opposed to anti-linear (or conjugate linear)
}%
) isometries defined below

\begin{itemize}
    \item $\Big(\Id \otimes F\Big) \Big(\Id_V \otimes \mathsf{Conj}\Big) \Big(\Id \otimes F^\dagger\Big) \cdot  Q_\rho^{\mathrm{Tab}} \cdot \Big(\Id \otimes F\Big) \Big(\Id_V \otimes \mathsf{Conj}\Big) \Big(\Id \otimes F^\dagger\Big)$,
    \item $\Big(\Id \otimes F\Big) \Big(\Id_V \otimes \mathsf{Transpose}\Big) \Big(\Id \otimes F^\dagger\Big) \cdot  Q_\rho^{\mathrm{Tab}} \cdot \Big(\Id \otimes F\Big) \Big(\Id_V \otimes \mathsf{Transpose}\Big) \Big(\Id \otimes F^\dagger\Big)$,
    \item $\Big(\Id \otimes F\Big) \Big(\Id_V \otimes \mathsf{Inv}\Big) \Big(\Id \otimes F^\dagger\Big) \cdot  Q_\rho^{\mathrm{Tab}} \cdot \Big(\Id \otimes F\Big) \Big(\Id_V \otimes \mathsf{Inv}\Big) \Big(\Id \otimes F^\dagger\Big)$.
\end{itemize}
Using the three identities from \cref{sec:fourier-unitary-identities}, we can see that
\begin{itemize}
    \item $\Big(\Id \otimes F\Big) \Big(\Id_V \otimes \mathsf{Conj}\Big) \Big(\Id \otimes F^\dagger\Big)$ performs the map 
    \[\sum_{\lambda, X,Y} \alpha_{\lambda, X,Y}\ket{\lambda, X, Y} \mapsto \sum_{\lambda, X,Y} \overline{\alpha}_{\lambda, X,Y}\ket{\lambda^*, X, Y}.
    \]
    \item $\Big(\Id \otimes F\Big) \Big(\Id_V \otimes \mathsf{Transpose}\Big) \Big(\Id \otimes F^\dagger\Big)$ performs the map
    \[\sum_{\lambda, X,Y} \alpha_{\lambda, X,Y}\ket{\lambda, X, Y} \mapsto \sum_{\lambda, X,Y} \overline{\alpha}_{\lambda, X,Y}\ket{\lambda, Y, X}.
    \]
    \item $\Big(\Id \otimes F\Big) \Big(\Id_V \otimes \mathsf{Inv}\Big) \Big(\Id \otimes F^\dagger\Big)$ performs the map 
    \[\sum_{\lambda, X,Y} \alpha_{\lambda, X,Y}\ket{\lambda, X, Y} \mapsto \sum_{\lambda, X,Y} \alpha_{\lambda, X,Y}\ket{\lambda^*, Y, X}.
    \]
\end{itemize}
Thus, we conclude that
\begin{itemize}
    \item Conjugate queries can be implemented using the \emph{complex conjugate} $\overline{Q_\rho^{\mathrm{Tab}}}$ of $Q_\rho^{\mathrm{Tab}}$ conjugated by the linear isometry $\ket{\lambda, X, Y}\mapsto \ket{\lambda^*, X, Y}$.

    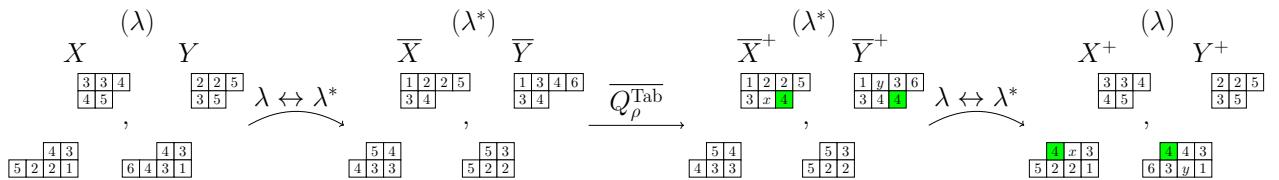
\begin{figure}[H]
        \centering
        \scalebox{.5}{
        \begin{tikzpicture}[baseline=(AX.base)]
            \node (CX) at (-9,0) [label=above:{\huge $\;\,X$}] {
                \Ytableau{
                    334,
                    45,
                }
                {
                        ,
                      43,
                    5221,
                }
            };
            \node (Ccomma) at (-7.5,0) {\huge $,$};
            \node (Clambda) at (-7.2,2.7) {\huge $(\lambda)$};
            \node (CY) at (-6,0) [label=above:{\huge $\;\,Y$}] {
                \Ytableau{
                    225,
                    35,
                }
                {
                        ,
                      43,
                    6431,
                }
            };
            \node (AX) at (0,0) [label=above:{\huge $\overline{X}$}] {
                \Ytableau{
                    1225,
                    34,
                    ,
                }
                {
                     54,
                    433
                }
            };
            \node (Acomma) at (1.5,0) {\huge $,$};
            \node (Alambda) at (1.7,2.7) {\huge $(\lambda^*)$};
            \node (AY) at (3,0) [label=above:{\huge $\overline{Y}$}] {
                \Ytableau{
                    1346,
                    34,
                    ,
                }
                {
                     53,
                    522
                }
            };
            \node (BX) at (9,0) [label=above:{\huge $\phantom{}^{\phantom +}\overline{X}^+$}] {
                \Ytableau{
                    1225,
                    3{x},
                    ,
                }
                {
                     54,
                    433
                }
            };
            \node at (BX) {
                \gyoung(:<>::::::<>,:::::!<\Yfillcolour{green}>;4,,,,:<>)
            };
            \node (Bcomma) at (10.5,0) {\huge $,$};
            \node (Blambda) at (10.7,2.7) {\huge $(\lambda^*)$};
            \node (BY) at (12,0) [label=above:{\huge $\phantom{}^{\phantom +}\overline{Y}^+$}] {
                \Ytableau{
                    1y36,
                    34,
                    ,
                }
                {
                     53,
                    522
                }
            };
            \node at (BY) {
                \gyoung(:<>::::::<>,:::::!<\Yfillcolour{green}>;{4},,,,:<>)
            };
            \node (DX) at (18,0) [label=above:{\huge $\phantom{}^{\phantom +}X^+$}] {
                \Ytableau{
                    334,
                    45,
                }
                {
                        ,
                    {x}3,
                    5221,
                }
            };
            \node at (DX) {
                \gyoung(:<>::::::<>,,,,:!<\Yfillcolour{green}>;{4},:<>)
            };
            \node (Dcomma) at (19.5,0) {\huge $,$};
            \node (Dlambda) at (19.8,2.7) {\huge $(\lambda)$};
            \node (DY) at (21,0) [label=above:{\huge $\phantom{}^{\phantom +}Y^+$}] {
                \Ytableau{
                    225,
                    35,
                }
                {
                          ,
                        43,
                    63{y}1,
                }
            };
            \node at (DY) {
                \gyoung(:<>::::::<>,,,,:!<\Yfillcolour{green}>;{4},:<>)
            };
            \draw[->] (AY.east) to node[midway,above] {{\huge $\overline{Q_\rho^{\mathrm{Tab}}}$}} (BX.west);
            \draw[->,bend left] (CY.east) to node[midway,above] {{\huge $\lambda \leftrightarrow \lambda^*$}} (AX.west);
            \draw[->,bend left] (BY.east) to node[midway,above] {{\huge $\lambda \leftrightarrow \lambda^*$}} (DX.west);
        \end{tikzpicture}
        }
        \caption{Example basis vectors in a complex conjugate query for the unitary group.}
    \end{figure}

    \item Transpose queries can be implemented using $\overline{Q_\rho^{\mathrm{Tab}}}$ conjugated by the unitary $\ket{\lambda, X, Y}\mapsto \ket{\lambda, Y, X}$

    \begin{figure}[H]
        \centering
        \scalebox{.5}{
        \begin{tikzpicture}[baseline=(AX.base)]
            \node (CX) at (-9,0) [label=above:{\huge $\;\,X$}] {
                \Ytableau{
                    334,
                    45,
                }
                {
                        ,
                      43,
                    5221,
                }
            };
            \node (Ccomma) at (-7.5,0) {\huge $,$};
            \node (Clambda) at (-7.2,2.7) {\huge $(\lambda)$};
            \node (CY) at (-6,0) [label=above:{\huge $\;\,Y$}] {
                \Ytableau{
                    225,
                    35,
                }
                {
                        ,
                      43,
                    6431,
                }
            };
            \node (AX) at (0,0) [label=above:{\huge $\;\,Y$}] {
                \Ytableau{
                    225,
                    35,
                }
                {
                        ,
                      43,
                    6431,
                }
            };
            \node (Acomma) at (1.5,0) {\huge $,$};
            \node (Alambda) at (1.8,2.7) {\huge $(\lambda)$};
            \node (AY) at (3,0) [label=above:{\huge $\;\,X$}] {
                \Ytableau{
                    334,
                    45,
                }
                {
                        ,
                      43,
                    5221,
                }
            };
            \node (BX) at (9,0) [label=above:{\huge $\;\,\phantom{}^{\phantom +}Y^+$}] {
                \gyoung(:<>:::::<>,::::::!<\Yfillcolour{green}>;{5},,,,<>:<>)
            };
            \node at (BX) {
                \Ytableau{
                    225,
                    3{x},
                }
                {
                        ,
                      43,
                    6431,
                }
            };
            \node (Bcomma) at (10.5,0) {\huge $,$};
            \node (Blambda) at (10.8,2.7) {\huge $(\lambda)$};
            \node (BY) at (12,0) [label=above:{\huge $\;\,\phantom{}^{\phantom +}X^+$}] {
                \gyoung(:<>:::::<>,::::::!<\Yfillcolour{green}>;{5},,,,<>:<>)
            };
            \node at (BY) {
                \Ytableau{
                    {y}33,
                    44,
                }
                {
                        ,
                      43,
                    5221,
                }
            };
            \node (DX) at (18,0) [label=above:{\huge $\;\,\phantom{}^{\phantom +}X^+$}] {
                \gyoung(:<>:::::<>,::::::!<\Yfillcolour{green}>;{5},,,,<>:<>)
            };
            \node at (DX) {
                \Ytableau{
                    {y}33,
                    44,
                }
                {
                        ,
                      43,
                    5221,
                }
            };
            \node (Dcomma) at (19.5,0) {\huge $,$};
            \node (Dlambda) at (19.8,2.7) {\huge $(\lambda)$};
            \node (DY) at (21,0) [label=above:{\huge $\;\,\phantom{}^{\phantom +}Y^+$}] {
                \gyoung(:<>:::::<>,::::::!<\Yfillcolour{green}>;{5},,,,<>:<>)
            };
            \node at (DY) {
                \Ytableau{
                    225,
                    3{x},
                }
                {
                        ,
                      43,
                    6431,
                }
            };
            \draw[->] (AY.east) to node[midway,above] {{\huge $\overline{Q_\rho^{\mathrm{Tab}}}$}} (BX.west);
            \draw[->,bend left] (CY.east) to node[midway,above] {{\huge $X \leftrightarrow Y$}} (AX.west);
            \draw[->,bend left] (BY.east) to node[midway,above] {{\huge $\;\;X^+ \!\!\leftrightarrow\! Y^+$}} (DX.west);
        \end{tikzpicture}
        }
        \caption{Example basis vectors in a transpose query for the unitary group.}
    \end{figure}
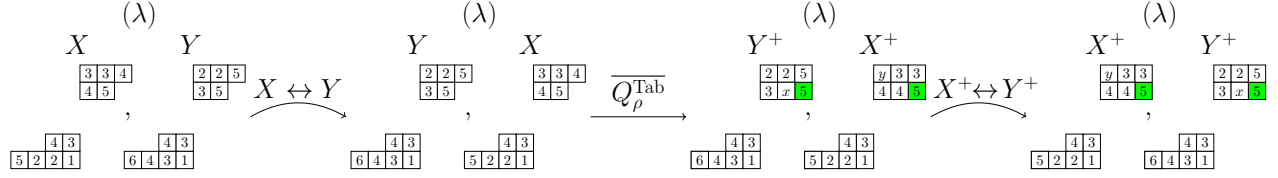

    \item Inverse queries can be implemented using  $Q_\rho^{\mathrm{Tab}}$ conjugated by the linear isometry $\ket{\lambda, X, Y}\mapsto \ket{\lambda^*, Y, X}$. 

    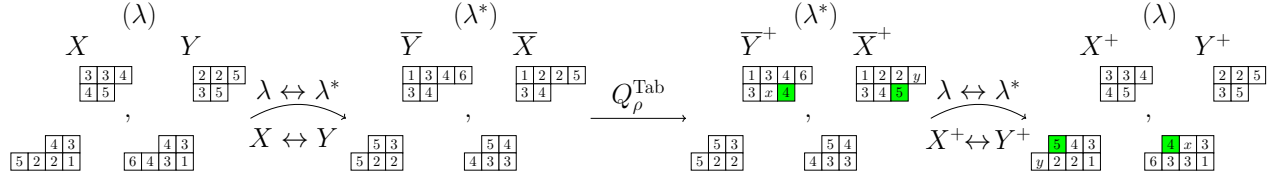
\begin{figure}[H]
        \centering
        \scalebox{.5}{
        \begin{tikzpicture}[baseline=(AX.base)]
            \node (CX) at (-9,0) [label=above:{\huge $\;\,X$}] {
                \Ytableau{
                    334,
                    45,
                }
                {
                        ,
                      43,
                    5221,
                }
            };
            \node (Ccomma) at (-7.5,0) {\huge $,$};
            \node (Clambda) at (-7.2,2.7) {\huge $(\lambda)$};
            \node (CY) at (-6,0) [label=above:{\huge $\;\,Y$}] {
                \Ytableau{
                    225,
                    35,
                }
                {
                        ,
                      43,
                    6431,
                }
            };
            \node (AX) at (0,0) [label=above:{\huge $\overline{Y}$}] {
                \Ytableau{
                    1346,
                    34,
                    ,
                }
                {
                     53,
                    522
                }
            };
            \node (Acomma) at (1.5,0) {\huge $,$};
            \node (Alambda) at (1.7,2.7) {\huge $(\lambda^*)$};
            \node (AY) at (3,0) [label=above:{\huge $\overline{X}$}] {
                \Ytableau{
                    1225,
                    34,
                    ,
                }
                {
                     54,
                    433
                }
            };
            \node (BX) at (9,0) [label=above:{\huge $\phantom{}^{\phantom +}\overline{Y}^+$}] {
                \Ytableau{
                    1346,
                    3{x},
                    ,
                }
                {
                     53,
                    522
                }
            };
            \node at (BX) {
                \gyoung(:<>::::::<>,:::::!<\Yfillcolour{green}>;4,,,,:<>)
            };
            \node (Bcomma) at (10.5,0) {\huge $,$};
            \node (Blambda) at (10.7,2.7) {\huge $(\lambda^*)$};
            \node (BY) at (12,0) [label=above:{\huge $\phantom{}^{\phantom +}\overline{X}^+$}] {
                \Ytableau{
                    122{y},
                    34,
                    ,
                }
                {
                     54,
                    433
                }
            };
            \node at (BY) {
                \gyoung(:<>::::::<>,:::::!<\Yfillcolour{green}>;{5},,,,:<>)
            };
            \node (DX) at (18,0) [label=above:{\huge $\phantom{}^{\phantom +}X^+$}] {
                \Ytableau{
                    334,
                    45,
                }
                {
                          ,
                        43,
                    {y}221,
                }
            };
            \node at (DX) {
                \gyoung(:<>::::::<>,,,,:!<\Yfillcolour{green}>;{5},:<>)
            };
            \node (Dcomma) at (19.5,0) {\huge $,$};
            \node (Dlambda) at (19.8,2.7) {\huge $(\lambda)$};
            \node (DY) at (21,0) [label=above:{\huge $\phantom{}^{\phantom +}Y^+$}] {
                \Ytableau{
                    225,
                    35,
                }
                {
                        ,
                    {x}3,
                    6331,
                }
            };
            \node at (DY) {
                \gyoung(:<>::::::<>,,,,:!<\Yfillcolour{green}>;{4},:<>)
            };
            \draw[->] (AY.east) to node[midway,above] {{\huge $Q_\rho^{\mathrm{Tab}}$}} (BX.west);
            \draw[->,bend left] (CY.east) to 
            node[midway,above] {{\huge $\lambda \leftrightarrow \lambda^*$}}
            node[midway,below=.5cm] {{\huge $\!\!X\leftrightarrow Y$}} 
            (AX.west);
            \draw[->,bend left] (BY.east) to 
            node[midway,above] {{\huge $\lambda \leftrightarrow \lambda^*$}}
            node[midway,below=.5cm] {{\huge $X^+ \!\!\leftrightarrow\! Y^+$}} (DX.west);
        \end{tikzpicture}
        }
        \caption{Example basis vectors in an inverse query for the unitary group.}
    \end{figure}
\end{itemize}
Thus, given any polynomial-size circuit for $Q_\rho^{\mathrm{Tab}}$ and efficient isomorphisms between $V_{\lambda^*}$ and $V_\lambda^*$, we have corresponding polynomial-size circuits for these other query types. 

\subsubsection{Efficient Implementation for the Unitary Group}\label{sec:tableau-efficient-unitary-group}

We now show how to efficiently implement the update rule, $Q_\rho^{\mathrm{Tab}}$, in~\Cref{thm:CG_updates} for the case of Haar random unitaries, $G = U(N)$. 
The Clebsch--Gordan transform for unitaries was shown to have an efficient algorithm in~\cite{BCH05,Har05,burchardt2025highdimensionalquantumschurtransforms}, and this algorithm was extended to the mixed unitary case, in which $\rho(U)$ could be either the standard irrep or the dual irrep, by~\cite{nguyen2023mixed,fei2023efficientquantumalgorithmportbased,grinko2023gelfand}. For the case of the irrep $\lambda$ consisting of a tableau containing at most $t$ boxes (either positive or negative), and the oracle representation being a single-box irrep (either the standard or dual irrep), they show how to implement the Clebsch--Gordan transform with error at most $\epsilon$ in time $\poly(\log(N), t, \log(\frac{1}{\epsilon}))$.

The recording update is exact when the underlying transforms are implemented exactly. Approximate implementations introduce only the error inherited from approximating these transforms. Each update uses one Clebsch--Gordan transform and one inverse dual Clebsch--Gordan transform. If each is implemented to operator-norm error at most $\delta$ on the relevant subspaces, the simulation error after $t$ queries is $O(t\delta)$ in trace distance. Consequently, the unitary simulator achieves total error at most $\varepsilon$ in time $\poly(\log(N), t, \log(\frac{1}{\epsilon}))$.

We show here that the Clebsch--Gordan transform implemented by these algorithms, or any similar algorithm that works in the Gelfand-Tsetlin basis, implies a corresponding dual Clebsch--Gordan algorithm with the same complexity which satisfies~\Cref{lemma:vilenkin-klimyk} and can be used in~\Cref{thm:CG_updates}.

\begin{definition}
    [Signed Flipping of a Young Tableau]
    \label{def:signed-flip-tableau}
    For every basis state $\ket{X}$ given by a semi-standard Young tableau $X$ (allowing negative boxes), let $\mathsf{Flip}\ket{X}$ be the operation that, up to a $\pm 1$ sign, flips the Young tableau $X$ by rotating it 180 degrees, for example, sending
    \begin{align}
        \begin{tikzinline}[baseline=(A.base)]
            \node (A) at (0,0) [label=above:{$X$}] {
                \Ytableau
                {
                    1225,
                    34,
                    ,
                }
                {
                     54,
                    433
                }
            };
            \node (B) at (5,0) [label=above:{$\barX$}] {
                \Ytableau
                {
                    334,
                    45,
                }
                {
                        ,
                      43,
                    5221,
                }
            };
            \draw[->] (A.east) -- node[above]{$\mathsf{Flip}$} (B.west);
        \end{tikzinline}
        \label{eq:flipping-tableau-definition}
    \end{align}
    with a phase of $(-1)^{Nt + p}$ where $p$ is the \emph{total parity} (even or odd) of all the entries in the tableau. 
    (Equivalently, the phase is $(-1)^{\sum_{x_i \in X} (N - x_i)}$ where the sum runs over all the entries in the boxes of the tableau.)

    For instance, in the example in \Cref{eq:flipping-tableau-definition} above, there are 6 odd entries (namely $1$, $5$, $3$, $5$, $3$, and $3$), so the sum is even, and $t = 11$ is odd, so in this case we would apply a $-1$ phase whenever $N$ is odd.

    When working with a semi-standard Young tableau $X$, we always apply a sign when flipping. Therefore, when clear from context, we refer to the signed flipping of a tableau as simply the \textquote{flipping operation}. 
\end{definition}

\begin{lemma} \label{claim:flip-gives-dual-CG}
    Let $G = U(N)$, and let $\mu = \boxed{\phantom{x}}$ be the standard irrep of $U(N)$.
    The dual Clebsch--Gordan transform implied by \Cref{lemma:cg-in-gt-basis} is equivalent to applying the signed flip of \Cref{def:signed-flip-tableau} to the Young tableau $X$ (and a corresponding unsigned flip to corresponding shape $\lambda$), applying the standard Clebsch--Gordan transform of \Cref{lemma:cg-in-gt-basis}, and then (signed) flipping it back.

    That is, we have that 
    \begin{align}
        \dcgcoeffmultfree{\lambda}{\mu}{\lambda^-}{X}{x}{X^-}
        =
        \langle
            \lambda,
            \mu,
            \lambda^-,
            X^-
        \vert
        \,
        \mathsf{Flip}
        \cdot
        C
        \cdot
        \mathsf{Flip}
        \,
        \vert
            \lambda,
            \mu,
            X,
            x
        \rangle
        \,.
    \end{align}
    Where the $\mathsf{Flip}$ applies to the $X$ register, and also flips $\lambda$ and $\lambda^-$ (without sign) accordingly.

    This means that every efficient algorithm for the Clebsch--Gordan transform implies a corresponding efficient dual Clebsch--Gordan transform algorithm where we $\mathsf{Flip}$, apply Clebsch--Gordan, and then $\mathsf{Flip}$ back.
\end{lemma}

\begin{proof}
    We can show that the reduced Wigner coefficients, or scalar factors, of \Cref{lemma:cg-in-gt-basis} match up term by term (up to a $\pm 1$ phase, which we will handle separately).
    Let $Z = \barX$ be the \emph{unsigned}%
    \footnote{
        We are starting with the unsigned flip, which we then use to compute the necessary sign. This is the only time we take the flip to be unsigned.
    }
    flip of $X$. Then $Z_{k}^{\le z} = -X_{z - k + 1}^{\le z}$ (a positive row turns into a negative row and vice versa, and the row counting is reversed to go from $z$ down to $1$ rather than $1$ to $z$).
    We can see in \Cref{lemma:cg-in-gt-basis} that the Clebsch--Gordan scalar factors for $Z$ are zero in the same places as for the dual Clebsch--Gordan, so it suffices to consider nonzero scalar factors.

    \paragraph{First scalar factor}
    The first scalar factor of the Clebsch--Gordan coefficient of adding 
    $
        (
            0, 
            \allowbreak
            0, 
            \allowbreak
            \dots, 
            \allowbreak
            0, 
            \allowbreak
            \Box_{\Delta_{x}}, 
            \allowbreak
            \Box_{\Delta_{x + 1}},
            \allowbreak
            \dots, 
            \allowbreak
            \Box_{\Delta_{N}}
        )
    $
    to $Z$
    is
    \begin{align}
        &
        \left\lvert
            \frac{
                \displaystyle\prod_{k = 1}^{x - 1}
                \left(
                    Z_{k}^{\le x - 1}
                    -
                    Z_{\Delta_{x}}^{\le x}
                    +
                    \Delta_{x}
                    -
                    k
                    -
                    1
                \right)
            }{
                \displaystyle\prod_{
                    \substack{
                        k = 1 
                        \\
                        k \ne \Delta_{x}
                    }
                }^{x}
                \left(
                    Z_{k}^{\le x}
                    -
                    Z_{\Delta_{x}}^{\le x}
                    +
                    \Delta_{x}
                    -
                    k
                \right)
            }
        \right\rvert
        ^{\frac{1}{2}}
        \allowdisplaybreaks
        \\
        =
        &
        \left\lvert
            \frac{
                \displaystyle\prod_{k = 1}^{x - 1}
                \left(
                    -
                    X_{x - k}^{\le x - 1}
                    +
                    X_{x - \Delta_{x} + 1}^{\le x}
                    +
                    \Delta_{x}
                    -
                    k
                    -
                    1
                \right)
            }{
                \displaystyle\prod_{
                    \substack{
                        k = 1 
                        \\
                        k \ne \Delta_{x}
                    }
                }^{x}
                \left(
                    -
                    X_{x - k +1}^{\le x}
                    +
                    X_{x - \Delta_{x} + 1}^{\le x}
                    +
                    \Delta_{x}
                    -
                    k
                \right)
            }
        \right\rvert
        ^{\frac{1}{2}}
        \allowdisplaybreaks
        \\
        =
        &
        \left\lvert
            \frac{
                \displaystyle\prod_{k = 1}^{x - 1}
                \left(
                    -
                    X_{k}^{\le x - 1}
                    +
                    X_{x - \Delta_{x} + 1}^{\le x}
                    +
                    \Delta_{x}
                    -
                    x
                    +
                    k
                    -
                    1
                \right)
            }{
                \displaystyle\prod_{
                    \substack{
                        k = 1 
                        \\
                        x - k + 1 \ne \Delta_{x}
                    }
                }^{x}
                \left(
                    -
                    X_{k}^{\le x}
                    +
                    X_{x - \Delta_{x} + 1}^{\le x}
                    +
                    \Delta_{x}
                    -
                    x 
                    + 
                    k 
                    - 
                    1
                \right)
            }
        \right\rvert
        ^{\frac{1}{2}}
        \tag{%
            by re-indexing 
            $k \to x-k$ in numerator and $k \to x - k + 1$ in denominator
        }
        \allowdisplaybreaks
        \\
        =
        &
        \left\lvert
            \frac{
                \displaystyle\prod_{k = 1}^{x - 1}
                \left(
                    X_{k}^{\le x - 1}
                    -
                    X_{x - \Delta_{x} + 1}^{\le x}
                    +
                    (
                        x
                        -
                        \Delta_{x}
                        +
                        1
                    )
                    -
                    k
                \right)
            }{
                \displaystyle\prod_{
                    \substack{
                        k = 1 
                        \\
                        k \ne (x - \Delta_{x} + 1)
                    }
                }^{x}
                \left(
                    X_{k}^{\le x}
                    -
                    X_{x - \Delta_{x} + 1}^{\le x}
                    +
                    (
                        x 
                        -
                        \Delta_{x}
                        + 
                        1
                    )
                    - 
                    k 
                \right)
            }
        \right\rvert
        ^{\frac{1}{2}}
        \,.
    \end{align}
    When we now relabel $\Delta_x \to (x - \Delta_x + 1)$ (since each row of the Gelfand-Tsetlin pattern is reversed when we apply the flipping operation), we can see that this becomes identical to the first scalar factor of the dual Clebsch--Gordan coefficient of subtracting 
    $
        (
            0, 
            \allowbreak
            0, 
            \allowbreak
            \dots, 
            \allowbreak
            0, 
            \allowbreak
            \Box_{\big(x - \Delta_{x} + 1\big)}, 
            \allowbreak
            \Box_{\big((x + 1) - \Delta_{(x + 1)} + 1\big)},
            \allowbreak
            \dots, 
            \allowbreak
            \Box_{\big(N - \Delta_{N} + 1\big)}
        )
    $
    from~$X$.

    \needspace{3\baselineskip}
    \paragraph{Remaining scalar factors}
    A similar argument holds for the other scalar factors. For the Clebsch--Gordan coefficient of adding 
    $
        (
            0, 
            \allowbreak
            0, 
            \allowbreak
            \dots, 
            \allowbreak
            0, 
            \allowbreak
            \Box_{\Delta_{x}}, 
            \allowbreak
            \Box_{\Delta_{x + 1}},
            \allowbreak
            \dots, 
            \allowbreak
            \Box_{\Delta_{N}}
        )
    $
    to the tableau $Z = \barX$, they are
    \begin{align}
        &
        \left\lvert
            \frac{
                \displaystyle\prod_{
                    \substack{
                        k = 1 
                        \\
                        k \ne \Delta_{(z - 1)}
                    }
                }^{z - 1}
                \left(
                    Z_{k}^{\le z - 1}
                    -
                    Z_{\Delta_{z}}^{\le z}
                    +
                    \Delta_{z}
                    -
                    k
                    -
                    1
                \right)
            }{
                \displaystyle\prod_{
                    \substack{
                        k = 1 
                        \\
                        k \ne \Delta_{z}
                    }
                }^{z}
                \left(
                    Z_{k}^{\le z}
                    -
                    Z_{\Delta_{z}}^{\le z}
                    +
                    \Delta_{z}
                    -
                    k
                \right)
            }
        \right\rvert
        ^{\frac{1}{2}}
        \left\lvert
            \frac{
                \displaystyle\prod_{
                    \substack{
                        k = 1 
                        \\
                        k \ne \Delta_{z}
                    }
                }^{z}
                \left(
                    Z_{k}^{\le z}
                    -
                    Z_{\Delta_{(z - 1)}}^{\le z - 1}
                    +
                    \Delta_{(z - 1)}
                    -
                    k
                \right)
            }{
                \displaystyle\prod_{
                    \substack{
                        k = 1 
                        \\
                        k \ne \Delta_{(z - 1)}
                    }
                }^{z - 1}
                \left(
                    Z_{k}^{\le z - 1}
                    -
                    Z_{\Delta_{(z - 1)}}^{\le z - 1}
                    +
                    \Delta_{(z - 1)}
                    -
                    k
                    -
                    1
                \right)
            }
        \right\rvert
        ^{\frac{1}{2}}
        {\scriptstyle 
            (-1)^{\Delta_{z} > \Delta_{(z - 1)}}
        }
        \allowdisplaybreaks
        \\
        &
        \hspace*{-1.5cm}
        =
        \left\lvert
            \frac{
                \displaystyle\prod_{
                    \substack{
                        k = 1 
                        \\
                        k \ne \Delta_{(z - 1)}
                    }
                }^{z - 1}
                \left(
                    -
                    X_{z - k}^{\le z - 1}
                    +
                    X_{z - \Delta_{z} + 1}^{\le z}
                    +
                    \Delta_{z}
                    -
                    k
                    -
                    1
                \right)
            }{
                \displaystyle\prod_{
                    \substack{
                        k = 1 
                        \\
                        k \ne \Delta_{z}
                    }
                }^{z}
                \left(
                    -
                    X_{z - k + 1}^{\le z}
                    +
                    X_{z - \Delta_{z} + 1}^{\le z}
                    +
                    \Delta_{z}
                    -
                    k
                \right)
            }
        \right\rvert
        ^{\frac{1}{2}}
        \left\lvert
            \frac{
                \displaystyle\prod_{
                    \substack{
                        k = 1 
                        \\
                        k \ne \Delta_{z}
                    }
                }^{z}
                \left(
                    -
                    X_{z - k + 1}^{\le z}
                    +
                    X_{z - \Delta_{(z - 1)}}^{\le z - 1}
                    +
                    \Delta_{(z - 1)}
                    -
                    k
                \right)
            }{
                \displaystyle\prod_{
                    \substack{
                        k = 1 
                        \\
                        k \ne \Delta_{(z - 1)}
                    }
                }^{z - 1}
                \left(
                    -
                    X_{z - k}^{\le z - 1}
                    +
                    X_{z - \Delta_{(z - 1)}}^{\le z - 1}
                    +
                    \Delta_{(z - 1)}
                    -
                    k
                    -
                    1
                \right)
            }
        \right\rvert
        ^{\frac{1}{2}}
        {\scriptstyle 
            (-1)^{\Delta_{z} > \Delta_{(z - 1)}}
        }
        \allowdisplaybreaks
        \\
        &
        \hspace*{-2cm}
        =
        \left\lvert
            \frac{
                \displaystyle\prod_{
                    \substack{
                        k = 1 
                        \\
                        z - k \ne \Delta_{(z - 1)}
                    }
                }^{z - 1}
                \!\!\!\!\!\!
                \left(
                    -
                    X_{k}^{\le z - 1}
                    +
                    X_{z - \Delta_{z} + 1}^{\le z}
                    +
                    \Delta_{z}
                    -
                    z
                    +
                    k
                    -
                    1
                \right)
            }{
                \displaystyle\prod_{
                    \substack{
                        k = 1 
                        \\
                        z - k + 1 \ne \Delta_{z}
                    }
                }^{z}
                \!\!\!\!\!\!
                \left(
                    -
                    X_{k}^{\le z}
                    +
                    X_{z - \Delta_{z} + 1}^{\le z}
                    +
                    \Delta_{z}
                    -
                    z 
                    + 
                    k 
                    - 
                    1
                \right)
            }
        \right\rvert
        ^{\frac{1}{2}}
        \hspace*{-.1cm}
        \left\lvert
            \frac{
                \displaystyle\prod_{
                    \substack{
                        k = 1 
                        \\
                        z - k + 1 \ne \Delta_{z}
                    }
                }^{z}
                \!\!\!\!\!\!
                \left(
                    -
                    X_{k}^{\le z}
                    +
                    X_{z - \Delta_{(z - 1)}}^{\le z - 1}
                    +
                    \Delta_{(z - 1)}
                    -
                    z 
                    + 
                    k 
                    - 
                    1
                \right)
            }{
                \displaystyle\prod_{
                    \substack{
                        k = 1 
                        \\
                        z - k \ne \Delta_{(z - 1)}
                    }
                }^{z - 1}
                \!\!\!\!\!\!
                \left(
                    -
                    X_{k}^{\le z - 1}
                    +
                    X_{z - \Delta_{(z - 1)}}^{\le z - 1}
                    +
                    \Delta_{(z - 1)}
                    -
                    z 
                    + 
                    k
                    -
                    1
                \right)
            }
        \right\rvert
        ^{\frac{1}{2}}
        {\scriptstyle 
            \hspace*{-.3cm}
            (-1)^{\Delta_{z} > \Delta_{(z - 1)}}
        }
        \allowdisplaybreaks
        \\
        &
        \hspace*{-2cm}
        =
        \left\lvert
            \frac{
                \displaystyle\prod_{
                    \substack{
                        k = 1 
                        \\
                        k \ne z - \Delta_{(z - 1)}
                    }
                }^{z - 1}
                \!\!\!\!\!\!
                \left(
                    X_{k}^{\le z - 1}
                    -
                    X_{z - \Delta_{z} + 1}^{\le z}
                    +
                    (
                        z
                        -
                        \Delta_{z}
                        +
                        1
                    )
                    -
                    k
                \right)
            }{
                \displaystyle\prod_{
                    \substack{
                        k = 1 
                        \\
                        k \ne z - \Delta_{z} + 1
                    }
                }^{z}
                \!\!\!\!\!\!
                \left(
                    X_{k}^{\le z}
                    -
                    X_{z - \Delta_{z} + 1}^{\le z}
                    +
                    (
                        z
                        -
                        \Delta_{z}
                        +
                        1
                    )
                    - 
                    k 
                \right)
            }
        \right\rvert
        ^{\frac{1}{2}}
        \hspace*{-.1cm}
        \left\lvert
            \frac{
                \displaystyle\prod_{
                    \substack{
                        k = 1 
                        \\
                        k \ne z - \Delta_{z} + 1
                    }
                }^{z}
                \!\!\!\!\!\!
                \left(
                    X_{k}^{\le z}
                    -
                    X_{z - \Delta_{(z - 1)}}^{\le z - 1}
                    +
                    (
                        z
                        -
                        \Delta_{(z - 1)}
                    )
                    - 
                    k 
                    +
                    1
                \right)
            }{
                \displaystyle\prod_{
                    \substack{
                        k = 1 
                        \\
                        z - k \ne \Delta_{(z - 1)}
                    }
                }^{z - 1}
                \!\!\!\!\!\!
                \left(
                    X_{k}^{\le z - 1}
                    -
                    X_{z - \Delta_{(z - 1)}}^{\le z - 1}
                    +
                    (
                        z
                        -
                        \Delta_{(z - 1)}
                    )
                    - 
                    k
                    +
                    1
                \right)
            }
        \right\rvert
        ^{\frac{1}{2}}
        {\scriptstyle 
            \hspace*{-.3cm}
            (-1)^{\Delta_{z} > \Delta_{(z - 1)}}
        }
        \,.
    \end{align}
    As before, we now relabel $\Delta_z \to (z - \Delta_z + 1)$ and $\Delta_{z-1} \to (z - \Delta_{z-1})$, and, up to a $\pm 1$ phase, it becomes identical to the corresponding scalar factor of the dual Clebsch--Gordan coefficient of subtracting 
    $
        (
            0, 
            \dots, 
            \Box_{\big(x - \Delta_{x} + 1\big)}, 
            \Box_{\big((x + 1) - \Delta_{(x + 1)} + 1\big)},
            \dots, 
            \Box_{\big(N - \Delta_{N} + 1\big)}
        )
    $
    from $X$.

    It now remains to handle the sign. For each scalar factor besides the first (which does not come with a sign), the attached sign 
    (after relabeling $\Delta_z \to (z - \Delta_z + 1)$ and $\Delta_{z-1} \to (z - \Delta_{z-1})$) is
    $
        (-1)^{
            (
                z 
                -
                \Delta_{z}
                +
                1
            ) 
            > 
            (
                z 
                -
                \Delta_{(z - 1)}
            ) 
        }
        =
        (-1)^{
            \Delta_{z}
            < 
            (
                \Delta_{(z - 1)}
                +
                1
            ) 
        }
        =
        (-1)^{
            \Delta_{z}
            \le 
            \Delta_{(z - 1)}
        }
    $ 
    for the (unsigned) flipped Clebsch--Gordan, but 
    $(-1)^{\Delta_{z} > \Delta_{(z - 1)}}$
    for the dual Clebsch--Gordan coefficient.
    Since the conditions
    $
        \Delta_{z}
        \le 
        \Delta_{(z - 1)}
    $
    and
    $
        \Delta_{z}
        >
        \Delta_{(z - 1)}
    $
    are mutually exclusive, they always disagree, meaning that the scalar factor for our unsigned flipped Clebsch--Gordan always has the wrong sign.

    The product of the scalar factors that makes up the Clebsch--Gordan coefficient has $N - x$ such scalar factors, which means that we are off by a phase of $(-1)^{N-x} = (-1)^{N + x}$.
    This is exactly the sign that we get by applying a phase of 
    $(-1)^{\sum_{x_i \in X} (N - x_i)}$ both times that we flip. The phase from each value $x_i$ in the tableau that is not added/removed will cancel between the flip before and after the  
    Clebsch--Gordan transform. The only phase remaining will be from the entries that are added or removed in between. 

    The Clebsch--Gordan transform of inserting $x$ either adds a positive box containing $x$, removes a negative box containing $x$, or replaces a negative box containing $x$ with $y$ and adds positive box containing $y$. In any of these cases, when performing the signed flip before and after the Clebsch--Gordan transform, the resulting phase is $(-1)^{N-x}$ as desired, completing the proof. 
\end{proof}

\begin{claim}
    Let $C$ be the Clebsch--Gordan transform implied by \Cref{lemma:cg-in-gt-basis}, and let $dC$ be the corresponding dual Clebsch--Gordan transform, given by applying $C$ on the flipped tableau as in \Cref{claim:flip-gives-dual-CG}. Then $C$ and $dC$ satisfy \Cref{lemma:vilenkin-klimyk}. Namely,
    \begin{align}
        \bra*{
            \lambda^+, 
            \lambda, 
            X 
        }
        dC 
        \ket*{
            \lambda^+, 
            X^+, 
            x
        } 
        = 
        \sqrt{
            \frac {
                \dim
                \left(
                    V_{U(N)}^{\lambda}
                \right)
            }
            {
                \dim
                \left(
                    V_{U(N)}^{\lambda^+}
                \right)
            }
        }
        \;
        \bra*{
            \lambda, 
            \lambda^+, 
            X^+
        }
        C 
        \ket*{
            \lambda, 
            X,
            x
        } 
    \end{align}
\end{claim}
This implies 
that the tableau recording update for random unitaries has an exact implementation relative to $C$ alone, and can be implemented efficiently to arbitrary accuracy using only and algorithms for the Clebsch--Gordan transform. No separate dual Clebsch–Gordan transform is required.

\begin{proof}
    We first observe that each Clebsch--Gordan coefficient and its dual Clebsch--Gordan coefficient are non-zero on the same support.
    Where they are both non-zero, we proceed by computing the ratio
    \begin{align}
        \frac
        {
            \bra*{
                \lambda, 
                \lambda^+, 
                X^+
            }
            C 
            \ket*{
                \lambda, 
                X,
                x
            } 
        }
        {
            \bra*{
                \lambda^+, 
                \lambda, 
                X 
            }
            dC 
            \ket*{
                \lambda^+, 
                X^+, 
                x
            } 
        }
    \end{align}
    by evaluating the ratios of each scalar factor individually.

    \paragraph{First scalar factor}
    The ratio of the first scalar factors of the two Clebsch--Gordan coefficients is
    \begin{align}
        &
        \left\lvert
            \frac{
                \displaystyle\prod_{k = 1}^{x - 1}
                \left(
                    X_{k}^{\le x - 1}
                    -
                    X_{\Delta_{x}}^{\le x}
                    +
                    \Delta_{x}
                    -
                    k
                    -
                    1
                \right)
            }{
                \displaystyle\prod_{
                    \substack{
                        k = 1 
                        \\
                        k \ne \Delta_{x}
                    }
                }^{x}
                \left(
                    X_{k}^{\le x}
                    -
                    X_{\Delta_{x}}^{\le x}
                    +
                    \Delta_{x}
                    -
                    k
                \right)
            }
        \right\rvert
        ^{\frac{1}{2}}
        \left\lvert
            \frac
            {
                \displaystyle\prod_{
                    \substack{
                        k = 1 
                        \\
                        k \ne \Delta_{x}
                    }
                }^{x}
                \left(
                    (X^+)_{k}^{\le x}
                    -
                    (X^+)_{\Delta_{x}}^{\le x}
                    +
                    \Delta_{x}
                    -
                    k
                \right)
            }
            {
                \displaystyle\prod_{k = 1}^{x - 1}
                \left(
                    (X^+)_{k}^{\le x - 1}
                    -
                    (X^+)_{\Delta_{x}}^{\le x}
                    +
                    \Delta_{x}
                    -
                    k
                \right)
            }
        \right\rvert
        ^{\frac{1}{2}}
        \allowdisplaybreaks
        \\
        &
        =
        \left\lvert
            \frac{
                \displaystyle\prod_{k = 1}^{x - 1}
                \left(
                    X_{k}^{\le x - 1}
                    -
                    X_{\Delta_{x}}^{\le x}
                    +
                    \Delta_{x}
                    -
                    k
                    -
                    1
                \right)
            }{
                \displaystyle\prod_{
                    \substack{
                        k = 1 
                        \\
                        k \ne \Delta_{x}
                    }
                }^{x}
                \left(
                    X_{k}^{\le x}
                    -
                    X_{\Delta_{x}}^{\le x}
                    +
                    \Delta_{x}
                    -
                    k
                \right)
            }
        \right\rvert
        ^{\frac{1}{2}}
        \left\lvert
            \frac
            {
                \displaystyle\prod_{
                    \substack{
                        k = 1 
                        \\
                        k \ne \Delta_{x}
                    }
                }^{x}
                \left(
                    X_{k}^{\le x}
                    -
                    X_{\Delta_{x}}^{\le x}
                    +
                    \Delta_{x}
                    -
                    k
                    -
                    1
                \right)
            }
            {
                \displaystyle\prod_{k = 1}^{x - 1}
                \left(
                    X_{k}^{\le x - 1}
                    -
                    X_{\Delta_{x}}^{\le x}
                    +
                    \Delta_{x}
                    -
                    k
                    -
                    1
                \right)
            }
        \right\rvert
        ^{\frac{1}{2}}
        \allowdisplaybreaks
        \\
        &
        =
        \left\lvert
            \frac
            {
                \displaystyle\prod_{
                    \substack{
                        k = 1 
                        \\
                        k \ne \Delta_{x}
                    }
                }^{x}
                \left(
                    X_{k}^{\le x}
                    -
                    X_{\Delta_{x}}^{\le x}
                    +
                    \Delta_{x}
                    -
                    k
                    -
                    1
                \right)
            }
            {
                \displaystyle\prod_{
                    \substack{
                        k = 1 
                        \\
                        k \ne \Delta_{x}
                    }
                }^{x}
                \left(
                    X_{k}^{\le x}
                    -
                    X_{\Delta_{x}}^{\le x}
                    +
                    \Delta_{x}
                    -
                    k
                \right)
            }
        \right\rvert
        ^{\frac{1}{2}}
        \allowdisplaybreaks
        \\
        &
        =
        \left\lvert
            \displaystyle\prod_{
                \substack{
                    k = 1 
                    \\
                    k \ne \Delta_{x}
                }
            }^{x}
            \frac
            {
                X_{k}^{\le x}
                -
                X_{\Delta_{x}}^{\le x}
                +
                \Delta_{x}
                -
                k
                -
                1
            }
            {
                X_{k}^{\le x}
                -
                X_{\Delta_{x}}^{\le x}
                +
                \Delta_{x}
                -
                k
            }
        \right\rvert
        ^{\frac{1}{2}}
    \end{align}

    \paragraph{Remaining scalar factors}
    The ratios of the remaining scalar factors of the two Clebsch--Gordan coefficients are
    \begin{align}
        &
        \hspace*{-1cm}
        \left\lvert
            \frac{
                \displaystyle\prod_{
                    \substack{
                        k = 1 
                        \\
                        k \ne \Delta_{z - 1}
                    }
                }^{z - 1}
                \left(
                    X_{k}^{\le z - 1}
                    -
                    X_{\Delta_{z}}^{\le z}
                    +
                    \Delta_{z}
                    -
                    k
                    -
                    1
                \right)
            }{
                \displaystyle\prod_{
                    \substack{
                        k = 1 
                        \\
                        k \ne \Delta_{z}
                    }
                }^{z}
                \left(
                    X_{k}^{\le z}
                    -
                    X_{\Delta_{z}}^{\le z}
                    +
                    \Delta_{z}
                    -
                    k
                \right)
            }
        \right\rvert
        ^{\frac{1}{2}}
        \left\lvert
            \frac{
                \displaystyle\prod_{
                    \substack{
                        k = 1 
                        \\
                        k \ne \Delta_{z}
                    }
                }^{z}
                \left(
                    X_{k}^{\le z}
                    -
                    X_{\Delta_{z - 1}}^{\le z - 1}
                    +
                    \Delta_{z - 1}
                    -
                    k
                \right)
            }{
                \displaystyle\prod_{
                    \substack{
                        k = 1 
                        \\
                        k \ne \Delta_{z - 1}
                    }
                }^{z - 1}
                \left(
                    X_{k}^{\le z - 1}
                    -
                    X_{\Delta_{z - 1}}^{\le z - 1}
                    +
                    \Delta_{z - 1}
                    -
                    k
                    -
                    1
                \right)
            }
        \right\rvert
        ^{\frac{1}{2}}
        (-1)^{\Delta_{z} > \Delta_{z - 1}}
        \allowdisplaybreaks
        \\
        \cdot
        &
        \left\lvert
            \frac
            {
                \displaystyle\prod_{
                    \substack{
                        k = 1 
                        \\
                        k \ne \Delta_{z}
                    }
                }^{z}
                \left(
                    (X^+)_{k}^{\le z}
                    -
                    (X^+)_{\Delta_{z}}^{\le z}
                    +
                    \Delta_{z}
                    -
                    k
                \right)
            }
            {
                \displaystyle\prod_{
                    \substack{
                        k = 1 
                        \\
                        k \ne \Delta_{z - 1}
                    }
                }^{z - 1}
                \left(
                    (X^+)_{k}^{\le z - 1}
                    -
                    (X^+)_{\Delta_{z}}^{\le z}
                    +
                    \Delta_{z}
                    -
                    k
                \right)
            }
        \right\rvert
        ^{\frac{1}{2}}
        \left\lvert
            \frac
            {
                \displaystyle\prod_{
                    \substack{
                        k = 1 
                        \\
                        k \ne \Delta_{z - 1}
                    }
                }^{z - 1}
                \left(
                    (X^+)_{k}^{\le z - 1}
                    -
                    (X^+)_{\Delta_{z - 1}}^{\le z - 1}
                    +
                    \Delta_{z - 1}
                    -
                    k
                    +
                    1
                \right)
            }
            {
                \displaystyle\prod_{
                    \substack{
                        k = 1 
                        \\
                        k \ne \Delta_{z}
                    }
                }^{z}
                \left(
                    (X^+)_{k}^{\le z}
                    -
                    (X^+)_{\Delta_{z - 1}}^{\le z - 1}
                    +
                    \Delta_{z - 1}
                    -
                    k
                    +
                    1
                \right)
            }
        \right\rvert
        ^{\frac{1}{2}}
        (-1)^{\Delta_{z} > \Delta_{z - 1}}
        \allowdisplaybreaks
        \\
        =
        &
        \left\lvert
            \frac{
                \displaystyle\prod_{
                    \substack{
                        k = 1 
                        \\
                        k \ne \Delta_{z - 1}
                    }
                }^{z - 1}
                \left(
                    X_{k}^{\le z - 1}
                    -
                    X_{\Delta_{z}}^{\le z}
                    +
                    \Delta_{z}
                    -
                    k
                    -
                    1
                \right)
            }{
                \displaystyle\prod_{
                    \substack{
                        k = 1 
                        \\
                        k \ne \Delta_{z}
                    }
                }^{z}
                \left(
                    X_{k}^{\le z}
                    -
                    X_{\Delta_{z}}^{\le z}
                    +
                    \Delta_{z}
                    -
                    k
                \right)
            }
        \right\rvert
        ^{\frac{1}{2}}
        \left\lvert
            \frac{
                \displaystyle\prod_{
                    \substack{
                        k = 1 
                        \\
                        k \ne \Delta_{z}
                    }
                }^{z}
                \left(
                    X_{k}^{\le z}
                    -
                    X_{\Delta_{z - 1}}^{\le z - 1}
                    +
                    \Delta_{z - 1}
                    -
                    k
                \right)
            }{
                \displaystyle\prod_{
                    \substack{
                        k = 1 
                        \\
                        k \ne \Delta_{z - 1}
                    }
                }^{z - 1}
                \left(
                    X_{k}^{\le z - 1}
                    -
                    X_{\Delta_{z - 1}}^{\le z - 1}
                    +
                    \Delta_{z - 1}
                    -
                    k
                    -
                    1
                \right)
            }
        \right\rvert
        ^{\frac{1}{2}}
        \allowdisplaybreaks
        \\
        \cdot
        &
        \left\lvert
            \frac
            {
                \displaystyle\prod_{
                    \substack{
                        k = 1 
                        \\
                        k \ne \Delta_{z}
                    }
                }^{z}
                \left(
                    X_{k}^{\le z}
                    -
                    X_{\Delta_{z}}^{\le z}
                    +
                    \Delta_{z}
                    -
                    k
                    -
                    1
                \right)
            }
            {
                \displaystyle\prod_{
                    \substack{
                        k = 1 
                        \\
                        k \ne \Delta_{z - 1}
                    }
                }^{z - 1}
                \left(
                    X_{k}^{\le z - 1}
                    -
                    X_{\Delta_{z}}^{\le z}
                    +
                    \Delta_{z}
                    -
                    k
                    -
                    1
                \right)
            }
        \right\rvert
        ^{\frac{1}{2}}
        \left\lvert
            \frac
            {
                \displaystyle\prod_{
                    \substack{
                        k = 1 
                        \\
                        k \ne \Delta_{z - 1}
                    }
                }^{z - 1}
                \left(
                    X_{k}^{\le z - 1}
                    -
                    X_{\Delta_{z - 1}}^{\le z - 1}
                    +
                    \Delta_{z - 1}
                    -
                    k
                \right)
            }
            {
                \displaystyle\prod_{
                    \substack{
                        k = 1 
                        \\
                        k \ne \Delta_{z}
                    }
                }^{z}
                \left(
                    X_{k}^{\le z}
                    -
                    X_{\Delta_{z - 1}}^{\le z - 1}
                    +
                    \Delta_{z - 1}
                    -
                    k
                \right)
            }
        \right\rvert
        ^{\frac{1}{2}}
        \allowdisplaybreaks
        \\
        =
        &
        \left\lvert
            \frac
            {
                \displaystyle\prod_{
                    \substack{
                        k = 1 
                        \\
                        k \ne \Delta_{z}
                    }
                }^{z}
                \left(
                    X_{k}^{\le z}
                    -
                    X_{\Delta_{z}}^{\le z}
                    +
                    \Delta_{z}
                    -
                    k
                    -
                    1
                \right)
            }
            {
                \displaystyle\prod_{
                    \substack{
                        k = 1 
                        \\
                        k \ne \Delta_{z}
                    }
                }^{z}
                \left(
                    X_{k}^{\le z}
                    -
                    X_{\Delta_{z}}^{\le z}
                    +
                    \Delta_{z}
                    -
                    k
                \right)
            }
        \right\rvert
        ^{\frac{1}{2}}
        \left\lvert
            \frac
            {
                \displaystyle\prod_{
                    \substack{
                        k = 1 
                        \\
                        k \ne \Delta_{z - 1}
                    }
                }^{z - 1}
                \left(
                    X_{k}^{\le z - 1}
                    -
                    X_{\Delta_{z - 1}}^{\le z - 1}
                    +
                    \Delta_{z - 1}
                    -
                    k
                \right)
            }
            {
                \displaystyle\prod_{
                    \substack{
                        k = 1 
                        \\
                        k \ne \Delta_{z - 1}
                    }
                }^{z - 1}
                \left(
                    X_{k}^{\le z - 1}
                    -
                    X_{\Delta_{z - 1}}^{\le z - 1}
                    +
                    \Delta_{z - 1}
                    -
                    k
                    -
                    1
                \right)
            }
        \right\rvert
        ^{\frac{1}{2}}
        \allowdisplaybreaks
        \\
        =
        &
        \left\lvert
            \displaystyle\prod_{
                \substack{
                    k = 1 
                    \\
                    k \ne \Delta_{z}
                }
            }^{z}
            \frac
            {
                X_{k}^{\le z}
                -
                X_{\Delta_{z}}^{\le z}
                +
                \Delta_{z}
                -
                k
                -
                1
            }
            {
                X_{k}^{\le z}
                -
                X_{\Delta_{z}}^{\le z}
                +
                \Delta_{z}
                -
                k
            }
        \right\rvert
        ^{\frac{1}{2}}
        \left\lvert
            \displaystyle\prod_{
                \substack{
                    k = 1 
                    \\
                    k \ne \Delta_{z - 1}
                }
            }^{z - 1}
            \frac
            {
                    X_{k}^{\le z - 1}
                    -
                    X_{\Delta_{z - 1}}^{\le z - 1}
                    +
                    \Delta_{z - 1}
                    -
                    k
            }
            {
                    X_{k}^{\le z - 1}
                    -
                    X_{\Delta_{z - 1}}^{\le z - 1}
                    +
                    \Delta_{z - 1}
                    -
                    k
                    -
                    1
            }
        \right\rvert
        ^{\frac{1}{2}}
    \end{align}

    \paragraph{Putting them together}
    We now combine these to get the ratio of the Clebsch--Gordan coefficients as the product of the ratios of the scalar factors:
    \begin{align}
        &
        \hspace*{-.5cm}
        \frac
        {
            \bra*{
                \lambda, 
                \lambda^+, 
                X^+
            }
            C 
            \ket*{
                \lambda, 
                X,
                x
            } 
        }
        {
            \bra*{
                \lambda^+, 
                \lambda, 
                X 
            }
            dC 
            \ket*{
                \lambda^+, 
                X^+, 
                x
            } 
        }
        \\
        =
        &
        \left\lvert
            \displaystyle\prod_{
                \substack{
                    k = 1 
                    \\
                    k \ne \Delta_{x}
                }
            }^{x}
            \frac
            {
                X_{k}^{\le x}
                -
                X_{\Delta_{x}}^{\le x}
                +
                \Delta_{x}
                -
                k
                -
                1
            }
            {
                X_{k}^{\le x}
                -
                X_{\Delta_{x}}^{\le x}
                +
                \Delta_{x}
                -
                k
            }
        \right\rvert
        ^{\frac{1}{2}}
        \\
        \cdot
        &
        \prod_{z = x + 1}^{N}
        \left\lvert
            \displaystyle\prod_{
                \substack{
                    k = 1 
                    \\
                    k \ne \Delta_{z}
                }
            }^{z}
            \frac
            {
                X_{k}^{\le z}
                -
                X_{\Delta_{z}}^{\le z}
                +
                \Delta_{z}
                -
                k
                -
                1
            }
            {
                X_{k}^{\le z}
                -
                X_{\Delta_{z}}^{\le z}
                +
                \Delta_{z}
                -
                k
            }
        \right\rvert
        ^{\frac{1}{2}}
        \left\lvert
            \displaystyle\prod_{
                \substack{
                    k = 1 
                    \\
                    k \ne \Delta_{z - 1}
                }
            }^{z - 1}
            \frac
            {
                X_{k}^{\le z - 1}
                -
                X_{\Delta_{z - 1}}^{\le z - 1}
                +
                \Delta_{z - 1}
                -
                k
            }
            {
                X_{k}^{\le z - 1}
                -
                X_{\Delta_{z - 1}}^{\le z - 1}
                +
                \Delta_{z - 1}
                -
                k
                -
                1
            }
        \right\rvert
        ^{\frac{1}{2}}
    \end{align}
    This is a telescoping product, since in each factor of the outside product, the second part cancels with the first part of the previous factor, ultimately giving
    \begin{align}
        \allowdisplaybreaks
        =
        &
        \left\lvert
            \displaystyle\prod_{
                \substack{
                    k = 1 
                    \\
                    k \ne \Delta_{N}
                }
            }^{N}
            \frac
            {
                X_{k}^{\le N}
                -
                X_{\Delta_{N}}^{\le N}
                +
                \Delta_{N}
                -
                k
                -
                1
            }
            {
                X_{k}^{\le N}
                -
                X_{\Delta_{N}}^{\le N}
                +
                \Delta_{N}
                -
                k
            }
        \right\rvert
        ^{\frac{1}{2}}
        \allowdisplaybreaks
        \\
        =
        &
        \left\lvert
            \displaystyle\prod_{
                \substack{
                    k = 1 
                    \\
                    k \ne \Delta_{N}
                }
            }^{N}
            \frac
            {
                \lambda_{k}
                -
                (
                    \lambda_{\Delta_{N}}
                    +
                    1
                )
                +
                \Delta_{N}
                -
                k
            }
            {
                \lambda_{k}
                -
                \lambda_{\Delta_{N}}
                +
                \Delta_{N}
                -
                k
            }
        \right\rvert
        ^{\frac{1}{2}}
        \allowdisplaybreaks
        \\
        =
        &
        \left\lvert
            \displaystyle\prod_{
                \substack{
                    k = 1 
                    \\
                    k \ne \Delta_{N}
                }
            }^{N}
            \frac
            {
                \lambda^+_{k}
                -
                \lambda^+_{\Delta_{N}}
                +
                \Delta_{N}
                -
                k
            }
            {
                \lambda_{k}
                -
                \lambda_{\Delta_{N}}
                +
                \Delta_{N}
                -
                k
            }
        \right\rvert
        ^{\frac{1}{2}}
        \allowdisplaybreaks
        \\
        =
        &
        \left\lvert
            \prod_{
                \substack{
                    k = 1
                    \\
                    k \ne \Delta_{N}
                }
            }^{N}
            \frac
            {
                (\lambda^+_k - k)
                -
                (\lambda^+_{\Delta_{N}} - {\Delta_{N}})
            }
            {
                (\lambda_k - k)
                -
                (\lambda_{\Delta_{N}} - {\Delta_{N}})
            }
        \right\rvert
        ^{\frac{1}{2}}
        \allowdisplaybreaks
        \\
        =
        &
        \left\lvert
            \prod_{i = 1}^{\Delta_{N} - 1}
            \frac
            {
                (\lambda^+_i - i)
                -
                (\lambda^+_{\Delta_{N}} - {\Delta_{N}})
            }
            {
                (\lambda_i - i)
                -
                (\lambda_{\Delta_{N}} - {\Delta_{N}})
            }
        \right\rvert
        ^{\frac{1}{2}}
        \left\lvert
            \prod_{j = \Delta_{N} + 1}^N
            \frac
            {
                (\lambda^+_{\Delta_{N}} - {\Delta_{N}})
                -
                (\lambda^+_j - j)
            }
            {
                (\lambda_{\Delta_{N}} - {\Delta_{N}})
                -
                (\lambda_j - j)
            }
        \right\rvert
        ^{\frac{1}{2}}
        \cdot
        \underbrace{
            \left\lvert
                \prod_{
                    \substack{
                        i < j
                        \\
                        i \ne \Delta_{N}
                        \\
                        j \ne \Delta_{N}
                    }
                }
                \frac
                {
                    (\lambda^+_i - i)
                    -
                    (\lambda^+_j - j)
                }
                {
                    (\lambda_i - i)
                    -
                    (\lambda_j - j)
                }
            \right\rvert
            ^{\frac{1}{2}}
        }_{
            = 1
            \,
            \text{
                since $\lambda$ and $\lambda^+$ only differ on row $\Delta_{N}$
            }
        }
        \allowdisplaybreaks
        \\
        =
        &
        \left\lvert
            \prod_{i = 1}^N
            \;
            \prod_{j = i+1}^N
            \frac
            {
                (\lambda^+_i - i)
                -
                (\lambda^+_j - j)
            }
            {
                (\lambda_i - i)
                -
                (\lambda_j - j)
            }
        \right\rvert
        ^{\frac{1}{2}}
        \allowdisplaybreaks
        \\
        =
        &
        \left\lvert
            \frac
            {
                \dim
                \left(
                    V_{U(N)}^{\lambda^+}
                \right)
            }
            {
                \dim
                \left(
                    V_{U(N)}^{\lambda}
                \right)
            }
        \right\rvert
        ^{\frac{1}{2}}
        \,.
        \qedhere
    \end{align}
\end{proof}

\section{The Path Recording Oracle}
\label{sec:path}

The tableau-recording oracle gives an exact simulation of queries to a Haar-random group representation by storing the purification in the Fourier basis of $L^2(G)$. Unfortunately, the tableau-recording oracle can be challenging to interpret
operationally: from the contents of the tableau register, it can be difficult to characterize in simple terms what information about the queried representation has been learned or fixed by the algorithm.

The goal of this section is to construct an equivalent recording in a path basis. Informally, after
$t$ queries, we would like the recording register to store a superposition of Feynman path information $\ket{x_1,\ldots,x_t,y_1,\hdots,y_t}$.
For a fixed group element $g$, this path corresponds to the matrix element
\[
    \prod_{i=1}^t \langle y_i|\rho(g)|x_i\rangle
    =
    \langle y_1,\ldots,y_t|
        \rho(g)^{\otimes t}
    |x_1,\ldots,x_t\rangle,
\]
and indeed, a non-adaptive algorithm that queries the uncompressed oracle $Q_\rho$ on inputs $\ket{x_1, \hdots, x_t}$ would result in the state 
\[ \sum_{y_1, \hdots, y_t} \ket{y_1, \hdots, y_t} \otimes \int_G \bra{\vec y} \rho^{\otimes t}\ket{\vec x} \cdot \ket{g} dg.
\]
How does this compare with recording the path $\ket{\vec x, \vec y}$? To make this precise, we introduce the matrix element map
\[
    \Theta_t:
    V^{\otimes t}\otimes \overline V^{\otimes t}
    \longrightarrow
    L^2(G)
\]
defined as
\begin{equation}
    \Theta_t|\vec x,\vec y\rangle
    :=
    \int_G
    \langle \vec y|\rho(g)^{\otimes t}|\vec x\rangle
    |g\rangle\,dg,
\end{equation}
so that 
\begin{equation} Q_\rho^{\otimes t} \ket{\vec x} = \sum_{\vec y \in \mathcal B(V)} \ket{\vec y} \otimes \Theta_t(\ket{\vec x}\ket{\vec y}).
\end{equation}
Using this, we can see that for any particular representation $\rho$, path states contain some redundancies. Let
\[
    \mathcal A_t=\End_G(V^{\otimes t})
\]
be the commutant algebra of the tensor-power representation. As discussed in the technical overview, since every \(A\in \mathcal A_t\) commutes
with \(\rho(g)^{\otimes t}\), the matrix-element map satisfies
\[
    \Theta_t(A\otimes \Id)
    =
    \Theta_t(\Id\otimes A^T).
\]
Therefore, it should be the case that $(A\otimes \Id)\ket{\psi} =(\Id\otimes A^T) \ket{\psi}$ on valid path-recording states. This identification is implemented by the commutant EPR projector \(\Omega_{\mathcal A_t}\).

Furthermore, path states may require different relative normalizations. These factors are encoded in the reweighting operator \(\Lambda_{\mathcal A_t}\) defined later.

With these ingredients, the path-recording update has the following form: append a new
input-output pair \((x_t,y_t)\), project the recording register onto the commutant-EPR subspace $\Omega_{\mathcal A_t}$,
and apply an appropriate Schur-Weyl subspace reweighting. The remainder of this section defines these
operators precisely, proves that the resulting path-recording oracle is isometric to the tableau
recording oracle, gives an interpretation of $\Omega_{\mathcal A_t}$ in terms of the algebra $\mathcal A_t$ (when it is semisimple), and derives forms of the update that are useful for adaptive query algorithms.

\subsection{Defining the Path Recording Oracle}
\label{sec:general-path-rec-update}
Let
\[
    \rho:G\to U(V)
\]
be a finite-dimensional unitary representation of a compact Lie group $G$. Fix an orthonormal basis $\calB(V) = \{\ket{x}\}$ of $V$. We use the same ket notation $\ket y$ for the corresponding conjugate basis of $\overline V$, specifying registers when needed.

\noindent For each $t$, let
\[
    \schur_t:
    V^{\otimes t}
    \longrightarrow
    \bigoplus_{\lambda}
    V_G^\lambda\otimes V_{\calA_t}^{\lambda}
\]
be the generalized Schur transform for $V^{\otimes t}$, so that
\[
    \rho^{\otimes t}(g)
    =
    \bigoplus_{\lambda}
    \rho_\lambda(g)\otimes \Id_{V_{\calA_t}^{\lambda}}.
\]
Define the ordered path space
\[
    \mathsf{Path}_t
    \coloneqq
    V^{\otimes t}\otimes\overline V^{\otimes t}.
\]
A path basis vector is written
\[
    \ket{\vec x,\vec y}
    \coloneqq
    \ket{x_1,\ldots,x_t}
    \otimes
    \ket{y_1,\ldots,y_t}.
\]
Note that $\ket{y_1, \ldots, y_t}$ implicitly corresponds to the dual basis element $\ket{y_1^*, \ldots, y_t^*} \in \overline V^{\otimes t}$. Next, we define the \emph{append} map
\[
    \label{eq:append_map}
    \App^{(t)}_{x,y}:
    \mathsf{Path}_t
    \longrightarrow
    \mathsf{Path}_{t+1}
\]
by
\begin{equation}
    \App^{(t)}_{x,y}
    \ket{\vec x,\vec y}
    =
    \ket{x_1,\ldots,x_t,x}
    \otimes
    \ket{y_1,\ldots,y_t,y}.
\end{equation}
Equivalently,
\[
    \App^{(t)}_{x,y}
    =
    \Id\otimes\ket{x,y}.
\]

We relate the path space $V^{\otimes t} \otimes \overline V^{\otimes t}$ to the (Fourier/tableau) compressed oracle via the double Schur transform
\[
    \schur_t\otimes\overline{\schur_t}:
    V^{\otimes t}\otimes\overline V^{\otimes t}
    \longrightarrow
    \bigoplus_{\lambda,\lambda'}
    \left(
        V_G^\lambda\otimes V_{\calA_t}^{\lambda}
    \right)
    \otimes
    \overline{
        \left(
            V_G^{\lambda'}\otimes V_{\calA_t}^{\lambda'}
        \right)
    }.
\]
We write double-Schur basis vectors as
\[
    \ket{\lambda,X,T}
    \otimes
    \ket{\lambda',Y,R},
\]
where
\[
    X\in\calB(V_G^\lambda),
    \qquad
    Y\in\calB(V_G^{\lambda'}),
    \qquad
    T\in\calB(V_{\calA_t}^{\lambda}),
    \qquad
    R\in\calB(V_{\calA_t}^{\lambda'}).
\]
This is again shorthand for the vector $\ket{\lambda,X,T}
    \otimes
    \overline{\ket{\lambda',Y,R}}$. 

Let $\widehat\Pi_{\lambda,t}$ be the projector in the Schur basis onto block $\lambda$, that is, onto $ V_G^\lambda\otimes V_{\calA_t}^{\lambda}$. We then define its path basis version as
\begin{equation}
    \Pi_{\lambda,t}
    := \schur_t^\dagger \cdot \widehat\Pi_{\lambda,t} \cdot\schur_t.
\end{equation}

In a slight abuse of notation, we use $\widehat\Pi_{\lambda,t}^{\otimes 2}$ to denote $\widehat\Pi_{\lambda,t} \otimes \overline{\widehat\Pi_{\lambda,t}}$, which is the projector in the
double-Schur basis onto the $(\lambda, \lambda)$ block
\[
    \left(
        V_G^\lambda\otimes V_{\calA_t}^{\lambda}
    \right)
    \otimes
    \overline{
        \left(
            V_G^\lambda\otimes V_{\calA_t}^{\lambda}
        \right)
    }.
\]
Equivalently,
\[
\begin{aligned}
    \widehat\Pi_{\lambda,t}^{\otimes 2}
    \left(
        \ket{\lambda_1,X,T}
        \otimes
        \ket{\lambda_2,Y,R}
    \right)
    =
    \begin{cases}
        \ket{\lambda,X,T}\otimes\ket{\lambda,Y,R},
        & \lambda_1=\lambda_2=\lambda,\\
        0,
        & \text{otherwise.}
    \end{cases}
\end{aligned}
\]
The corresponding path basis projector is then
\[
    \Pi_{\lambda,t}^{\otimes 2}
    :=
    \left(
        \schur_t^\dagger
        \otimes
        \schur_t^\top
    \right)
    \widehat\Pi_{\lambda,t}^{\otimes 2}
    \left(
        \schur_t
        \otimes
        \overline{\schur_t}
    \right).
\]

For each $\lambda$, define the normalized EPR state on the pair of
$\calA_t$-registers by
\[
    \ket{\mathsf{EPR}_{\calA_t}^{\lambda}}
    :=
    \frac{1}{\sqrt{\dim(V_{\calA_t}^{\lambda})}}
    \sum_{T\in\calB(V_{\calA_t}^{\lambda})}
    \ket T\otimes\ket T. 
\]

Let $\widehat{\Pi}^{\calA_t}_{\mathrm{EPR},t}$ be the double-Schur-basis
projector that acts as zero on blocks $\lambda\neq\lambda'$, and on the
$\lambda=\lambda'$ block projects the two $\calA_t$-registers onto
$\ket{\mathsf{EPR}_{\calA_t}^{\lambda}}$ while acting as identity on the
two $G$-registers. Equivalently,
\[
\begin{aligned}
    \widehat{\Pi}^{\calA_t}_{\mathrm{EPR},t}
    \left(
        \ket{\lambda,X,T}
        \otimes
        \ket{\lambda',Y,R}
    \right)
    =
    \begin{cases}
    \displaystyle
    \frac{\delta_{T,R}}{\dim(V_{\calA_t}^{\lambda})}
    \sum_{S\in\calB(V_{\calA_t}^{\lambda})}
    \ket{\lambda,X,S}
    \otimes
    \ket{\lambda,Y,S},
    & \lambda=\lambda',\\[2ex]
    0,
    & \lambda\neq\lambda'.
    \end{cases}
\end{aligned}
\]
Via the double Schur transform, this induces the following projector on the path space.
\begin{equation}
    \Omega_{\calA_t}
    :=
    \left(
        \schur_t^\dagger
        \otimes
        \schur_t^\top
    \right)
    \widehat{\Pi}^{\calA_t}_{\mathrm{EPR},t}
    \left(
        \schur_t
        \otimes
        \overline{\schur_t}
    \right).
\end{equation}
Finally, define the subspace-reweighting operator
\begin{equation}
    W_t
    :=
    \sum_\lambda
    \sqrt{
        \frac{
            \dim(V_{\calA_t}^{\lambda})
        }{
            \dim(V_G^\lambda)
        }
    }
    \Pi_{\lambda,t}^{\otimes 2}.
\end{equation}
This operator reweights each double-Schur subspace with matching irrep label $\lambda$ on the two sides. 

On the equal-irrep subspace
\[
    \Pi_{\mathrm{diag},t}
    :=
    \sum_\lambda
    \Pi_{\lambda,t}^{\otimes 2},
\]
the inverse reweighting is
\begin{equation}
    W_t^{+}
    :=
    \sum_{\lambda}
    \sqrt{
        \frac{
            \dim(V_G^\lambda)
        }{
            \dim(V_{\calA_t}^{\lambda})
        }
    }
    \Pi_{\lambda,t}^{\otimes 2}.
\end{equation}
Thus
\[
    W_t^{+}W_t
    =
    W_tW_t^{+}
    =
    \Pi_{\mathrm{diag},t}.
\]
Since
\[
    \Omega_{\calA_t}
    \preceq
    \Pi_{\mathrm{diag},t},
\]
this is a valid inverse restricted to the legal path subspace. 

\subsection{Path Recording Theorem Statement}

With these definitions set up, we are ready to state and analyze our path recording oracle. 

\begin{definition}[Path Recording Oracle]\label{def:path-recording}
    The path recording oracle is defined to be the following linear operator. The input space of the operator is $V \otimes \bigoplus_t \mathsf{Path}_t$.

    On subspace $V\otimes \mathsf{Path}_t$, the operator is defined as 

    \begin{equation}
    Q_{\rho,t+1}^{\mathrm{Path}}
    =
    \sum_{x,y\in\calB(V)}
    \ketbra{y}{x}
    \otimes
    W_{t+1} \cdot 
    \Omega_{\calA_{t+1}} \cdot 
    \App^{(t)}_{x,y} \cdot W_t^{+} \cdot 
    \Omega_{\calA_t},
\end{equation}
which outputs an element of $V \otimes \mathsf{Path}_{t+1}$. 

When restricted to the legal path-recording subspace $\Omega_{\calA_t}$, the operator is given by
\begin{equation}
    Q_{\rho,t+1}^{\mathrm{Path}}
    =
    \sum_{x,y\in\calB(V)}
    \ketbra{y}{x}
    \otimes
    W_{t+1}
    \Omega_{\calA_{t+1}}
    \App^{(t)}_{x,y}
    W_t^{+}.
\end{equation}
\end{definition}

\begin{remark}[Equivalent definition]\label{def:path-recording-equivalent}
    Another definition of the path recording oracle is the following:
    \begin{equation}
        Q_{\rho,t+1}^{\mathrm{Path}}
    =
    \sum_{x,y\in\calB(V)}
    \ketbra{y}{x}
    \otimes
    \Lambda_{\calA_{t+1}} \cdot 
    \Omega_{\calA_{t+1}} \cdot 
    \App^{(t)}_{x,y} \cdot \Lambda_{\calA_t}^{+} \cdot 
    \Omega_{\calA_t},
    \end{equation}
    where
    \begin{equation} \Lambda_{\calA_t} = \sum_\lambda
    \sqrt{
        \frac{
            \dim(V_{\calA_t}^{\lambda})
        }{
            \dim(V_G^\lambda)
        }
    }
    \cdot \Big(\Pi_{\lambda,t} \otimes \Id\Big)
    \end{equation}
    and $\Lambda_{\calA_t}^+$ is inverse to $\Lambda_{\calA_t}$ on $\im(\schur_t)$. This is equivalent because $\Omega_{\mathcal A_t}$ in particular projects onto $\Pi_{\mathrm{diag},t}$, on which $\Lambda_{\calA_t}$ and $W_t$ are equivalent (as are their pseudoinverses). 
\end{remark}

We will prove that $Q_{\rho,t+1}^{\mathrm{Path}}$ is equivalent to $Q_\rho$ on an appropriate subspace of legal ``reachable'' states. Let
\[
    \widehat{G}_t
    :=
    \mathrm{Irr}(\rho^{\otimes t})
    =
    \left\{
        \lambda\in\widehat{G}:
        \dim(V_{\calA_t}^{\lambda})>0
    \right\}
\]
be the set of irreps appearing in the Schur--Weyl decomposition of
$V^{\otimes t}$, and let
\[
    \mathsf{Tab}_t
    :=
    \bigoplus_{\lambda\in\widehat{G}_t}
    V_G^\lambda\otimes \overline{V_G^\lambda}
\]
be the $t$-query reachable tableau space. 
\begin{theorem}[Correctness of the path-recording oracle]\label{thm:path-recording}
There is a sequence of partial isometries $\Gamma_t: \bigoplus_\lambda V^\lambda_G \otimes \overline V^\lambda_G \rightarrow \mathsf{Path}_t$ such that the following holds.  For every $t$, 

\begin{equation} Q_{\rho,t+1}^{\mathrm{Path}} \cdot (\Id_V \otimes \Gamma_t \cdot F )= (\Id_V \otimes \Gamma_{t+1} \cdot F ) \cdot Q_\rho,
\end{equation}
when inputs are restricted to $V\otimes F^{-1}(\mathsf{Tab}_t)$. 

Therefore, for every $t$-query quantum algorithm $A$ whose query register is $V$, for all input states $\ket{\psi}$, we have that 
\[ A^{Q_\rho}(\ket{\psi}) \simeq A^{Q_\rho^{\mathrm{Path}}}(\ket{\psi})
\]
via a fixed partial isometry $\Gamma_t \cdot F$ on the recording register. 
\end{theorem}

\paragraph{Setup} With this notation, the path-recording update is given by
\[
    Q_{\rho,t+1}^{\mathrm{Path}}
    =
    \sum_{x,y\in\calB(V)}
    \ketbra{y}{x}
    \otimes
    W_{t+1}
    \Omega_{\calA_{t+1}}
    \App^{(t)}_{x,y}
    W_t^{+}\Omega_{\calA_t}.
\]
We will prove
\[
    Q_{\rho,t+1}^{\mathrm{Path}}
    (\Id_V\otimes \Gamma_t F)
    =
    (\Id_V\otimes \Gamma_{t+1} F)
    Q_\rho
\]
on the appropriately restricted input states. 

\subsection{Defining the partial isometry $\Gamma_t$}
To begin proving \cref{thm:path-recording}, we first define the partial isometries
\[
    \Gamma_t:
    \bigoplus_\lambda
    V_G^\lambda\otimes \overline{V_G^\lambda}
    \longrightarrow
    \mathsf{Path}_t.
\]
For every irrep $\lambda$, we write the basis
states of $V_G^\lambda \otimes \overline V_G^\lambda$ as $\ket{\lambda,X,Y}$. 

With this setup, we have that

\[
    \im\Big(\schur_t\otimes\overline{\schur_t} \Big) = 
    \bigoplus_{\lambda,\lambda'\in\widehat{G}_t}
    \left(
        V_G^\lambda\otimes V_{\calA_t}^{\lambda}
    \right)
    \otimes
    \overline{
    \left(
        V_G^{\lambda'}\otimes V_{\calA_t}^{\lambda'}
    \right)
    }.
\]
Since $\schur_t\otimes\overline{\schur_t}$ is an isometry, we thus conclude that
\[
    \Big(\schur_t\otimes\overline{\schur_t}\Big)^\dagger \Big(\schur_t\otimes\overline{\schur_t}\Big) = \Id_{\mathsf{Path}_t},
    \qquad
    \Big(\schur_t\otimes\overline{\schur_t}\Big)\Big(\schur_t\otimes\overline{\schur_t}\Big)^\dagger =\Pi_{\im(\schur_t\otimes\overline{\schur_t} )}.
\]

Next, we define the $\calA_t$ EPR isometry
\[
    E_t:
    \bigoplus_\lambda V_G^\lambda \otimes \overline V_G^\lambda
    \longrightarrow
    \bigoplus_\lambda
    \left(
        V_G^\lambda\otimes V_{\calA_t}^{\lambda}
    \right)
    \otimes
    \overline{
    \left(
        V_G^\lambda\otimes V_{\calA_t}^{\lambda}
    \right)
    } \subset \bigoplus_{\lambda, \lambda'}
    \left(
        V_G^\lambda\otimes V_{\calA_t}^{\lambda}
    \right)
    \otimes
    \overline{
    \left(
        V_G^{\lambda'}\otimes V_{\calA_t}^{\lambda'}
    \right)
    }
\]
by
\[
    E_t\ket{\lambda,X,Y}
    :=
    \frac{1}{
        \sqrt{\dim(V_{\calA_t}^{\lambda})}
    }
    \sum_{R\in\calB(V_{\calA_t}^{\lambda})}
    \ket{\lambda,X,R}
    \otimes
    \ket{\lambda,Y,R},
    \qquad
    \lambda\in\widehat{G}_t
\]
and $E_t\ket{\lambda,X,Y} = 0$ otherwise. Since this operation is simply tensoring with
a normalized EPR state on the two $\calA_t$-registers, $E_t$ is an isometry when restricted to $\mathsf{Tab}_t$. Moreover, by construction,
\[ \im(E_t) \subset \im(\schur_t\otimes\overline{\schur_t})
\]
The adjoint of $E_t$ is given on double-Schur basis states by
\[
    E_t^\dagger
    \left(
        \ket{\lambda,X,T}\otimes\ket{\lambda',Y,R}
    \right)
    =
    \begin{cases}
    \displaystyle
    \frac{1}{\sqrt{\dim(V_{\calA_t}^{\lambda})}}
    \ket{\lambda,X,Y},
    &
    \lambda=\lambda'\in\widehat{G}_t
    \text{ and } T=R,
    \\[2ex]
    0,
    &
    \text{otherwise.}
    \end{cases}
\]
With these definitions, it follows directly that
\[
    E_t^\dagger E_t=\Pi_{\mathsf{Tab}_t}, \qquad 
    E_tE_t^\dagger
    =
    \widehat{\Pi}^{\calA_t}_{\mathrm{EPR},t}.
\]

Finally, we are ready to define the operators
\[
    \Gamma_t
    :=
    \left(
        \schur_t^\dagger
        \otimes
        \schur_t^\top
    \right)
    E_t.
\]
To understand the domain and image on which $\Gamma_t$ is an isometry, we calculate

\[
    \Gamma_t^\dagger \Gamma_t
    =
    E_t^\dagger \Pi_{\im(\schur_t\otimes\overline{\schur_t})} E_t
    =
    E_t^\dagger E_t
    =
    \Pi_{\mathsf{Tab}_t},
\]
while
\begin{align*}
     \Gamma_t \Gamma_t^\dagger
    &=
    (\schur_t\otimes\overline{\schur_t})^\dagger E_tE_t^\dagger (\schur_t\otimes\overline{\schur_t}) \\
    &=
    (\schur_t\otimes\overline{\schur_t})^\dagger
    \widehat{\Pi}^{\calA_t}_{\mathrm{EPR},t}
    (\schur_t\otimes\overline{\schur_t}) \\
    &=
    \Omega_{\calA_t},
\end{align*}
where the last equation holds by definition of $\Omega_{\calA_t}$.  Thus, we have proved the following fact.

\begin{claim}
    $\Gamma_t$ is an isometry between $\mathsf{Tab}_t$ and the legal path subspace $\im(\Omega_{\mathcal A_t})\subset \mathsf{Path}_t$. 
\end{claim}

As a result, we see that $\Gamma_t \cdot F$ is an isometry between $F^{-1}(\mathsf{Tab}_t)$ and $\im(\Omega_{\mathcal A_t})$. We next give a characterization of the subspace $F^{-1}(\mathsf{Tab}_t)$. 

\subsection{Path Matrix Elements} 

To properly understand $\Gamma_t$, we consider the matrix element map

\[
    \Theta_t:
    \mathsf{Path}_t
    \longrightarrow
    L^2(G)
\]
defined as
\begin{equation}
    \Theta_t\ket{\vec x,\vec y}
    :=
    \int_G
    \bra{\vec y}\rho^{\otimes t}(g)\ket{\vec x}
    \ket g\,dg.
\end{equation}
We make the following claims about the map $\Theta_t$.

\begin{lemma}\label{lemma:path-matrix-exlement}
    $\im(F\cdot \Theta_t \cdot \Omega_{\calA_t}) = \im(F\cdot \Theta_t) = \mathsf{Tab}_t$. Moreover, as operators defined on $\mathsf{Path}_t$, 
    \begin{align}
    \Gamma_t \cdot F \cdot \Theta_t
    &=
    W_t\Omega_{\calA_t}. \label{eq:matrix-element-identity}
\end{align}
\end{lemma}
\begin{proof}
We first prove that $\mathsf{Tab}_t\subset \im(F\cdot \Theta_t \cdot \Omega_{\calA_t})$. To prove this, let $\ket{\lambda, X, Y} \in \mathsf{Tab}_t$ be a tableau basis vector. Let $\ket{T} \in  V_{\mathcal A_t}^\lambda$ be an arbitrary basis state, which we also identify with a basis state in $\overline V_{\mathcal A_t}^\lambda$. Then, we observe that
\begin{align} \Theta_t\Big( \schur^\dagger \ket{\lambda, X, T} \otimes \schur^\top \ket{\lambda, Y, T} \Big) &= \int_G \bra{\lambda, Y, T}  \schur  \cdot \rho^{\otimes t}(g) \cdot \schur^\dagger \ket{\lambda, X, T} \cdot \ket{g} dg \nonumber \\&= \int_G \bra{\lambda, Y, T} \rho_\lambda(g) \ket{\lambda, X, T} \cdot \ket{g} dg \\ &= \frac 1 {\sqrt{\dim(V_G^\lambda)}} \cdot F^{-1} \ket{\lambda, X, Y}.
\end{align}
Since this holds for every basis state $\ket{T}$, we can initialize the $V^\lambda_{\calA_t} \otimes \overline{V}^\lambda_{\calA_t}$ to be an EPR state (instead of a fixed basis vector) and also conclude that $F^{-1} \ket{\lambda, X, Y} \in \im(\Theta_t \Omega_{\calA_t})$. 

Next, we prove that $\im(F \cdot \Theta_t)\subset \mathsf{Tab}_t$ as well as \cref{eq:matrix-element-identity}. It suffices to prove both of these claims on basis vectors $\ket{\vec x,\vec y} \in \mathsf{Path}_t$. Expand the Schur transform as
\[
    \schur_t\ket{\vec x}
    =
    \sum_{\lambda,X,T}
    S_{\vec x}^{\lambda,X,T}
    \ket{\lambda,X,T},
\]
where
\[
    X\in\calB(V_G^\lambda),
    \qquad
    T\in\calB(V_{\calA_t}^{\lambda}).
\]
Since Schur-Weyl duality exactly tells us that the representation $\rho^{\otimes t}$ decomposes as
\[
    \rho^{\otimes t}(g)
    =
    \bigoplus_{\lambda}
    \rho_\lambda(g)\otimes \Id_{V_{\calA_t}^{\lambda}},
\]
we have that
\[
\begin{aligned}
    \bra{\vec y}\rho^{\otimes t}(g)\ket{\vec x}
    =
    \sum_{\lambda,X,Y,T}
    \overline{
        S_{\vec y}^{\lambda,Y,T}
    }
    S_{\vec x}^{\lambda,X,T}
    \bra{Y}\rho_\lambda(g)\ket{X}.
\end{aligned}
\]
Therefore, by the definition of the Fourier transform, 
\[
\begin{aligned}
    F\Theta_t\ket{\vec x,\vec y}
    =
    \sum_{\lambda,X,Y,T}
    \frac{
        \overline{
            S_{\vec y}^{\lambda,Y,T}
        }
        S_{\vec x}^{\lambda,X,T}
    }{
        \sqrt{\dim(V_G^\lambda)}
    }
    \ket{\lambda,X,Y}.
\end{aligned}
\]
Since the coefficient of $\ket{\lambda, X, Y}$ is nonzero only if both $\ket{\lambda, X, T}$ and $\ket{\lambda, Y, T}$ are in the image of the Schur transform for some $T$, this tells us that $F \Theta_t \ket{\vec x, \vec y} \in \mathsf{Tab}_t$. 

Next, applying $\Gamma_t=(\schur_t^\dagger\otimes\schur_t^\top)E_t$ gives
\begin{equation}
\begin{aligned}
    \Gamma_tF\Theta_t\ket{\vec x,\vec y}
    =
    \sum_{\lambda,X,Y,T,R}
    \frac{
        \overline{
            S_{\vec y}^{\lambda,Y,T}
        }
        S_{\vec x}^{\lambda,X,T}
    }{
        \sqrt{
            \dim(V_G^\lambda)
            \dim(V_{\calA_t}^{\lambda})
        }
    }
    \left(
        \schur_t^\dagger
        \otimes
        \schur_t^\top
    \right)
    \left(
        \ket{\lambda,X,R}
        \otimes
        \ket{\lambda,Y,R}
    \right).
\end{aligned}
\end{equation}

On the other hand,
\[
\begin{aligned}
    \left(
        \schur_t\otimes\overline{\schur_t}
    \right)
    \ket{\vec x,\vec y}
    =
    \sum_{\lambda,\lambda',X,Y,T,R}
    S_{\vec x}^{\lambda,X,T}
    \overline{
        S_{\vec y}^{\lambda',Y,R}
    }
    \ket{\lambda,X,T}
    \otimes
    \ket{\lambda',Y,R}.
\end{aligned}
\]
Projecting onto the EPR subspace gives
\[
\begin{aligned}
    \Omega_{\calA_t}\ket{\vec x,\vec y}
    =
    \left(
        \schur_t^\dagger
        \otimes
        \schur_t^\top
    \right)
    \sum_{\lambda,X,Y,T,R}
    \frac{
        S_{\vec x}^{\lambda,X,T}
        \overline{
            S_{\vec y}^{\lambda,Y,T}
        }
    }{
        \dim(V_{\calA_t}^{\lambda})
    }
    \ket{\lambda,X,R}
    \otimes
    \ket{\lambda,Y,R}.
\end{aligned}
\]
The operator $W_t$ multiplies the $\lambda$-subspace by
\[
    \sqrt{
        \frac{
            \dim(V_{\calA_t}^{\lambda})
        }{
            \dim(V_G^\lambda)
        }
    }.
\]
Hence
\begin{equation}
\begin{aligned}
    W_t\Omega_{\calA_t}\ket{\vec x,\vec y}
    =
    \sum_{\lambda,X,Y,T,R}
    \frac{
        \overline{
            S_{\vec y}^{\lambda,Y,T}
        }
        S_{\vec x}^{\lambda,X,T}
    }{
        \sqrt{
            \dim(V_G^\lambda)
            \dim(V_{\calA_t}^{\lambda})
        }
    }
    \left(
        \schur_t^\dagger
        \otimes
        \schur_t^\top
    \right)
    \left(
        \ket{\lambda,X,R}
        \otimes
        \ket{\lambda,Y,R}
    \right).
\end{aligned}
\end{equation}
We conclude that
\[
    \Gamma_tF\Theta_t\ket{\vec x,\vec y}
    =
    W_t\Omega_{\calA_t}\ket{\vec x,\vec y},
\]
as desired. 
\end{proof}
\subsection{Proof of \cref{thm:path-recording}}
To prove \cref{thm:path-recording}, we derive an expression that is simultaneously equal to both $(\Id_V \otimes \Gamma_{t+1} F) Q_\rho$ and $Q_{\rho,t+1}^{\mathrm{Path}} (\Id_V \otimes \Gamma_t F) $.

Beginning with the former, we claim the following operator identity holds on inputs $\ket{x} \otimes \ket{\vec x}\ket{\vec y}$.

\begin{claim}

\begin{align}
    Q_\rho \cdot (\ket{x} \otimes \Theta_t) = 
    \sum_y
    \ket y\otimes
    \Theta_{t+1}\App^{(t)}_{x,y}. \label{eq:representation-append}
\end{align}

\end{claim}

\noindent To see this, we simply observe that the left-hand side is equal to 

\[
    Q_\rho
    \left(
        \ket x\otimes
        \int_G \bra{\vec y}\rho^{\otimes t}(g) \ket{\vec x} \ket g\,dg
    \right)
    =
    \sum_{y\in\calB(V)}
    \ket y\otimes
    \int_G
    \bra y\rho(g)\ket x
    \bra{\vec y}{\rho^{\otimes t}(g)}\ket{\vec x}\ket g\,dg.
\]
and the second tensor factor exactly contains the $\ket{\vec x, x,\vec y, y}$ matrix element function. 

Applying $\Id_V\otimes \Gamma_{t+1}F$ to the recording register and invoking \cref{eq:matrix-element-identity}, we see that
\[
\begin{aligned}
    (\Id_V\otimes \Gamma_{t+1}F)
    Q_\rho
    \left(
        \ket{x} \otimes
        \Theta_t
    \right)
    &=
    \sum_y
    \ket y\otimes
     \Gamma_{t+1}F
    \Theta_{t+1}
    \App^{(t)}_{x,y}
    \\
    &=
    \sum_y
    \ket y\otimes
    W_{t+1}
    \Omega_{\calA_{t+1}}
    \App^{(t)}_{x,y}.
\end{aligned}
\]
Next, for every $x\in V$, we calculate (again invoking \cref{eq:matrix-element-identity})

\begin{align}
    Q_{\rho,t+1}^{\mathrm{Path}} (\Id_V \otimes \Gamma_t F)
    \left(
        \ket x\otimes
        \Theta_t
    \right)
    &=
    Q_{\rho,t+1}^{\mathrm{Path}}
    \left(
        \ket x\otimes
        W_t\Omega_{\calA_t}
    \right)
    \\
    &=
    \sum_y
    \ket y
    \otimes
    W_{t+1}
    \Omega_{\calA_{t+1}}
    \App^{(t)}_{x,y}
    W_t^{+} \Omega_{\calA_t}
    W_t
    \Omega_{\calA_t}
    \\
    &=
    \sum_y
    \ket y
    \otimes
    W_{t+1}
    \Omega_{\calA_{t+1}}
    \App^{(t)}_{x,y}
    \Omega_{\calA_t},
\end{align}
where the last equation holds because $\Omega_{\calA_t}$ and $W_t^+$ commute and $W_t^+ W_t \Omega_{\calA_t} = \Omega_{\calA_t}$. 

Inserting an additional $\Omega_{\calA_t}$ on the right, we conclude that 

\[ Q_{\rho,t+1}^{\mathrm{Path}} (\Id_V \otimes \Gamma_t F )(\Id_V \otimes \Theta_t \Omega_{\calA_t}) = (\Id_V \otimes \Gamma_{t+1} F) Q_\rho (\Id_V \otimes \Theta_t \Omega_{\calA_t}),
\]
and thus
\[ Q_{\rho,t+1}^{\mathrm{Path}} (\Id_V \otimes \Gamma_t F) = (\Id_V \otimes \Gamma_{t+1} F) Q_\rho
\]
when restricted to inputs in $V\otimes \im(\Theta_t \Omega_{\calA_t}) = V\otimes F^{-1} (\mathsf{Tab}_t)$, which is the first claim of \cref{thm:path-recording}. 

\paragraph{Quantum query algorithms} To obtain the second claim of \cref{thm:path-recording}, we note that all recording states obtainable by making $t$ queries to $Q_\rho$ lie in $F^{-1}(\mathsf{Tab}_t)$. 

\begin{lemma}[Reachable states are in $\im(\Theta_t)$]
Fix any adaptive quantum algorithm whose query register is $V$ and queries $Q_\rho$. Then, its recording register always remains in $\im(\Theta_t) = F^{-1}(\mathsf{Tab}_t)$. 
\end{lemma}

\begin{proof}
This follows immediately from \cref{eq:representation-append} and an induction on $t$.
\end{proof}

Finally, we observe that in this context, the one-query identity implies the $t$-query statement by induction. In particular, suppose that immediately before the $(t+1)$-st query, the path-recorded
simulation state is obtained from the uncompressed representation oracle by
applying
\[
    \Id\otimes \Gamma_t F
\]
to the recording register. The adversary's next local unitary commutes with the recording-register
isometry. The next oracle call is handled by the identity
\[
    Q_{\rho,t+1}^{\mathrm{Path}}
    (\Id_V\otimes \Gamma_t F)
    =
    (\Id_V\otimes \Gamma_{t+1} F)
    Q_\rho.
\]
Therefore the invariant is preserved with $t+1$ in place of $t$.

Starting from the trivial depth-$0$ recording register, induction over the
$t$ queries gives
\[
    A^{Q_\rho}(\ket\psi)
    \simeq
    A^{Q_\rho^{\mathrm{Path}}}(\ket\psi)
\]
via the final isometry $\Gamma_t F$. This proves the second claim of \cref{thm:path-recording}. 

\subsection{Transpose queries}
The same argument also gives a path-recording update for transpose queries.
Here transpose is taken with respect to the fixed basis $\mathcal B(V)$.

\begin{theorem}[Transpose path-recording update]
Let $Q_{\rho^\top}$ denote the purified oracle which applies $\rho(g)^\top$
to the adversary's query register:
\[
    Q_{\rho^\top}
    \left(
        |x\rangle\otimes \int_G f(g)|g\rangle\,dg
    \right)
    =
    \int_G f(g)\,\rho(g)^\top |x\rangle\otimes |g\rangle\,dg .
\]
Then, for every $x\in \mathcal B(V)$,
\[
    Q_{\rho^\top}\cdot (|x\rangle\otimes \Theta_t)
    =
    \sum_{y\in \mathcal B(V)}
    |y\rangle\otimes
    \Theta_{t+1}\operatorname{App}^{(t)}_{y,x}.
\]
Consequently, if we define
\[
    Q^{\mathrm{Path},\top}_{\rho,t+1}
    :=
    \sum_{x,y\in \mathcal B(V)}
    \ketbra{y}{x}
    \otimes
    W_{t+1}\Omega_{\calA_{t+1}}
    \operatorname{App}^{(t)}_{y,x}
    W_t^+\Omega_{\mathcal A_t},
\]
then
\[
    Q^{\mathrm{Path},\top}_{\rho,t+1}
    \cdot
    (\Id_V\otimes \Gamma_tF)
    =
    (\Id_V\otimes \Gamma_{t+1}F)
    \cdot
    Q_{\rho^\top}
\]
on inputs restricted to $V\otimes F^{-1}(\mathsf{Tab}_t)$.
\end{theorem}

\begin{proof}
First we prove the matrix-element identity. By
definition,
\[
    \Theta_t|\vec x,\vec y\rangle
    =
    \int_G
    \langle \vec y|\rho(g)^{\otimes t}|\vec x\rangle
    |g\rangle\,dg .
\]
Therefore
\begin{align}
    Q_{\rho^\top}
    \left(
        |x\rangle\otimes \Theta_t|\vec x,\vec y\rangle
    \right)
    &=
    \sum_{y\in \mathcal B(V)}
    |y\rangle\otimes
    \int_G
    \langle y|\rho(g)^\top|x\rangle
    \langle \vec y|\rho(g)^{\otimes t}|\vec x\rangle
    |g\rangle\,dg  \\
    &=
    \sum_{y\in \mathcal B(V)}
    |y\rangle\otimes
    \int_G
    \langle x|\rho(g)|y\rangle
    \langle \vec y|\rho(g)^{\otimes t}|\vec x\rangle
    |g\rangle\,dg  \\
    &=
    \sum_{y\in \mathcal B(V)}
    |y\rangle\otimes
    \Theta_{t+1}
    \operatorname{App}^{(t)}_{y,x}
    |\vec x,\vec y\rangle .
\end{align}
Thus a transpose query appends the reversed pair $(y,x)$. We now prove the path-recording identity. By \cref{eq:matrix-element-identity},
\[
    \Gamma_tF\Theta_t = W_t\Omega_{\mathcal A_t},
    \qquad
    \Gamma_{t+1}F\Theta_{t+1}=W_{t+1}\Omega_{\calA_{t+1}}.
\]
Applying $\Id_V\otimes \Gamma_{t+1}F$ to the transpose matrix-element identity gives
\[
    (\Id_V\otimes \Gamma_{t+1}F)
    Q_{\rho^\top}
    (|x\rangle\otimes \Theta_t)
    =
    \sum_y
    |y\rangle\otimes
    W_{t+1}\Omega_{\calA_{t+1}}
    \operatorname{App}^{(t)}_{y,x}.
\]
On the other hand,
\begin{align}
    Q^{\mathrm{Path},\top}_{\rho,t+1}
    (|x\rangle\otimes W_t\Omega_{\mathcal A_t})
    &=
    \sum_y
    |y\rangle\otimes
    W_{t+1}\Omega_{\calA_{t+1}}
    \operatorname{App}^{(t)}_{y,x}
    W_t^+\Omega_{\mathcal A_t}W_t\Omega_{\mathcal A_t} \\
    &=
    \sum_y
    |y\rangle\otimes
    W_{t+1}\Omega_{\calA_{t+1}}
    \operatorname{App}^{(t)}_{y,x}
    \Omega_{\mathcal A_t},
\end{align}
using $W_t^+W_t\Omega_{\mathcal A_t}=\Omega_{\mathcal A_t}$. Therefore the two expressions agree
when restricted to inputs in
\[
    V\otimes \operatorname{Im}(\Theta_t\Omega_{\mathcal A_t})
    =
    V\otimes F^{-1}(\mathsf{Tab}_t),
\]
which proves the theorem.
\end{proof}

Thus arbitrary adaptive algorithms making forward and transpose queries are
simulated by using $\operatorname{App}^{(t)}_{x,y}$ for a forward query and
$\operatorname{App}^{(t)}_{y,x}$ for a transpose query.
    \subsection{Schur Orthogonality for the Commutant Algebras}
\label{sec:schur-orthogonality}

Our next step is to \emph{interpret} the commutant EPR projector $\Omega_{\mathcal A_t}$. To do this, we first show that whenever the commutant algebra is semisimple, it satisfies the Schur orthogonality relations for the matrix elements of its simple modules.   

\begin{theorem}[Schur Orthogonality]
    \label{thm:schur-orthogonality}
    Let $\calA$ be a semisimple algebra, and let $\calB(\calA)$ be a basis for $\calA$. 
    For $a \in \calA$, let $\tr(a)$ be the trace of the regular representation of $\calA$.
    For each $a \in \calA$, let $a^*$ be its dual element under this trace. 
    Let $\lambda, \mu \in \Lambda(\calA)$ be the labels of two simple modules, $V_{\calA}^{\lambda}$ and $V_{\calA}^{\mu}$, of $\calA$.
    Let $\ket{a}, \ket{b} \in \mathcal B(V_{\calA}^{\lambda})$ and $\ket{c}, \ket{d} \in \mathcal B(V_{\calA}^{\mu})$.
    Then%
    \footnote{
        Note that compared with the Schur orthogonality theorem for irrep matrix elements of \emph{groups}, the second matrix element is transposed rather than complex conjugated. As we mention below in \Cref{rmk:schur-orthogonality-comparison}, this is because for groups, $\sigma^*$ is proportional to its inverse, which of course conjugate-transposes the irrep.
    }
    \begin{align}
        \sum_{
            \sigma 
            \in 
            \calB(\calA)
        }
        \bra{a}
        \rho_{\lambda}(\sigma)
        \ket{b}
        \;
        \bra{c}
        \rho_{\mu}(\sigma^*)
        \ket{d}
        =
        \frac{
            1
        }{
            \dim(V_{\calA}^{\lambda})
        }
        \delta_{\lambda\mu}
        \delta_{ad}
        \delta_{bc}
    \end{align}
\end{theorem}

\begin{proof}
    Consider the matrix
    \begin{equation*}
        T_{bc} \coloneqq \sum_{\sigma \in \calB(\calA)} \rho_{\lambda}(\sigma) \cdot \ketbra{b}{c} \cdot \rho_{\mu}(\sigma^*).
    \end{equation*}
    Our goal is to show that
    \begin{equation*}
        T_{bc} = 
        \begin{cases}
            \frac{
                1
            }{
                \dim(V_{\calA}^{\lambda})
            }
            \Id_{V^{\lambda}_{\calA}} & \text{if $\lambda = \mu$ and $b = c$},\\
            0 & \text{otherwise}.
        \end{cases}
    \end{equation*}
    If we can show this, then 
    \begin{equation*}
        \sum_{
            \sigma 
            \in 
            \calB(\calA)
        }
        \bra{a}
        \rho_{\lambda}(\sigma)
        \ket{b}
        \;
        \bra{c}
        \rho_{\mu}(\sigma^*)
        \ket{d}
        = \bra{a} T_{bc} \ket{d}
    \end{equation*}
    is equal to 
    $
        \frac{
            1
        }{
            \dim(V_{\calA}^{\lambda})
        }
        \bra{a} 
        \Id_{M_{\lambda}}
        \ket{d} 
        = 
        \frac{
            1
        }{
            \dim(V_{\calA}^{\lambda})
        }
        \braket{a}{d} 
        = 
        \frac{
            1
        }{
            \dim(V_{\calA}^{\lambda})
        }
        \delta_{ad}
    $ 
    if $\lambda = \mu$ and $b = c$ and is equal to $0$ otherwise.
    This gives the desired equality.

    To prove this, we observe that by~\cite[Proposition 5.2(b)]{halverson2005partition}, applied with
\(M=V_{\calA}^{\mu}\), \(N=V_{\calA}^{\lambda}\), and
\(\phi=\ketbra{b}{c}\), the map $T_{b,c}$ is a \(\calA\)-module homomorphism, meaning that for every $a \in \calA$,
    \[
    T_{bc}\rho_\mu(a)=\rho_\lambda(a)T_{bc}.
    \]
    We now consider two cases.

    \paragraph{Case 1: $\lambda \neq \mu$} If $\lambda \neq \mu$, Schur's lemma (\cite[Proposition 5.3(2)]{halverson2005partition}) states that $T_{bc} = 0$. 

    \paragraph{Case 2: $\lambda = \mu$} In this case, Schur's lemma (\cite[Proposition 5.3(1)]{halverson2005partition}) states that $T_{bc} = \alpha \cdot \Id_{V^\lambda_{\calA}}$ for some $\alpha \in \mathbb C$. 
    ~Taking the (linear algebraic) trace of both sides, we have that
    \begin{equation*}
        \tr(T_{bc}) = \alpha \cdot \dim(V_{\calA}^{\lambda}),
    \end{equation*}
    from which we see that $\alpha = \tr(T_{bc}) / \dim(V^\lambda_{\calA})$.
    It therefore suffices to compute the trace of $T_{bc}$.

    To compute this, we write
    \begin{align*}
        \tr(T_{bc}) &= \sum_{\sigma \in \mathcal B(\mathcal A)} \tr(\rho_\lambda(\sigma) \cdot \ketbra{b}{c} \cdot \rho_\lambda(\sigma^*)) \\
        &=\sum_{\sigma \in \mathcal B(\mathcal A)} \tr(  \ketbra{b}{c} \cdot \rho_\lambda(\sigma^*) \cdot \rho_\lambda(\sigma)) \\
        &=  \tr(  \ketbra{b}{c} \cdot \rho_\lambda\Big(\sum_{\sigma \in \mathcal B(\mathcal A)} \sigma^*\sigma\Big)) \\
        &= \tr(\ketbra{b}{c}) \\
        &= \delta_{bc},
    \end{align*}
    where we have invoked the identity $\sum_{\sigma \in \mathcal B(\mathcal A)} \sigma^* \sigma = 1$ \cite[Proof of Theorem 5.8]{halverson2005partition}.
\end{proof}
An alternative proof of \Cref{thm:schur-orthogonality} also appears in~\cite[Corollary 2.12]{foxman2026efficient}.

\begin{remark} \label{rmk:schur-orthogonality-comparison}
Note that this generalizes the Schur orthogonality for groups, which states that 
\begin{align}
    \frac{
        1
    }{
        \abs{G}
    }
    \sum_{
        g 
        \in 
        G
    }
    \bra{a}
    \rho_{\lambda}(g)
    \ket{b}
    \;
    \overline{
        \bra{d}
        \rho_{\mu}(g)
        \ket{c}
    }
    =
    \sum_{
        g 
        \in 
        \calB(\C[G])
    }
    \bra{a}
    \rho_{\lambda}(g)
    \ket{b}
    \;
    \bra{c}
    \rho_{\mu}
    \left(
        \frac{
            1
        }{
            \abs{G}
        }
        g^{-1}
    \right)
    \ket{d}
    =
    \frac{
        1
    }{
        \dim(V_{\calA}^{\lambda})
    }
    \delta_{\lambda\mu}
    \delta_{ad}
    \delta_{bc}
\end{align}
since note that whenever $\calA$ is a group algebra $\C[G]$, then we have that, $g^* = \frac{1}{\abs{G}} g^{-1} \quad \forall g \in G$ (since $\frac{1}{\abs{G}} \tr(h^{-1} g) = \delta_{hg}$).
\end{remark}

\subsection{Interpretation of the Commutant EPR projector}

\todo{Add intuition: applying the EPR projector is just teleporting the commutant tableau. Either here or in the intro}

\begin{lemma}
    \label{thm:commutant-epr-proj-as-symmetrization}
    Let $\rho: G\rightarrow \mathrm{End}(V)$ be a representation, and consider the tensor product representation $\rho^{\otimes t}$ on $V^{\otimes t}$. Let $\mathcal A_t$ denote a semisimple algebra that is isomorphic to the commutant algebra $\mathrm{End}_{G}(V^{\otimes t})$; for any $a\in \mathcal A_t$, let $S(a)\in \mathrm{End}_{G}(V^{\otimes t})$ denote its action. 

    Finally, let $\mathsf{Schur}$ be the corresponding generalized Schur transform. Let $V_{\calA_t}^{\lambda}$ and $V_{G}^{\lambda}$ be the corresponding simple $\calA_t$-modules and $G$-representations with bases $\calB(V_{\calA_t}^{\lambda})$ and $\calB(V_{G}^{\lambda})$ implied by the Schur transform.

    Then, 
    \begin{align}
        \Omega_{\calA_t} 
        = 
        \sum_{\sigma \in \mathcal B(\mathcal A_t) }
        S(\sigma) 
        \otimes 
        S(\sigma^*)^T. 
    \end{align} 
\end{lemma}

\begin{proof}
    We calculate 
    \begin{align}
        &
        \left(
            \mathsf{Schur} 
            \ot 
            \overline{\mathsf{Schur}}
        \right)
        \left(
            \sum_{
                \sigma \in \calB(\calA_t)
            } 
            S(\sigma) 
            \otimes 
            S(\sigma^*)^{T}
        \right)
        \left(
            \mathsf{Schur}^{\dagger} 
            \ot 
            \mathsf{Schur}^{T}
        \right)
        \\
        &
        =
        \sum_{
            \substack{
                \lambda,
                \mu,
                \in 
                \widehat{\calA}_t
            }
        }
        \proj{\lambda, \mu}
        \ot
        \left(
            \sum_{
                \sigma \in \calB(\calA_t)
            } 
            \rho_{\lambda}(\sigma)
            \otimes 
            \rho_{\mu}(\sigma^*)^{T}
        \right)
        \ot
        \Id_{V_{G}^{\lambda}}\otimes \Id_{V_{G}^{\mu}}
        \tag{definition of Schur transform (\Cref{def:schur-transform})}
        \allowdisplaybreaks
        \\
        &
        =
        \sum_{
            \substack{
                \lambda,
                \mu,
                \in 
                \widehat{\calA}_t
            }
        }
        \proj{\lambda, \mu}
        \ot
        \left(
            \sum_{
                \ket{a}, 
                \ket{b} 
                \ket{c}, 
                \ket{d}, 
                \in
                \calB(V_{\calA_t}^{\lambda})
            } 
            \ketbra{a}{b}
            \otimes 
            \ketbra{d}{c}
        \right)
        \ot
        \Id_{V_{G}^{\lambda}\otimes V_{G}^{\mu}}
        \cdot
        \sum_{
            \sigma \in \calB(\calA_t)
        }
        \bra{a}
        \rho_{\lambda}(\sigma)
        \ket{b} \cdot 
        \bra{d}
        \rho_{\mu}(\sigma^*)^{T}
        \ket{c}
        \tag{$\sum_{a} \proj{a} = \Id$}
        \allowdisplaybreaks
        \\
        &
        =
        \sum_{
            \substack{
                \lambda,
                \mu,
                \in 
                \widehat{\calA}_t
            }
        }
        \proj{\lambda, \mu}
        \ot
        \left(
            \sum_{
                \ket{a}, 
                \ket{b} 
                \ket{c}, 
                \ket{d}, 
                \in
                \calB(V_{\calA_t}^{\lambda})
            } 
            \ketbra{a}{b}
            \otimes 
            \ketbra{d}{c}
        \right)
        \ot
        \Id_{V_{G}^{\lambda}}\otimes \Id_{V_{G}^{\mu}}
        \cdot
        \frac{
            1
        }{
            \dim(V_{\lambda})
        }
        \delta_{\lambda \mu}
        \delta_{ad}
        \delta_{bc}
        \tag{\Cref{thm:schur-orthogonality}}
        \allowdisplaybreaks
        \\
        &
        =
        \sum_{
            \substack{
                \lambda
                \in 
                \widehat{\calA}_t
            }
        }
        \frac{
            1
        }{
            \dim(V_{\calA_t}^{\lambda})
        }
        \proj{\lambda, \lambda}
        \ot
        \left(
            \sum_{
                \ket{a}, 
                \ket{b} 
                \in
                \calB(V_{\calA_t}^{\lambda})
            } 
            \ketbra{a}{b}
            \otimes 
            \ketbra{a}{b}
        \right)
        \ot
        \Id_{V_{G}^{\lambda}}\otimes \Id_{V_{G}^{\mu}}
        \\
        &
        =
        \sum_{
            \substack{
                \lambda 
                \in 
                \widehat{\calA}_t
                \\
                \ket{a}, 
                \ket{b} 
                \in
                \calB(V_{\calA_t}^{\lambda})
            }
        }
        \frac{
            1
        }{
            \dim(V_{\calA_t}^{\lambda})
        }
        \proj{\lambda, \lambda}
        \ot
        \ketbra{
            a, a
        }{
            b, b
        }
        \ot
        \Id_{V_{G}^{\lambda}}\otimes \Id_{V_{G}^{\mu}}
        \\
        &
        =
        \left(
            \mathsf{Schur} 
            \ot 
            \overline{\mathsf{Schur}}
        \right)
        \Omega_{\calA_t}
        \left(
            \mathsf{Schur}^{\dagger} 
            \ot 
            \mathsf{Schur}^{T}
        \right)
        \tag*{\qedhere}
    \end{align}
\end{proof}

\begin{example}[Group Algebra Commutant EPR Projector]\label{ex:group-algebra-EPR-projector}
    In the case of $\calA_t = \mathbb C[H]$ for some finite group $H$, we have that $[h]^* = \frac 1 {|H|} [h^{-1}]$ for all $h\in H$ (\cref{ex:group-algebra-dual}); since $S([h^{-1}])^T = \overline{S([h])}$ for all $h$, this means that the commutant EPR projector formula simplifies to 
\[ \Omega_{\mathbb C[H]} = \frac 1 {|H|}\sum_{h\in H} S([h]) \otimes \overline{S([h])}.
\]
This further simplifies to 
\[ \Omega_{\mathbb C[H]} = \frac 1 {|H|}\sum_{h\in H} S([h]) \otimes S([h])
\]
whenever the Schur representation $S(\cdot)$ has real entries, such as in the case of $H = S_t$. 
\end{example}

    \subsection{Adaptive Queries to the Update Rule}\label{subsec:update_rule_for_adversaries}

In this section, we derive an equivalent form for the general update rule in \Cref{sec:general-path-rec-update}, which we will use for the main argument in~\cref{subsec:putting_together}. Specifically, we are interested in analyzing (potentially adaptive) quantum query algorithms that have oracle access to one of two different isometries:
\begin{itemize}
    \item The path basis update $Q_\rho^{\mathrm{Path}}$, or
    \item $Q_\rho^{\mathrm{Path}}\cdot C$, which applies $Q_\rho^{\mathrm{Path}}$ after first applying a unitary $C$ to the adversary's query register.
\end{itemize}

We will use $\reg{A}$, $\reg{B}$, and $\reg{R}$ to denote the query, workspace, and database registers as defined in~\cref{def:oracle_adversary}. We will also subdivide $\reg{R}$ as 
\begin{equation}
    \reg{R}  = (\reg{R_0}, \reg{R_1} = (\reg{R_{X, 1}, R_{Y, 1}}), \reg{R_2} = (\reg{R_{X, 2}, R_{Y, 2}}), \dots, \reg{R_t} = (\reg{R_{X, t}, R_{Y, t}}))
\end{equation}
where $\reg{R_i}$ denotes the register to which the update operator adds the $i$th query, and $\reg{R_{X, i}}$ and $\reg{R_{Y, i}}$ denote the registers for recording the input and output for the $i$'th query. We also write $\reg{R_X} := (\reg{R_{X, 1}}, \reg{R_{X, 2}}, \dots \reg{R_{X, t}})$ and $\reg{R_Y} := (\reg{R_{Y, 1}}, \reg{R_{Y, 2}}, \dots \reg{R_{Y, t}})$. 
We will also use $\reg{R_0}$ to denote register containing the initial database, before making any queries (which we will assume to be properly formatted as a list of pairs of registers like the rest of $\reg{R}$). Most often, we will take $\reg{R_0}$ to be the empty ($0$ qubit) register (with associated Hilbert space $\C^{2^0} = \C$). We write such an empty register as $\ket{\emptyset}_{\reg R}$. However it can be non-empty if, for example, we are not starting with a clean slate and a previous algorithm has already made queries.
We will also write $\reg{R}_{\le i} := (\reg{R}_0, \reg{R}_1, \dots, \reg{R}_i)$.

We will use $\calA_i$ to denote the commutant algebra of the $k$'th tensor product of the group representation, where $k = |\reg{R}_{\le i}| = i + |\reg{R}_0|$ (the size of $\reg{R}_0$ being the number of pairs in the initial database).
When the database is empty, then this is the commutant algebra of the trivial representation of the group, which is simply the one-dimensional algebra $\C$.

\paragraph{Notation for the Update Rule}
We will frequently drop the subscript $t$ from the Schur transform $\schur_t$ when the tensor power is unambiguous. For ease of working with the update rule from \Cref{sec:general-path-rec-update}, we will write it in terms of the following operators:

Let
\begin{align}
    \Pi_{\lambda, i}
    &
    := \schur^\dagger \Big( \proj{\lambda} \otimes \Id_{V_{\calA_i}^{\lambda}}
        \ot
        \Id_{V_{G}^{\lambda}} \Big) \schur \otimes \Id_{\overline V^{\otimes t}}
    \,,
    \allowdisplaybreaks
    \\
    \Lambda_{\lambda, i} 
    &
    \coloneqq 
    \beta_{\lambda, i} 
    \;
    \Pi_{\lambda, i}
    \,,
    \qquad
    \Lambda_{\lambda, i}^+ 
    \coloneqq 
    \beta_{\lambda, i}^{-1}
    \;
    \Pi_{\lambda, i}
    \,,
    \qquad
    \text{where }
    \beta_{\lambda, i} 
    \coloneqq 
    \sqrt{
        \frac{
            \dim(V^{\lambda}_{\calA_i})
        }{
            \dim(V^{\lambda}_{G})
        }
    }
    \,,
    \allowdisplaybreaks
    \\
    \Lambda_{\calA_i} 
    &
    \coloneqq 
    \sum_{\lambda \in \wh{\calA}_i}
    \Lambda_{\lambda, i} 
    \,,
    \qquad
    \Lambda_{\calA_i}^+
    \coloneqq 
    \sum_{\lambda \in \wh{\calA}_i}
    \Lambda_{\lambda, i}^+ 
    \allowdisplaybreaks
    \\
    \Omega_{\calA_i} 
    &
    := 
    \left(
        \mathsf{Schur}^{\dagger} 
        \ot 
        \mathsf{Schur}^{T}
    \right)
    \left(
        \sum_{
            \substack{
                \lambda 
                \in 
                \widehat{\calA}_i
                \\
                \ket{a}, 
                \ket{b} 
                \in
                \calB(V_{\calA_i}^{\lambda})
            }
        }
        \frac{
            1
        }{
            \dim(V_{\calA_i}^{\lambda})
        }
        \proj{\lambda, \lambda}
        \ot
        \ketbra{
            a, a
        }{
            b, b
        }
        \ot
        \Id_{V_{G}^{\lambda}}^{\ot 2}
    \right)
    \left(
        \mathsf{Schur} 
        \ot 
        \overline{\mathsf{Schur}}
    \right)
    \allowdisplaybreaks
    \\
    &
    =
    \sum_{
        \sigma \in \calB(\calA_i)
    } 
    S(\sigma) 
    \otimes 
    S(\sigma^*)^{T}
    \tag{by \Cref{thm:commutant-epr-proj-as-symmetrization}, where $S$ is the Schur representation}
    \,.
    \allowdisplaybreaks
    \\
    \App_{\reg{A}\,\reg{R}_i} 
    &
    \coloneqq 
    \sum_{x, y \in [N]} 
    \left(
    \ketbra{y}{x}
    \right)
    _{\reg{A}}
    \otimes
    \ket{x, y}
    _{\reg{R_i}}
\end{align}

When the database register is of size $i-1$, we can write the path recording update rule (\Cref{def:path-recording-equivalent}), when restricted to input databases in $\im(\Omega_{\mathcal A_{i-1}})$, as
\begin{align}
    \label{eq:update-rule-path-simplified}
    Q_\rho^{\mathrm{Path}} = V
    :=
    \Big(
        \Lambda_{\calA_i}
    \Big)_{\reg R_{\le i}}
    \Big(
        \Omega_{\calA_i}
    \Big)_{\reg{R}_{\le i}}
    \;
    \Big(
        \App
    \Big)_{\reg{A}\,\reg{R}_i} 
    \Big(
        \Lambda_{\calA_{i-1}}^+
    \Big)_{\reg R_{\le i - 1}}
\end{align}

\paragraph{Rewriting the Update Operator for Adaptive Queries}

\begin{lemma}[Telescoping product of irrep ratio reweighting]
    \label{lemma:telescoping-irrep-ratios}
    Let $V$ be the isometry defined in~\cref{eq:update-rule-path-simplified}, and define the reduced update operator (which without the rescaling is not an isometry) to be
    \begin{align}
        \label{eq:update-rule-path-reduced}
        \wt{V}
        &
        :=
        \Big(
            \Omega_{\calA_i}
        \Big)_{\reg{R}_{\le i}}
        \Big(
            \App
        \Big)
        _{\reg{A}\reg{R}_i} 
        \\
        &
        =
        \sum_{
            x_i, y_i \in [N]
        } 
        \ketbra{y_i}{x_i}_{\reg{A}} 
        \otimes 
        \Omega_{\calA_i}
        \Big(
            \Id_{\reg{R}_{< i}}
            \otimes 
            \ket{x_i, y_i}_{\reg{R}_{i}}
        \Big)
        \,.
    \end{align}
    whenever the database register $\reg{R}$ is of size $i-1$.
    Let $\mathsf{Adv}_t$ be any $t$-query algorithm with workspace register $\reg{A}$, ancilla register $\reg{B}$, and that makes queries to the update rule $V$ on workspace register $\reg{A}$ and database register $\reg{R}$. Then
    \begin{equation}
        \mathsf{Adv}_t^{V}
        = 
        \Big(
            \Lambda_{\calA_t}
        \Big)
        _{\reg{R}_{\le t}}
        \;
        \mathsf{Adv}_t^{\wt{V}}
        \;
        \Big(
            \Lambda_{\calA_0}^+
        \Big)
        _{\reg{R}_0}
    \end{equation}

    In other words, if we make $t$ queries to the path recording oracle, then this is the same as if we were to apply the path recording update rule without the reweighting, and then apply a reweighting just once at the end.%
    \footnote{
        \label{foot:no-rescaling-when-empty}
        As described in \Cref{lem:no-rescaling-when-empty}, when beginning with an empty database, the first reweighting operator always occurs on the one-dimensional irreps corresponding to $|\lambda| = 0$, and is thus just the identity operator.
    }
\end{lemma}

\begin{proof}
    Let $(A_1)_{\reg{A}\reg{B}}, \dots, (A_t)_{\reg{A}\reg{B}}$ be the unitaries that $\mathsf{Adv}$ applies in between queries.
    We can write 
    \begin{align}
        &
        \mathsf{Adv}_t^{V}
        \\
        &
        =
        \prod_{i = 1}^t
        \left(
            V_{\reg{AR}}
            \cdot
            A_{i\, \reg{AB}}
        \right)
        \\
        &
        =
        \prod_{i = 1}^t
        \left(
            \Big(
                \Lambda_{\calA_i}
            \Big)
            _{\reg{R}_{\le i}}
            \Big(
                \Omega_{\calA_i}
            \Big)_{\reg{R}_{\le i}}
            \Big(
                \App
            \Big)
            _{\reg{A}\reg{R}_i} 
            \Big(
                \Lambda_{\calA_{i - 1}}^+
            \Big)
            _{\reg{R}_{\le i - 1}}
            \big(
                A_i
            \big)_{\reg{AB}}
        \right)
        \tag{by the definition of $V$}
        \allowdisplaybreaks
        \\
        &
        = 
        \prod_{i = 1}^t
        \left(
            \left(
                \sum_{\lambda^+ \in \wh{\calA}_{i}}
                \beta_{\lambda^+, i}
                \;
                \Pi_{\lambda^+, i}
            \right)_{\reg{R}_{\le i}}
            \Big(
                \Omega_{\calA_i}
            \Big)_{\reg{R}_{\le i}}
            \Big(
                \App
            \Big)
            _{\reg{A}\reg{R}_i} 
            \left(
                \sum_{\lambda \in \wh{\calA}_{i-1}}
                \beta_{\lambda, i-1}^{-1}
                \;
                \Pi_{\lambda, i-1}
            \right)_{\reg{R_{\le i - 1}}}
            \big(
                A_i
            \big)_{\reg{AB}}
        \right)
        \tag{by the definition of $\Lambda_{\calA_i}$}
        \allowdisplaybreaks
        \\
        &
        =
        \sum_{
            \substack{
                \lambda_1, \dots, \lambda_t
                \\
                \lambda^+_1, \dots, \lambda^+_t 
                \\
                \lambda_i \in \wh{\calA}_{i-1}
                ,
\;                \lambda^+_i \in \wh{\calA}_{i}
            }
        }
        \prod_{i = 1}^t
        \left(
            \left(
                \beta_{\lambda^+_i, i}
                \;
                \Pi_{\lambda^+_i, i}
            \right)_{\reg{R}_{\le i}} 
            \Big(
                \Omega_{\calA_i}
            \Big)_{\reg{R}_{\le i}}
            \Big(
                \App
            \Big)
            _{\reg{A}\reg{R}_i} 
            \left(
                \beta_{\lambda_i, i-1}^{-1}
                \;
                \Pi_{\lambda_i, i-1}
            \right)_{\reg{R_{\le i - 1}}}
            \big(
                A_i
            \big)_{\reg{AB}}
        \right)
        \allowdisplaybreaks
        \\
        &
        =
        \sum_{
            \substack{
                \lambda_1, \dots, \lambda_t
                \\
                \lambda^+_1, \dots, \lambda^+_t 
                \\
                \lambda_i \in \wh{\calA}_{i-1}
                ,
\;                \lambda^+_i \in \wh{\calA}_{i}
            }
        }
        \beta_{\lambda^+_{t}, t}
        \prod_{i = 1}^t
        \left(
            \left(
                \beta_{\lambda_{i+1}, i}^{-1}
                \;
                \Pi_{\lambda_{i+1}, i}
            \right)_{\reg{R}_{\le i}}
            \left(
                \beta_{\lambda^+_i, i}
                \;
                \Pi_{\lambda^+_i, i}
            \right)_{\reg{R}_{\le i}} 
            \Big(
                \Omega_{\calA_i}
            \Big)_{\reg{R}_{\le i}}
            \Big(
                \App
            \Big)
            _{\reg{A}\reg{R}_i} 
            \big(
                A_i
            \big)_{\reg{AB}}
        \right)
        \left(
            \beta_{\lambda_1, 0}^{-1}
            \;
            \Pi_{\lambda_1, 0}
        \right)_{\reg{R_{0}}}
        \tag{by moving each 
        $
            \beta_{\lambda_i, i-1}^{-1}
            \;
            \Pi_{\lambda_i, i-1}
        $ to the previous factor in the product\footnotemark}
        \allowdisplaybreaks
        \\
        &
        =
        \sum_{
            \substack{
                \lambda_1, \dots, \lambda_t
                \\
                \lambda^+_1, \dots, \lambda^+_t 
                \\
                \lambda_i \in \wh{\calA}_{i-1}
                ,\;
                \lambda^+_i \in \wh{\calA}_{i}
            }
        }
        \prod_{i = 1}^t
        \left(
            \frac{
                \beta_{\lambda^+_i, i}
            }{
                \beta_{\lambda_{i+1}, i}
            }
            \;
            \left(
                \Pi_{\lambda_{i+1}, i}
            \right)_{\reg{R}_{\le i}}
            \left(
                \Pi_{\lambda^+_i, i}
            \right)_{\reg{R}_{\le i}} 
            \Big(
                \Omega_{\calA_i}
            \Big)_{\reg{R}_{\le i}}
            \Big(
                \App
            \Big)
            _{\reg{A}\reg{R}_i} 
            \big(
                A_i
            \big)_{\reg{AB}}
        \right)
        \left(
            \Pi_{\lambda_{1},0}
        \right)_{\reg{R_{0}}}
        \frac{
            \beta_{\lambda^+_{t}, t}
        }{
            \beta_{\lambda_{1}, 0}
        }
        \allowdisplaybreaks
        \\
        &
        =
        \sum_{
            \substack{
                \lambda_1, \dots, \lambda_t
                \\
                \lambda^+_1, \dots, \lambda^+_t 
                \\
                \lambda_i \in \wh{\calA}_{i-1}
                ,\;
                \lambda^+_i \in \wh{\calA}_{i}
            }
        }
        \prod_{i = 1}^t
        \left(
            \frac{
                \beta_{\lambda^+_i, i}
            }{
                \beta_{\lambda_{i+1}, i}
            }
            \;
            \delta_{\lambda_{i+1},\, \lambda_i^+}
            \left(
                \Pi_{\lambda^+_i, i}
            \right)_{\reg{R}_{\le i}} 
            \Big(
                \Omega_{\calA_i}
            \Big)_{\reg{R}_{\le i}}
            \Big(
                \App
            \Big)
            _{\reg{A}\reg{R}_i} 
            \big(
                A_i
            \big)_{\reg{AB}}
        \right)
        \left(
            \Pi_{\lambda_{1}, 0}
        \right)_{\reg{R_{0}}}
        \frac{
            \beta_{\lambda^+_{t}, t}
        }{
            \beta_{\lambda_{1}, 0}
        }
        \tag*{%
            $
                \left(
                \text{since}
                \;\;
                \Pi_{\lambda_{i+1}}
                \Pi_{\lambda^+_{i}}
                = 
                \delta_{\lambda_{i+1}, \lambda^+_{i}}
                \;
                \Pi_{\lambda^+_{i}}
                \right)
            $
        }
        \allowdisplaybreaks
        \\
        &
        =
        \sum_{
            \substack{
                \lambda_1, \lambda^+_1, \dots, \lambda^+_t 
                \\
                \lambda_1 \in \wh{\calA}_{0}
                ,\;
                \lambda^+_i \in \wh{\calA}_{i}
            }
        }
        \prod_{i = 1}^t
        \left(
            \left(
                \Pi_{\lambda^+_i, i}
            \right)_{\reg{R}_{\le i}} 
            \Big(
                \Omega_{\calA_i}
            \Big)_{\reg{R}_{\le i}}
            \Big(
                \App
            \Big)
            _{\reg{A}\reg{R}_i} 
            \big(
                A_i
            \big)_{\reg{AB}}
        \right)
        \left(
            \Pi_{\lambda_{1}, 0}
        \right)_{\reg{R_{0}}}
        \frac{
            \beta_{\lambda^+_{t}, t}
        }{
            \beta_{\lambda_{1}, 0}
        }
        \allowdisplaybreaks
        \\
        &
        =
        \prod_{i = 1}^t
        \left(
            \left(
                \sum_{
                    \substack{
                        \lambda^+_i \in \wh{\calA}_{i}
                    }
                }
                \Pi_{\lambda^+_i, i}
            \right)_{\reg{R}_{\le i}} 
            \Big(
                \Omega_{\calA_i}
            \Big)_{\reg{R}_{\le i}}
            \Big(
                \App
            \Big)
            _{\reg{A}\reg{R}_i} 
            \big(
                A_i
            \big)_{\reg{AB}}
        \right)
        \left(
            \sum_{
                \substack{
                    \lambda_1 \in \wh{\calA}_{0}
                }
            }
            \Pi_{\lambda_{1}, 0}
        \right)_{\reg{R_{0}}}
        \frac{
            \beta_{\lambda^+_{t}, t}    
        }{
            \beta_{\lambda_{1}, 0}    
        }
        \allowdisplaybreaks
        \\
        &
        =
        \left(
            \sum_{
                \substack{
                    \lambda^+_{t} \in \wh{\calA}_t
                }
            }
            \beta_{\lambda^+_{t}, t}
            \;
            \Pi_{\lambda^+_t, t}
        \right)_{\reg{R}_{\le t}}
        \prod_{i = 1}^t
        \left(
            \Big(
                \Omega_{\calA_i}
            \Big)_{\reg{R}_{\le i}}
            \Big(
                \App
            \Big)
            _{\reg{A}\reg{R}_i} 
            \big(
                A_i
            \big)_{\reg{AB}}
        \right)
        \left(
            \sum_{
                \substack{
                    \lambda_1 \in \wh{\calA}_{0}
                }
            }
            \beta_{\lambda_{1}, 0}^{-1}
            \;
            \Pi_{\lambda_{1}, 0}
        \right)_{\reg{R_{0}}}
        \tag{Using that $\sum_{\lambda} \Pi_{\lambda} = \Id$ in all except the last factor of the product}
        \allowdisplaybreaks
        \\
        &
        =
        \Big(
            \Lambda_{\calA_t}
        \Big)
        _{\reg{R}_{\le t}}
        \;
        \mathsf{Adv}_t^{\wt{V}}
        \;
        \Big(
            \Lambda_{\calA_0}^+
        \Big)
        _{\reg{R}_{0}}
        \tag{by the definitions of $\Lambda_{\calA_i}$, $\wt{V}$}
    \end{align}
\end{proof}

\footnotetext{For convenience of not having to write the last $t$'th factor separately, we take $\lambda_{t+1} := \lambda^+_{t}$.}

\begin{lemma}
    \label{rem:telescoping-ratios}
    The same holds for queries to $V \cdot C$ for any unitary $C \in U(N)$ acting on the $\reg{A}$ register. That is,
    \begin{equation}
        \mathsf{Adv}_t^{V \cdot C}
        = 
        \Big(
            \Lambda_{\calA_t}
        \Big)
        _{\reg{R}_{\le t}}
        \;
        \mathsf{Adv}_t^{\wt{V} \cdot C}
        \;
        \Big(
            \Lambda_{\calA_0}^+
        \Big)
        _{\reg{R}_{0}}
        \,.
    \end{equation}
\end{lemma}
\noindent
The proof is the same as that of \Cref{lemma:telescoping-irrep-ratios} above, but where each $A_{i\, \reg{AB}}$ has an extra $C$ attached to the left.

\begin{lemma}[Ricochet Property of the Append Operator]
    \label{lem:ricochet}
    For any $N$-dimensional unitary $C \in U(N)$, 
    \begin{equation}
        \App_{\reg {A\, R_i}} \cdot \;C_{\reg{A}} =  C_{{\reg{R_{X, i}}}} \cdot \App_{\reg {A\, R_i}} 
    \end{equation}
\end{lemma}
\begin{proof}
    \begin{align}
        \App_{\reg {A\, R_i}} 
        \cdot 
        \;
        C_{\reg{A}} 
        &
        =
        \sum_{x, y \in [N]} 
        \ketbra{y}{x}_{\reg{A}}
        C_{\reg{A}}
        \otimes
        \ket{x, y}
        _{\reg{R}_{i}}
        \tag{Definition of $\App_{\reg {A\, R_i}}$}
        \\
        &= 
        \sum_{x, y, z \in [N]} 
        \ket{y}_{\reg{A}} 
        \inner{x|C|z}
        \bra{z}_{\reg{A}} 
        \otimes
        \ket{x, y}
        _{\reg{R}_{i}}
        \tag{$\sum_z \ketbra{z}{z} = \Id$}
        \allowdisplaybreaks
        \\
        &= 
        \sum_{y, z \in [N]} 
        \ketbra{y}{z}_{\reg{A}} 
        \otimes 
        \sum_{x \in [N]} 
        \inner{x|C|z}
        \ket{x, y}
        _{\reg{R}_{i}}
        \allowdisplaybreaks
        \\
        &= 
        \sum_{y, z \in [N]} 
        \ketbra{y}{z}_{\reg{A}} 
        \otimes 
        \sum_{x \in [N]} 
        \ket{x}_{\reg{R_{X, i}}}
        \inner{x|C|z} 
        \ket{y}_{\reg{R_{Y, i}}}
        \allowdisplaybreaks
        \\
        &= 
        \sum_{y, z \in [N]} \ketbra{y}{z}_{\reg{A}} \otimes C\ket{z}_{\reg{R_{X, i}}} \otimes \ket{y}_{\reg{R_{Y, i}}}
        \tag{$\sum_x \ketbra{x}{x} = \Id$}
        \\
        &= 
        C_{{\reg{R_{X, i}}}} 
        \cdot 
        \App_{\reg {A\, R_i}}
        \tag*{\qedhere}
    \end{align}
\end{proof}

\begin{lemma}[Alternate Form of $\mathsf{Adv}_{t}^{\wt{V} \cdot C}$]
    \label{lem:reduced_adversary_expansion}
    \begin{equation}
        \mathsf{Adv}_{t}^{\wt{V} \cdot C} 
        = \left(
            \prod_{i=1}^t
            \left(
                \Omega_{\calA_i}
            \right)
            _{\reg{R}_{\le i}}
        \right) 
        \cdot 
        \left(
            C^{\otimes t}
        \right)
        _{\reg{R_X}}
        \cdot 
        \mathsf{Adv}_{t}^{\App} 
    \end{equation}
\end{lemma}
In other words, the following process is identical to querying $\wt{V} \cdot C$: during each query, only the $\App$ operator is applied. Then, after all queries have been made, $C^{\otimes t}$ is applied to $\reg{R_X}$, the input half of the database in the oracle register, and then for every $i \in [t]$, $\Omega_{\calA_i}$ is applied to the first $i$ registers of $\reg{R}$.
\begin{proof}
    \begin{align}
        \mathsf{Adv}_{t}^{\wt{V} \cdot C} 
        &
        = 
        \prod_{i=1}^t 
        \left( 
            \wt{V}
            _\reg{A\, R} 
            \cdot C_{\reg{A}} 
            \cdot A_{i\, \reg{AB}}
        \right) 
        \\
        &
        = 
        \prod_{i=1}^t 
        \left( 
            \left(
                \Omega_{\calA_i}
            \right)
            _{\reg{R}_{\le i}}
            \cdot 
            \App_{\reg{A\, R_i}} 
            \cdot C_{\reg{A}} 
            \cdot A_{i\, \reg{AB}}
        \right) 
        \tag{Definition of $\wt{V}$}
        \allowdisplaybreaks
        \\
        &= \prod_{i=1}^t \left( \left(\Omega_{\calA_i}\right)_{\reg{R}_{\le i}}  \cdot C_{\reg{R_{X, i}}} \cdot \App_{\reg{A\, R_i}} \cdot A_{i\, \reg{AB}}\right) 
        \tag{\cref{lem:ricochet}}
        \allowdisplaybreaks
        \\
    \intertext{
        Next, notice that for all $j < i$, $\App_{\reg{A\, R_i}} \cdot A_{i\, \reg{AB}}$ commutes with both $C_{\reg{R_{X, j}}}$ and $\left(\Omega_{\calA_j}\right)_\reg{R_{\le j}}$, since the former acts only on the $\reg{A}$, $\reg{B}$, and $\reg{R_i}$ registers, while both of the latter operators act only on $(\reg{R_1} \dots \reg{R_j})$. Therefore, we can commute them all past:
    }
        &=
        \prod_{i=1}^t 
        \left(
            \left(
                \Omega_{\calA_i}
            \right)
            _{\reg{R}_{\le i}}
            \cdot 
            C_{\reg{R_{X, i}}} 
        \right)
        \cdot 
        \prod_{i=1}^t 
        \left(
            \App_{\reg{A\, R_i}}
            \cdot 
            A_{i\, \reg{AB}}
        \right) 
        \allowdisplaybreaks
        \\
        &
        =
        \left(
            \prod_{i=1}^t
            \left(
                \Omega_{\calA_i}
            \right)
            _{\reg{R}_{\le i}}
        \right) 
        \cdot 
        \left(
            \prod_{i=1}^t 
            C_{\reg{R_{X, i}}} 
        \right)
        \cdot 
        \prod_{i=1}^t 
        \left(
            \App_{\reg{A\, R_i}} 
            \cdot 
            A_{i\, \reg{AB}} 
        \right)
        \tag{$C_{\reg{R_{X, i}}}$ similarly commutes with  $\left(\Omega_{\calA_j}\right)_\reg{R_{\le j}}$, for all $j < i$}
        \\
        &
        =
        \left(
            \prod_{i=1}^t
            \left(
                \Omega_{\calA_i}
            \right)
            _{\reg{R}_{\le i}}
        \right) 
        \cdot 
        \Big(
            C^{\otimes t}
        \Big)
        _{\reg{R_{X}}} 
        \cdot 
        \mathsf{Adv}_{t}^{\App} 
        \qedhere
    \end{align}
\end{proof}

\begin{lemma}
    \label{lem:t-fold-omega-collapses}
    \begin{align}
        \prod_{k=1}^t
        \left(
            \Omega_{\calA_k}
        \right)
        _{\reg{R}_{\le k}}
        =
        \Omega_{\calA_t}
    \end{align}
\end{lemma}
\begin{proof}
For each $k$, write $S_k:\calA_k\to \End(V^{\otimes k})$ for the Schur
representation action of $\calA_k$ on the first $k$ path registers. When
$k\le t$, we view $\Omega_{\calA_k}$ as an operator on $\reg R_{\le t}$ by
tensoring with $\Id$ on the later registers $\reg R_{k+1},\ldots,\reg R_t$.

We first recall the subspace onto which $\Omega_{\calA_k}$ projects. Namely,
$\Omega_{\calA_k}$ is the orthogonal projector onto the subspace of states
$\ket{\psi}$ satisfying
\[
    \bigl(S_k(a)\bigr)_{\reg X_{\le k}}\ket{\psi}
    =
    \bigl(S_k(a)^T\bigr)_{\reg Y_{\le k}}\ket{\psi}
    \qquad
    \forall a\in \calA_k .
\]
Here the $\reg X_{\le k}$ registers are identified with $V^{\otimes k}$,
whereas the $\reg Y_{\le k}$ registers are identified with
$\overline V^{\otimes k}$, using the conjugate basis labeled by the same
basis elements.

To see the characterization, pass to the double Schur basis for
$V^{\otimes k}\ot \overline V^{\otimes k}$. In this basis,
\[
    V^{\otimes k}\ot \overline V^{\otimes k}
    \cong
    \bigoplus_{\lambda,\mu}
    \left(V_G^\lambda\ot V_{\calA_k}^\lambda\right)
    \ot
    \left(\overline{V_G^\mu}\ot \overline{V_{\calA_k}^\mu}\right).
\]
The algebra $\calA_k$ acts on the $\reg X$ half by
\[
    S_k(a)
    =
    \bigoplus_{\lambda}
    \Id_{V_G^\lambda}\ot \rho_\lambda(a),
\]
and the transposed action on the $\reg Y$ half is
\[
    S_k(a)^T
    =
    \bigoplus_{\mu}
    \Id_{\overline{V_G^\mu}}\ot \rho_\mu(a)^T.
\]

Taking $a$ to be central idempotents of $\calA_k$ first forces any state
satisfying the displayed relations to have support only on the
$(\lambda,\lambda)$ blocks. On such a block, after treating the two
$G$-registers as spectators, the relations become
\[
    \bigl(\rho_\lambda(a)\ot \Id\bigr)\ket{\psi_{\lambda,\lambda}}
    =
    \bigl(\Id\ot \rho_\lambda(a)^T\bigr)\ket{\psi_{\lambda,\lambda}}
    \qquad
    \forall a\in \calA_k .
\]
Since $\rho_\lambda(\calA_k)=\End(V_{\calA_k}^\lambda)$, this is equivalent to
\[
    (M\ot \Id)\ket{\psi_{\lambda,\lambda}}
    =
    (\Id\ot M^T)\ket{\psi_{\lambda,\lambda}}
    \qquad
    \forall M\in \End(V_{\calA_k}^\lambda).
\]
The solutions are precisely arbitrary states on the two $G$-registers tensored
with the EPR state on the two commutant registers,
\[
    \ket{\mathsf{EPR}_{\calA_k}^{\lambda}}
    :=
    \frac{1}{\sqrt{\dim(V_{\calA_k}^{\lambda})}}
    \sum_{T\in \mathcal B(V_{\calA_k}^{\lambda})}
    \ket{T}\ot\ket{T},
\]
where the second $\ket{T}$ denotes the corresponding conjugate-basis vector
in $\overline{V_{\calA_k}^{\lambda}}$. This is exactly the image of
$\Omega_{\calA_k}$ by definition.

The lemma now follows from the fact that the $\Omega_{\mathcal A_k}$ are all orthogonal projectors onto subspaces that form a linear chain. This is because the commutant algebras satisfy $S(\calA_k) \subset S(\calA_t)$ for all $k\leq t$, so the constraints
\[
    \bigl(S_k(a)\bigr)_{\reg X_{\le k}}\ket{\psi}
    =
    \bigl(S_k(a)^T\bigr)_{\reg Y_{\le k}}\ket{\psi}
    \qquad
    \forall a\in \calA_k
\]
are contained within the corresponding set of constraints for $\calA_t$. This implies that
\[ \Omega_{\calA_t} \Omega_{\calA_k} = \Omega_{\calA_t}
\]
for all $k\leq t$, which proves the lemma. 
\end{proof}

Combining~\cref{lem:reduced_adversary_expansion,rem:telescoping-ratios,lem:t-fold-omega-collapses}, we obtain an equivalent expanded form for an adversary querying $V \cdot C$: 
\begin{corollary}
    \label{eq:fully_expanded_adversary_action}
    \begin{equation}
        \mathsf{Adv}_t^{V \cdot C}
        =
        \underbrace{
            \vphantom{
                \left(
                    \prod_{i=1}^t
                    \left(
                        \Omega_{\calA_i}
                    \right)_{\reg{R}_{\le i}}
                \right)
            }
        \Big(
            \Lambda_{\calA_t}
        \Big)
        _{\reg{R}_{\le t}}
        }_{\operatorname{ReweightEnd}}
        \cdot
        \underbrace{
            \vphantom{
                \left(
                    \prod_{i=1}^t
                    \left(
                        \Omega_{\calA_i}
                    \right)_{\reg{R}_{\le i}}
                \right)
            }
        \Big(
            \Omega_{\calA_t}
        \Big)
        _{\reg{R}_{\le t}}
        }_{\operatorname{Symmetrize}}
        \cdot
        \underbrace{
            \vphantom{
                \left(
                    \prod_{i=1}^t
                    \left(
                        \Omega_{\calA_i}
                    \right)_{\reg{R}_{\le i}}
                \right)
            }
        \Big(
            C^{\otimes t}
        \Big)
        _{\reg{R_{X}}} 
        }_{C^{\otimes t}}
        \cdot
        \underbrace{
            \vphantom{
                \left(
                    \prod_{i=1}^t
                    \left(
                        \Omega_{\calA_i}
                    \right)_{\reg{R}_{\le i}}
                \right)
            }
            \mathsf{Adv}_{t}^{\App} 
        }_{\operatorname{Append}}
        \cdot
        \underbrace{
            \vphantom{
                \left(
                    \prod_{i=1}^t
                    \left(
                        \Omega_{\calA_i}
                    \right)_{\reg{R}_{\le i}}
                \right)
            }
        \Big(
            \Lambda_{\calA_0}^+
        \Big)
        _{\reg{R}_{0}}
        }_{\operatorname{ReweightStart}}
    \end{equation}
\end{corollary}
Both reweighting steps, as well as the $\operatorname{Symmetrize}$ step, depend on the group $G$ (in particular, they depend on the commutant algebra of its tensor power representation), while $C^{\otimes t}$ and $\operatorname{Append}$ are independent of the group and its representation. 

When we start with $\reg{R_0}$ containing the empty database, this simplifies further, since initial reweighting step disappears:

\begin{remark}
    \label{lem:no-rescaling-when-empty}
    When $R$ contains the empty database, $\ket{\emptyset}$, then
    $
        \Lambda_{\calA_0}^{+}
        \ket{\emptyset}_{\reg{R}} 
        = 
        \ket{\emptyset}_{\reg{R}}
        \,.
    $
\end{remark}
\begin{proof}
    Since the database register $\mathsf{R}$ is empty, it corresponds to the 1-dimensional space $(\C^N)^{\otimes 0} = \C$, on which $G$ acts via the trivial irrep ($\rho^{\otimes 0} = 1$), and its commutant $\calA_0 = \C$ is a simple module (its only irrep is the one that acts via scalar multiplication). Since the irreps are both 1-dimensional, the ratio $\beta_{\lambda}$ is 1, and the rescaling has no effect.
\end{proof}

\section{Approximation and Specialization to Specific Groups}
\label{sec:approximation-specialization} Our general update rule can recover an interpretable recording oracle for any group representation. To demonstrate its generality, in this section, we show how it recovers existing recording oracles, including Zhandry's original recording oracle for functions~\cite{C:Zhandry19} and Ma and Huang's path-recording oracle~\cite{STOC:MaHua25} for unitaries as special cases.
For the latter, since the path recording oracle of \cite{STOC:MaHua25} is only approximate, we describe two explicit approximations that relate our path-recording oracle to theirs.

We also analyze the \emph{unitary Haar cipher} model, in which unitaries $U_1,\, \dots,\, U_K$ are each chosen i.i.d. from the Haar measure on $U(N)$ and queried in a controlled manner as $\sum_{k \in [K]} \proj{k} \otimes U_k$.

\subsection{The Unitary Path-Recording Oracle of~\cite{STOC:MaHua25}}

We will use the following lemma from \cite{STOC:MaHua25}:

\begin{lemma}[{\cite[Lemma 2.2]{STOC:MaHua25}}]
    \label{lem:proj_and_trace_distance}
Let $\rho_{\reg{ST}}$ be a density matrix on registers $\reg{S}, \reg{T}$ and let $\Pi_\reg{T}$ be a projector that acts on register $\reg{T}$. Then
\begin{equation}
    \left\|
        \operatorname{Tr}_\reg{T}(\rho_{\reg{ST}})
        -
        \operatorname{Tr}_\reg{T}(\Pi_\reg{T} \rho_{\reg{ST}} \Pi_\reg{T})
    \right\|_1
    =
    1 - \operatorname{Tr}(\Pi_\reg{T} \rho_{\reg{{ST}}}).
\end{equation}
\end{lemma}

\subsubsection{The Distinct Subspace}
\begin{definition}
    For $1 \le t \le N$, define 
\begin{equation}
    [N]_{\text{dist}}^t \coloneqq \{(x_1, \dots, x_t) \in [N]^t: \text{all $x_i$'s distinct} \}
\end{equation}
as the set of all distinct ordered $t$-tuples over $[N]$. 
\end{definition}
\begin{definition}\label{def:distinct-subspace}
    The \textit{distinct subspace} $\mathsf{Dist}_{N, t}  \subseteq (\C^N)^{\otimes t}$ is defined as the span of all distinct strings, i.e.
\begin{equation}
   \mathsf{Dist}_{N, t} \coloneqq \text{span}\{\ket{x}: x \in [N]_{\text{dist}}^t\}
\end{equation}
\end{definition}

\subsubsection{The \cite{STOC:MaHua25} oracle}
The \cite{STOC:MaHua25} path-recording oracle acts on an adversary input register $\reg A$ and recording register $\reg R_{MH}$ whose basis consists of \emph{sets} of input-output pairs $\{(x_1, y_1), \hdots, (x_t, y_t)\}$ with the property that $(y_1, \hdots, y_t)$ are distinct. 

To compare this with our compressed oracle, we consider the following simple isometry mapping $\reg R_{MH}$ to our recording space $\reg R$:
\[ \ket{\{(x_1, y_1), \hdots, (x_t, y_t)\}} \mapsto \frac 1 {\sqrt{t!}} \sum_{\pi\in S_t} \ket{x_{\pi^{-1}(1)}, \hdots, x_{\pi^{-1}(t)}, y_{\pi^{-1}(1)}, \hdots, y_{\pi^{-1}(t)}} = \sqrt{t!} \cdot \Omega_{S_t} \ket{x_1, \hdots, x_t, y_1, \hdots, y_t}.
\]
Via this isometry, we identify the Ma-Huang path recording update with the following update rule
\begin{align} \ket{x_{t}}_{\reg A} \otimes \ket{x_1, \hdots, x_{t-1}, y_1, \hdots, y_{t-1}} &\mapsto \Omega_{S_{t}} \cdot \frac {\sqrt t} {\sqrt{N-t+1}} \sum_{y_{t}\notin \{y_1, \hdots, y_{t-1}\}} \ket{y_{t}}_{\reg A} \otimes \ket{x_1, \hdots, x_{t}, y_1, \hdots, y_{t}} \\
&=\frac {\sqrt t} {\sqrt{N-t+1}} \cdot \Big(\Omega_{S_{t}}\Big)_{\reg R} \cdot \Big(\Pi_{\mathsf{Dist}}\Big)_{\reg R_Y} \mathsf{App}_{\reg A \reg R}.
\end{align}

Thus, the statement that the Ma-Huang oracle is indistinguishable from our oracle amounts to two informal statements:

\begin{itemize}
    \item The application of $\Pi_{\mathsf{Dist}}$ is approximately the identity map, and
    \item The overall effect of inverse subspace reweighting $\Lambda_{S_{t-1}}^+$ and reweighting $\Lambda_{S_t}$ is approximately constant, with a scalar factor of $\sqrt{\frac{t}{N-t+1}}$. 
\end{itemize}

Next, we will give a formal justification for the approximate equivalence of $V_{MH}$ and $V_{U(N)}$ for $t$-query quantum algorithms.

\subsubsection{Proof of approximate equivalence}
Using the fact that $(\Pi_{\mathsf{Dist}})_{\reg R_{Y, \leq t}}$ commutes with $\Omega_{S_{t}}$, we can then express the isometry implemented by a $t$-query adaptive adversary $\mathsf{Adv}^{V_{MH}}$ as

\begin{align}
        \label{eq:fully_expanded_adversary_MH}
        \mathsf{Adv}_t^{V_{MH}}
        &= \sqrt{t!} \prod_{i=1}^t \frac 1 {\sqrt{N-i+1}} \cdot 
        \Big(
            \Pi_{\mathsf{Dist}}
        \Big)
        _{\reg{R}_{Y}}
        \cdot
        \Big(
            \Omega_{S_t}
        \Big)
        _{\reg{R}}
        \cdot
        \mathsf{Adv}_{t}^{\App}
        \\
        &= 
        \frac{1}{\sqrt{{N\choose t}}} \cdot 
        \Big(
            \Pi_{\mathsf{Dist}}
        \Big)
        _{\reg{R}_{Y}}
        \cdot
        \Big(
            \Omega_{S_t}
        \Big)
        _{\reg{R}}
        \cdot
        \mathsf{Adv}_{t}^{\App}
        \,.
    \end{align}
On the other hand, our own path-recording oracle implements

    \begin{align}
        \mathsf{Adv}_t^{V_{U(N)}}
        &
        =
        \Big(
            \Lambda_{S_t}
        \Big)
        _{\reg{R}}
        \cdot
        \Big(
            \Omega_{S_t}
        \Big)
        _{\reg{R}}
        \cdot
        \mathsf{Adv}_{t}^{\App} 
        \cdot
        \Big(
            \Lambda_{S_0}^+
        \Big)
        _{\reg{R}_{0}}
        \\
        &
        =
        \Big(
            \Lambda_{S_t}
        \Big)
        _{\reg{R}}
        \cdot
        \Big(
            \Omega_{S_t}
        \Big)
        _{\reg{R}}
        \cdot
        \mathsf{Adv}_{t}^{\App}.\label{eq:fully_expanded_adversary_UN}
    \end{align}
To compare these, define the quantum states
\begin{equation} \ket{\mathsf{Adv}_t^{V_{MH}}} = \mathsf{Adv}_t^{V_{MH}} \ket{0}_{\reg A \reg B} \ket{\emptyset}_{\reg R}, \qquad \ket{\mathsf{Adv}_t^{V_{U(N)}}} = \mathsf{Adv}_t^{V_{U(N)}} \ket{0}_{\reg A \reg B} \ket{\emptyset}_{\reg R}. 
\end{equation}
We wish to derive the fact that
\alex{John B points out that the next equation is not really what we want to say here, as this is already proved by MH + Section 5
\\
\textbf{\underline{Barak}}:
Yeah, what we want to say is that it's actually close without any Uhlmann unitary, right? 
\\
\textbf{\underline{Alex}}:
Slightly more complicated than that: if you don't trace out the purification the error is actually $t/\sqrt{N}$ because of Gentle measurement error. Honestly we could just leave it for now... I changed the wording to ``we wish to derive the fact that''
}
\begin{align}
        \left\lVert
            \Tr_{\reg{R}}^{\phantom{R}}
            \!\!
            \left[
                \Proj{\mathsf{Adv}_t^{V_{MH}}}
            \right]
            \!
            -
            \Tr_{\reg{R}}^{\phantom{R}}
            \!\!
            \left[
                \Proj{\mathsf{Adv}_t^{V_{U(N)}}}
            \right]
        \right\rVert_1
        \le
        O
        \left(
            \frac{t^2}{N}
        \right)
    \end{align}
We prove this via the following sequence of steps. First, we show that
\begin{align}
    \Big|\Big| (\Pi_{\mathsf{Dist}})_{\reg R_Y} \ket{\mathsf{Adv}_t^{V_{U(N)}}} \Big|\Big|^2 \geq 1-O\left(\frac{t^2}{N}\right).
\end{align}
Then, by \cref{lem:proj_and_trace_distance}, we have that 
\begin{align}
        \left\lVert
            \Tr_{\reg{R}}^{\phantom{R}}
            \!\!
            \left[ (\Pi_{\mathsf{Dist}})_{\reg R_Y}
                \Proj{\mathsf{Adv}_t^{V_{U(N)}}}(\Pi_{\mathsf{Dist}})_{\reg R_Y}
            \right]
            \!
            -
            \Tr_{\reg{R}}^{\phantom{R}}
            \!\!
            \left[
                \Proj{\mathsf{Adv}_t^{V_{U(N)}}}
            \right]
        \right\rVert_1
        \le
        O
        \left(
            \frac{t^2}{N}
        \right)
    \end{align}
    Finally, we prove that 
    \begin{align}
        \left\lVert
            (\Pi_{\mathsf{Dist}})_{\reg R_Y}
                \Proj{\mathsf{Adv}_t^{V_{U(N)}}}
                (\Pi_{\mathsf{Dist}})_{\reg R_Y}
            \!
            -
                \Proj{\mathsf{Adv}_t^{V_{MH}}}
        \right\rVert_1
        \le
        O
        \left(
            \frac{t^2}{N}
        \right)
    \end{align}
For the remainder of this subsection, write
\[
    \Pi_{\mathsf{Dist}}^Y
    :=
    \left(\Pi_{\mathsf{Dist}_{N,t}}\right)_{\reg R_Y},
    \qquad
    \Pi_{\mathsf{Dist},t-1}^Y
    :=
    \left(\Pi_{\mathsf{Dist}_{N,t-1}}\right)_{\reg R_{Y,<t}},
    \qquad
    \alpha_t
    :=
    \frac{1}{\sqrt{\binom{N}{t}}}.
\]
We also abbreviate the append-only state by
\[
    \ket{\psi_t}
    :=
    \mathsf{Adv}_t^{\App}
    \ket{0}_{\reg A\reg B}
    \ket{\emptyset}_{\reg R}.
\]
With this notation, \cref{eq:fully_expanded_adversary_MH} says
\[
    \ket{\mathsf{Adv}_t^{V_{MH}}}
    =
    \alpha_t \cdot
    \Pi_{\mathsf{Dist}}^Y \cdot 
    \Omega_{S_t}
    \ket{\psi_t}.
\]
while \cref{eq:fully_expanded_adversary_UN} says
\[
    \ket{\mathsf{Adv}_t^{V_{U(N)}}}
    =
    \Lambda_{S_t} \cdot 
    \Omega_{S_t}
    \ket{\psi_t}.
\]

\begin{lemma}[Analyzing $\Pi_{\mathsf{Dist}}^Y \ket{\psi_t}$]
\label{lem:mh-distinct-append-state}
For every non-identity \(\pi\in S_t\),
\[
    \bra{\psi_t}
        \Pi_{\mathsf{Dist}}^Y\Big(S(\pi)\otimes S(\pi)\Big)\Pi_{\mathsf{Dist}}^Y
    \ket{\psi_t}
    =
    0.
\]
Moreover,
\[
    \bra{\psi_t}\Pi_{\mathsf{Dist}}^Y\ket{\psi_t}
    =
    t!\binom{N}{t}.
\]
Consequently,
\[
    \left\|
        \Omega_{S_t}\Pi_{\mathsf{Dist}}^Y\ket{\psi_t}
    \right\|^2
    =
    \binom{N}{t}
    =
    \frac{1}{\alpha_t^2}.
\]
\end{lemma}

\begin{proof}
We prove the first two claims by induction on \(t\). The case $t=1$ is trivial (there are no non-identity permutations and $\Pi_{\mathsf{Dist}}^Y = \Id$).

For the inductive step, we write
\[ \ket{\psi_t} = \mathsf{App}_{\reg A \reg R_t} \cdot (A_t)_{\reg A \reg B} \cdot \ket{\psi}_{t-1} = \mathsf{App}_{\reg A \reg R_t} \ket{\psi_t^{\mathrm{pre}}}.
\]
Moreover, since $\Pi_{\mathsf{Dist}}^Y = \Pi_{\mathsf{Dist}}^Y \cdot \Pi_{\mathsf{Dist},t-1}^Y$ and $\Pi_{\mathsf{Dist},t-1}^Y$ commutes with $\mathsf{App}_{\reg A \reg R_t}$, we have that
\[ \Pi_{\mathsf{Dist}}^Y \ket{\psi_t} = \Pi_{\mathsf{Dist}}^Y \cdot \mathsf{App}_{\reg A \reg R_t} \cdot \Pi_{\mathsf{Dist},t-1}^Y \ket{\psi_t^{\mathrm{pre}}}.
\]
We now write explicitly in the standard basis
\[\Pi_{\mathsf{Dist},t-1}^Y \ket{\psi_t^{\mathrm{pre}}} = \sum_{\substack{x_1, \hdots, x_t \\ \text{distinct } y_1, \hdots, y_{t-1}}} \ket{x_t}_{\reg A} \ket{\phi_{x_{1:t}, y_{1:t-1}}}_{\reg B} \ket{x_1,\hdots x_{t-1}, y_1, \hdots, y_{t-1}}_{\reg R_{<t}}.
\]
On this state, the effect of $\Pi_{\mathsf{Dist}}^Y \cdot \mathsf{App}_{\reg A \reg R_t}$ is appending with respect to an (unnormalized) EPR state over $y_t\notin \{y_1, \hdots, y_{t-1}\}$ (controlled on $\{y_1, \hdots, y_{t-1}\}$), which tells us that
\[\Pi_{\mathsf{Dist}}^Y \ket{\psi_t} = \sum_{\substack{x_1, \hdots, x_t \\ \text{distinct } y_1, \hdots, y_{t}}} \ket{y_t}_{\reg A} \ket{\phi_{x_{1:t}, y_{1:t-1}}}_{\reg B} \ket{x_1,\hdots x_{t}, y_1, \hdots, y_{t}}_{\reg R_{\leq t}}.
\]
Given that $\Pi_{\mathsf{Dist},t-1}^Y$ also commutes with $(A_t)_{\reg A \reg B}$, this immediately tells us that
\[\Big|\Big| \Pi_{\mathsf{Dist}}^Y \ket{\psi_t} \Big|\Big|^2 = (N-t+1)\Big|\Big| \Pi_{\mathsf{Dist},t-1}^Y \ket{\psi_{t-1}} \Big|\Big|^2,
\]
completing the second part of the induction. For the first part, we have that
\[\Big(S(\pi) \otimes S(\pi) \Big) \Pi_{\mathsf{Dist}}^Y \ket{\psi_t} = \sum_{\substack{x_1, \hdots, x_t \\ \text{distinct } y_1, \hdots, y_{t}}} \ket{y_t}_{\reg A} \ket{\phi_{x_{1:t}, y_{1:t-1}}}_{\reg B} \ket{x_{\pi^{-1}(1)},\hdots x_{\pi^{-1}(t)}, y_{\pi^{-1}(1)}, \hdots, y_{\pi^{-1}(t)}}_{\reg R_{\leq t}}.
\]
Now, suppose that $\pi(t)\neq t$. Then, the inner product of these two states is zero because a nonzero term in the expansion of this inner product would require terms $y_t = y'_t$ on the $\reg A$ register but also $y_t = y'_{\pi^{-1}(t)}$ on the $\reg R_{Y, t}$ register.

On the other hand, suppose that $\pi(t) = t$ and identify $\pi$ as a non-identity element of $S_{t-1}$; in particular, $S(\pi)$ commutes with $\mathsf{App}_{\reg A \reg R_t} \cdot (A_t)_{\reg A \reg B}$. Then, we write
\[ \bra{\psi_t} \Pi_{\mathsf{Dist}}^Y \Big(S(\pi) \otimes S(\pi) \Big) \Pi_{\mathsf{Dist}}^Y \ket{\psi_t} = \bra{\psi_{t}^{\mathrm{pre}}}\Pi_{\mathsf{Dist},t-1}^Y \mathsf{App}_{\reg A, \reg R_t}^\dagger \Pi_{\mathsf{Dist}}^Y \mathsf{App}_{\reg A, \reg R_t} \Pi_{\mathsf{Dist},t-1}^Y  \Big(S(\pi) \otimes S(\pi) \Big) \ket{\psi_{t}^{\mathrm{pre}}}
\]

Next, we make use of the identity
\[\Pi_{\mathsf{Dist},t-1}^Y \mathsf{App}_{\reg A, \reg R_t}^\dagger \Pi_{\mathsf{Dist}}^Y \mathsf{App}_{\reg A, \reg R_t} \Pi_{\mathsf{Dist},t-1}^Y = (N-t+1) \Pi_{\mathsf{Dist},t-1}^Y,
\]
which again holds because the effect of $\Pi_{\mathsf{Dist}}^Y \mathsf{App}_{\reg A, \reg R_t}$ on the image of $\Pi_{\mathsf{Dist},t-1}^Y$ is to append an un-normalized EPR state over $y_t \notin \{y_1, \hdots, y_{t-1}\}$:
\[\Pi_{\mathsf{Dist}}^Y \mathsf{App}_{\reg A, \reg R_t} \Pi_{\mathsf{Dist},t-1}^Y = \sum_{\substack{x_1, \hdots x_t\\ \text{distinct }y_1, \hdots, y_t}} \ketbra{y_t}{x_t}_{\reg A} \otimes \ketbra{x_1, \hdots, x_t, y_1, \hdots, y_t}{x_1, \hdots, x_{t-1}, y_1, \hdots, y_{t-1}}.
\]
The claimed identity follows from this by a direct computation.

We conclude that
\[
\begin{aligned}
    \bra{\psi_t} \Pi_{\mathsf{Dist}}^Y \Big( S(\pi) \otimes S(\pi)\Big) \Pi_{\mathsf{Dist}}^Y \ket{\psi_t}
    &=
    (N-t+1)
    \bra{\psi_t^{\mathrm{pre}}}
        \Pi_{\mathsf{Dist},t-1}^Y \Big( S(\pi) \otimes S(\pi)\Big)
    \ket{\psi_t^{\mathrm{pre}}} \\
    &=
    (N-t+1)
    \bra{\psi_{t-1}}
        \Pi_{\mathsf{Dist},t-1}^Y (A_t^\dagger)_{\reg A \reg B} \Big( S(\pi) \otimes S(\pi)\Big) (A_t)_{\reg A \reg B}  \Pi_{\mathsf{Dist},t-1}^Y
    \ket{\psi_{t-1}} \\
    &= (N-t+1)
    \bra{\psi_{t-1}}
        \Pi_{\mathsf{Dist},t-1}^Y  \Big( S(\pi) \otimes S(\pi)\Big)  \Pi_{\mathsf{Dist},t-1}^Y
    \ket{\psi_{t-1}}.
\end{aligned}
\]
This completes the induction.

Finally, using
\[
    \Omega_{S_t}
    =
    \frac{1}{t!}
    \sum_{\pi\in S_t}
    S(\pi)\otimes S(\pi),
\]
the cross terms below vanish and so
\[
    \left\|
        \Omega_{S_t}\Pi_{\mathsf{Dist}}^Y\ket{\psi_t}
    \right\|^2
    =
    \frac{1}{t!}
    \left\|\Pi_{\mathsf{Dist}}^Y\ket{\psi_t}\right\|^2
    =
    \binom{N}{t}. \qedhere 
\]
\end{proof}

\begin{lemma}[Reweighting is approximately scalar]
\label{lem:mh-lambda-scalar}
Assume \(t^2=O(N)\). Then
\[
    \left\|
        \Lambda_{S_t}
        -
        \alpha_t \Id
    \right\|
    \le
    \alpha_t \cdot 
    O\left(\frac{t^2}{N}\right).
\]
Equivalently,
\begin{align}
    \label{eq:lambda-symmetric-identity}
    \alpha_t^2
    \left(
        1-O\left(\frac{t^2}{N}\right)
    \right)\Id
    \preceq
    \Lambda_{S_t}^{\dagger}\Lambda_{S_t}
    \preceq
    \alpha_t^2
    \left(
        1+O\left(\frac{t^2}{N}\right)
    \right)\Id .
\end{align}
\end{lemma}

\begin{proof}
We equivalently calculate
    \begin{align}
        \left\lVert
            \frac{N^t}{t!}
            \Lambda_{S_t}^\dagger \Lambda_{S_t}
            -
            \Id
        \right\rVert
        &
        =
        \left\lVert
            \frac{N^t}{t!}
            \left(
                \sum_{\lambda \in \wh{S}_t}
                \beta_{\lambda, t}^2
                \Pi_{\lambda, t}
            \right)
            -
            \Id
        \right\rVert
        \allowdisplaybreaks
        \\
        &
        =
        \left\lVert
            \sum_{\lambda \in \wh{S}_t}
            \left(
                \frac{N^t}{t!}
                \beta_{\lambda, t}^2
                -
                1
            \right)
            \Pi_{\lambda, t}
        \right\rVert
        \allowdisplaybreaks
        \\
        &
        \le
        \max_{\lambda \in \wh{S}_t}
        \left\lvert
            \frac{N^t}{t!}
            \beta_{\lambda, t}^2
            -
            1
        \right\rvert
        \allowdisplaybreaks
        \\
        &
        \le
        O
        \left(
            \frac{t^2}{N}
        \right)
        \tag{\Cref{lem:irrep-ratios}}
        \,.
    \end{align}
\end{proof}

\begin{lemma}[The exact unitary recording is almost distinct on \(\reg R_Y\)]
\label{lem:unitary-mostly-distinct}
\[
    \left\|
        \Pi_{\mathsf{Dist}}^Y
        \ket{\mathsf{Adv}_t^{V_{U(N)}}}
    \right\|^2
    \ge
    1
    -
    O\left(\frac{t^2}{N}\right).
\]
\end{lemma}

\begin{proof}
We have that
\[
    \Pi_{\mathsf{Dist}}^Y
    \ket{\mathsf{Adv}_t^{V_{U(N)}}}
    =
    \Lambda_{S_t}
    \Omega_{S_t}
    \Pi_{\mathsf{Dist}}^Y
    \ket{\psi_t}.
\]
By \cref{lem:mh-lambda-scalar},
\[
    \Lambda_{S_t}^{\dagger}\Lambda_{S_t}
    \succeq
    \alpha_t^2
    \left(
        1-O\left(\frac{t^2}{N}\right)
    \right)\Id,
\]
and by \cref{lem:mh-distinct-append-state},
\[
    \left\|
        \Omega_{S_t}\Pi_{\mathsf{Dist}}^Y\ket{\psi_t}
    \right\|^2
    =
    \frac{1}{\alpha_t^2}.
\]
Together, these prove the lemma. 
\end{proof}

\begin{theorem}
\label{thm:compare-mh-distinct-path-recording}
For any \(t\)-query adversary with \(t^2=O(N)\),
\[
    \left\lVert
            \Proj{\mathsf{Adv}_t^{V_{MH}}}
        -
            \Pi_{\mathsf{Dist}}^Y\Proj{\mathsf{Adv}_t^{V_{U(N)}}}\Pi_{\mathsf{Dist}}^Y
    \right\rVert_1
    \le
    O\left(\frac{t^2}{N}\right).
\]
\end{theorem}

\begin{proof}
We write
\[
    \Pi_{\mathsf{Dist}}^Y
    \ket{\mathsf{Adv}_t^{V_{U(N)}}}-\ket{\mathsf{Adv}_t^{V_{MH}}}
    =
    \left(
        \Lambda_{S_t}
        -
        \alpha_t \Id
    \right)
    \Omega_{S_t}
    \Pi_{\mathsf{Dist}}^Y
    \ket{\psi_t}.
\]
By \cref{lem:mh-lambda-scalar,lem:mh-distinct-append-state},
\[
    \left\|
        \Pi_{\mathsf{Dist}}^Y
    \ket{\mathsf{Adv}_t^{V_{U(N)}}}-\ket{\mathsf{Adv}_t^{V_{MH}}}
    \right\|
    \le
    \alpha_t
    \cdot O\left(\frac{t^2}{N}\right)
    \cdot
    \frac{1}{\alpha_t}
    =
    O\left(\frac{t^2}{N}\right).
\]
Therefore,
\begin{align}
        &\left\lVert
            \Proj{\mathsf{Adv}_t^{V_{MH}}}
        -
            \Pi_{\mathsf{Dist}}^Y\Proj{\mathsf{Adv}_t^{V_{U(N)}}}\Pi_{\mathsf{Dist}}^Y
    \right\rVert_1 \\
    \le &
    2\cdot 
    \left\|
        \Pi_{\mathsf{Dist}}^Y
    \ket{\mathsf{Adv}_t^{V_{U(N)}}}-\ket{\mathsf{Adv}_t^{V_{MH}}}
    \right\| \\
    \le &
    O\left(\frac{t^2}{N}\right).
\end{align}
\end{proof}
This completes the full comparison between the \cite{STOC:MaHua25} path-recording oracle and our exact unitary path recording oracle.
    \subsection{The Unitary Haar Cipher}
\label{sec:unitary_haar_cipher}
We now consider the case of the unitary Haar cipher, which consists of a collection of keyed unitaries $U_1,\, \dots,\, U_K$, each sampled independently from the Haar measure on $U(N)$.
The group is therefore $G = U(N)^K$, and the representation is its block-diagonal embedding:
\begin{align}
    \rho(U_1,\, \dots,\, U_K)
    &
    =
    \sum_{k \in [K]}
    \proj{k}
    \otimes
    U_k
    \,.
    \label{eq:haar-cipher-rep}
\end{align}
This is the idealization of a pseudorandom unitary, in which the unitary for each key is chosen independently from the Haar measure.

As the reader might rightfully guess, we can get a recording oracle for the unitary Haar cipher by simply maintaining $K$ different recordings of each of the $K$ underlying unitaries. And in fact, we can compress it to size that grows only with the number of queries by only explicitly recording the keys whose individual recording has a non-zero number of entries.
This is in fact the correct intuition, and for practical purposes, it might suffice to stop there.
But the beauty of our framework is that it eliminates the guesswork. We can in fact just blindly take the representation above and plug it into our framework to recover this outcome.

For this, we must first consider the commutant algebra of the tensor power of the representation in \Cref{eq:haar-cipher-rep}. It turns out to be the Schur representation of the diagram algebra that comes out of the wreath product $\Z_K \wr S_t$ of the cyclic group mod $K$ and the symmetric group.

\begin{definition}
    The \emph{$K$-colored permutation group} is the wreath product $\Z_K \wr S_t$. That is, it is the group whose group elements $(f, \pi) \in \Z_K \wr S_t$ are parameterized by
    \begin{itemize}
        \item a ``function'' $f \in \Z_K^t$ (we often write group elements of $\Z_K^t$ as functions $f \, : \, [t] \to \Z_K$, or as vectors $\vec f = (f(1), \, \dots, \, f(t))$), and
        \item a permutation $\pi \in S_t$, 
    \end{itemize}
    with a group operation that is given by entry-wise addition in $\Z_K^t$, composing permutations in $S_t$, and permuting a function $f \in \Z_K^t$ by a permutation $\pi \in S_t$ whenever they commute past each other.
    Specifically,
    \begin{align}
        \Z_K \wr S_t
        =
        \left\langle
        (f, \pi)
        \;\middle|\;
        f \in \Z_K^t,\,
        \pi \in S_t,\,
        (
            f_1,
            \pi_1
        )
        \cdot
        (
            f_2,
            \pi_2
        )
        =
        \big(
            f_1
            +
            \pi_1(f_2) 
            ,\, 
            \pi_1 
            \circ 
            \pi_2
        \big)
        \right\rangle
    \end{align}
This group is also often called 
\begin{itemize}
    \item the $K$-phased permutation group, since the $\Z_K$ can be viewed as phases that come from $K$'th roots of unity,
    \item the monomial unitary group with modulus $K$, or
    \item the complex reflection group $G(K, 1, t)$.\todo{add citations} 
\end{itemize}
When specializing to $K = 2$, it is also known as the hyperoctahedral group.

Note that when it is clear from context we will often write the group elements $(f, e_{S_t})$ and $(e_{\Z_K^t}, \pi)$ more simply as $f$ and $\pi$, respectively,%
\footnote{
    $e_{S_t}$ and $e_{\Z_K^t}$ are the identity elements of the respective groups.
}
so that we can write any group element $(f, \pi)$  as a product $f\, \pi$, or equivalently, as $\pi\, f'$, for $f' = \pi^{-1}(f)$.
\end{definition}

The defining representation of the colored permutation group is often taken to be as permutation matrices with nonzero entries taken from the $K$'th roots of unity. 
We can also view group elements as colored permutation diagrams, where each strand has an attached ``color'' in $\Z_K$:
\begin{figure}[H]
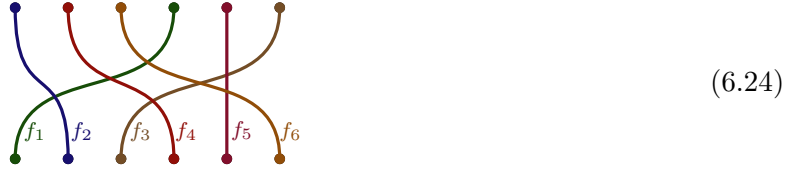

    \centering
    \begin{equation}
    \begin{tikzinline}[scale=1,baseline]
        \makenodes{6}{1cm}{-1cm}{.7cm}

        \connectset[
            label={$f_1$}, 
            color=darkgreen!80!darkbrown
        ]
        {
            B1, 
            T4
        }
        \connectset[
            label={$f_2$}, 
            color=darkblue!80!darkbrown
        ]
        {
            B2, 
            T1
        }
        \connectset[
            label={$f_3$}, 
            color=darkbrown!80!darkbrown
        ]
        {
            B3, 
            T6
        }
        \connectset[
            label={$f_4$}, 
            color=darkred!80!darkbrown
        ]
        {
            B4, 
            T2
        }
        \connectset[
            label={$f_5$}, 
            color=darkpurple!80!darkbrown
        ]
        {
            B5, 
            T5
        }
        \connectset[
            label={$f_6$}, 
            color=darkorange!80!darkbrown
        ]
        {
            B6, 
            T3
        }
    \end{tikzinline}
    \end{equation}
    \caption{Colored permutation diagram on $6$ elements with colors $f_1, \dots, f_6 \in \Z_K$ attached~to~the~strands.} 
    \label{fig:colored-permutation-diagram}
\end{figure}
\noindent Diagram multiplication occurs by concatenation of the underlying permutations and summing (mod $K$) the colors along each merged strand.

We will be interested here in the Schur representation of $\Z_K \wr S_t$ on $(\C^{KN})^{\otimes t}$, which generalizes the Schur representation of $S_t$ (\cref{def:schur_rep}) by adding a phase of $\omega_K^{f_i k_i}$ whenever a register containing $\ket{k_i,\, x_i}$ passes through a strand with color $f_i$.

For example,
\begin{equation}
    S\!\left(
    \vcenter{\hbox{
    \scalebox{0.85}{
    \begin{tikzpicture}[scale=1]
        \makenodes{6}{1cm}{-1cm}{.7cm}

        \connectset[
            label={$f_1$}, 
            color=darkgreen!80!darkbrown
        ]
        {
            B1, 
            T4
        }
        \connectset[
            label={$f_2$}, 
            color=darkblue!80!darkbrown
        ]
        {
            B2, 
            T1
        }
        \connectset[
            label={$f_3$}, 
            color=darkbrown!80!darkbrown
        ]
        {
            B3, 
            T6
        }
        \connectset[
            label={$f_4$}, 
            color=darkred!80!darkbrown
        ]
        {
            B4, 
            T2
        }
        \connectset[
            label={$f_5$}, 
            color=darkpurple!80!darkbrown
        ]
        {
            B5, 
            T5
        }
        \connectset[
            label={$f_6$}, 
            color=darkorange!80!darkbrown
        ]
        {
            B6, 
            T3
        }
    \end{tikzpicture}
    }}}
    \right)
    = 
    \sum_{
        \substack{
            \vec{k} \in [K]^6
            \\
            \vec{x} \in [N]^6
        }
    }
    \omega_{K}^{f_1 k_1 + \dots + f_6 k_6}
    \;
    \substack{
        \phantom{\big(}
        \\
        \ket
        {
            \vphantom{\big(}
            (k_2, x_2), 
            (k_4, x_4), 
            (k_6, x_6),
            (k_1, x_1), 
            (k_5, x_5), 
            (k_3, x_3)
        }
        \\
        \bra
        {
            \vphantom{\big(}
            (k_1, x_1), 
            (k_2, x_2), 
            (k_3, x_3), 
            (k_4, x_4), 
            (k_5, x_5), 
            (k_6, x_6)
        }
    }
    \,.
\end{equation}

Formally,
\begin{align}
    \label{eq:wreath-schur-rep}
    S
    \big(
        \pi\,
        f 
    \big)
    :=
    \sum_{
        \substack{
            \vec{k} \in [K]^t
            \\
            \vec{x} \in [N]^t
        }
    }
    \;
    \omega_{K}^{\vec f \,\cdot\, \vec k}
    \ketbra
    {
        (k_{\pi^{-1}(1)}, x_{\pi^{-1}(1)}), 
        \dots, 
        (k_{\pi^{-1}(t)}, x_{\pi^{-1}(t)}) 
    }
    {
        (k_1, x_1), 
        \dots, 
        (k_t, x_t) 
    }
\end{align}

The \emph{colored permutation algebra} is the group algebra $\C[\Z_K \wr S_t]$, which is the diagram algebra spanned by colored permutation diagrams.

\begin{lemma}
    The commutant of $\rho(\vec U)^{\otimes t}$ ove all $\vec U\in U(N)^K$ (see \Cref{eq:haar-cipher-rep}) is the image of the colored permutation algebra $\C[\Z_K \wr S_t]$ under the Schur representation.
\end{lemma}
\begin{proof}
    First, we show that $S(\pi\, f)$ commutes with $\rho(\vec U)^{\otimes t}$.
    \begin{align}
        S(\pi\, f)
        \;
        \rho(\vec U)^{\otimes t}
        &
        =
        S(\pi)
        \;
        S(f)
        \;
        \rho(\vec U)^{\otimes t}
        \allowdisplaybreaks
        \\
        &
        =
        S(\pi)
        \;
        \left(
            \sum_{
                \substack{
                    \vec{k} \in [K]^t
                }
            }
            \omega_{K}^{\vec f \,\cdot\, \vec k}
            \proj{
                \vec k
            }
            \otimes 
            \Id
        \right)
        \left(
            \sum_{
                \substack{
                    \vec{k} \in [K]^t
                }
            }
            \proj{\vec k}
            \otimes
            U_{k_1} 
            \otimes 
            \cdots 
            \otimes 
            U_{k_t}
        \right)
        \allowdisplaybreaks
        \\
        &
        =
        S(\pi)
        \;
        \left(
            \sum_{
                \substack{
                    \vec{k} \in [K]^t
                }
            }
            \omega_{K}^{\vec f \,\cdot\, \vec k}
            \proj{\vec k}
            \otimes
            U_{k_1} 
            \otimes 
            \cdots 
            \otimes 
            U_{k_t}
        \right)
        \allowdisplaybreaks
        \\
        &
        =
        S(\pi)
        \;
        \left(
            \sum_{
                \substack{
                    \vec{k} \in [K]^t
                }
            }
            \proj{\vec k}
            \otimes
            U_{k_1} 
            \otimes 
            \cdots 
            \otimes 
            U_{k_t}
        \right)
        \left(
            \sum_{
                \substack{
                    \vec{k} \in [K]^t
                }
            }
            \omega_{K}^{\vec f \,\cdot\, \vec k}
            \proj{
                \vec k
            }
            \otimes 
            \Id
        \right)
        \allowdisplaybreaks
        \\
        &
        =
        S(\pi)
        \;
        \rho(\vec U)^{\otimes t}
        \;
        S(f)
        \allowdisplaybreaks
        \\
        &
        =
        \rho(\vec U)^{\otimes t}
        \;
        S(\pi)
        \;
        S(f)
    \end{align}
    This shows that the image of $\C[\Z_K \wr S_t]$ under the Schur representation is contained in the commutant.

    It now remains to prove that every operator that commutes with $\rho(\vec U)^{\otimes t}$ must have the form $S(a)$ for some $a \in \C[\Z_K \wr S_t]$.
    Let $T \in \End((\C^{KN})^{\otimes t})$ be a generic operator on this space, and let's say that it commutes with $\rho(\vec U)^{\otimes t}$ for all tuples of unitaries $\vec U \in U(N)^K$.
    We first write $T$ in terms of its blocks as
    \begin{align}
        T
        =
        \sum_{
            \vec p
            ,\, 
            \vec q 
            \in 
            [K]^t
        }
        \vert
            \vec p
        \rangle
        \!
        \langle
            \vec q
        \vert
        \otimes
        T_{
            \vec p,
            \vec q
        }
    \end{align}
    where the
    $
        \vert
            \vec p
        \rangle
        \!
        \langle
            \vec q
        \vert
    $
    acts on all the control (or key) registers,
    and each 
    $
        T_{
            \vec p,
            \vec q
        }
        \in
        \End((\C^{N})^{\otimes t})
    $
    acts on the rest of the registers (where each of the unitaries $U_{k_i}$ acts).

    We know that for all $\vec U$, we have
    $
        [
            \rho(\vec U)^{\otimes t}
            ,\, 
            T
        ]
        =
        \rho(\vec U)^{\otimes t}
        \,
        T
        -
        T
        \,
        \rho(\vec U)^{\otimes t}
        =
        0
    $.
    For brevity, we write 
    $
        U_{\vec k}
        \coloneqq
        U_{k_1} 
        \otimes 
        \cdots 
        \otimes 
        U_{k_t}
    $.
    The commutator is then
    \begin{align}
        &
        \rho(\vec U)^{\otimes t}
        \,
        T
        -
        T
        \,
        \rho(\vec U)^{\otimes t}
        \allowdisplaybreaks
        \\
        &
        =
        \left(
            \sum_{
                \substack{
                    \vec{k} \in [K]^t
                }
            }
            \proj{\vec k}
            \otimes
            U_{\vec k}
        \right)
        \,
        \left(
            \sum_{
                \vec p
                ,\, 
                \vec q 
                \in 
                [K]^t
            }
            \vert
                \vec p
            \rangle
            \!
            \langle
                \vec q
            \vert
            \otimes
            T_{
                \vec p,
                \vec q
            }
        \right)
        -
        \left(
            \sum_{
                \vec p
                ,\, 
                \vec q 
                \in 
                [K]^t
            }
            \vert
                \vec p
            \rangle
            \!
            \langle
                \vec q
            \vert
            \otimes
            T_{
                \vec p,
                \vec q
            }
        \right)
        \,
        \left(
            \sum_{
                \substack{
                    \vec{k} \in [K]^t
                }
            }
            \proj{\vec k}
            \otimes
            U_{\vec k}
        \right)
        \allowdisplaybreaks
        \\
        &
        =
        \left(
            \sum_{
                \substack{
                    \vec{k}
                    ,\,
                    \vec p
                    ,\, 
                    \vec q 
                    \in 
                    [K]^t
                }
            }
            \proj{\vec k}
            \cdot
            \vert
                \vec p
            \rangle
            \!
            \langle
                \vec q
            \vert
            \otimes
            U_{\vec k}
            \cdot
            T_{
                \vec p,
                \vec q
            }
        \right)
        -
        \left(
            \sum_{
                \substack{
                    \vec{k}
                    ,\,
                    \vec p
                    ,\, 
                    \vec q 
                    \in 
                    [K]^t
                }
            }
            \vert
                \vec p
            \rangle
            \!
            \langle
                \vec q
            \vert
            \cdot
            \proj{\vec k}
            \otimes
            T_{
                \vec p,
                \vec q
            }
            \cdot
            U_{\vec k}
        \right)
        \allowdisplaybreaks
        \\
        &
        =
        \left(
            \sum_{
                \substack{
                    \vec p
                    ,\, 
                    \vec q 
                    \in 
                    [K]^t
                }
            }
            \vert
                \vec p
            \rangle
            \!
            \langle
                \vec q
            \vert
            \otimes
            U_{\vec p}
            \cdot
            T_{
                \vec p,
                \vec q
            }
        \right)
        -
        \left(
            \sum_{
                \substack{
                    \vec p
                    ,\, 
                    \vec q 
                    \in 
                    [K]^t
                }
            }
            \vert
                \vec p
            \rangle
            \!
            \langle
                \vec q
            \vert
            \otimes
            T_{
                \vec p,
                \vec q
            }
            \cdot
            U_{\vec q}
        \right)
        \allowdisplaybreaks
        \\
        &
        =
        \sum_{
            \substack{
                \vec p
                ,\, 
                \vec q 
                \in 
                [K]^t
            }
        }
        \vert
            \vec p
        \rangle
        \!
        \langle
            \vec q
        \vert
        \otimes
        \left(
            U_{\vec p}
            \cdot
            T_{
                \vec p,
                \vec q
            }
            -
            T_{
                \vec p,
                \vec q
            }
            \cdot
            U_{\vec q}
        \right)
        \,.
    \end{align}
    Now, if this operator is $0$, then it must be $0$ on each block. That is, for every $\vec p, \vec q \in [K]^t$ (and every $\vec U$), we have that
    $
        U_{\vec p}
        \cdot
        T_{
            \vec p,
            \vec q
        }
        -
        T_{
            \vec p,
            \vec q
        }
        \cdot
        U_{\vec q}
        =
        0
    $.

    Since this holds for all $\vec U \in U(N)^K$, it must also hold for the case where each unitary is simply $U_k = e^{i\, \theta_k} \, \Id$. In this case, we have that 
    \begin{align}
        U_{\vec p}
        \cdot
        T_{
            \vec p,
            \vec q
        }
        -
        T_{
            \vec p,
            \vec q
        }
        \cdot
        U_{\vec q}
        &
        =
        e^{
            i 
            \sum_{j \in [t]} 
            \theta_{p_j}
        }
        T_{
            \vec p,
            \vec q
        }
        -
        e^{
            i 
            \sum_{j \in [t]} 
            \theta_{q_j}
        }
        T_{
            \vec p,
            \vec q
        }
        \allowdisplaybreaks
        \\
        &
        =
        \left(
            e^{
                i 
                \sum_{j \in [t]} 
                \theta_{p_j}
            }
            -
            e^{
                i 
                \sum_{j \in [t]} 
                \theta_{q_j}
            }
        \right)
        T_{
            \vec p,
            \vec q
        }
        \,.
    \end{align}
    Therefore, $T_{\vec p,\, \vec q}$ must be $0$ unless $\vec p$ is a permutation of $\vec q$ (that is, they have the same entries in a different order).

    Fix $\vec p,\vec q$ with the same key multiplicities,
    and choose $\tau\in S_t$ such that $\tau(\vec q)=\vec p$.
    In the following, permutation operators act only on
    the data registers.
    Since
    \begin{align}
        U_{\vec p}S(\tau)=S(\tau)U_{\vec q},
    \end{align}
    the operator
    \begin{align}
        \widetilde T_{\vec p,\vec q}
        \coloneqq S(\tau)^{-1}T_{\vec p,\vec q}
    \end{align}
    commutes with $U_{\vec q}$ for every $\vec U\in U(N)^K$.

    Let
    \begin{align}
        H_{\vec q}
        \coloneqq
        \{\sigma\in S_t:\sigma(\vec q)=\vec q\},
    \end{align}
    the subgroup that permutes positions having the same key.
    Writing $m_k=|\{j:q_j=k\}|$ and grouping the data registers
    by key, the representation $U_{\vec q}$ becomes
    $\bigotimes_{k=1}^K U_k^{\otimes m_k}$.
    Because the $U_k$ vary independently, its commutant is
    \begin{align}
        \bigotimes_{k=1}^K
        \End_{U(N)}\bigl((\C^N)^{\otimes m_k}\bigr).
    \end{align}
    Applying classical Schur--Weyl duality to each factor
    therefore gives
    \begin{align}
        \widetilde T_{\vec p,\vec q}
        \in \operatorname{span}
        \{S(\sigma):\sigma\in H_{\vec q}\}.
    \end{align}
    Consequently,
    \begin{align}
        T_{\vec p,\vec q}
        \in \operatorname{span}
        \{S(\tau\sigma):\sigma\in H_{\vec q}\}
        =
        \operatorname{span}
        \{S(\pi):\pi(\vec q)=\vec p\}.
    \end{align}
    Thus we may choose coefficients
    $c^{\vec p,\vec q}_{\pi}$ such that
    \begin{align}
        T_{\vec p,\vec q}
        =
        \sum_{\substack{\pi\in S_t\\\pi(\vec q)=\vec p}}
        c^{\vec p,\vec q}_{\pi}\,S(\pi).
    \end{align}

    We thus have that 
    \begin{align}
        T
        &
        =
        \sum_{
            \substack{
                \vec q 
                \in 
                [K]^t
                \\
                \pi
                \in S_t
            }
        }
        c^{\pi(\vec q), \vec q}_{\pi}
        \,
        \vert
            \pi(\vec q)
        \rangle
        \!
        \langle
            \vec q
        \vert
        \otimes
        S(\pi)
        \allowdisplaybreaks
        \\
        &
        =
        \sum_{
            \substack{
                \vec q 
                \in 
                [K]^t
                \\
                \pi
                \in S_t
            }
        }
        c^{\pi(\vec q), \vec q}_{\pi}
        \,
        S(\pi)
        \left(
            \vert
                \vec q
            \rangle
            \!
            \langle
                \vec q
            \vert
            \otimes
            \Id
        \right)
        \,,
    \end{align}
    where in the first line, $S(\pi)$ acts on only the second register, while in the second line $S(\pi)$ acts on both registers together.
    Thus, $T$ is in the span of operators of the form 
    $
        M_{\pi, \vec q}
        \coloneqq
        S(\pi)
        \left(
            \vert
                \vec q
            \rangle
            \!
            \langle
                \vec q
            \vert
            \otimes
            \Id
        \right)
    $.
    It thus remains to show that $M_{\pi, \vec q}$ is in the span of $S(\pi f)$ for $(\pi,f) \in \Z_K \wr S_t$. To this end, observe that 
    \begin{align}
        \label{eq:haar-cipher-commutant-change-of-basis}
        \vert
            \vec q
        \rangle
        \!
        \langle
            \vec q
        \vert
        \otimes
        \Id
        &
        =
        \frac 1 {K^t} \sum_{
            \substack{
                \vec f \in \Z_K^t
            }
        }
        \!
        \omega_{K}^{-\vec f \cdot \vec q}
        \;
        \left(
            \sum_{
                \substack{
                    \vec r \in [K]^t
                }
            }
            \omega_{K}^{\vec f \cdot \vec r}
            \;
            \proj{
                \vec r
            }
            \otimes
            \Id
        \right)
        \allowdisplaybreaks
        \\
        &
        =
        \frac 1 {K^t} \sum_{
            \substack{
                \vec f \in \Z_K^t
            }
        }
        \!
        \omega_{K}^{-\vec f \cdot \vec q}
        \;
        S(f)
        \,.
    \end{align}
    Thus $T$ is in the span of $S(\pi f)$, which completes the proof.
\end{proof}

\begin{lemma}
    \barak{\underline{\textbf{ToDo:}} Should this be moved to the preliminaries? Maybe with the other irrep dims?}
    The irreps of both the group $G = U(N)^K$ and the commutant algebra $\C[\Z_K \wr S_t]$ are parameterized by $K$-tuples of Young diagrams, 
    $
        \left(
            \lambda^{(1)},\, \lambda^{(2)},\, \dots,\, \lambda^{(K)}
        \right)
    $, 
    where each $\lambda^{(i)} \in \wh{S}_{t_i} \cap \wh{U(N)}$, and $\sum_{i \in [K]} t_i = t$.

    Furthermore, the dimensions of the irreps are as follows:

    Since the group $U(N)^K$ is a direct product, its irreps are also direct products of irreps of $U(N)$, and so the dimension of an irrep 
    $
        \vec{\lambda} 
        \coloneqq
        \left(
            \lambda^{(1)},\, \lambda^{(2)},\, \dots,\, \lambda^{(K)}
        \right)
    $ 
    is the product of the individual dimensions 
    \begin{align}
        \dim\left(
            V_{U(N)^K}^{\vec{\lambda}}
        \right)
        &
        =
        \prod_{i \in [K]}
        \dim\left(
            V_{U(N)}^{\lambda^{(i)}}
        \right)
        \,.
    \end{align}
    Similarly, the dimension of the corresponding irrep of $\Z_K \wr S_t$ is~\cite{strahov2024generalizedregularrepresentationsbig,macdonald1998symmetric}
    \begin{align}
        \dim\left(
            V_{\Z_K \wr S_t}^{\vec{\lambda}}
        \right)
        &
        =
        \frac{
            t!
        }{
            \prod_{i \in [K]}
            (t_i!)
        }
        \prod_{i \in [K]}
        \dim\left(
            V_{S_{t_i}}^{\lambda^{(i)}}
        \right)
        =
        \frac{
            t!
        }{
            \prod_{i \in [K]}
            (t_i!)
        }
        \prod_{i \in [K]}
        f^{\lambda^{(i)}}
        \,.
    \end{align}
    Therefore, the irrep ratios are 
    \label{lem:wreath-irrep-ratios}
    \begin{align}
        \label{eq:wreath-irrep-ratios}
        \frac{
            \dim\left(
                V_{\Z_K \wr S_t}^{\vec{\lambda}}
            \right)
        }{
            \dim\left(
                V_{U(N)^K}^{\vec{\lambda}}
            \right)
        }
        &
        =
        \frac{
            t!
        }{
            \prod_{i \in [K]}
            (t_i!)
        }
        \prod_{i \in [K]}
        \frac
        {
            \dim\left(
                V_{S_{t_i}}^{\lambda^{(i)}}
            \right)
        }
        {
            \dim\left(
                V_{U(N)}^{\lambda^{(i)}}
            \right)
        }
        \\
        &
        =
        \frac{
            t!
        }{
            \prod_{i \in [K]}
            (t_i!)
        }
        \prod_{i \in [K]}
        \frac
        {
            t_i!
        }
        {
            N^{t_i}
        }
        \left(
            1
            \pm
            O
            \left(
                \frac{t_i^2}{N}
            \right)
        \right)
        \tag{\Cref{lem:irrep-ratios}}
        \\
        &
        =
        \frac{
            t!
        }{
            N^t
        }
        \left(
            1
            \pm
            O
            \left(
                \frac{t^2}{N}
            \right)
        \right)
        \tag{$\sum_{i} t_i^2 \le \left(\sum_{i} t_i\right)^2 = t^2$}
    \end{align}
\end{lemma}

Note that since the colored permutation algebra is a group algebra, its dual basis (\cref{ex:group-algebra-dual}) is proportional to the inverse of each group element:
\begin{align}
    [
        \pi
        \, 
        f
    ]^{*}
    = 
    \frac{1}{t! \, K^t}
    [f^{-1}
    \pi^{-1}]
    \,.
    \label{eq:colored-permutation-dual}
\end{align}

\subsubsection{Recording Oracle for the Unitary Haar Cipher}

We will use the following notation for the registers:
Let the adversary's query register be split as  $\reg{A} \coloneqq (\reg{A_k}, \reg{A_u})$, corresponding respectively to the key/control register, and the register on which the unitary $U_k$ acts. That is, the query representation acts as $\rho(\vec U) = \sum_{k \in [K]} \proj{k}_{\reg{A_k}} \otimes \left(U_k\right)_{\reg{A_u}}$.

Similarly, let the purification register $\reg{R}$ of the recording be split into $t$ registers $\reg{R}_1, \dots, \reg{R}_t$, and let each $\reg{R}_i$ contain a 4-tuple $(\reg{R}_{\reg{X_k},i},\reg{R}_{\reg{X_u},i},\reg{R}_{\reg{Y_k},i},\reg{R}_{\reg{Y_u},i})$, with $\reg{R}_{\reg{X}, i} \coloneqq (\reg{R}_{\reg{X_k},i},\reg{R}_{\reg{X_u},i})$, and $\reg{R}_{\reg{Y}, i} \coloneqq (\reg{R}_{\reg{Y_k},i},\reg{R}_{\reg{Y_u},i})$. Let 
$\reg{R}_{\reg{X_k}} := (\reg{R}_{\reg{X_k},1}, \dots, \reg{R}_{\reg{X_k},t})$
$\reg{R}_{\reg{X_u}} := (\reg{R}_{\reg{X_u},1}, \dots, \reg{R}_{\reg{X_u},t})$, and so forth.

We have from \Cref{eq:update-rule-path-simplified}
that the path recording update rule for the unitary Haar cipher, $G = U(N)^K$, has the form
\begin{align}
    V_{U(N)^K}
    :=
    \Big(
        \Lambda_{\Z_K \wr S_i}
    \Big)_{\reg R_{\le i}}
    \Big(
        \Omega_{\Z_K \wr S_i}
    \Big)_{\reg{R}_{\le i}}
    \;
    \Big(
        \App
    \Big)_{\reg{A}\,\reg{R}_i} 
    \Big(
        \Lambda_{\Z_K \wr S_{i-1}}^+
    \Big)_{\reg R_{\le i - 1}}
    \;,
\end{align}
and from \Cref{eq:fully_expanded_adversary_action} that a $t$-query adaptive algorithm $\mathsf{Adv}_t$ querying $V_{U(N)^K}$ has the form%
\footnote{
    when starting with an empty recording. Otherwise, we of course also have an additional rescaling of 
    $
        \left(
            \Lambda_{\Z_K \wr S_{|\reg{R_0}|}}^+
        \right)_{\reg R_{0}}
    $
    at the beginning.
}
\begin{align}
    \mathsf{Adv}_t^{V_{U(N)^K}}
    &
    =
    \Big(
        \Lambda_{\Z_K \wr S_t}
    \Big)
    _{\reg{R}}
    \cdot
    \Big(
        \Omega_{\Z_K \wr S_t}
    \Big)
    _{\reg{R}}
    \cdot
    \mathsf{Adv}_{t}^{\App} 
    \,.
\end{align}

To understand the path recording oracle for the unitary Haar cipher, we therefore only need to look at the form of $\Lambda_{\Z_K \wr S_t}$ and 
$\Omega_{\Z_K \wr S_t}$.

\begin{lemma}[Factoring $\Omega$ along the wreath product]
    \begin{align}
        \Omega_{\Z_K \wr S_t}
        =
        \Omega_{S_t}
        \,
        \Omega_{\Z_K^t}
    \end{align}
\end{lemma}
\begin{proof}
    We have that 
    \begin{align}
        S
        \big(
            \pi
            \, 
            f
        \big)
        &
        =
        S
        \big(
            \pi
        \big)
        \, 
        S
        \big(
            f
        \big)
        \,,
    \end{align}
    and
    \begin{align}
        S
        \big(
            (
                \pi
                \, 
                f
            )^{*}
        \big)^T
        &
        = 
        \frac{1}{t! \, K^t}
        S
        \left(
            f^{-1}
            \pi^{-1}
        \right)^T
        \tag{by \Cref{eq:colored-permutation-dual}}
        \\
        &
        = 
        \frac{1}{t! \, K^t}
        \left(
        S
        \left(
            \pi^{-1}
        \right)
        \,
        S
        \left(
            f^{-1}
        \right)
        \right)^T
        \\
        &
        = 
        \frac{1}{t! \, K^t}
        S
        \left(
            \pi^{-1}
        \right)^T
        \,
        S
        \left(
            f^{-1}
        \right)^T
        \\
        &
        = 
        \frac{1}{t! \, K^t}
        S
        \left(
            \pi
        \right)
        \,
        \overline{
            S
            \left(
                f
            \right)
        }
        \,,
        \tag{since $S(\pi)$ is real, and $S(f)$ is diagonal}
    \end{align}
    and therefore
    \begin{align}
        \Omega_{\Z_K \wr S_t}
        &
        =
        \sum_{\pi f \in \Z_K \wr S_t}
        S(\pi f) 
        \otimes 
        S((\pi f)^*)^T
        \tag{\Cref{thm:commutant-epr-proj-as-symmetrization}}
        \\
        &
        =
        \frac{1}{t! \, K^t}
        \sum_{\pi f \in \Z_K \wr S_t}
        S
        \big(
            \pi
        \big)
        \, 
        S
        \big(
            f
        \big)
        \otimes 
        S
        \left(
            \pi
        \right)
        \,
        \overline{
            S
            \left(
                f
            \right)
        }
        \\
        &
        =
        \left(
            \frac{1}{t!}
            \sum_{\pi \in S_t}
            S
            \big(
                \pi
            \big)
            \otimes 
            S
            \left(
                \pi
            \right)
        \right)
        \left(
            \frac{1}{K^t}
            \sum_{f \in \Z_K^t}
            S
            \big(
                f
            \big)
            \otimes 
            \overline{
                S
                \left(
                    f
                \right)
            }
        \right)
        \\
        &
        =
        \Omega_{S_t}
        \,
        \Omega_{\Z_K^t}
        \qedhere
    \end{align}
\end{proof}
\noindent Note that the same argument also shows that 
$
    \Omega_{\Z_K \wr S_t}
    =
    \Omega_{\Z_K^t}
    \,
    \Omega_{S_t}
$ (just write the elements of $\Z_K \wr S_t$ as $f' \pi$ instead of $\pi f$),
implying that $\Omega_{\Z_K^t}$ and~$\Omega_{S_t}$~commute.

\begin{lemma}
    \begin{align}
        \left(
            \Omega_{\Z_K^t}
        \right)
        _{
            \reg{R}_{\reg{X_k}},
            \reg{R}_{\reg{X_u}},
            \reg{R}_{\reg{Y_k}},
            \reg{R}_{\reg{Y_u}}
        }
        &
        =
        \;\;
        \Big(
            \Pi_{\reg{X_k=Y_k}}
        \Big)
        _{
            \reg{R}_{\reg{X_k}},
            \reg{R}_{\reg{Y_k}}
        }
    \end{align}
    where
    $
        \Pi_{\reg{X_k=Y_k}}
    $
    is the projector onto the registers $\reg{R}_{\reg{X_k}}$ and $\reg{R}_{\reg{Y_k}}$ being equal in the standard basis.
\end{lemma}
\begin{proof}
    \begin{align}
        \Omega_{\Z_K^t}
        &
        =
        \frac{1}{K^t}
        \sum_{f \in \Z_K^t}
        S
        \big(
            f
        \big)
        \otimes 
        \overline{
            S
            \left(
                f
            \right)
        }
        \\
        &
        =
        \frac{1}{K^t}
        \sum_{f \in \Z_K^t}
        \sum_{
            \substack{
                \vec{k} \in [K]^t
                \\
                \vec{x} \in [N]^t
            }
        }
        \omega_{K}^{\vec f \,\cdot\, \vec k}
        \ketbra
        {
            \vec k,
            \vec x
        }
        {
            \vec k,
            \vec x
        }
        _{\reg{R_{X_k}}, \reg{R_{X_u}}}
        \otimes
        \sum_{
            \substack{
                \vec{k'} \in [K]^t
                \\
                \vec{y} \in [N]^t
            }
        }
        \omega_{K}^{-\vec f \,\cdot\, \vec k'}
        \ketbra
        {
            \vec k',
            \vec y
        }
        {
            \vec k',
            \vec y
        }
        _{\reg{R_{Y_k}}, \reg{R_{Y_u}}}
        \\
        &
        =
        \sum_{
            \substack{
                \vec{k},
                \vec{k'} \in [K]^t
                \\
                \vec{x},
                \vec{y} \in [N]^t
            }
        }
        \underbrace{
            \left(
                \frac{1}{K^t}
                \sum_{f \in \Z_K^t}
                \omega_{K}^{\vec f \,\cdot\, (\vec k -\vec k')}
            \right)
        }_{\delta_{\vec k, \vec k'}}
        \ketbra
        {
            \vec k,
            \vec x
        }
        {
            \vec k,
            \vec x
        }
        _{\reg{R_{X_k}}, \reg{R_{X_u}}}
        \otimes
        \ketbra
        {
            \vec k',
            \vec y
        }
        {
            \vec k',
            \vec y
        }
        _{\reg{R_{Y_k}}, \reg{R_{Y_u}}}
        \\
        &
        =
        \sum_{
            \substack{
                \vec{k} \in [K]^t
                \\
                \vec{x},
                \vec{y} \in [N]^t
            }
        }
        \ketbra
        {
            \vec k,
            \vec x
        }
        {
            \vec k,
            \vec x
        }
        _{\reg{R_{X_k}}, \reg{R_{X_u}}}
        \otimes
        \ketbra
        {
            \vec k,
            \vec y
        }
        {
            \vec k,
            \vec y
        }
        _{\reg{R_{Y_k}}, \reg{R_{Y_u}}}
        \\
        &
        =
        \sum_{
            \substack{
                \vec{k} \in [K]^t
            }
        }
        \ketbra
        {
            \vec k,
            \vec k,
        }
        {
            \vec k,
            \vec k,
        }
        _{
            \reg{R_{X_k}}, 
            \reg{R_{Y_k}}
        }
        \otimes
        \Id
        _{
            \reg{R_{X_u}},
            \reg{R_{Y_u}}
        }
        \qedhere
    \end{align}
\end{proof}

We see, therefore, that $\Omega_{\Z_K^t}$ plays the role of ensuring that the control register remains the same before and after the query, and moreover, that the same key is recorded in both the $\reg{R_X}$ and the $\reg{R_Y}$ halves of the recording.
That is, the recording will always have pairs of the form $(k_i, x_i, k_i, y_i)$.
Specifically, by the following lemma, we have that when the $\reg{R}$ register is already correctly formatted in a state of this form, 
the update rule without the rescaling proceeds by appending a new tuple of the form $(k_i, x_i, k_i, y_i)$, and symmetrizing over all permutations with the existing tuples.

\begin{lemma}
    \begin{align}
        \Big(
            \Omega_{\Z_K \wr S_i}
        \Big)_{\reg{R}_{\le i}}
        \;
        \Big(
            \App
        \Big)_{\reg{A}\,\reg{R}_i} 
        =
        \Big(
            \Omega_{S_i}
        \Big)_{\reg{R}_{\le i}}
        \;
        \Big(
            \App_{
                \left(
                    \reg{X_k}
                    =
                    \reg{Y_k}
                \right)
            }
        \Big)_{\reg{A}\,\reg{R}_i}
        \Big(
            \Pi_{\reg{X_k=Y_k}}
        \Big)_{\reg{R}_{\le {i-1}}}
    \end{align}
    where
    \begin{align}
        \App_{
            \left(
                \reg{X_k}
                =
                \reg{Y_k}
            \right)
        }
        \coloneqq
        \sum_{k \in [K],\, x, y \in [N]} 
        \ketbra{k, y}{k, x}
        _{\reg{A}}
        \otimes
        \ket{k, x, k, y}
        _{\reg{R}_i}
    \end{align}
\end{lemma}
\begin{proof}
    \begin{align}
        \Big(
            \Omega_{\Z_K \wr S_i}
        \Big)_{\reg{R}_{\le i}}
        \Big(
            \App
        \Big)_{\reg{A}\,\reg{R}_i} 
        &
        =
        \Big(
            \Omega_{S_i}
        \Big)_{\reg{R}_{\le i}}
        \Big(
            \Pi_{\reg{X_k=Y_k}}
        \Big)
        _{
            \reg{R}_{\le i}
        }
        \Big(
            \App
        \Big)_{\reg{A}\,\reg{R}_i} 
        \allowdisplaybreaks
        \\
        &
        =
        \Big(
            \Omega_{S_i}
        \Big)_{\reg{R}_{\le i}}
        \Big(
            \Pi_{\reg{X_k=Y_k}}
        \Big)
        _{
            \reg{R}_{i}
        }
        \Big(
            \Pi_{\reg{X_k=Y_k}}
        \Big)
        _{
            \reg{R}_{\le i-1}
        }
        \Big(
            \App
        \Big)_{\reg{A}\,\reg{R}_i} 
        \allowdisplaybreaks
        \\
        &
        =
        \Big(
            \Omega_{S_i}
        \Big)_{\reg{R}_{\le i}}
        \Big(
            \Pi_{\reg{X_k=Y_k}}
        \Big)
        _{
            \reg{R}_{i}
        }
        \Big(
            \App
        \Big)_{\reg{A}\,\reg{R}_i} 
        \Big(
            \Pi_{\reg{X_k=Y_k}}
        \Big)
        _{
            \reg{R}_{\le i-1}
        }
        \allowdisplaybreaks
        \\
        &
        =
        \Big(
            \Omega_{S_i}
        \Big)_{\reg{R}_{\le i}}
        \left(
            \sum_{
                \substack{
                    k, k' \in [K]
                    \\
                    x, y \in [N]
                }
            } 
            \ketbra{k', y}{k, x}
            _{\reg{A}}
            \otimes
            \Pi_{\reg{X_k=Y_k}}
            \ket{k, x, k', y}
            _{\reg{R}_i}
        \right)
        \Big(
            \Pi_{\reg{X_k=Y_k}}
        \Big)
        _{
            \reg{R}_{\le i-1}
        }
        \allowdisplaybreaks
        \\
        &
        =
        \Big(
            \Omega_{S_i}
        \Big)_{\reg{R}_{\le i}}
        \left(
            \sum_{
                \substack{
                    k \in [K]
                    \\
                    x, y \in [N]
                }
            } 
            \ketbra{k, y}{k, x}
            _{\reg{A}}
            \otimes
            \ket{k, x, k, y}
            _{\reg{R}_i}
        \right)
        \Big(
            \Pi_{\reg{X_k=Y_k}}
        \Big)
        _{
            \reg{R}_{\le i-1}
        }
        \allowdisplaybreaks
        \\
        &
        =
        \Big(
            \Omega_{S_i}
        \Big)_{\reg{R}_{\le i}}
        \Big(
            \App_{
                \left(
                    \reg{X_k}
                    =
                    \reg{Y_k}
                \right)
            }
        \Big)_{\reg{A}\,\reg{R}_i}
        \Big(
            \Pi_{\reg{X_k=Y_k}}
        \Big)_{\reg{R}_{\le {i-1}}}
        \tag*{\qedhere}
    \end{align}
\end{proof}

\begin{corollary}
    \label{lem:haar-cipher-update}
    Assuming we start with a validly formatted recording register, we can write the path recording update rule for the unitary Haar cipher as
    \begin{align}
        V_{U(N)^K}
        =
        \Big(
            \Lambda_{\Z_K \wr S_i}
        \Big)_{\reg R_{\le i}}
        \Big(
            \Omega_{S_i}
        \Big)_{\reg{R}_{\le i}}
        \;
        \Big(
            \App_{
                \left(
                    \reg{X_k}
                    =
                    \reg{Y_k}
                \right)
            }
        \Big)_{\reg{A}\,\reg{R}_i} 
        \Big(
            \Lambda_{\Z_K \wr S_{i-1}}^+
        \Big)_{\reg R_{\le i - 1}}
        \;,
    \end{align}
    and we can write a $t$-query adaptive algorithm $\mathsf{Adv}_t$ querying $V_{U(N)^K}$ starting with an empty recording as
    \begin{align}
        \mathsf{Adv}_t^{V_{U(N)^K}}
        &
        =
        \Big(
            \Lambda_{\Z_K \wr S_t}
        \Big)
        _{\reg{R}}
        \cdot
        \Big(
            \Omega_{S_t}
        \Big)
        _{\reg{R}}
        \cdot
        \mathsf{Adv}_{t}^{
            \App_{
                \left(
                    \reg{X_k}
                    =
                    \reg{Y_k}
                \right)
            }
        } 
        \,.
    \end{align}
\end{corollary}

Thus, we have that up to the reweighting operator $\Lambda_{\Z_K \wr S_t}$ (which for the sake of simplicity, we can safely set aside, as it is $O\left(\frac{t^2}{N}\right)$-close to a scalar multiple of the identity, see \Cref{lem:wreath-irrep-ratios}\footnote{
    A more careful treatment can show that even the $\Lambda_{\Z_K \wr S_t}$ operator can be properly decomposed, and it is therefore not necessary to incur any error in transforming this path recording to a $K$-tuple of individual path recordings. Since this is slightly more involved and offers little additional clarity or intuition at this point, we will not do so here. However, we do work this out explicitly in the special case of diagonal unitaries ($N=1$) below.
}), our recording oracle is encoding multiset states of the form 
\begin{align}
    \ket{
        \left\{
            (k_1, x_1, y_1),
            (k_2, x_2, y_2),
            \,\dots,
            (k_t, x_t, y_t)
        \right\}
    }
    &
    \coloneqq
    \Omega_{S_t}
    \ket{
        (k_1, x_1, k_1, y_1),
        (k_2, x_2, k_2, y_2),
        \,\dots,
        (k_t, x_t, k_t, y_t)
    }
    \,,
    \label{eq:haar-cipher-recording-states}
\end{align}
and each $i$'th query adds $(k_i, x_i, y_i)$ to the set (by appending $(\App)$ and then reshuffling the order $(\Omega_{S_t})$).
We can, of course, alternatively associate each such multiset state with a $K$-tuple $(R_1, \dots, R_K)$ of multisets of the form $R_k = \{(x_i,y_i) \;|\; k_i = k\}$, where the $k$'th multiset in the $K$-tuple contains the $(x,y)$ pairs corresponding to querying the $k$'th unitary. 

\subsubsection{Diagonal Unitaries}
\label{sec:diagonal-unitaries}
Specializing further, we can take $N = 1$, to get the group $U(1)^K$ of $K$-dimensional diagonal unitaries (the $K$-torus). In this case, each $U_k$ in \Cref{eq:haar-cipher-rep} is simply a unit-norm complex phase, and $x_i$ and $y_i$ are 0-qubit registers, and therefore disappear from the update rule, dropping out of \Cref{eq:haar-cipher-recording-states} to give recording states of the form 
\begin{align}
    \ket{
        \left\{
            k_1,
            k_2,
            \,\dots,
            k_t
        \right\}
    }
    &
    :\propto
    \Omega_{S_t}
    \ket{
        (k_1, k_1),
        (k_2, k_2),
        \,\dots,
        (k_t, k_t)
    }
    \,.
\end{align}

Furthermore, the commutant of $\rho(\vec U)^{\otimes t}$ is still a representation of $\C[\Z_K \wr S_t]$, but it is no longer a faithful representation: as noted in \Cref{lem:wreath-irrep-ratios} since $t > N = 1$, the irreps $\vec \lambda$ that survive are those where each $\lambda^{(i)}$ indexes an irrep of $U(1)$. That is, we have that each $\lambda^{(i)}$ is a Young diagram with a single row. Let $\calA_t$ be this quotient algebra, that is $\calA_t \coloneqq \C[\Z_K \wr S_t] / \ker(S)$ (where $S(\cdot)$ is the Schur representation of $\Z_K \wr S_t$, see \Cref{eq:wreath-schur-rep}).

\begin{lemma}
    Let $G = U(1)^K$ be the group of $K$ dimensional diagonal unitaries, with representation $\rho(\theta_1, \dots, \theta_K) = \sum_{k \in [K]} e^{-i\theta_k} \proj{k}$. Then the path recording update rule for this group simplifies to a multiset insertion 
    \begin{align}
        V_{U(1)^K}
        =
        \sum_{
            \substack{
                k \in [K]
                \\
                R \textnormal{ multiset from } [K]
            }
        }
        \proj{k}_{\reg{A}}
        \otimes
        \ketbra{R \uplus \{k\}}{R\,}_{\reg{R}}
    \end{align}
    on multiset states of the form
    $
        \ket{
            \left\{
                k_1,
                k_2,
                \dots,
                k_t
            \right\}
        }
        \coloneqq
        \frac{
            1
        }{
            \sqrt{
                t!
                \prod_{i \in [K]}
                (t_i!)
            }
        }
        \sum_{\pi \in S_t}
        \ket{
            \pi
            (
                k_1,
                k_2,
                \dots,
                k_t
            )
        }
    $,
    where $t_i$ is the number of times $i$ appears in the multiset.%
    \footnote{
        Note that when the multiset contains collisions/repeats, different permutations can give the same basis string, and so the prefactor provides the proper normalization to make this a normalized state.
    }
\end{lemma}

\begin{proof}
We start with the recording oracle for $U(N)^K$ from \Cref{lem:haar-cipher-update}, specialized to the case where $N = 1$:
\begin{align}
    V_{U(1)^K}
    =
    \Big(
        \Lambda_{\calA_i}
    \Big)_{\reg R_{\le i}}
    \Big(
        \Omega_{S_i}
    \Big)_{\reg{R}_{\le i}}
    \;
    \Big(
        \App_{
            \left(
                \reg{X_k}
                =
                \reg{Y_k}
            \right)
        }
    \Big)_{\reg{A}\,\reg{R}_i} 
    \Big(
        \Lambda_{\calA_{i-1}}^+
    \Big)_{\reg R_{\le i - 1}}
    \;,
\end{align}

Note that the purification register here contains both an $\reg{R_X} = (\reg{R_X}_1, \dots, \reg{R_X}_t)$ and a corresponding $\reg{R_Y} = (\reg{R_Y}_1, \dots, \reg{R_Y}_t)$. But we know by \Cref{lem:haar-cipher-update} that $\reg{R_X}$ and $\reg{R_Y}$ will contain identical strings (recall that we have thrown out the part of $\reg{R}$ that was not identical when we took $N=1$). 
So of course, there is redundancy. 
We can therefore write the update rule here as if it were acting on just a single purification register $\reg{R} = (\reg{R}_1, \dots, \reg{R}_t)$ containing a string of $t$ $k$-values, 
These are off by a simple relabeling on the purification register, so this distinction will not matter, but it will simplify notation, so we will write it here as if it were a single register $\reg{R} = (\reg{R}_1, \dots, \reg{R}_t)$.
We thus replace
$
    \Big(
        \App_{
            \left(
                \reg{X_k}
                =
                \reg{Y_k}
            \right)
        }
    \Big)_{\reg{A}\,\reg{R}_i} 
    =
    \sum_{k \in [K]}
    \proj{k}_{\reg{A}}
    \otimes
    \ket{k, k}_{\reg{R_X}_i\, \reg{R_Y}_i}
$
with
$
    \Big(
        \App
    \Big)_{\reg{A}\,\reg{R}_i} 
    =
    \sum_{k \in [K]}
    \proj{k}_{\reg{A}}
    \otimes
    \ket{k}_{\reg{R}_i}
$
and write 
$
    \Omega_{S_t}
$
as
$
    \frac{1}{t!}
    \sum_{\pi \in S_t}
    S(\pi)
$.

Now, we argue about the reweighting operators $\Lambda_{\calA_i}$ and $\Lambda_{\calA_{i-1}}^+$:
Consider the irrep ratios in \Cref{lem:wreath-irrep-ratios}. The group $U(1)$ is Abelian, which implies that all the irreps in the denominator of \Cref{eq:wreath-irrep-ratios} have dimension 1, and furthermore, the irreps of $U(1)$ are parameterized by Young diagrams of a single-row, which means that each $\lambda^{(i)}$ in the numerator is the trivial irrep of the corresponding symmetric group $S_{t_i}$, all of which have dimension 1.
We thus have that the irrep ratios simplify to
$
    \frac{
        t!
    }{
        \prod_{i \in [K]}
        (t_i!)
    }
$, where $t_i$ is the size of $\lambda^{(i)}$, or the number of times that $i$ appears in the recording register.
Therefore, the reweighting operator (for this case when $N = 1$) produces a type-dependent scalar: 
\begin{align}
    \Lambda_{\calA_t} 
    &
    = 
    \sum_{
        \substack{
            (
                t_1, 
                \dots, 
                t_K
            )
        }
    }
    \sqrt{
        \frac{
            t!
        }{
            \prod_{i \in [K]}
            (t_i!)
        }
    }
    \;
    \Pi_{
        (
            t_1, 
            \dots, 
            t_K
        )
    }
    \,,
    \;\;
    \text{ and }
    \;\;
    \Lambda_{\calA_{t-1}} ^+
    = 
    \sum_{
        \substack{
            (
                t_1, 
                \dots, 
                t_K
            )
        }
    }
    \sqrt{
        \frac{
            \prod_{i \in [K]}
            (t_i!)
        }{
            t!
        }
    }
    \;
    \Pi_{
        (
            t_1, 
            \dots, 
            t_K
        )
    }
    \,,
    \label{eq:wreath-lambda-N-is-1}
\end{align}
where 
$
    \Pi_{
        (
            t_1, 
            \dots, 
            t_K
        )
    }
$
is the projector onto strings with a type of the form $(t_1, \dots, t_K)$.

We thus have that
\begin{align}
    \Lambda_{\calA_t}
    \Omega_{S_t}
    &
    =
    \sum_{
        \substack{
            (
                t_1, 
                \dots, 
                t_K
            )
        }
    }
    \sqrt{
        \frac{
            t!
        }{
            \prod_{i \in [K]}
            (t_i!)
        }
    }
    \;
    \Pi_{
        (
            t_1, 
            \dots, 
            t_K
        )
    }
    \cdot
    \frac{1}{t!}
    \sum_{\pi \in S_t}
    S(\pi)
    \allowdisplaybreaks
    \\
    &
    =
    \sum_{
        \substack{
            (
                t_1, 
                \dots, 
                t_K
            )
        }
    }
    \sqrt{
        \frac{
            t!
        }{
            \prod_{i \in [K]}
            (t_i!)
        }
    }
    \;
    \cdot
    \frac{1}{t!}
    \sum_{\pi \in S_t}
    S(\pi)
    \;
    \Pi_{
        (
            t_1, 
            \dots, 
            t_K
        )
    }
    \tag{%
        $
            \Pi_{
                (
                    t_1, 
                    \dots, 
                    t_K
                )
            }
        $
        is invariant under permutations
    }
    \allowdisplaybreaks
    \\
    &
    =
    \sum_{
        \substack{
            (
                t_1, 
                \dots, 
                t_K
            )
        }
    }
    \frac{
        1
    }{
        \sqrt{
            t!
            \prod_{i \in [K]}
            (t_i!)
        }
    }
    \;
    \sum_{\pi \in S_t}
    S(\pi)
    \;
    \Pi_{
        (
            t_1, 
            \dots, 
            t_K
        )
    }
    \allowdisplaybreaks
    \\
    &
    =
    \sum_{
        \substack{
            k_1,
            k_2,
            \dots,
            k_t
            \in
            [K]
        }
    }
    \ketbra
    {
        \left\{
            k_1,
            k_2,
            \dots,
            k_t
        \right\}
    }
    {
        k_1,
        k_2,
        \dots,
        k_t
    }
    \allowdisplaybreaks
    \\
    &
    =:
    \sum_{
        \substack{
            \vec k \in [K]^t
        }
    }
    \lvert
        \{
            \vec k
        \}
    \rangle
    \langle
        \vec k
    \rvert
    \label{eq:wreath-norm-sym1}
    \,.
\end{align}
So this is the normalized symmetrization operator.
Similarly, we have that
\begin{align}
    \Omega_{S_{t-1}}
    \Lambda_{\calA_{t-1}}^+
    &
    =
    \frac{1}{(t-1)!}
    \sum_{\pi \in S_{t-1}}
    S(\pi)
    \sum_{
        \substack{
            (
                t_1, 
                \dots, 
                t_K
            )
        }
    }
    \sqrt{
        \frac{
            \prod_{i \in [K]}
            (t_i!)
        }{
            (t-1)!
        }
    }
    \;
    \Pi_{
        (
            t_1, 
            \dots, 
            t_K
        )
    }
    \allowdisplaybreaks
    \\
    &
    =
    \sum_{
        \substack{
            (
                t_1, 
                \dots, 
                t_K
            )
        }
    }
    \sqrt{
        \frac{
            \prod_{i \in [K]}
            (t_i!)
        }{
            (t-1)!
        }
    }
    \;
    \cdot
    \frac{1}{(t-1)!}
    \sum_{\pi \in S_{t-1}}
    \Pi_{
        (
            t_1, 
            \dots, 
            t_K
        )
    }
    \;
    S(\pi)
    \tag{%
        $
            \Pi_{
                (
                    t_1, 
                    \dots, 
                    t_K
                )
            }
        $
        is invariant under permutations
    }
    \allowdisplaybreaks
    \\
    &
    =
    \sum_{
        \substack{
            (
                t_1, 
                \dots, 
                t_K
            )
        }
    }
    \frac{
        \prod_{i \in [K]}
        (t_i!)
    }
    {
        (t-1)!
    }
    \sqrt{
        \frac{
            1
        }{
            (t-1)!
            \prod_{i \in [K]}
            (t_i!)
        }
    }
    \sum_{\pi \in S_{t-1}}
    \Pi_{
        (
            t_1, 
            \dots, 
            t_K
        )
    }
    \;
    S(\pi)
    \allowdisplaybreaks
    \\
    &
    =
    \sum_{
        \substack{
            k_1,
            k_2,
            \dots,
            k_t
            \in
            [K]
        }
    }
    \frac{
        \prod_{i \in [K]}
        \left(
            t_i^{(\vec k)}!
        \right)
    }
    {
        (t-1)!
    }
    \ketbra
    {
        k_1,
        k_2,
        \dots,
        k_t
    }
    {
        \left\{
            k_1,
            k_2,
            \dots,
            k_t
        \right\}
    }
    \allowdisplaybreaks
    \\
    &
    =:
    \sum_{
        \substack{
            \vec k \in [K]^t
        }
    }
    \frac{
        \prod_{i \in [K]}
        \left(
            t_i^{(\vec k)}!
        \right)
    }
    {
        (t-1)!
    }
    \lvert
        \vec k
    \rangle
    \langle
        \{
            \vec k
        \}
    \rvert
    \label{eq:wreath-norm-sym2}
    \,.
\end{align}
where we take $t_i^{(\vec k)}!$ to be the number of times that a value $i \in [K]$ appears inside $\vec k$.

We now combine everything to get the result:
\begin{align}
    V_{U(1)^K}
    &
    =
    \Big(
        \Lambda_{\calA_i}
    \Big)_{\reg R_{\le i}}
    \Big(
        \Omega_{S_i}
    \Big)_{\reg{R}_{\le i}}
    \Big(
        \App
    \Big)_{\reg{A}\,\reg{R}_i} 
    \Big(
        \Lambda_{\calA_{i-1}}^+
    \Big)_{\reg R_{\le i - 1}}
    \allowdisplaybreaks
    \\
    &
    =
    \Big(
        \Lambda_{\calA_i}
    \Big)_{\reg R_{\le i}}
    \Big(
        \Omega_{S_i}
    \Big)_{\reg{R}_{\le i}}
    \Big(
        \Omega_{S_{i-1}}
    \Big)_{\reg{R}_{\le {i-1}}}
    \Big(
        \App
    \Big)_{\reg{A}\,\reg{R}_i} 
    \Big(
        \Lambda_{\calA_{i-1}}^+
    \Big)_{\reg R_{\le i - 1}}
    \tag{by \Cref{lem:t-fold-omega-collapses}}
    \allowdisplaybreaks
    \\
    &
    =
    \Big(
        \Lambda_{\calA_i}
    \Big)_{\reg R_{\le i}}
    \Big(
        \Omega_{S_i}
    \Big)_{\reg{R}_{\le i}}
    \Big(
        \App
    \Big)_{\reg{A}\,\reg{R}_i} 
    \Big(
        \Omega_{S_{i-1}}
    \Big)_{\reg{R}_{\le {i-1}}}
    \Big(
        \Lambda_{\calA_{i-1}}^+
    \Big)_{\reg R_{\le i - 1}}
    \tag{$\Omega_{S_{i-1}}$ and $\App$ act on different registers}
    \allowdisplaybreaks
    \\
    &
    =
    \Big(
        \Lambda_{\calA_i}
    \Big)_{\reg R_{\le i}}
    \Big(
        \Omega_{S_i}
    \Big)_{\reg{R}_{\le i}}
    \left(
        \!\!\!\!\!\!
        \sum_{
            \substack{
                k \in [K]
                \\
                \phantom{}^{\phantom{i-1}}
                \vec k \in [K]^{i-1}
            }
        }
        \!\!\!\!\!\!
        \proj{k}_{\reg{A}}
        \otimes
        \lvert
            \vec k,
            k
        \rangle
        _{\reg{R}_{\le i}}
        \langle
            \vec k\,
        \rvert
        _{\reg{R}_{\le {i-1}}}
    \right)_{\reg{A}\,\reg{R}_i} 
    \!\!\!\!\!\!
    \Big(
        \Omega_{S_{i-1}}
    \Big)_{\reg{R}_{\le {i-1}}}
    \Big(
        \Lambda_{\calA_{i-1}}^+
    \Big)_{\reg R_{\le i - 1}}
    \tag{definition of $\App$}
    \allowdisplaybreaks
    \\
    &
    =
    \left(
        \!\!\!\!\!\!
        \sum_{
            \substack{
                \phantom{}^{\phantom{i-1}}
                \vec k' \in [K]^{i-1}
            }
        }
        \!\!\!\!\!\!
        \lvert
            \{
                \vec k'
            \}
        \rangle
        \langle
            \vec k'
        \rvert
    \right)_{\!\!\!\reg{R}_{\le i}}
    \left(
        \!\!\!\!\!\!
        \sum_{
            \substack{
                k \in [K]
                \\
                \phantom{}^{\phantom{i-1}}
                \vec k \in [K]^{i-1}
            }
        }
        \!\!\!\!\!\!
        \proj{k}_{\reg{A}}
        \otimes
        \lvert
            \vec k,
            k
        \rangle
        _{\reg{R}_{\le i}}
        \langle
            \vec k\,
        \rvert
        _{\reg{R}_{\le {i-1}}}
    \right)_{\!\!\!\!\reg{A}\,\reg{R}_i} 
    \!\!\!\!
    \left(
        \!\!\!\!\!\!
        \sum_{
            \substack{
                \phantom{}^{\phantom{i-1}}
                \vec k'' \in [K]^{i-1}
            }
        }
        \!\!\!\!\!\!
        \frac{
            \prod_{i \in [K]}
            \left(
                t_i^{(\vec k'')}!
            \right)
        }
        {
            (t-1)!
        }
        \lvert
            \vec k''
        \rangle
        \langle
            \{
                \vec k''
            \}
        \rvert
    \right)_{\!\!\!\reg R_{\le i - 1}}
    \tag{by \Cref{eq:wreath-norm-sym1,eq:wreath-norm-sym2}}
    \allowdisplaybreaks
    \\
    &
    =
    \sum_{
        \substack{
            k \in [K]
            \\
            \phantom{}^{\phantom{i-1}}
            \vec k \in [K]^{i-1}
        }
    }
    \frac{
        \prod_{i \in [K]}
        \left(
            t_i^{(\vec k)}!
        \right)
    }
    {
        (t-1)!
    }
    \proj{k}_{\reg{A}}
    \otimes
    \lvert
        \{
            \vec k
        \}
        \uplus
        \{
            k
        \}
    \rangle
    _{\reg{R}_{\le i}}
    \langle
        \{
            \vec k
        \}
    \rvert
    _{\reg{R}_{\le {i-1}}}
    \allowdisplaybreaks
    \\
    &
    =
    \sum_{
        \substack{
            k \in [K]
            \\
            R \textnormal{ multiset from } [K]
        }
    }
    \!\!\!\!\!\!
    \proj{k}_{\reg{A}}
    \otimes
    \ket{
        R
        \uplus
        \{
            k
        \}
    }
    _{\reg{R}_{\le i}}
    \bra{
        R\,
    }
    _{\reg{R}_{\le {i-1}}}
\end{align}
Where the last equality follows since each multiset $R = \{\vec k\}$ corresponds to 
$
    \frac{
        (t-1)!
    }
    {
        \prod_{i \in [K]}
        \left(
            t_i^{(\vec k)}!
        \right)
    }
$ 
distinct strings. This completes the proof.
\end{proof}

We can see, therefore, that the path recording oracle for $U(1)^K$ is simply keeping a running total that keeps track of how many times each key $k$ has been queried.
This is, of course, the recording oracle of Zhandry~\cite{C:Zhandry19} for the case of queries to a Haar random diagonal unitary (that is, a function $f : [K] \to U(1)$).

    \subsection{Zhandry's Compressed Phase Oracle}

We can, of course, specialize this further to random Boolean functions, $f : \bit^n \to \bit$. Without loss of generality, we can view these functions as being applied in the phase: 
\begin{align}
    \rho(f)
    =
    \sum_{x \in \bit^n} 
    (-1)^{f(x)} 
    \proj{x}
    \,.
    \label{eq:rep-of-boolean-functions}
\end{align}
The group of such Boolean functions is thus $G = \Z_2^{N}$ (again, taking $N = 2^n$), with the group operation given by an XOR on the truth tables, and a representation as diagonal unitaries with $\pm 1$ values. 

\begin{lemma}\label{lemma:path-recording-zhandry}
    Let $G = \Z_2^N$ be the group of Boolean functions, with the representation in \Cref{eq:rep-of-boolean-functions} above. Then the path recording update rule for this group simplifies to a set insertion mod 2 (that is, add to the set if it does not yet appear, and remove it if it does): 
    \begin{align}
        V_{\Z_2^N}
        =
        \sum_{
            \substack{
                x \in [N]
                \\
                R \subseteq [N]
            }
        }
        \proj{x}_{\reg{A}}
        \otimes
        \ketbra{R \oplus \{x\}}{R\,}_{\reg{R}}
        \,.
    \end{align}
    This is the compressed oracle of~\cite{C:Zhandry19} for Boolean functions applied in the phase. 
\end{lemma}

We remark that Zhandry's compressed oracle can also be derived from the tableau-recording oracle: since all irreducible representations of a finite Abelian group are one-dimensional (i.e., characters of the group), the relevant Clebsch-Gordan transforms all simplify to scalars. Nevertheless, we prove \cref{lemma:path-recording-zhandry} to demonstrate how the path-recording oracle can be used to derive a wide variety of existing compressed oracles, and to improve our general-purpose understanding of the path-recording oracle.

\begin{proof}
We can easily compute the commutant of $\rho(f)^{\otimes t}$ as follows.
Let $\vec f$ be the bit string encoding of $f$ as $(f(1), f(2), \dots, f(N)) \in \bit^{N}$, and similarly let $m_2(x) \in \bit^{N}$ be the one-hot encoding of $x \in [N]$. Then $f(x) = \vec f \cdot m_2(x)$.
We write
\begin{align}
    \rho(f)^{\otimes t}
    &
    =
    \sum_{
        x_1, 
        \dots, 
        x_t 
        \in 
        [N]
    } 
    (-1)^{
        f(x_1)
        \,
        \oplus
        \,
        \cdots
        \,
        \oplus
        \,
        f(x_t) 
    } 
    \proj{
        x_1, 
        \dots, 
        x_t
    }
    \\
    &
    =
    \sum_{
        x_1, 
        \dots, 
        x_t 
        \in 
        [N]
    } 
    (-1)^{
        \vec f
        \,
        \cdot
        \,
        \left(
            m_2(x_1) 
            \,
            \oplus
            \,
            \cdots
            \,
            \oplus
            \,
            m_2(x_t)
        \right)
    } 
    \proj{
        x_1, 
        \dots, 
        x_t
    }
    \\
    &
    =
    \sum_{
        \boldsymbol{x} 
        \in 
        [N]^t
    } 
    (-1)^{
        \vec f
        \,
        \cdot
        \,
        \left(
            m_2(\boldsymbol{x})
        \right)
    } 
    \proj{
        \boldsymbol{x}
    }
    \,.
\end{align}
where $\boldsymbol{x} \coloneqq (x_1, \dots, x_t)$, and $m_2(\boldsymbol{x}) \coloneqq m_2(x_1) \oplus \dots \oplus m_2(x_t)$ is the count mod 2 of $\boldsymbol{x}$.

By extending it linearly, we can view the tensor power representation 
$\rho(f)^{\otimes t}$
as a representation of the corresponding group algebra $\C[\Z_2^{N}]$: For each algebra element 
$
    a 
    \in 
    \C[\Z_2^{N}]
$,
where
$
    a 
    = 
    \sum_{f \in \Z_2^{N}} 
    \; 
    a_f 
    \; 
    f 
$, 
we have that 
\begin{align}
    R(a)
    \coloneqq
    \sum_{f \in \Z_2^{N}} 
    \; 
    a_f 
    \; 
    \rho(f)^{\otimes t}
    \,.
\end{align}
Now consider the Fourier basis of the group algebra:
\begin{align}
    \bigg\{
        \hat \mu
        \coloneqq
        \frac{
            1
        }{
            2^{N}
        }
        \sum_{f \in \Z_2^{N}} 
        (-1)^{
            \vec f
            \,
            \cdot
            \,
            \vec \mu
        }
        \;
        f
    \bigg\}
    _{
        \mu \in \Z_2^{N}
    }
    \,.
\end{align}
The representations $R(\hat \mu)$ of the Fourier basis form another basis for the span of the tensor power operators. We can compute that
\begin{align}
    R(\hat \mu)
    &
    =
    \frac{
        1
    }{
        2^{N}
    }
    \sum_{f \in \Z_2^{N}} 
    (-1)^{
        \vec f
        \,
        \cdot
        \,
        \vec \mu
    }
    \;
    \rho(f)^{\otimes t}
    \\
    &
    =
    \frac{
        1
    }{
        2^{N}
    }
    \sum_{
        \substack{
            f \in \Z_2^{N}
            \\
            \boldsymbol{x}
            \in
            [N]^t
        }
    } 
    (-1)^{
        \vec f
        \,
        \cdot
        \,
        \left(
            \vec \mu
            \;
            \oplus
            \;
            m_2(\boldsymbol{x})
        \right)
    } 
    \proj{
        \boldsymbol{x}
    }
    \\
    &
    =
    \sum_{
        \substack{
            \boldsymbol{x} 
            \in 
            [N]^t
            \\
            m_2(\boldsymbol{x})
            =
            \vec \mu
        }
    } 
    \proj{
        \boldsymbol{x}
    }
    \label{eq:mod-2-couning-projection}
\end{align}

Define the set of occurring parity labels by
$\mathcal M_t:=\{\mu\subseteq[N]:|\mu|\le t,\ |\mu|\equiv t\pmod 2\},$
where we identify bit strings with subsets of $[N]$. For a string of length $t$, the number of labels occurring an odd number of times is at most $t$ and has the same parity as $t$. Consequently, $R(\hat\mu)$ is nonzero exactly when $\mu\in\mathcal M_t$. For such $\mu$, it projects onto strings in which each label in $\mu$ occurs an odd number of times and each label outside $\mu$ occurs an even number of times. 
The subspaces corresponding to these projectors are mutually orthogonal: $R(\hat \mu) R(\hat \nu) = \delta_{\mu \nu} R(\hat \mu)$.

Their commutant $\calA_t$ is therefore the algebra of operators that preserve these subspaces, and is spanned by operators of the form 
\begin{align}
    \Big\{
        E^{\mu}_{\boldsymbol{a}, \boldsymbol{b}}
        \coloneqq
        \ketbra{
            \boldsymbol{a}
        }{
            \boldsymbol{b}
        }
    \;
    \Big\vert
    \;
        m_2(\boldsymbol{a})
        =
        m_2(\boldsymbol{b})
        =
        \mu
    \Big\}
    _{
        \boldsymbol{a}, \boldsymbol{b} \in [N]^t,
        \,
        \mu \subseteq [N],
    }
    \label{eq:basis-zandry-commutant}
\end{align}
This basis of the commutant algebra already presents it as a direct sum of matrix algebras. That is, this is the Fourier basis of the algebra. We can thus directly see that 
the irreps of this algebra are parameterized by $\mu\in\mathcal M_t$,
and furthermore, that the dimension of irrep $\mu$ is the total number of ways to choose $\boldsymbol{a} \in [N]^t$ such that $m_2(a) = \mu$:
\begin{align}
    \dim(V^{\mu}_{\calA_t}) 
    &
    =
    \sum_{
        \substack{
            t_1,
            \dots,
            t_N
            \ge
            0
            \\
            t_i 
            \text{ odd iff }
            i \in \mu
            \\
            \sum_{i} t_i = t
        }
    }
    \frac{t!}{t_1!\dots t_N!}
    \\
    &
    =
    \sum_{
        \substack{
            \vec t
            \in
            \Z_{\ge 0}^N
            ,\,\,
            |\,\vec t\,| = t
            \\
            \vec \mu
            = 
            \vec t 
            \mod 2
        }
    }
    \frac{t!}{t_1!\dots t_N!}
    \,.
\end{align}

By \Cref{lem:properties-of-Fourier-basis}, \Cref{item:Fourier-basis-self-dual},
we have that
$
    S
    \bigg(
    \Big(
        E^{\mu}_{\boldsymbol{a}, \boldsymbol{b}}
    \Big)^*
    \bigg)^T
    =
    \frac{1}{\dim(V^{\mu}_{\calA_t})}
    S
    \Big(
        E^{\mu}_{\boldsymbol{a}, \boldsymbol{b}}
    \Big)
$.

We thus have that the reweighting and symmetrization operators have the following form:

\begin{align}
    \Omega_{\calA_t}
    &
    =
    \sum_{
        \substack{
            \mu 
            \subseteq
            [N],
            \,
            \boldsymbol{a}, 
            \boldsymbol{b} 
            \in
            [N]^t
            \\
            m_2(\boldsymbol{a})
            =
            m_2(\boldsymbol{b})
            =
            \mu
        }
    }
    \frac{1}{
        \dim(V^{\mu}_{\calA_t})
    }
    \;
    S
    \Big(
        E^{\mu}_{\boldsymbol{a}, \boldsymbol{b}}
    \Big)
    \otimes
    S
    \Big(
        E^{\mu}_{\boldsymbol{a}, \boldsymbol{b}}
    \Big)
    \\
    &
    =
    \sum_{
        \substack{
            \boldsymbol{a}, 
            \boldsymbol{b} 
            \in
            [N]^t
            \\
            m_2(\boldsymbol{a})
            =
            m_2(\boldsymbol{b})
        }
    }
    \frac{1}{
        \dim
        \left(
            V^{
                m_2(\boldsymbol{a})
            }
            _{
                \calA_t
            }
        \right)
    }
    \;
    \ketbra{
        \boldsymbol{a}
    }{
        \boldsymbol{b}
    }
    \otimes
    \ketbra{
        \boldsymbol{a}
    }{
        \boldsymbol{b}
    }
    \,,
\end{align}
and as before, since we now have that the two registers are always equal in the standard basis, we will abuse notation and write it as if it were a single register 
\begin{align}
    \Omega_{\calA_t}
    =
    \sum_{
        \substack{
            \boldsymbol{a}, 
            \boldsymbol{b} 
            \in
            [N]^t
            \\
            m_2(\boldsymbol{a})
            =
            m_2(\boldsymbol{b})
        }
    }
    \frac{1}{
        \dim
        \left(
            V^{
                m_2(\boldsymbol{a})
            }
            _{
                \calA_t
            }
        \right)
    }
    \;
    \ketbra{
        \boldsymbol{a}
    }{
        \boldsymbol{b}
    }
    \,.
\end{align}

\begin{align}
    \Lambda_{\calA_t}
    &
    =
    \sum_{
        \substack{
            \mu\in\mathcal M_t
        }
    }
    \sqrt{
        \frac{
            \dim(V^{\mu}_{\calA_t})
        }{
            \dim(V^{\mu}_{\Z_2^N})
        }
    }
    \;
    \Pi_{\mu}
    &
    \Lambda_{\calA_t}^+
    &
    =
    \sum_{
        \substack{
            \mu\in\mathcal M_t
        }
    }
    \sqrt{
        \frac{
            \dim(V^{\mu}_{\Z_2^N})
        }{
            \dim(V^{\mu}_{\calA_t})
        }
    }
    \;
    \Pi_{\mu}
    \\
    &
    =
    \sum_{
        \substack{
            \mu\in\mathcal M_t
        }
    }
    \sqrt{
        \dim(V^{\mu}_{\calA_t})
    }
    \;
    \Pi_{\mu}
    &
    &
    =
    \sum_{
        \substack{
            \mu\in\mathcal M_t
        }
    }
    \frac{1}{
        \sqrt{
            \dim(V^{\mu}_{\calA_t})
        }
    }
    \Pi_{\mu}
\end{align}
where
$
    \;
    \Pi_{\mu}
    \coloneqq
    \sum_{
        \substack{
            \boldsymbol{x} 
            \in 
            [N]^t
            ,\,
            m_2(\boldsymbol{x})
            =
            \vec \mu
        }
    } 
    \proj{
        \boldsymbol{x}
    }
$
are the projectors from \Cref{eq:mod-2-couning-projection},
and the second line follows since $\Z_2^N$ is an Abelian group, so its irreps are all 1-dimensional.

For $\boldsymbol{x} \in [N]^t$, let 
$
    \{
        \boldsymbol{x}
    \}_{\textrm{mod } 2}
    \subseteq
    [N]
$
be the set corresponding to $m_2(\boldsymbol{x})$. That is, it is the set of values appearing an odd number of times in $\boldsymbol{x}$.
Let 
\begin{align}
    \ket{
        \{
            \boldsymbol{x}
        \}_{\textrm{mod } 2}
    }
    \coloneqq
    \frac{1}{
        \sqrt{
            \dim
            \left(
                V^{
                    m_2(\boldsymbol{x})
                }
                _{
                    \calA_t
                }
            \right)
        }
    }
    \sum_{
        \substack{
            \boldsymbol{y}
            \in
            [N]^t
            \\
            m_2(\boldsymbol{y})
            =
            m_2(\boldsymbol{x})
        }
    }
    \ket{\boldsymbol{y}}
\end{align}
be the uniform superposition over strings of the same parity.

We then have that
\begin{align}
    \Lambda_{\calA_t}
    \Omega_{\calA_t}
    &
    =
    \sum_{
        \substack{
            \boldsymbol{a}, 
            \boldsymbol{b} 
            \in
            [N]^t
            \\
            m_2(\boldsymbol{a})
            =
            m_2(\boldsymbol{b})
        }
    }
    \sqrt{
        \dim
        \left(
            V^{
                m_2(\boldsymbol{a})
            }
            _{
                \calA_t
            }
        \right)
    }
    \;
    \frac{1}{
        \dim
        \left(
            V^{
                m_2(\boldsymbol{a})
            }
            _{
                \calA_t
            }
        \right)
    }
    \;
    \ketbra{
        \boldsymbol{a}
    }{
        \boldsymbol{b}
    }
    \allowdisplaybreaks
    \\
    &
    =
    \sum_{
        \substack{
            \boldsymbol{a}, 
            \boldsymbol{b} 
            \in
            [N]^t
            \\
            m_2(\boldsymbol{a})
            =
            m_2(\boldsymbol{b})
        }
    }
    \;
    \frac{1}{
        \sqrt{
            \dim
            \left(
                V^{
                    m_2(\boldsymbol{a})
                }
                _{
                    \calA_t
                }
            \right)
        }
    }
    \;
    \ketbra{
        \boldsymbol{a}
    }{
        \boldsymbol{b}
    }
    \allowdisplaybreaks
    \\
    &
    =
    \sum_{
        \substack{
            \boldsymbol{x} 
            \in
            [N]^t
        }
    }
    \lvert
        \{
            \boldsymbol{x}
        \}_{\textrm{mod } 2}
    \rangle
    \langle
        \boldsymbol{x}
    \rvert
    \label{eq:zhandry-norm-sym1}
    \,,
\end{align}
and similarly,
\begin{align}
    \Omega_{\calA_{t-1}}
    \Lambda_{\calA_{t-1}}^+
    &
    =
    \sum_{
        \substack{
            \boldsymbol{a}, 
            \boldsymbol{b} 
            \in
            [N]^{t-1}
            \\
            m_2(\boldsymbol{a})
            =
            m_2(\boldsymbol{b})
        }
    }
    \frac{1}{
        \sqrt{
            \dim
            \left(
                V^{
                    m_2(\boldsymbol{a})
                }
                _{
                    \calA_{t-1}
                }
            \right)
        }
    }
    \;
    \frac{1}{
        \dim
        \left(
            V^{
                m_2(\boldsymbol{a})
            }
            _{
                \calA_{t-1}
            }
        \right)
    }
    \;
    \ketbra{
        \boldsymbol{a}
    }{
        \boldsymbol{b}
    }
    \allowdisplaybreaks
    \\
    &
    =
    \sum_{
        \substack{
            \boldsymbol{x} 
            \in
            [N]^{t-1}
        }
    }
    \;
    \frac{1}{
        \dim
        \left(
            V^{
                m_2(\boldsymbol{x})
            }
            _{
                \calA_{t-1}
            }
        \right)
    }
    \;
    \lvert
        \boldsymbol{x}
    \rangle
    \langle
        \{
            \boldsymbol{x}
        \}_{\textrm{mod } 2}
    \rvert
    \label{eq:zhandry-norm-sym2}
    \,.
\end{align}

Let 
$
    d_{\boldsymbol{x}} 
    \coloneqq 
    \dim
    \left(
        V^{
            m_2(\boldsymbol{x})
        }
        _{
            \calA_{i-1}
        }
    \right)
$
for brevity.
As before,
we now combine these to conclude:
\begin{align}
    V_{\Z_2^N}
    &
    =
    \Big(
        \Lambda_{\calA_i}
    \Big)_{\reg R_{\le i}}
    \Big(
        \Omega_{\calA_i}
    \Big)_{\reg{R}_{\le i}}
    \Big(
        \App
    \Big)_{\reg{A}\,\reg{R}_i} 
    \Big(
        \Lambda_{\calA_{i-1}}^+
    \Big)_{\reg R_{\le i - 1}}
    \allowdisplaybreaks
    \\
    &
    =
    \Big(
        \Lambda_{\calA_i}
    \Big)_{\reg R_{\le i}}
    \Big(
        \Omega_{\calA_i}
    \Big)_{\reg{R}_{\le i}}
    \Big(
        \Omega_{\calA_{i-1}}
    \Big)_{\reg{R}_{\le {i-1}}}
    \Big(
        \App
    \Big)_{\reg{A}\,\reg{R}_i} 
    \Big(
        \Lambda_{\calA_{i-1}}^+
    \Big)_{\reg R_{\le i - 1}}
    \tag{by \Cref{lem:t-fold-omega-collapses}}
    \allowdisplaybreaks
    \\
    &
    =
    \Big(
        \Lambda_{\calA_i}
    \Big)_{\reg R_{\le i}}
    \Big(
        \Omega_{\calA_i}
    \Big)_{\reg{R}_{\le i}}
    \Big(
        \App
    \Big)_{\reg{A}\,\reg{R}_i} 
    \Big(
        \Omega_{\calA_{i-1}}
    \Big)_{\reg{R}_{\le {i-1}}}
    \Big(
        \Lambda_{\calA_{i-1}}^+
    \Big)_{\reg R_{\le i - 1}}
    \tag{$\Omega_{\calA_{i-1}}$ and $\App$ act on different registers}
    \allowdisplaybreaks
    \\
    &
    =
    \Big(
        \Lambda_{\calA_i}
    \Big)_{\reg R_{\le i}}
    \Big(
        \Omega_{\calA_i}
    \Big)_{\reg{R}_{\le i}}
    \left(
        \!\!\!\!\!\!
        \sum_{
            \substack{
                x \in [N]
                \\
                \phantom{}^{\phantom{i-1}}
                \boldsymbol{x} \in [N]^{i-1}
            }
        }
        \!\!\!\!\!\!
        \proj{x}_{\reg{A}}
        \otimes
        \lvert
            \boldsymbol{x},
            x
        \rangle
        _{\reg{R}_{\le i}}
        \langle
            \boldsymbol{x}\,
        \rvert
        _{\reg{R}_{\le {i-1}}}
    \right)_{\reg{A}\,\reg{R}_i} 
    \!\!\!\!\!\!
    \Big(
        \Omega_{\calA_{i-1}}
    \Big)_{\reg{R}_{\le {i-1}}}
    \Big(
        \Lambda_{\calA_{i-1}}^+
    \Big)_{\reg R_{\le i - 1}}
    \tag{definition of $\App$}
    \allowdisplaybreaks
    \\
    &
    =
    \left(
        \!\!
        \sum_{
            \substack{
                \phantom{}^{\phantom{i}}
                \boldsymbol{x}' \in [N]^{i}
            }
        }
        \!\!
        \lvert
            \{
                \boldsymbol{x}'
            \}_{\textrm{mod } 2}
        \rangle
        \langle
            \boldsymbol{x}'
        \rvert
    \right)_{\!\!\!\reg{R}_{\le i}}
    \left(
        \!\!\!\!\!\!
        \sum_{
            \substack{
                x \in [N]
                \\
                \phantom{}^{\phantom{i-1}}
                \boldsymbol{x} \in [N]^{i-1}
            }
        }
        \!\!\!\!\!\!
        \proj{x}_{\reg{A}}
        \otimes
        \lvert
            \boldsymbol{x},
            x
        \rangle
        _{\reg{R}_{\le i}}
        \langle
            \boldsymbol{x}\,
        \rvert
        _{\reg{R}_{\le {i-1}}}
    \right)_{\!\!\!\!\reg{A}\,\reg{R}_i} 
    \!\!\!\!
    \left(
        \!\!\!\!\!\!
        \sum_{
            \substack{
                \phantom{}^{\phantom{i-1}}
                \boldsymbol{x}'' \in [N]^{i-1}
            }
        }
        \!
        \frac{1}{
            d_{\boldsymbol{x}''}
        }
        \lvert
            \boldsymbol{x}''
        \rangle
        \langle
            \{
                \boldsymbol{x}''
            \}_{\textrm{mod } 2}
        \rvert
    \right)_{\!\!\!\reg R_{\le i - 1}}
    \tag{by \Cref{eq:zhandry-norm-sym1,eq:zhandry-norm-sym2}}
    \allowdisplaybreaks
    \\
    &
    =
    \sum_{
        \substack{
            x \in [N]
            \\
            \phantom{}^{\phantom{i-1}}
            \boldsymbol{x} \in [N]^{i-1}
        }
    }
    \frac{1}{
        d_{\boldsymbol{x}}
    }
    \proj{x}_{\reg{A}}
    \otimes
    \lvert
        \{
            \boldsymbol{x}
        \}_{\textrm{mod } 2}
        \oplus
        \{
            x
        \}
    \rangle
    _{\reg{R}_{\le i}}
    \langle
        \{
            \boldsymbol{x}
        \}_{\textrm{mod } 2}
    \rvert
    _{\reg{R}_{\le {i-1}}}
    \allowdisplaybreaks
    \\
    &
    =
    \sum_{
        \substack{
            x \in [N]
            \\
            R\in\mathcal M_{i-1}
        }
    }
    \proj{x}_{\reg{A}}
    \otimes
    \ket{
        R
        \oplus
        \{
            x
        \}
    }
    _{\reg{R}_{\le i}}
    \bra{
        R\,
    }
    _{\reg{R}_{\le {i-1}}}
\end{align}
Where the last equality follows since each parity set $R = \{\boldsymbol{x}\}_{\textrm{mod } 2}$ corresponds to 
$d_{\boldsymbol{x}}$ 
distinct strings. This completes the proof.
\end{proof}

It turns out that there is another characterization of this commutant algebra in terms of colored even-partition diagrams.

\begin{definition}
    [Colored Even-Partition Algebra]
    \label{def:colored-even-partition-algebra}
    The \emph{even-partition algebra} is the algebra of partition diagrams in which every component has even number of vertices. 
    This is equivalent to the Tanabe algebra $T_t(N, 2)$~\cite{tanabe1997centralizer}.

    The $\Z_N$-\emph{colored even-partition algebra} is the algebra that has a basis of even-partition diagrams with each component assigned a color in $\Z_N$ (see \Cref{fig:colored-even-partition-diagram} for an example of such a colored diagram).%
    \footnote{
        This diagram algebra is closely related to the $G$-colored partition algebras studied by~\cite{bloss2003g}. 
    }
    Diagram multiplication is given by concatenation, with the colors
    of merged components added in $\Z_N$.
    Each component contained entirely in the identified middle row
    is removed and contributes a scalar factor of $N$ if its
    accumulated color is zero, and $0$ otherwise.

    \begin{figure}[H]
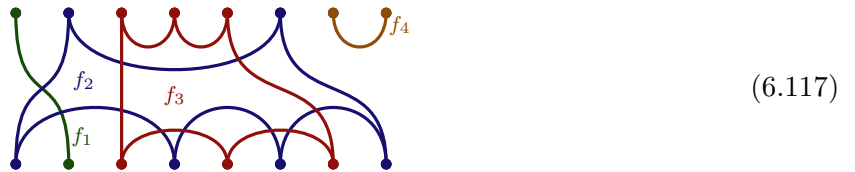

    \centering
    \begin{equation}
    \begin{tikzinline}[scale=1,baseline]
        \makenodes{8}{1cm}{-1cm}{.7cm}

        \connectset[
            label={$f_1$}, 
            color=darkgreen!80!darkbrown
        ]
        {
            T1, 
            B2
        }
        \connectset[
            height=1,
            color=darkblue!80!darkbrown
        ]
        {
            T2,
            T6,
            B1, 
            B4,
            B6,
            B8
        }
        \connectset[
            color=darkblue!80!darkbrown
        ]
        {
            T6,
            B8
        }
        \node [darkblue!80!darkbrown, font=\scriptsize] at (1.6,0.1) {$f_2$};
        \connectset[
            color=darkred!80!darkbrown
        ]
        {
            T3, 
            T4,
            T5,
            B3,
            B5,
            B7
        }
        \connectset[
            color=darkred!80!darkbrown
        ]
        {
            T5,
            B7
        }
        \node [darkred!80!darkbrown, font=\scriptsize] at (2.8,-0.1) {$f_3$};
        \connectset[
            label={$f_4$}, 
            color=darkorange!80!darkbrown
        ]
        {
            T7, 
            T8
        }
    \end{tikzinline}
    \end{equation}
    \caption{Colored even-partition diagram on $8$ elements with colors $f_1, f_2, f_3, f_4 \in \Z_N$ attached to the four components.} 
    \label{fig:colored-even-partition-diagram}
\end{figure}

    The Schur representation on $(\C^{N})^{\otimes t}$ differs from that of the partition algebra by attaching a phase of $\omega_N^{f_i x_i}$ to the sum over $x_i$ of a component of color $f_i$
    (similar to Schur representation of $\Z_K \wr S_t$ on $(\C^{KN})^{\otimes t}$).

    For example, 
    \begin{equation}
        S\!\left(
        \vcenter{\hbox{
        \scalebox{0.85}{
        \begin{tikzpicture}[scale=1]
            \makenodes{8}{1cm}{-1cm}{.7cm}

            \connectset[
                label={$f_1$}, 
                color=darkgreen!80!darkbrown
            ]
            {
                T1, 
                B2
            }
            \connectset[
                height=1,
                color=darkblue!80!darkbrown
            ]
            {
                T2,
                T6,
                B1, 
                B4,
                B6,
                B8
            }
            \connectset[
                color=darkblue!80!darkbrown
            ]
            {
                T6,
                B8
            }
            \node [darkblue!80!darkbrown, font=\scriptsize] at (1.6,0.1) {$f_2$};
            \connectset[
                color=darkred!80!darkbrown
            ]
            {
                T3, 
                T4,
                T5,
                B3,
                B5,
                B7
            }
            \connectset[
                color=darkred!80!darkbrown
            ]
            {
                T5,
                B7
            }
            \node [darkred!80!darkbrown, font=\scriptsize] at (2.8,-0.1) {$f_3$};
            \connectset[
                label={$f_4$}, 
                color=darkorange!80!darkbrown
            ]
            {
                T7, 
                T8
            }
        \end{tikzpicture}
        }}}
        \right)
        = 
        \sum_{
            \substack{
                x_1, x_2, x_3, x_4 \in [N]
            }
        }
        \omega_{N}^{f_1 x_1 + \dots + f_4 x_4}
        \;
        \substack{
            \phantom{\big(}
            \\
            \ket
            {
                \vphantom{\big(}
                x_1, 
                x_2, 
                x_3, 
                x_3, 
                x_3, 
                x_2,
                x_4,
                x_4
            }
            \\
            \bra
            {
                \vphantom{\big(}
                x_2,
                x_1, 
                x_3,
                x_2, 
                x_3, 
                x_2, 
                x_3, 
                x_2
            }
        }
        \;
        \,.
    \end{equation}
\end{definition}

\begin{lemma}
    The commutant algebra of $\rho(f)^{\otimes t}$
    is the is the image of the \emph{colored even-partition algebra} under the Schur representation. 
    A basis%
    \footnote{
        As in most of this paper, we assume here that $t \ll N$.
    }
    consists of the Schur images of colored even-partition diagrams, where, under some canonical ordering,%
    \footnote{
        For instance, ordering the blocks by their smallest vertex.
    }
    the $j$th block's color is restricted to $[N+1-j]$.
\end{lemma}

\begin{proof}[Proof sketch]
    This comes from performing a change of basis on the basis of \Cref{eq:basis-zandry-commutant}, similarly to \Cref{eq:haar-cipher-commutant-change-of-basis}, but with a block-triangular structure.
    We omit the details, as this characterization is not
    used elsewhere in the paper.
\end{proof}

\section{PC is a Pseudorandom Unitary}
\label{sec:prus}
In this section, we prove that any algorithm making $t$ queries cannot distinguish a Haar-random unitary from a random Clifford circuit (or arbitrary $2$-design) composed with a random permutation with advantage better than $O(t^2/N)$. Throughout this section, we assume \(t\le c\sqrt N\) for a sufficiently small
absolute constant \(c>0\). In particular, this guarantees that the partition algebra \(P_t(N)\) is semisimple and all \(O(t^2/N)\) error terms below are smaller than an absolute constant.
We also note that we consider standard PRU security in this section.%
\footnote{
    This PC construction, much like the PFC construction of~\cite{STOC:MaHua25} does not satisfy strong PRU security, since inverting it allows one to query on the distinct nonplussed subspace.
    If we want strong PRU security, we conjecture that arguments along these lines can show that $C_1 \cdot P \cdot C_2$ satisfies the stronger security notions.
}

In order to prove this, we carefully analyze the path-recording oracles $Q_{U(N)}^{\mathrm{Path}}$, $Q_{S_N}^{\mathrm{Path}}$ for the defining representation of $U(N)$ and the in-place representation of $S_N$. In particular, we prove that they are close in trace distance on the distinct, nonplussed subspace $\mathsf{DNP}$ (\cref{def:dist-nonplussed-subspace}), and prove that a random $2$-design $C$ causes almost all of the recording state to land in $\mathsf{DNP}$. Proving the first of these statements, in turn, requires two main steps: an analysis of the dual elements of permutations $[\pi]^*\in P_t(N)$ within the partition algebra, and an analysis of $\lambda$-irrep dimensions in $(U(N), S_t)$ and $(S_N, P_t(N))$ Schur-Weyl duality. 

\subsection{The Distinct, Nonplussed Subspace}
\subsubsection{Preliminaries}

\begin{lemma}[\cite{FOCS:MPSY24,STOC:MaHua25}]
    \label{lem:distinct-twirling}
    Let $C$ be any unitary $2$-design on $\C^N$. Then, 
    \begin{equation}
        \mathbb{E}_C\!\left[\Tr\left(
               \Pi_{\mathsf{Dist}_{N,t}}C^{\otimes t} \rho (C^\dagger)^{\otimes t}  
            \right)\right]  \ge 1 - \frac{t(t-1)}{N + 1}
    \end{equation}
\end{lemma}

\begin{definition}
    Let $\ket{+}$ denote the uniform superposition on the standard basis of $\C^N$, and define $W = \Id - \ketbra{+}{+}$, i.e.\ the projector away from $\ket{+}$. For a system of $t$ registers, the \textit{nonplussed subspace}\footnote{In other words, we are completely unfazed by states in this subspace.} $\mathsf{NoPlus}_{N, t} =  W^{\otimes t}$ is obtained by projecting away the uniform superposition on each register.

    An orthonormal basis for $\mathsf{NoPlus}_{N, t}$ is given by 
    \begin{equation}
        \label{eq:nonplussed_basis_phases}
        \left\{
        \frac{1}{{N}^{t/2}}
        \sum_{\vec x \in [N]^t}
        \omega_N^{\left\langle{\vec x, \vec y}\right\rangle}
        \ket{\vec x}
        \right\}_{
            (y_1, \dots, y_t) 
            \in 
            \left(
                \Z_N \setminus \{0\}
            \right)^t
        }
    \end{equation}

    Another basis for $\mathsf{NoPlus}_{N, t}$ that will often be convenient for us (but not orthonormal) is given by 
    \begin{equation}
        \label{eq:nonplussed_basis}
        \left\{
            \bigotimes_{i=1}^t 
            \big(
                \ket{1} - \ket{x_i}
            \big)
        \right\}_{
            (x_1, \dots, x_t) 
            \in 
            \left(
                [N] \setminus \{1\}
            \right)^t
        }
    \end{equation}
    Note that, while not orthonormal, these basis states are linearly independent, since only the state corresponding to $\vec x$ has support on $\ket{\vec x}$.
\end{definition}

\begin{lemma}
\label{lem:uniform_twirling}
Let $C$ be sampled from any unitary $1$-design on $\C^N$. Then, 
\begin{equation}
   \normalfont  \mathbb{E}_C\!\left[\Tr\left(
           \Pi_{\mathsf{NoPlus}_{N,t}}C^{\otimes t} \rho (C^\dagger)^{\otimes t}  
        \right)\right]  \ge 1 - \frac{t}{N}
\end{equation}
\end{lemma}

\begin{proof}
Since $\{\ketbra{+}{+}_i\}_{i=1}^t$ are commuting projectors, we can use the operator form of the union bound:
\begin{lemma}[\cite{Khabbazi_Oskouei_2019}]
    \label{lem:commuting_union_bound}
    Let $\rho$ be a density operator, and let
    $\Pi_1, \ldots, \Pi_m$ be commuting orthogonal projectors. Then
    \begin{equation}
        1
        -
        \operatorname{Tr}
        \left(
            \Pi_m \cdots \Pi_1
            \rho
            \Pi_1 \cdots \Pi_m
        \right)
        \leq
        \sum_{i = 1}^{m}
        \operatorname{Tr}
        \left(
            (\Id - \Pi_i)\rho
        \right).
\end{equation}
\end{lemma}
\noindent which implies that 
\begin{equation}
    W^{\otimes t}=(\Id-\ketbra{+}{+})^{\otimes t} \succeq \Id - \sum_{i=1}^t \ketbra{+}{+}
\end{equation}
Therefore, 
\begin{align}
    \Tr( \Pi_{\mathsf{NoPlus}_{N,t}}C^{\otimes t} \rho (C^\dagger)^{\otimes t})
    \geq
    1 - \sum_{i=1}^t \Tr(\ketbra{+}{+}_iC^{\otimes t} \rho (C^\dagger)^{\otimes t}) \\ 
    \implies  \mathbb{E}_C\!\left[\Tr\left(
           \Pi_{\mathsf{NoPlus}_{N,t}}C^{\otimes t} \rho (C^\dagger)^{\otimes t}  
        \right)\right]
    \geq
    1 - \sum_{i=1}^t \mathbb{E}_C\!\left[\Tr(\ketbra{+}{+}_iC^{\otimes t} \rho (C^\dagger)^{\otimes t})\right]
    \label{eq:non_uniform_subspace_bound}
\end{align}
Now, consider the summand on the right hand side. By the cyclic property of trace and linearity of expectation, the summand on the right hand side is equivalent to 
\begin{equation}
   \Tr( \mathbb{E}_C[(C^\dagger)^{\otimes t}\ketbra{+}{+}_iC^{\otimes t} \rho ]) = \Tr( \mathbb{E}_C[(C^\dagger\ketbra{+}{+}C)_i] \rho ) 
   \label{eq:simplified_summand_uniform_bound}
\end{equation}
For any $1$-design, $\mathbb{E}_C[C^\dagger \sigma C] = \frac{\Tr(\sigma)}{N}\Id$.\ Since $\sigma =\ketbra{+}{+}$ has trace $1$,~\cref{eq:simplified_summand_uniform_bound} simplifies to $\Tr(\rho)/N = 1/N$. Plugging this into the summand in~\cref{eq:non_uniform_subspace_bound} completes the proof.
\end{proof}

\begin{definition}
    \label{def:dist-nonplussed-subspace}
    The \textit{distinct nonplussed subspace} $\mathsf{DNP}_{N, t} := \mathsf{Dist}_{N, t} \cap \mathsf{NoPlus}_{N, t}$ is the intersection of the distinct and nonplussed subspaces.
\end{definition}

\begin{lemma}
    The distinct nonplussed subspace $\mathsf{DNP}_{N, t}$ has dimension 
    $
        N^t 
        \left(
            1 - O 
            \left( 
                \frac{t^2}{N}
            \right) 
        \right)
    $.
    In particular, this means that the fully mixed state has probability 
    at least 
    $
        1 - O 
        \left( 
            \frac{t^2}{N}
        \right) 
    $
    of being in the distinct nonplussed subspace.
\end{lemma}

\begin{proof}
We can take the following to be a partial and non-orthogonal basis for the distinct nonplussed subspace:

\begin{equation}
    \label{eq:dist_nonplussed_basis}
    S = \left\{
        \bigotimes_{i=1}^t 
        \big(
            \ket{i} - \ket{x_i}
        \big)
    \right\}_{
        (x_1, \dots, x_t) 
        \in 
        \left(
            [N] \setminus \{1,..., t\}
        \right)^t_{\text{dist}}
    }
\end{equation}
As in \Cref{eq:nonplussed_basis}, these basis states are linearly independent (though not orthogonal) because only the state corresponding to $\vec x$ has support on $\ket{\vec x}$.
Furthermore, $S$ is contained in the distinct nonplussed subspace:
Fix $\ket{\psi} \in S$. Clearly, $\ket{\psi} \in \mathsf{Dist}_{N, t}$, since all $t$ registers are supported on pairwise disjoint sets of computational basis vectors. Moreover, 
\begin{align}
    W(\ket{i} - \ket{x_i}) 
    &=
    \ket{i} - \ket{x_i} + \left(\inner{+|i} - \inner{+|x_i}\right)\ket{+}
    \\ 
    &= 
    \ket{i} - \ket{x_i} + \left(\frac{1}{\sqrt{N}} - \frac{1}{\sqrt{N}}\right)\ket{+}
    \\ 
    &= 
    \ket{i} - \ket{x_i}
    \\ 
    &\implies 
    W^{\otimes t}\ket{\psi} = \ket{\psi}
\end{align}
and so $\ket{\psi} \in \mathsf{NoPlus}_{N, t}$ as well. 
So the size of $S$ is a lower bound on the dimension of the distinct nonplussed subspace.
This gives us that 
\begin{align}
    \dim(\mathsf{DNP}_{N, t}) 
    \ge
    |S| 
    &
    = 
    (N - t)(N - t - 1) \dots (N - 2t + 1) 
    \\
    &
    \ge  
    N^t 
    \left(
        1 - O 
        \left( 
            \frac{t^2}{N}
        \right) 
    \right).
    \qedhere
\end{align}
\end{proof}

\subsubsection{Unitary 2-designs and the Distinct Nonplussed Subspace}

In this section, we will show that applying a unitary $2$-design with high probability maps an arbitrary state $\rho \in (\C^N)^{\otimes t}$ to a state that has high overlap with the distinct nonplussed subspace. More precisely, we will prove the following theorem: 
\begin{theorem}
    \label{thm:proj_to_dist_cap_nu}
    Let $\Pi_{\mathsf{DNP}_{N, t}}$ be the projection onto the distinct nonplussed subspace, and $C$ be any unitary $2$-design on $\C^N$. Then, 
    \begin{equation}
        \mathbb{E}_C\!\left[\Tr\!\left(
            \Pi_{\mathsf{DNP}_{N, t}} C^{\otimes t} \rho (C^\dagger)^{\otimes t}
        \right)\right]
    \geq
    1 - O\left(\frac{t^2}{N}\right)
\end{equation}
for any state $\rho \in (\C^N)^{\otimes t}$.
\end{theorem}
\noindent To prove~\cref{thm:proj_to_dist_cap_nu}, we first collect some facts about projectors: 

\begin{lemma}[\cite{Gao15}]
    \label{lemma:gao}
    Let $\rho$ be a density operator, and let
    $\Pi_1, \ldots, \Pi_m$ be orthogonal projectors. Then
    \begin{equation}
        1
        -
        \operatorname{Tr}
        \left(
            \Pi_m \cdots \Pi_1
            \rho
            \Pi_1 \cdots \Pi_m
        \right)
        \leq
        4
        \sum_{i = 1}^{m}
        \operatorname{Tr}
        \left(
            (\Id - \Pi_i)\rho
        \right).
    \end{equation}
    Equivalently,
    \begin{equation}
        \operatorname{Tr}
        \left(
            \Pi_m \cdots \Pi_1
            \rho
            \Pi_1 \cdots \Pi_m
        \right)
        \geq
        1
        -
        4
        \sum_{i = 1}^{m}
        \operatorname{Tr}
        \left(
            (\Id - \Pi_i)\rho
        \right).
    \end{equation}
\end{lemma}

\begin{definition}
    Assume that $A$ and $B$ are subspaces of $\C^N$, and let $C \coloneqq A \cap B$ be the intersection. The \textit{Friedrichs angle} $c(A, B)$ is defined as 
    \begin{equation}
        c(A, B) \coloneqq \sup\limits_{\substack{\ket{x} \in A \cap C^{\perp} \\ \ket{y} \in B \cap C^\perp \\ \norm{\ket{x}} = \norm{\ket{y}} = 1}} \abs{\inner{x|y}}
    \end{equation}
\end{definition}
\noindent Informally, the Friedrichs angle is the highest overlap between two unit vectors in $A$ and $B$, excluding the intersection. For example, when $\Pi_A$ and $\Pi_B$ are commuting projectors, $c(A, B) = 0$.  Notice that since $\ket{y} \in B \cap C^\perp$,
\begin{align}
    \norm{\Pi_A\ket{y}} = \sup\limits_{\substack{\ket{a} \in A \\ \norm{\ket{a}} = 1}} |\inner{a|y}| &= \sup\limits_{\substack{\ket{a} \in A \cap C^\perp \\ \norm{\ket{a}} = 1}} |\inner{a|y}| \\ 
    \implies \norm{\Pi_A\mid_{B \cap C^\perp}\!}^2 &= \sup\limits_{\substack{\ket{y} \in B \cap C^\perp 
    \\ \norm{\ket{y}} = 1}} |\inner{y|\Pi_{A}|y}|
    \\
    &= 
    \sup\limits_{
        \substack{
            \ket{x} \in A \cap C^{\perp} 
            \\ 
            \ket{y} \in B \cap C^\perp 
            \\ 
            \norm{\ket{x}} = 
            \norm{\ket{y}} = 1
        }
    } 
    |\inner{y|x}\inner{x|y}|  = c(A, B)^2
    \label{eq:alternate_friedrich}
\end{align}
We make use of two more characterizations of $c(A,B)$. First,
\begin{claim}\label{claim:friedrich-singular-value}
$c(A,B)$ is the largest singular value of $\Pi_A \Pi_B$ aside from $1$. 
\end{claim}

\begin{proof}
Since $A$ decomposes into $C \oplus (A\cap C^\bot)$, we have that 
\[
    \Pi_A \Pi_B = \Pi_C \Pi_B + \Pi_{A\cap C^\bot} \Pi_{B} = \Pi_C + \Pi_{A\cap C^\bot} \Pi_{B}.
\]
Moreover, decomposing $B = C\oplus (B\cap C^\bot)$, we can write
\[
    \Pi_A \Pi_B = \Pi_C + \Pi_{A\cap C^\bot} \Pi_{C} + \Pi_{A\cap C^\bot} \Pi_{B\cap C^\bot} = \Pi_C + \Pi_{A\cap C^\bot} \Pi_{B\cap C^\bot}.
\]
Thus, $\Pi_A \Pi_B$ block-diagonalizes according to the $(C, A\cap C^\bot)$, $(C, B\cap C^\bot)$ vector space decompositions. $c(A,B)$ is defined to be the top singular value of the $(A\cap C^\bot, B\cap C^\bot)$ block, so this proves the claim. 
\end{proof}

\noindent Kayalar and Weinert proved that the Friedrichs angle captures the difference between first projecting onto $A$, then projecting onto $B$, as opposed to a single projection onto $A \cap B$:
\begin{theorem}[\cite{kayalar1989oblique}]
    \label{thm:alternating_projection}
    \begin{equation}
        ||\Pi_A\Pi_B - \Pi_{A \cap B}|| = c(A, B).
    \end{equation}
\end{theorem}
From~\cref{thm:alternating_projection}, we can derive a related bound which will be useful in our setting: 
\begin{lemma}
    \label{lem:squared_projection}
    \begin{equation}
        \norm{\Pi_B\Pi_A\Pi_B - \Pi_{A \cap B}} 
        \le 
        c(A, B)^2 
    \end{equation}
\end{lemma}
\begin{proof}
    \begin{align}
        \left\lVert
            \Pi_B\Pi_A\Pi_B - \Pi_{A \cap B} 
        \right\rVert
        &= 
        \left\lVert
            \Pi_B\Pi_A\Pi_A\Pi_B - \Pi_{A \cap B}
        \right\rVert
        \tag{since $\Pi_A^2 = \Pi_A$}
        \\
        &= 
        \left\lVert
            \Pi_B\Pi_A\Pi_A\Pi_B - \Pi_{A \cap B} - \Pi_{A \cap B}+  \Pi_{A \cap B}
        \right\rVert
        \\
        &=
        \left\lVert
            \Pi_B\Pi_A\Pi_A\Pi_B - \Pi_B\Pi_A\Pi_{A \cap B} - \Pi_{A \cap B}\Pi_A\Pi_B +  \Pi_{A \cap B}
        \right\rVert
        \tag{since $\Pi_A \Pi_{A \cap B} = \Pi_{A \cap B}$}
        \\
         &=
        \left\lVert
            (\Pi_B\Pi_A - \Pi_{A \cap B} )(\Pi_A\Pi_B - \Pi_{A \cap B} ) 
        \right\rVert
        \\
        &
        \le
        \left\lVert
            \Pi_B\Pi_A - \Pi_{A \cap B} 
        \right\rVert
        \;
        \left\lVert
            \Pi_A\Pi_B - \Pi_{A \cap B} 
        \right\rVert
        \tag{submultiplicativity}
        \\
        &
        =
        \left\lVert
            \Pi_A\Pi_B - \Pi_{A \cap B} 
        \right\rVert^2
        \\
        &
        \le 
        c(A, B)^2 
        \tag{\Cref{thm:alternating_projection}}
    \end{align}
\end{proof}
\begin{corollary}
    \label{corollary:squared_projector_overlap}
    \begin{align}
        \left\lvert
        \Tr(\Pi_A\Pi_B\rho\Pi_B\Pi_A) 
        - 
        \Tr(\Pi_{A \cap B}\rho \Pi_{A \cap B}) 
        \right\rvert
        \le 
        c(A, B)^2.
    \end{align}
\end{corollary}
\begin{proof}
    \begin{align}
        \left\lvert
        \Tr(\Pi_A\Pi_B\rho\Pi_B\Pi_A) - \Tr(\Pi_{A \cap B}\rho\Pi_{A \cap B})
        \right\rvert
        &=
        \left\lvert
        \Tr(\Pi_B\Pi_A\Pi_A\Pi_B\rho) - \Tr(\Pi_{A \cap B}\Pi_{A \cap B}\rho) 
        \right\rvert
        \\
        & 
        = 
        \left\lvert
        \Tr(\Pi_B\Pi_A\Pi_B\rho) - \Tr(\Pi_{A \cap B}\rho)
        \right\rvert
        \\
        & 
        = 
        \left\lvert
        \Tr\big((\Pi_B\Pi_A\Pi_B - \Pi_{A \cap B})\rho\big)
        \right\rvert
        \\
        &  
        \le 
        ||\Pi_B\Pi_A\Pi_B - \Pi_{A \cap B}||\; \norm{\rho}_1
        \tag{Hölder's inequality}
        \\
        &  
        = 
        ||\Pi_B\Pi_A\Pi_B - \Pi_{A \cap B}||
        \tag{$\norm{\rho}_1$ = 1}
        \\
        &
        = 
        c(A, B)^2 
        \tag{\Cref{lem:squared_projection}}
    \end{align}
\end{proof}
Finally, we will make use of one more fact about the Friedrichs angle:
\begin{lemma}\label{lem:friedrichs-jordan}
    $\Pi_A \Pi_B + \Pi_B \Pi_A \succeq -c(A,B) \Big( \Pi_A + \Pi_B \Big)$
\end{lemma}
\begin{proof}
    By Jordan's lemma, $\Pi_A$ and $\Pi_B$ are simultaneously block-diagonalizable into one- and two-dimensional subspaces where on the two dimensional subspaces we have
\[
\Pi_A=\begin{pmatrix}1&0\\0&0\end{pmatrix},
\qquad
\Pi_B=\begin{pmatrix}
\cos^2\theta&\cos\theta\sin\theta\\
\cos\theta\sin\theta&\sin^2\theta
\end{pmatrix}
\]
for some angle $0< \theta < \frac \pi 2$. The one-dimensional subspaces together span $A\cap B$, $A^\bot \cap B$, $A\cap B^\bot$, and $A^\bot \cap B^\bot$. Thus, in this language, the Friedrichs angle is given by
\[ c(A,B) = \max \cos(\theta),
\]
where the maximum is taken over two-dimensional Jordan blocks. But on any fixed Jordan block, one can see that 
\[
\Pi_A \Pi_B + \Pi_B \Pi_A+\cos(\theta)(\Pi_A+\Pi_B) =\cos(\theta) \cdot \begin{pmatrix}
(1+\cos(\theta))^2&\sin\theta (1+\cos(\theta)) \\
\sin\theta (1+\cos(\theta))&\sin^2\theta
\end{pmatrix} \succeq 0,
\]
which proves the claim. 
\end{proof}

\noindent Our first goal will be to show that $c({\mathsf{Dist}_{N, t}}, {\mathsf{NoPlus}_{N,t}})^2$ is small when $t \ll N$:
\begin{lemma}
    \label{lem:friedrich_angle_small}
    \begin{equation}
        \normalfont c({\mathsf{Dist}_{N,t}}, {\mathsf{NoPlus}_{N,t}})^2 \le \frac{t(t-1)}{N - t + 1}.
    \end{equation}
\end{lemma}
\begin{proof}
    Define the averaging operator $A_i$ as follows:
    \begin{equation}       
        A_i\ket{\vec{x}}
        =
        \begin{cases}
        \dfrac{1}{N-t+1}
        \displaystyle\sum_{a\notin \{x_j:\, j\neq i\}}
        \ket{x_1, \dots, x_{i-1}, a, x_{i+1}, \dots, x_t}
        & \text{if } \vec{x} \in [N]_{\text{dist}}^t, \\[2ex]
        \ket{\vec{x}},
        & \text{otherwise.}
        \end{cases}
    \end{equation}
From the definition, it is clear that $A_i$ is an orthogonal projector, whose normalized $+1$ eigenvectors within the distinct subspace are given by 
\begin{equation}
\label{eq:basis_for_image}
    \frac{1}{\sqrt{N - t + 1}}\sum_{a\notin \{z_j:\, j\neq i\}} \ket{z_1, \dots, z_{i-1}, a, z_{i+1}, \dots, z_t},\;\; \vec{z} = (z_1, \dots, z_{i-1}, z_{i+1}, \dots, z_t) \in [N]_{\text{dist}}^{t-1}
\end{equation}
and a (overcomplete) basis for $\ker A_i$ (which is within the distinct subspace) is given by 
\begin{equation}
    \label{eq:basis_for_kernel}
    \frac{1}{\sqrt{2}} (\ket{z_1, \dots, z_{i-1}, a, z_{i+1}  \dots, z_t} - \ket{z_1, \dots, z_{i-1}, b, z_{i+1}  \dots, z_t}),\;\; a \ne b
\end{equation}
We will be interested in the (Hermitian, PSD) operator
\[ A = \sum_i A_i.
\]
We first show that the kernel of $A$ is exactly the intersection $\mathsf{Dist}_{N,t} \cap \mathsf{NoPlus}_{N,t}$:
\begin{claim}
    \label{lem:intersection_characterization}
\begin{equation}
    \normalfont \mathsf{Dist}_{N,t} \cap \mathsf{NoPlus}_{N,t} = \bigcap_{i=1}^t \ker A_i = \ker A
\end{equation}

\end{claim}
\begin{proof}
We know that $\bigcap_{i=1}^t \ker A_i = \ker A$ because the $A_i$ are all PSD.

For the main claim, by definition of $\mathsf{NoPlus}_{N,t}$, $\mathsf{NoPlus}_{N,t} = \bigcap_{i=1}^t \im W^{(i)}$.
So it is sufficient for us to show that for each $i$, 
\begin{equation}
    \ker A_i = \mathsf{Dist} \cap \im W^{(i)}
\end{equation}
For the forward inclusion, it is clear by definition of $A_i$ that $\ker A_i \subseteq \mathsf{Dist}$. Moreover, for any state $\ket{\phi}$ in~\cref{eq:basis_for_kernel},  $\Tr_{[t] \setminus i}(\ketbra{\phi}{\phi}) = \frac{1}{\sqrt{2}}(\ket{a} - \ket{b}) $ is orthogonal to $\ket{+}$, and therefore in $\im W$. Thus
\begin{equation}
    \ker A_i \subseteq \mathsf{Dist} \cap \im W^{(i)}.
\end{equation}
\noindent Conversely, if $\ket{\psi} \notin \ker A_i$, then either $\ket{\psi}$ has support on a non-distinct $t$-tuple (in which case $\ket{\psi} \notin \mathsf{Dist}$), or has support on a basis state $\ket{\phi}$ in~\cref{eq:basis_for_image}, which satisfies 
\begin{equation}
    \Tr_{[t] \setminus i}(\ketbra{\phi}{\phi}) =   \frac{1}{\sqrt{N - t + 1}}\sum_{a\notin \{z_j:\, j\neq i\}} \ket{a}
\end{equation}
which has nonzero overlap with $\ket{+} \in \ker W$. Hence, $\ket{\phi} \notin \im W^{(i)}$, and so 
\begin{equation}
    \ket{\psi} \notin \ker A_i \implies \ket{\psi} \notin  \mathsf{Dist} \cap \im W^{(i)}.
\end{equation}
\end{proof}
\noindent Next, we calculate the pairwise Friedrichs angles $c(\im(A_i), \im(A_j))$ for $i\neq j$.
\begin{claim}\label{claim:pairwise-friedrichs}
    $c(\im(A_i), \im(A_j)) \leq \frac 1 {N-t+1}$ for all $i<j$. 
\end{claim}
\begin{proof}
    The projections $A_i$ and $A_j$ are simultaneously block-diagonal in the coordinates $\ell \notin \{i, j\}$. Thus, it suffices to prove the claim restricting to states of the \barak{The notation $\ket{\psi_{x_{\neq i,j}}}$ is somewhat confusing here.}
    \[ \bigotimes_{\ell\notin \{i,j\}} \ket{x_\ell} \otimes \ket{\psi_{x_{\neq i,j}}}
    \]
    where $\ket{\psi_{x_{\neq i,j}}}$ is on the $i$th and $j$th registers. Restricted to these states and looking only at registers $i$ and $j$, $\im(A_i)$ is spanned by states of the form
    \[ \ket{\psi_{x_j}^{(i)}}=\frac 1 {\sqrt{N-t+1}}\sum_{x'_i \notin \{x_\ell\}_{\ell \neq i}} \ket{x'_i} \ket{x_j}
    \]
    for all $x_j\notin \{x_\ell\}_{\ell \notin \{i,j\}}$, while $\im(A_j)$ is similarly spanned by 
    \[ \ket{\psi_{x_i}^{(j)}}=\frac 1 {\sqrt{N-t+1}}\sum_{x'_j \notin \{x_\ell\}_{\ell \neq j}} \ket{x_i} \ket{x'_j}
    \]
    for all $x_i\notin \{x_\ell\}_{\ell \notin \{i,j\}}$. The corresponding Gram matrix of all such inner products $\braket{\psi_{x_i}^{(j)}}{\psi_{x_j}^{(i)}}$ is given by
    \[ \frac 1 {N-t+1} \Big(J_{N-t+2} - \Id_{N-t+2} \Big),
    \]
    where $J$ denotes an all ones matrix, as the diagonal terms are indeed zero while the off-diagonal terms are $\frac 1 {N-t+1}$. By \cref{claim:friedrich-singular-value}, the $c(\im(A_i), \im(A_j))$ is the largest singular value of this matrix aside from $1$. Because $J$ has rank $1$, all remaining singular values of the Gram matrix are $\frac{1}{N-t+1}$, so the claim follows.  
\end{proof}

\noindent In order to finish proving~\cref{lem:friedrich_angle_small}, we will need to establish one more key claim.

\begin{claim}
    $A \succeq (1-\frac {t-1}{N-t+1}) (\Id - \Pi_{\ker(A)})$.
\end{claim}
\begin{proof}
    By \cref{claim:pairwise-friedrichs} and \cref{lem:friedrichs-jordan}, we have that
    \[ A_i A_j + A_j A_i \succeq - \frac 1 {N-t+1} \Big(A_i + A_j\Big)
    \]
    for all $i\neq j$. Thus, we can calculate
    \begin{align}
        A^2 &= A + \sum_{i<j} (A_i A_j + A_j A_i) \\
        &\succeq A - \frac 1 {N-t+1} \sum_{i<j}(A_i + A_j) \\
        &= (1-\frac{t-1}{N-t+1}) A.
    \end{align}
    Since $A$ is Hermitian PSD, this means that all nonzero eigenvalues of $A$ are at least $(1-\frac{t-1}{N-t+1})$ and so
    \[ A \succeq (1-\frac{t-1}{N-t+1}) (\Id - \Pi_{\ker A}). \qedhere
    \]
\end{proof}
We are finally ready to prove \cref{lem:friedrich_angle_small}. Fix any $\ket{\psi} \in \mathsf{NoPlus}_{N,t} \cap (\mathsf{DNP}_{N, t})^\perp$, and define $\ket{\phi} \coloneqq \Pi_{\mathsf{Dist}_{N,t}}\ket{\psi}$ and $\ket{\eta}\coloneqq (\Id-\Pi_{\mathsf{Dist}_{N,t}})\ket{\psi}$. By~\cref{eq:alternate_friedrich}, bounding $c({\mathsf{Dist}_{N,t}}, {\mathsf{NoPlus}_{N,t}})^2$ reduces to bounding
 \begin{equation}
    \label{eq:reduce_to_ai}
 \inner{\psi|\Pi_{\mathsf{Dist}_{N,t}}|\psi} = \inner{\phi| \phi} = \inner{\phi|\Id - \Pi_{\mathsf{DNP}_{N, t}}|\phi} \le \frac{N-t+1}{N-2t+2}\sum_{i=1}^t \inner{\phi|A_i|\phi} = \frac{N-t+1}{N-2t+2}\sum_{i=1}^t \norm{A_i\ket{\phi}}^2
 \end{equation}
\noindent where the second equality holds because $\ket{\phi}$ is orthogonal to $\mathsf{DNP}_{N, t}$: for any $\ket{\gamma} \in \mathsf{DNP}_{N, t}$, 
\begin{equation}
    \inner{\gamma|\phi} = \inner{\gamma|\Pi_{\mathsf{Dist}_{N,t}}|\psi} =  \inner{\gamma|\psi} = 0
\end{equation}
\noindent Now, we turn to bounding $\norm{A_i\ket{\phi}}^2$. For any $\vec{z} \in [N]^{t-1}$ and $a \in [N]$, let $\alpha_{\vec{z}, a}$ be the coefficient of $\ket{z_1, \dots, z_{i-1}, a, z_{i+1}, \dots, z_t}$ when expanding $\ket{\psi}$ in the computational basis. Since $\ket{\psi} \in \mathsf{NoPlus}_{N,t}$,
\begin{equation}
    \label{eq:flip_for_cauchy}
    \sum_{a \in [N]} \alpha_{\vec{z}, a} = 0 \implies -\sum_{a \in \{z_j: j \ne i\}} \alpha_{\vec{z}, a} = \sum_{a \notin \{z_j: j \ne i\}} \alpha_{\vec{z}, a} 
\end{equation}
But then, 
\begin{align}
    A_i\ket{\phi} 
    &
    = 
    A_i\left(\Pi_{\mathsf{Dist}_{N,t}}\ket{\psi} \right) 
    \\
    &
    = 
    A_i\left(\sum_{\vec{z} \in [N]^{t-1}_{\text{dist}}} \sum_{a \notin \{z_j: j \ne i\}} \alpha_{\vec{z}, a}\ket{z_1, \dots, z_{i-1}, a, z_{i+1}, \dots, z_t} \right) 
    \\
    &
    =  
    \frac{1}{N - t + 1}\sum_{\vec{z} \in [N]^{t-1}_{\text{dist}}} \sum_{b \notin \{z_j: j \ne i\}}\sum_{a \notin \{z_j: j \ne i\}} \alpha_{\vec{z}, a} \ket{z_1, \dots, z_{i-1}, b, z_{i+1}, \dots, z_t} 
    \\
    &
    =  
    -\frac{1}{N - t + 1}\sum_{\vec{z} \in [N]^{t-1}_{\text{dist}}} \sum_{b \notin \{z_j: j \ne i\}}\sum_{a \in \{z_j: j \ne i\}} \alpha_{\vec{z}, a} \ket{z_1, \dots, z_{i-1}, b, z_{i+1}, \dots, z_t}
    \\ 
    \implies 
    \norm{A_i\ket{\phi}}^2 
    &
    = 
    \frac{1}{(N - t + 1)^2}  \sum_{\vec{z} \in [N]^{t-1}_{\text{dist}}} \sum_{b \notin \{z_j: j \ne i\}} \left| \sum_{a \in \{z_j: j \ne i\}} \alpha_{\vec{z}, a} \right|^2
    \\ 
    &
    = 
    \frac{1}{N - t + 1}  \sum_{\vec{z} \in [N]^{t-1}_{\text{dist}}} \left| \sum_{a \in \{z_j: j \ne i\}} \alpha_{\vec{z}, a} \right|^2
    \\
    &
    \le 
    \frac{t - 1}{N - t + 1}  \sum_{\vec{z} \in [N]^{t-1}_{\text{dist}}} \sum_{a \in \{z_j: j \ne i\}} \left| \alpha_{\vec{z}, a} \right|^2
    \\
    &
    \le 
    \frac{t - 1}{N - t + 1}\norm{\ket{\eta}}^2
    =
    \frac{t - 1}{N - t + 1}
    \left(1-\norm{\ket{\phi}}^2\right)
    \label{eq:bound-on-a_i}
\end{align}
On the first line we expanded out the definition of $\ket{\phi}$, and on the second line we expanded out the application of $A_i$. On the third line, we plug in~\cref{eq:flip_for_cauchy}, the fourth is by the definition of the Euclidean norm, and the fifth uses that $\left|\{z_j: j \ne i\}\right| = t - 1$. The sixth line follows by an application of Cauchy Schwarz, and on the last line, we use that the remaining sum is over non-distinct basis states and is therefore bounded by $\norm{\ket{\eta}}^2=1-\norm{\ket{\phi}}^2$.

We therefore conclude that
\begin{align}
    \norm{\ket{\phi}}^2
    &
    \le 
    \frac{N-t+1}{N-2t+2}\sum_{i=1}^t \norm{A_i\ket{\phi}}^2
    \tag{by \Cref{eq:reduce_to_ai}}
    \\
    &
    \le 
    \frac{t(t-1)}{N-2t+2}
    (1-\norm{\ket{\phi}}^2)
    \tag{by \Cref{eq:bound-on-a_i}}
\end{align}
Rearranging gives
\[
    \norm{\ket{\phi}}^2
    \le
    \frac{t(t-1)}{N-2t+2+t(t-1)}
    \le
    \frac{t(t-1)}{N-t+1}.
\]
Since this holds for every $\ket{\psi}$ in the optimization in~\cref{eq:alternate_friedrich}, this gives the claimed bound,
which completes the proof of \Cref{lem:friedrich_angle_small}.
\end{proof}

We can now prove~\cref{thm:proj_to_dist_cap_nu}:
\begin{proof}[Proof of~\cref{thm:proj_to_dist_cap_nu}]
    First,
    \begin{align}
    & 
    \mathbb{E}_C
    \!
    \left[
        \Tr
        \left(
            \Pi_{\mathsf{DNP}_{N, t}} 
            C^{\otimes t} 
            \rho 
            (C^\dagger)^{\otimes t}
        \right)
    \right] 
    \\ 
    &
    =
    \mathbb{E}_C
    \!
    \left[
        \Tr
        \left(
            \Pi_{\mathsf{DNP}_{N, t}} 
            C^{\otimes t} 
            \rho 
            (C^\dagger)^{\otimes t}
            \Pi_{\mathsf{DNP}_{N, t}} 
        \right)
    \right] 
    \\ 
    &
    \ge
    \mathbb{E}_C
    \!
    \left[
        \Tr
        \left(
            \Pi_{\mathsf{Dist}_{N,t}}
            \Pi_{\mathsf{NoPlus}_{N,t}}
            C^{\otimes t} 
            \rho 
            (C^\dagger)^{\otimes t}  
            \Pi_{\mathsf{NoPlus}_{N,t}}
            \Pi_{\mathsf{Dist}_{N,t}}
        \right)
    \right] 
    - 
    c({\mathsf{Dist}_{N,t}}, {\mathsf{NoPlus}_{N,t}})^2
    \tag{by \Cref{corollary:squared_projector_overlap}}
    \\ 
    &
    \ge
    \mathbb{E}_C
    \!
    \left[
        \Tr
        \left(
            \Pi_{\mathsf{Dist}_{N,t}}
            \Pi_{\mathsf{NoPlus}_{N,t}}
            C^{\otimes t} 
            \rho 
            (C^\dagger)^{\otimes t}  
            \Pi_{\mathsf{NoPlus}_{N,t}}
            \Pi_{\mathsf{Dist}_{N,t}}
        \right)
    \right] 
    - 
    \frac{t(t-1)}{N - t + 1} 
    \tag{by \Cref{lem:friedrich_angle_small}}
    \\ 
    &
    \ge
    1 
    - 
    4\cdot 
    \mathbb{E}_C\!\left[\Tr\left(
    (\Id  - \Pi_{\mathsf{Dist}_{N,t}})C^{\otimes t} \rho (C^\dagger)^{\otimes t}
    \right)\right] 
    \\
    & 
    \qquad\qquad 
    -\; 
    4\cdot 
    \mathbb{E}_C\!\left[\Tr\left(
    (\Id  - \Pi_{\mathsf{NoPlus}_{N,t}})C^{\otimes t} \rho (C^\dagger)^{\otimes t}
    \right)\right] - \frac{t(t-1)}{N - t + 1}
    \tag{by \Cref{lemma:gao}}
    \\ 
    &
    =
    1 - 4 
    \left( 
        1
        -
        \mathbb{E}_C\!\left[\Tr\left(
        \Pi_{\mathsf{Dist}_{N,t}}
        C^{\otimes t} \rho (C^\dagger)^{\otimes t}  
        \right)\right] 
    \right) 
    \\
    & 
    \qquad\qquad 
    -\; 
    4 
    \left(
        1
        -
        \mathbb{E}_C
        \!
        \left[
            \Tr
            \left(
                \Pi_{\mathsf{NoPlus}_{N,t}}
                C^{\otimes t} 
                \rho 
                (C^\dagger)^{\otimes t}  
            \right)
        \right] 
    \right) 
    - 
    \frac{t(t-1)}{N - t + 1}
    \tag{%
        since 
        $\Tr
        \left(
            C^{\otimes t} 
            \rho 
            (C^\dagger)^{\otimes t}  
        \right)$
        =
        1
    }
    \\
    &
    \ge 
    1 - \frac{4t(t - 1)}{N + 1} - \frac{4t}{N} - \frac{t(t-1)}{N - t + 1}
    \tag{by \Cref{lem:distinct-twirling,lem:uniform_twirling}}
    \\
    &
    \ge 
    1 - \frac{6t^2}{N - t + 1}
    \end{align}
    which is $1 - O(t^2/N)$.
\end{proof}

\subsection{The Action of the Partition Algebra on the Distinct Nonplussed Subspace}

\begin{lemma}
    \label{lem:non_permutation_mapping_to_DNP}
    Let $D \in \mathcal{B}(P_t(N))$ be a partition diagram, and let $S(\cdot)$ denote the Schur representation (\cref{def:schur_rep}) of $P_t(N)$. 
    Then 
    $
        S(D)
        \,
        \Pi_{\mathsf{DNP}_{N, t}}
        =
        0
    $
    if and only if $D$ is not a permutation diagram.
\end{lemma}
\begin{proof}
        The ``only if'' direction is straightforward, since $S(D)$ is unitary if $D$ is a permutation. For the ``if'' direction, assuming that $D$ is not a permutation diagram, there are one of two possible cases to consider: 
    \begin{enumerate}
        \item[a)] $D$ has a connected component with at least two vertices in the input row, and/or
        \item[b)] $D$ has a singleton in the input row. 
    \end{enumerate}
    This is because if no two input vertices are in the same connected component, and every input vertex is propagating (no singletons), then the input vertices must each connect to a distinct output vertex, producing a permutation.

    First, assume that $D$ has a connected component $C_1$ containing two input vertices $i_{\operatorname{in}}$ and $j_{\operatorname{in}}$. By definition of the Schur representation, all summands in~\cref{eq:schur_rep} will have $\bra{y_{i}} = \bra{y_j}$, which implies that $S(D)\ket{\vec{z}} = 0$ unless $z_{i} = z_j$. Therefore, $S(D)\ket{\psi} = 0$ for all $\ket{\psi} \in \mathsf{Dist}_{N, t}$.

    Next, assume that $D$ has a singleton $i_{\operatorname{in}}$ in the input row. Again using~\cref{eq:schur_rep}, this implies that $S(D)$ factors as $V \otimes \bra{+}_i$, where $V: (\C^{N})^{\otimes t - 1} \rightarrow (\C^{N})^{\otimes t}$. $V \otimes \bra{+}_i$ annihilates all states that are orthogonal to $\ket{+}_i$, including all states of the form $\ket{\psi} \otimes (\ket{1} - \ket{z_t})$ (\cref{eq:nonplussed_basis}). Therefore, $S(D)\ket{\psi} = 0$ for all $\ket{\psi} \in \mathsf{NoPlus}_{N, t}$. 
    Combining the two cases, it follows that 
    \begin{equation}
        S(D)
        \,
        \Pi_{\mathsf{DNP}_{N, t}}
        =
        0 
    \end{equation}
\end{proof}

\begin{lemma}\label{lem:DNP-fullbox}
    \label{lem:all_on_t}
    $\schur_{P_t(N)}$ maps the distinct nonplussed subspace 
    $\mathsf{DNP}_{N, t}$ to the subspace spanned by $(S_N, P_t(N))$-Schur basis states $\ket{\lambda, X, Y}$ with $|\lambda| = t$. 
\end{lemma}
In order to prove~\cref{lem:all_on_t}, we first show that an irrep of $P_t(N)$ which annihilates all non-permutation diagrams must correspond to a full-box irrep, i.e. $|\lambda| = t$: 
\begin{lemma} 
    \label{lem:full_box_irrep_quotient}
     Let $J \trianglelefteq P_t(N)$ be the two-sided ideal spanned by all non-permutation diagrams in $\mathcal{B}(P_t(N))$.\footnote{That $J$ is in fact a two-sided ideal is a consequence of \cite{halverson2005partition}, Equation 1.3.}  Then, $\rho_{\lambda}(J) = 0$ if and only if $|\lambda| = t$. 
\end{lemma}
\begin{proof}
 The ``if'' direction follows from~\cite{foxman2026efficient} Theorem 4.17, which proves that if $|\lambda| = t$, then $\rho_\lambda(D) = 0$ for all diagrams $D$ with fewer than $t$ propagating components.\footnote{A connected component is \textit{propagating} if it intersects both the input and output rows of the diagram.} The only diagrams in $\calB(P_t(N))$ with $t$ propagating components are permutation diagrams, which implies that $\rho_{\lambda}(j) = 0$ for all $j \in J$. 

 To prove the reverse direction, note that there is a bijection between the irreps of $P_t(N)/J$ and the irreps of $P_t(N)$ which annihilate $J$ (\cite{erdmann_holm_2018_algebras}, Lemma 2.37, Lemma 3.5):
 \begin{equation}
    \begin{array}{c}
    \{\lambda: \lambda \in \widehat{P_t(N)},\;\; \ \rho_{\lambda}(j)=0\; \text{for all} \;j \in J\}
    \quad
    \Longleftrightarrow
    \quad
    \{\lambda': \lambda' \in \widehat{P_t(N)/J}\}
    \end{array}
\end{equation}
Since $P_t(N)/J \simeq \mathbb{C}[S_t]$ (\cite{halverson2005partition}, Equation 2.11), the number of irreducible representations of $P_t(N)$ which annihilate $J$ is equal to the number of irreducible representations of $\mathbb{C}[S_t]$. These are indexed by partitions of $t$. Since the $t$-box irreducible representations of $P_t(N)$ already annihilate $J$, it follows that no other irreducible representation of $P_t(N)$ (with $|\lambda| < t$) can annihilate $J$.
\end{proof}

    \begin{proof}[Proof of \cref{lem:all_on_t}]
    We begin with the forward direction, and let $j \in J$, where $J$ is defined as in~\cref{lem:full_box_irrep_quotient}. For any $\ket{\psi} \in (\C^N)^{\otimes t}$, we can write the state $S(j)\ket{\psi}$ in the $(S_N, P_t(N))$-Schur basis as follows: 

    \begin{align}
        \ket{\psi} &= \sum_{\substack{\lambda \in \widehat{P_t(N)}, \\ X \in \mathcal{B}(V_{S_N}^{(N - |\lambda|, \lambda)}), \\ Y \in \mathcal{B}(V_{P_t(N)}^{\lambda})}  } \alpha_{\lambda, X, Y}\ket{\lambda, X, Y}
        \\
        \implies 
        S(j)\ket{\psi} &= \sum_{\substack{\lambda \in \widehat{P_t(N)}, \\ X \in \mathcal{B}(V_{S_N}^{(N - |\lambda|, \lambda)}), \\ Y \in \mathcal{B}(V_{P_t(N)}^{\lambda})}  } \alpha_{\lambda, X, Y} \ket{\lambda, X} \otimes \rho_{\lambda}(j)\ket{Y}
        \label{eq:partition_schur_plus_schur_rep}
    \end{align}
     Next, fix some $\lambda$ with $|\lambda| < t$. By~\cref{lem:full_box_irrep_quotient}, there is some $k \in J$ such that $\rho_\lambda(k) \ne 0$. Therefore, $\rho_\lambda(J)$ is a non-zero ideal of $\rho_\lambda(A)$. But since $\lambda$ is an irrep, $\rho_\lambda(A)$ is a simple algebra, which has no non-zero proper ideals. Therefore, $\rho_\lambda(J) = \rho_\lambda(A)$, and so there exists $j_{\lambda} \in J$ such that $\rho_{\lambda}(j_{\lambda}) = \rho_\lambda(\mathbf{1}) = \Id_{V_{P_t(N)}^{\lambda}}$, where $\mathbf{1}$ is the identity diagram in $A$.

     Setting $j = j_\lambda$ in~\cref{eq:partition_schur_plus_schur_rep} implies that the coefficient on $\ket{\lambda, X, Y}$ in $S(j_\lambda)\ket{\psi}$ equal to $\alpha_{\lambda, X, Y}$. But, if $\ket{\psi} \in \mathsf{DNP}_{N, t}$, then \Cref{lem:non_permutation_mapping_to_DNP} implies that $S(j_\lambda)\ket{\psi} = 0$, so $\alpha_{\lambda, X, Y} = 0$. Since $\lambda$ was chosen arbitrarily among all irreps with fewer than $t$ boxes, it follows that in the $(S_N, P_t(N))$-Schur basis, $\ket{\psi}$ is supported entirely on full-box irreps.

    For the reverse direction, assume that in the $(S_N, P_t(N))$-Schur basis, $\ket{\psi}$ is supported entirely on states $\ket{\lambda, X, Y}$ with $|\lambda| = t$. By~\cref{lem:full_box_irrep_quotient}, $\rho_{\lambda}(J) = 0$, and so the block-diagonal decomposition on the right hand side of~\cref{eq:partition_schur_plus_schur_rep} implies that $S(j)\ket{\psi} = 0$ for all $j \in J$. In particular, $S(B_{ij})\ket{\psi} = 0$, where $B_{ij}$ is the diagram with a single block containing the input and output vertices in columns $i$ and $j$, with all other blocks propagating vertically along a single column. From~\cref{def:schur_rep}, it follows $S(B_{ij})$ is the equality projector between registers $i$ and $j$, and therefore
    \begin{equation}
        \forall i, j,\;S(B_{ij})\ket{\psi} = 0 \implies \ket{\psi} \in \mathsf{Dist}_{N, t}.
    \end{equation}
    Similarly, we also have that $S(P_i)\ket{\psi} = 0$, where $P_i$ is the diagram containing singleton in the $i$th column, again all other blocks propagating vertically along a single column. In this case, $S(P_i)$ is proportional to the projector onto $\Id \otimes \ket{+}_i$, and so
    \begin{equation}
        \forall i, \; S(P_i)\ket{\psi} = 0 \implies \ket{\psi} \in \mathsf{NoPlus}_{N, t}
    \end{equation}
    Combining the equations above, we conclude that $\ket{\psi} \in \mathsf{Dist}_{N, t} \cap \mathsf{NoPlus}_{N, t} = \mathsf{DNP}_{N, t}$.
\end{proof}

\subsubsection{The Dual of a Permutation Element}
Next, we prove a lemma which relates the dual of a permutation $\pi \in \C[S_t]$ with the dual of the same permutation, but in the partition algebra $P_t(N)$:
\begin{lemma}
    \label{lem:dual_element_of_permutation_in_partition_algebra}
   For any $\pi \in \mathcal{B}(\mathbb{C}[S_t])$, let $\pi^{*_{P_t(N)}}$ denote the dual element of $\pi$ with respect to $P_t(N)$, and let $\pi^{*_{\mathbb{C}[S_t]}}$ denote the dual element of $\pi$ with respect to $\mathbb{C}[S_t]$. For any $\sigma \in \mathcal{B}(\mathbb{C}[S_t])$, 
    \begin{equation}
        \inner{\pi^{*_{\mathbb{C}[S_t]}}, \sigma}_2 = \inner{\pi^{*_{P_t(N)}}, \sigma}_2
    \end{equation}
    Equivalently, the expansions of $\pi^{*_{\mathbb{C}[S_t]}}$ and $\pi^{*_{P_t(N)}}$ in the basis $\mathcal{B}(P_t(N))$ agree on all permutations $\sigma \in \mathcal{B}(\C[S_t])$. (Consequently, among all elements of $S_t$ only $\pi^{-1}$ appears in this expansion. )
\end{lemma}
\begin{proof}
    Since $\C[S_t]$ is a group algebra, $\pi^{*_{\C[S_t]}} = \pi^{-1}/|S_t|$. So, it suffices to show that 
    \begin{equation}
    \label{eq:dual_basis_restrict_main_claim}
        \inner{\pi^{*_{P_t(N)}}, \sigma}_2 = \frac{\delta_{\pi^{-1}, \sigma}}{|S_t|} = \frac{\delta_{\pi^{-1}, \sigma}}{t!}
    \end{equation}
    As in~\cref{lem:all_on_t}, we let $J  \trianglelefteq P_t(N)$ be the two-sided ideal spanned by all non-permutation diagrams, so that we have the canonical quotient map
    \begin{equation}
        \label{eq:canonical_quotient}
        \begin{aligned}
            q : P_t(N) &\longrightarrow P_t(N)/J, \\
            a &\longmapsto a + J.
        \end{aligned}
    \end{equation}
    Moreover, since $P_t(N)$ is semisimple, there exists a central idempotent\footnote{A central idempotent $e \in A$ is an element satisfying $ea = ae$ for all $a \in A$, and $e^2 = e$. If a semisimple algebra $A$ decomposed into a direct sum $I \oplus J$ of two ideals, then both $I$ and $J$ must be of the form $eA$ for some central idempotent $e$. See the discussion at the start of Chapter 22 in~\cite{Lam2001FirstCourse}.} $e$ such that~\cite{Lam2001FirstCourse}:
    \begin{equation}
        \label{eq:central_idempotent}
        P_t(N) \simeq eP_t(N) \oplus J
    \end{equation}
    Therefore, $eP_t(N) \simeq P_t(N)/J$, with one possible isomorphism given by restricting the quotient map in~\cref{eq:canonical_quotient}:
    \begin{equation}
        \label{eq:quotient_isomorphism}
        \begin{aligned}
            q_e : eP_t(N) &\xrightarrow{\sim}P_t(N)/J, \\
            ea &\longmapsto ea + J.
        \end{aligned}
    \end{equation}
    By definition, $q_e(e) = q_e(e \cdot \mathbf{1}) =  e + J$.
    Moreover, $e$ is a unit in $eP_t(N)$, so $e$ also maps to $\mathbf{1} + J$, where $\mathbf{1}$ is the identity permutation. Therefore, $\mathbf{1} + J = e + J$, which implies that $e = \mathbf{1} + j$ for some $j \in J$. 

     Using that $e = \mathbf{1} + j$, we now show that $\pi^{*_{P_t(N)}} = \pi^{-1}e/t!$:
    \begin{align}
        \inner{D, \pi^{*_{P_t(N)}}}_{\text{reg}_{P_t(N)}} &= \frac{1}{t!}\inner{D, \pi^{-1}e}_{\text{reg}_{P_t(N)}}
        \\
         & = \frac{1}{t!}\sum_{E \in \mathcal{B}(P_t(N))} \inner{E, D\pi^{-1}e \cdot E}_2
         \tag{Definition of $\inner{\cdot, \cdot}_{\text{reg}_{P_t(N)}}$ }
          \\ 
        & = \frac{1}{t!}\sum_{E \in \mathcal{B}(P_t(N))} \inner{E, eD\pi^{-1}E}_2 \tag{$e$ is central in $P_t(N)$}
        \\
        &=\frac{1}{t!} \cdot \delta_{D \in \mathcal{B}(\C[S_t])}\sum_{\sigma \in \mathcal{B}(\C[S_t])} \inner{\sigma, eD\pi^{-1}\sigma}_2 \tag{$eP_t(N) \cap J = \{0\}$,~\cref{eq:central_idempotent}}
        \\
        &=\frac{1}{t!} \cdot \delta_{D \in \mathcal{B}(\C[S_t])} \sum_{\sigma \in \mathcal{B}(\C[S_t])} \inner{\sigma, (\mathbf{1} + j)D\pi^{-1}\sigma}_2
         \\
        &=\frac{1}{t!} \cdot \delta_{D \in \mathcal{B}(\C[S_t])}\sum_{\sigma \in \mathcal{B}(\C[S_t])} \inner{\sigma, D\pi^{-1}\sigma}_2
        \tag{$\inner{\sigma, k} = 0$ for all $k \in J$} 
        \\
        &=\frac{1}{t!} \cdot \delta_{D \in \mathcal{B}(\C[S_t])} \cdot (\delta_{D, \pi} \cdot t!)
        \tag{\cref{eq:reg_trace_for_group}}
        \\ 
        &= \delta_{D, \pi}
    \end{align}
    Finally, since $\pi^{*_{P_t(N)}} = \pi^{-1}e/t!$, 
    \begin{align}
        \inner{\pi^{*_{P_t(N)}}, \sigma}_2 &= \frac{1}{t!}\inner{\pi^{-1}e, \sigma}_2 
        \\
         &=  \frac{1}{t!}\inner{\pi^{-1}(\mathbf{1} + j), \sigma}_2
         \\
         &=  \frac{1}{t!}\inner{\pi^{-1}, \sigma}_2 \tag{$\inner{\sigma, k} = 0$ for all $k \in J$}
         \\
        & = \frac{1}{t!} \cdot \delta_{\pi^{-1}, \sigma}
    \end{align}
    This matches~\cref{eq:dual_basis_restrict_main_claim}, concluding the proof.   
\end{proof}

\subsection{Additional Properties of the Update Operators}

We give some useful properties of the update operators in the path-recording basis of \Cref{sec:path}.

\begin{lemma}[$\Lambda_{S_t}$ commutes with permutation-respecting maps on $\reg R_X$ or $\reg R_Y$]
\label{lem:lambda-sym-commutes-with-pi-dnp}
Let $M_{\reg R_X}$ be a linear transformation such that $M \cdot S(\pi) = S(\pi) \cdot M$ for every permutation $\pi \in S_t$. Then, $M_{\reg R_X}\otimes \Id_{\reg R_Y}$ commutes with $\Lambda_{S_t}$. This applies to $M \in \{\Pi_{\mathsf{Dist}}, \Pi_{\NoPlus}, \Pi_{\DNP}, \Pi_{\NoPlus}^{\mathsf{FDist}} \}$ (the latter is defined below). 

The analogous statement holds for $M_{\reg R_Y}$.
\end{lemma}
\begin{proof}
    By \cref{thm:Schur-Weyl}, we know that $M$ (as an operator on $(\mathbb C^N)^{\otimes t}$) is in the span of $\{U^{\otimes t}\}_{U\in U(N)}$. As a result, in the Schur basis, $M$ is block-diagonal with respect to the irrep $V_{U(N)}^\lambda \otimes V^\lambda_{\mathcal A_t}$. Since $\Lambda_{S_t}$ is block-diagonal on the $\ket{\lambda}\otimes \ket{\lambda'}$ irrep spaces and acts as a scalar on each of them, we conclude that $\Lambda_{S_t}$ commutes with $M_{\reg R_X}\otimes \Id$ and $\Id \otimes M_{\reg R_Y}$. 
\end{proof}

\begin{lemma}[$\Lambda_{P_t(N)}$ commutes with $\Pi_{\DNP}^X$ and $\Pi_{\DNP}^Y$]
\label{lem:lambda-partition-commutes-with-pi-dnp}
    $\Lambda_{P_t(N)}$ commutes with $\Pi_{\DNP}^X$ and $\Pi_{\DNP}^Y$. 
\end{lemma}
\begin{proof}
    For $\Pi_{\DNP}^X$, this follows directly from the fact that $\Pi_{\DNP}$ projects, in the $(S_N, P_t(N))$-Schur basis, onto irrep spaces with $|\lambda| = t$ (\cref{lem:all_on_t}). $\Pi_{\DNP}^Y$ and $\Lambda_{P_t(N)}$ act on disjoint registers.
\end{proof}

\begin{lemma}[$\Lambda_{P_t(N)}$ is approximately $\sqrt{\frac{t!}{N^t}} \Id$ on full-box irreps]
\label{lem:lambda-approx-identity-partition}
    \begin{align}
        \label{eq:lambda-partition-identity}
        \frac{t!}{N^t}
        \left(
            1
            -
            O
            \left(
                \frac{t^2}{N}
            \right)
        \right) \cdot 
        \Pi_{\DNP}^{X,Y}
        \preceq
        \Pi_{\DNP}^{X,Y}\Lambda_{P_t(N)}^\dagger \Lambda_{P_t(N)}\Pi_{\DNP}^{X,Y}
        \preceq
        \frac{t!}{N^t}
        \left(
            1
            +
            O
            \left(
                \frac{t^2}{N}
            \right)
        \right) \cdot 
        \Pi_{\DNP}^{X,Y}
    \end{align}
\end{lemma}
\begin{proof}
    We have 
    \begin{align}
        \left\lVert
            \frac{N^t}{t!}
            \Lambda_{P_t(N)}^\dagger 
            \Lambda_{P_t(N)}
            \Pi_{\DNP}
            -
            \Pi_{\DNP}
        \right\rVert
        &
        =
        \left\lVert
            \frac{N^t}{t!}
            \left(
                \sum_{\lambda \in \wh{P}_t(N)}
                \beta_{\lambda, t}^2
                \Pi_{\lambda, t}
            \right)
            \Pi_{\DNP}
            -
            \Pi_{\DNP}
        \right\rVert
        \allowdisplaybreaks
        \\
        &
        =
        \left\lVert
            \frac{N^t}{t!}
            \left(
                \sum_{
                    \substack{
                        \lambda \in \wh{P}_t(N)
                        \\
                        |\lambda| = t
                    }
                }
                \beta_{\lambda, t}^2
                \Pi_{\lambda, t}
            \right)
            -
            \left(
                \sum_{
                    \substack{
                        \lambda \in \wh{P}_t(N)
                        \\
                        |\lambda| = t
                    }
                }
                \Pi_{\lambda, t}
            \right)
        \right\rVert
        \tag{\Cref{lem:DNP-fullbox}}
        \allowdisplaybreaks
        \\
        &
        =
        \left\lVert
            \sum_{
                \substack{
                    \lambda \in \wh{P}_t(N)
                    \\
                    |\lambda| = t
                }
            }
            \left(
                \frac{N^t}{t!}
                \beta_{\lambda, t}^2
                -
                1
            \right)
            \Pi_{\lambda, t}
        \right\rVert
        \allowdisplaybreaks
        \\
        &
        \le
        \max_{
            \substack{
                \lambda \in \wh{P}_t(N)
                \\
                |\lambda| = t
            }
        }
        \left\lvert
            \frac{N^t}{t!}
            \beta_{\lambda, t}^2
            -
            1
        \right\rvert
        \allowdisplaybreaks
        \\
        &
        \le
        O
        \left(
            \frac{t^2}{N}
        \right)
        \tag*{(\Cref{lem:irrep-ratios}) \qedhere}
        \,.
    \end{align}
\end{proof}

\begin{lemma}[$\Omega_{\mathcal A_t}$ on the Distinct Nonplussed Subspace]
\label{lem:omega-on-dnp}
Let
\[
    \Pi_{\mathsf{DNP}}^X
    :=
    \left(\Pi_{\DNP_{N,t}}\right)_{R_X},
    \qquad
    \Pi_{\mathsf{DNP}}^Y
    :=
    \left(\Pi_{\DNP_{N,t}}\right)_{R_Y},
    \qquad
    \Pi_{\mathsf{DNP}}^{X,Y}
    :=
    \Pi_{\mathsf{DNP}}^X \Pi_{\mathsf{DNP}}^Y .
\]
Let \(\Omega_{P_t(N)}\) denote the commutant EPR projector for the partition
algebra \(P_t(N)\), and let
\[
    \Omega_{S_t}:=\Omega_{\mathbb C[S_t]}
\]
denote the commutant EPR projector for the symmetric group algebra. Then:
\begin{equation}
    [\Pi_{\mathsf{DNP}}^X,\Omega_{P_t(N)}]=0,
    \qquad
    [\Pi_{\mathsf{DNP}}^Y,\Omega_{P_t(N)}]=0,
\end{equation}
\begin{equation}
    [\Pi_{\mathsf{DNP}}^X,\Omega_{S_t}]=0,
    \qquad
    [\Pi_{\mathsf{DNP}}^Y,\Omega_{S_t}]=0,
\end{equation}
and
\begin{equation}
    \Omega_{S_t}
    \Pi_{\mathsf{DNP}}^{X,Y}
    =
    \Omega_{P_t(N)}
    \Pi_{\mathsf{DNP}}^{X,Y} .
\end{equation}
\end{lemma}

\begin{proof}
We prove the three identities in order.

First, let
\[
    \Pi_{|\lambda|=t}
    :=
    \sum_{\substack{\lambda\in \widehat{P_t(N)}\\ |\lambda|=t}}
    \Pi_\lambda
\]
be the projector onto the full-box irreps in the \((S_N,P_t(N))\)-Schur decomposition
of \((\mathbb C^N)^{\otimes t}\). By \cref{lem:DNP-fullbox},
\[
    \Pi_{|\lambda|=t}=\Pi_{\DNP_{N,t}} .
\]
Thus
\[
    \Pi_{\mathsf{DNP}}^X=(\Pi_{|\lambda|=t})_{\reg R_X},
    \qquad
    \Pi_{\mathsf{DNP}}^Y=(\Pi_{|\lambda|=t})_{\reg R_Y}.
\]
In the double Schur basis, \(\Omega_{P_t(N)}\) is block-diagonal with respect to the irrep labels $(\lambda, \lambda')$. Since $(\Pi_{|\lambda|=t})_{\reg R_X}$ and $(\Pi_{|\lambda|=t})_{\reg R_Y}$ are just projections onto a subset of these blocks, we immediately obtain that they commute with $\Omega_{P_t(N)}$. 

Second, the symmetric-group commutant EPR projector (\cref{ex:group-algebra-EPR-projector}) has the form
\[
    \Omega_{S_t}
    =
    \frac1{t!}
    \sum_{\pi\in S_t}
    S(\pi)_{\reg R_X}\otimes S(\pi)_{\reg R_Y},
\]
For every
\(\pi\in S_t\), the operator \(S(\pi)\) only permutes the \(t\) tensor factors.
The subspace \(\Dist_{N,t}\) is invariant under tensor-factor permutations, and
\[
    \NoPlus_{N,t}=(\Id - \ketbra{+})^{\otimes t}
\]
is also invariant under tensor-factor permutations. Hence
\[
    \DNP_{N,t}=\Dist_{N,t}\cap\NoPlus_{N,t}
\]
is invariant under every \(S(\pi)\). Therefore
\[
    S(\pi)\Pi_{\DNP_{N,t}}
    =
    \Pi_{\DNP_{N,t}}S(\pi)
    \qquad
    \forall \pi\in S_t .
\]
Commuting term by term in the displayed formula for \(\Omega_{S_t}\), we obtain
\[
    [\Pi_{\mathsf{DNP}}^X,\Omega_{S_t}]=0,
    \qquad
    [\Pi_{\mathsf{DNP}}^Y,\Omega_{S_t}]=0 .
\]
For the third claim, we have the following calculation 

\begin{align}
        & \quad  
        \left(
                \Omega^{(t)}_{P_t(N)}
            \right)
            _{\reg{R}}
        \cdot 
        \Pi_{\DNP}^{X,Y}
        \\
        & = 
        \left(
            \sum_{
                D \in \calB(P_t(N))
            } 
            S(D) 
            \otimes 
            S(D^{*_{P_t(N)}})^{T}
        \right)
        \cdot
        \Pi_{\DNP}^{X,Y}
        \tag{\cref{thm:commutant-epr-proj-as-symmetrization}}
        \\
        & = 
        \left(
            \sum_{
                \pi \in S_t
            } 
            S(\pi) 
            \otimes 
            S([\pi]^{*_{P_t(N)}})^{T}
        \right)
        \cdot
        \Pi_{\DNP}^{X,Y}
        \tag{\cref{lem:non_permutation_mapping_to_DNP}}
    \end{align}
    Next, we expand $[\pi]^{*_{P_t(N)}}$ in the basis of $P_t(N)$ as $\sum_{E \in \calB(P_t(N))} \alpha_{\pi,E} \; E$: 
    \begin{align}
        & \quad 
        \left(
            \sum_{
                \pi \in S_t
            } 
            S(\pi) 
            \otimes 
            \sum_{
            E \in \calB(P_t(N))
            }
            \alpha_{\pi,E}\; S(E)^T
        \right)
        \cdot
        \Pi_{\DNP}^{X,Y}
        \\
        & = 
        \left(
            \sum_{
                \pi \in S_t
            } 
            S(\pi) 
            \otimes 
            \sum_{
            \pi' \in \C[S_t]
            }
            \alpha_{\pi, \pi'}\; S(\pi')^T
        \right)
        \cdot
        \Pi_{\DNP}^{X,Y}
        \tag{\cref{lem:non_permutation_mapping_to_DNP}}
        \\
        &= 
        \left(
            \sum_{
                \pi \in S_t
            } 
            S(\pi) 
            \otimes 
            S([\pi]^{*_{\C[S_t]}})^T
        \right)
        \cdot
        \Pi_{\DNP}^{X,Y}
        \tag{\cref{lem:dual_element_of_permutation_in_partition_algebra}}
        \\
        &=
        \left(
                \Omega^{(t)}_{\C[S_t]}
            \right)
            _{\reg{R}}
        \cdot 
        \Pi_{\DNP}^{X,Y}. \qedhere
    \end{align}

\end{proof}

\subsection{Updates in the Distinct Nonplussed Subspace}\label{subsec:adversary-updates-DNP}

Let $V_{S_N}$ and $V_{U(N)}$ be the update operators when $G = S_N$ or $U(N)$, respectively. In this case, the commutant algebra $\calA_t$ is equal to respectively either $P_t(N)$ or $\C[S_t]$, so \Cref{eq:fully_expanded_adversary_action} becomes 

\begin{align}
    \label{eq:permutation_query_adv}
    \mathsf{Adv}_t^{V_{S_N} \,\cdot\, C}
    &=
    \underbrace{
        \vphantom{
            \left(
                \prod_{i=1}^t
                \left(
                    \Omega_{\calA_i}
                \right)_{\reg{R}_{\le i}}
            \right)
        }
    \Big(
        \Lambda_{P_t(N)}
    \Big)
    _{\reg{R}}
    }_{\operatorname{ReweightEnd}}
    \cdot
    \underbrace{
        \vphantom{
            \left(
                \prod_{i=1}^t
                \left(
                    \Omega_{\calA_i}
                \right)_{\reg{R}_{\le i}}
            \right)
        }
    \Big(
        \Omega_{P_t(N)}
    \Big)
    _{\reg{R}}
    }_{\operatorname{Symmetrize}}
    \cdot
    \underbrace{
        \vphantom{
            \left(
                \prod_{i=1}^t
                \left(
                    \Omega_{\calA_i}
                \right)_{\reg{R}_{\le i}}
            \right)
        }
    \Big(
        C^{\otimes t}
    \Big)
    _{\reg{R_{X}}} 
    }_{C^{\otimes t}}
    \cdot
    \underbrace{
        \vphantom{
            \left(
                \prod_{i=1}^t
                \left(
                    \Omega_{\calA_i}
                \right)_{\reg{R}_{\le i}}
            \right)
        }
        \mathsf{Adv}_{t}^{\App} 
    }_{\operatorname{Append}}
    \cdot
    \underbrace{
        \vphantom{
            \left(
                \prod_{i=1}^t
                \left(
                    \Omega_{\calA_i}
                \right)_{\reg{R}_{\le i}}
            \right)
        }
    \Big(
        \Lambda_{P_0(N)}^+
    \Big)
    _{\reg{R}_{0}}
    }_{\operatorname{ReweightStart}}
    \\
    \mathsf{Adv}_t^{V_{U(N)} \,\cdot\, C}
    &=
    \underbrace{
        \vphantom{
            \left(
                \prod_{i=1}^t
                \left(
                    \Omega_{\calA_i}
                \right)_{\reg{R}_{\le i}}
            \right)
        }
    \Big(
        \Lambda_{\C[S_t]}
    \Big)
    _{\reg{R}}
    }_{\operatorname{ReweightEnd}}
    \cdot
    \underbrace{
        \vphantom{
            \left(
                \prod_{i=1}^t
                \left(
                    \Omega_{\calA_i}
                \right)_{\reg{R}_{\le i}}
            \right)
        }
    \Big(
        \Omega_{\C[S_t]}
    \Big)
    _{\reg{R}}
    }_{\operatorname{Symmetrize}}
    \cdot
    \underbrace{
        \vphantom{
            \left(
                \prod_{i=1}^t
                \left(
                    \Omega_{\calA_i}
                \right)_{\reg{R}_{\le i}}
            \right)
        }
    \Big(
        C^{\otimes t}
    \Big)
    _{\reg{R_{X}}} 
    }_{C^{\otimes t}}
    \cdot
    \underbrace{
        \vphantom{
            \left(
                \prod_{i=1}^t
                \left(
                    \Omega_{\calA_i}
                \right)_{\reg{R}_{\le i}}
            \right)
        }
        \mathsf{Adv}_{t}^{\App} 
    }_{\operatorname{Append}}
    \cdot
    \underbrace{
        \vphantom{
            \left(
                \prod_{i=1}^t
                \left(
                    \Omega_{\calA_i}
                \right)_{\reg{R}_{\le i}}
            \right)
        }
    \Big(
        \Lambda_{P_0(N)}^+
    \Big)
    _{\reg{R}_{0}}
    }_{\operatorname{ReweightStart}}
\end{align}

Let 
\begin{align}
    \ket{\mathsf{Adv}_t^{V_{S_N} \,\cdot\, C}} 
    &
    := 
    \mathsf{Adv}_t^{V_{S_N} \,\cdot\, C}\left(\ket{0}_{\reg{AB}} \otimes \ket{\emptyset}_{\reg R}\right)
    \\
    \ket{\mathsf{Adv}_t^{V_{U(N)} \,\cdot\, C}} 
    &
    := 
    \mathsf{Adv}_t^{V_{U(N)} \,\cdot\, C}\left(\ket{0}_{\reg{AB}} \otimes \ket{\emptyset}_{\reg R}\right)
\end{align}

\begin{lemma}
    \label{lem:C-commutes-with-compressed-unitary}
    We can rearrange the terms in 
    $\mathsf{Adv}_t^{V_{U(N)} \,\cdot\, C}$ so that $C^{\otimes t}$ appears at the end, as
    \begin{align}
        \mathsf{Adv}_t^{V_{U(N)} \,\cdot\, C}
        &
        =
        \Big(
            C^{\otimes t}
        \Big)
        _{\reg{R_{X}}} 
        \cdot
        \Big(
            \Lambda_{\C[S_t]}
        \Big)
        _{\reg{R}}
        \cdot
        \Big(
            \Omega_{\C[S_t]}
        \Big)
        _{\reg{R}}
        \cdot
        \mathsf{Adv}_{t}^{\App} 
        \cdot
        \Big(
            \Lambda_{\C[S_0]}^+
        \Big)
        _{\reg{R}_{0}}
        \\
        &
        =
        \Big(
            C^{\otimes t}
        \Big)
        _{\reg{R_{X}}} 
        \mathsf{Adv}_t^{V_{U(N)}}
    \end{align}
\end{lemma}
\begin{proof}
    To see this, it suffices to show that $C^{\otimes t}$ commutes with both $\Omega_{\C[S_t]}$ and $\Lambda_{\C[S_t]}$. This is clear when viewing the three operators in the Schur basis. All three are controlled by the irrep label, and $C^{\otimes t}$ acts only on the unitary (group) register, while $\Omega_{\C[S_t]}$ acts only on the commutant register, and $\Lambda_{\C[S_t]}$ is proportional to the identity within each block.
    \todo{Should we spell this out more formally? It's quite simple.}
\end{proof}

\begin{lemma}[Analyzing $\Pi_{\mathsf{Dist}}^Y \AdvApp \ket{0}$]
\label{lem:permute-distinct}
Let
\[
    \ket{\psi_t}_{\reg A \reg B \reg R} := \AdvApp\ket{0}
\]
be the unnormalized append-only state,%
\footnote{
    For brevity, we will write $\ket{0}$ to represent $\ket{0}_{\reg{AB}}\ket{\emptyset}_{\reg{R}}$.
}
and let
\[
    D_t^Y := (\Pi_{\Dist_{N,t}})_{\reg R_Y}.
\]
For every non-identity \(\pi\in S_t\),
\[
    \bra{\psi_t}
        D_t^Y\Big( S(\pi) \otimes S(\pi) \Big) D_t^Y
    \ket{\psi_t}
    =0.
\]
Also, 
\[
    \bra{\psi_t}
     D_t^Y
    \ket{\psi_t} 
    = 
    N^t 
    \cdot 
    \prod_{i=1}^t 
    \left(
        1
        -
        \frac{i-1}{N}
    \right) 
    = 
    N^t 
    \cdot 
    \left( 
        1
        - 
        O
        \left(
            \frac{t^2}{N}
        \right)
    \right).
\]
\end{lemma}

\begin{proof}
We prove the claim by induction on \(t\). The case $t=1$ is trivial (there are no non-identity permutations and $D_1^Y = \Id$).

For the inductive step, we write
\[ \ket{\psi_t} = \mathsf{App}_{\reg A \reg R_t} \cdot (A_t)_{\reg A \reg B} \cdot \ket{\psi}_{t-1} = \mathsf{App}_{\reg A \reg R_t} \ket{\psi_t^{\mathrm{pre}}}.
\]
Moreover, since $D_t^Y = D_t^Y \cdot D_{t-1}^Y$ and $D_{t-1}^Y$ commutes with $\mathsf{App}_{\reg A \reg R_t}$, we have that
\[ D_t^Y \ket{\psi_t} = D_t^Y \cdot \mathsf{App}_{\reg A \reg R_t} \cdot D_{t-1}^Y \ket{\psi_t^{\mathrm{pre}}}.
\]

We now write explicitly in the standard basis
\[D_{t-1}^Y \ket{\psi_t^{\mathrm{pre}}} = \sum_{\substack{x_1, \hdots, x_t \\ \text{distinct } y_1, \hdots, y_{t-1}}} \ket{x_t}_{\reg A} \ket{\phi_{x_{1:t}, y_{1:t-1}}}_{\reg B} \ket{x_1,\hdots x_{t-1}, y_1, \hdots, y_{t-1}}_{\reg R_{<t}}.
\]
On this state, the effect of $D_t^Y \cdot \mathsf{App}_{\reg A \reg R_t}$ is appending with respect to an (unnormalized) EPR state over $y_t\notin \{y_1, \hdots, y_{t-1}\}$ (controlled on $\{y_1, \hdots, y_{t-1}\}$), which tells us that 
\[D_t^Y \ket{\psi_t} = \sum_{\substack{x_1, \hdots, x_t \\ \text{distinct } y_1, \hdots, y_{t}}} \ket{y_t}_{\reg A} \ket{\phi_{x_{1:t}, y_{1:t-1}}}_{\reg B} \ket{x_1,\hdots x_{t}, y_1, \hdots, y_{t}}_{\reg R_{\leq t}}.
\]
Given that $D_{t-1}^Y$ also commutes with $(A_t)_{\reg A \reg B}$, this immediately tells us that
\[\Big|\Big| D_t^Y \ket{\psi_t} \Big|\Big|^2 = \Big(N-(t-1)\Big)\Big|\Big| D_{t-1}^Y \ket{\psi_{t-1}} \Big|\Big|^2,
\]
completing the first part of the induction. For the second part, we have that
\[\Big(S(\pi) \otimes S(\pi) \Big) D_t^Y \ket{\psi_t} = \sum_{\substack{x_1, \hdots, x_t \\ \text{distinct } y_1, \hdots, y_{t}}} \ket{y_t}_{\reg A} \ket{\phi_{x_{1:t}, y_{1:t-1}}}_{\reg B} \ket{x_{\pi^{-1}(1)},\hdots x_{\pi^{-1}(t)}, y_{\pi^{-1}(1)}, \hdots, y_{\pi^{-1}(t)}}_{\reg R_{\leq t}}.
\]
Now, suppose that $\pi(t)\neq t$. Then, the inner product of these two states is zero because a nonzero term in the expansion of this inner product would require terms $y_t = y'_t$ for the $\reg A$ register but also $y_{\pi^{-1}(t)} = y'_{\pi^{-1}(t)}$ for the $\reg R_{Y, t}$ register. 

On the other hand, suppose that $\pi(t) = t$ and identify $\pi$ as an element of $S_{t-1}$; in particular, $S(\pi)$ commutes with $\mathsf{App}_{\reg A \reg R_t} \cdot (A_t)_{\reg A \reg B}$. Then, we write
\[ \bra{\psi_t} D_t^Y \Big(S(\pi) \otimes S(\pi) \Big) D_t^Y \ket{\psi_t} = \bra{\psi_{t}^{\mathrm{pre}}}D_{t-1}^Y \mathsf{App}_{\reg A, \reg R_t}^\dagger D_t^Y \mathsf{App}_{\reg A, \reg R_t} D_{t-1}^Y  \Big(S(\pi) \otimes S(\pi) \Big) \ket{\psi_{t}^{\mathrm{pre}}}
\]

Next, we make use of the identity
\[D_{t-1}^Y \mathsf{App}_{\reg A, \reg R_t}^\dagger D_t^Y \mathsf{App}_{\reg A, \reg R_t} D_{t-1}^Y = (N-t+1) D_{t-1}^Y,
\]
which again holds because the effect of $D_t^Y \mathsf{App}_{\reg A, \reg R_t}$ on the image of $D_{t-1}^Y$ is to append an un-normalized EPR state over $y_t \notin \{y_1, \hdots, y_{t-1}\}$:
\[D_t^Y \mathsf{App}_{\reg A, \reg R_t} D_{t-1}^Y = \sum_{\substack{x_1, \hdots x_t\\ \text{distinct }y_1, \hdots, y_t}} \ketbra{y_t}{x_t}_{\reg A} \otimes \ketbra{x_1, \hdots, x_t, y_1, \hdots, y_t}{x_1, \hdots, x_{t-1}, y_1, \hdots, y_{t-1}}.
\]
The claimed identity follows from this by a direct computation.

We conclude that
\[
\begin{aligned}
    \bra{\psi_t} D_t^Y \Big( S(\pi) \otimes S(\pi)\Big) D_t^Y \ket{\psi_t}
    &=
    (N-t+1)
    \bra{\psi_t^{\mathrm{pre}}}
        D_{t-1}^Y \Big( S(\pi) \otimes S(\pi)\Big)
    \ket{\psi_t^{\mathrm{pre}}} \\
    &=
    (N-t+1)
    \bra{\psi_{t-1}}
        D_{t-1}^Y (A_t^\dagger)_{\reg A \reg B} \Big( S(\pi) \otimes S(\pi)\Big) (A_t)_{\reg A \reg B}  D_{t-1}^Y
    \ket{\psi_{t-1}} \\
    &= (N-t+1)
    \bra{\psi_{t-1}}
        D_{t-1}^Y  \Big( S(\pi) \otimes S(\pi)\Big)  D_{t-1}^Y
    \ket{\psi_{t-1}}.
\end{aligned}
\]
This completes the induction. 
\end{proof}

In order to argue about $\Pi_{\NoPlus}$, we pass to a ``smaller'' projection,
\begin{align} \Big(\Pi_{\NoPlus}^{\mathsf{FDist}}\Big)_{\reg R_Y} &:= \sum_{\substack{\widehat y_1, \hdots, \widehat y_t \\ \text{distinct and }\neq 0}}  \ketbra{\widehat y_1, \hdots, \widehat y_t}  \\
&= F_N^{\otimes t}\cdot (\Id - \ketbra{0})^{\otimes t} \cdot \Pi_{\mathsf{Dist}}^Y \cdot  (F_N^{\dagger})^{\otimes t},
\end{align}
where $\ket{\widehat y} \coloneqq \frac{1}{\sqrt{N}} \sum_{z \in \Z_N} \omega_{N}^{-yz} \ket{z}$ denotes the $y$th $\mathbb Z_N$ Fourier basis element of $\mathbb C^N$ and $F_N$ denotes the $\mathbb Z_N$ quantum Fourier transform. In other words, $\Pi_{\NoPlus}^{\mathsf{FDist}}$ projects, in the $\mathbb Z_N$ Fourier basis, onto distinct tuples of nonzero basis states. 

\begin{lemma}[Analyzing $\Pi_{\NoPlus}^{\mathsf{FDist}} \AdvApp \ket{0}$]\label{lem:permute-noplus}
Let
\[
    \ket{\psi_t}_{\reg A \reg B \reg R} := \AdvApp\ket{0}
\]
be the unnormalized append-only state, and let
\[
    D_t^Y := (\Pi_{\NoPlus}^{\mathsf{FDist}})_{\reg R_Y}.
\]
For every non-identity \(\pi\in S_t\),
\[
    \bra{\psi_t}
        D_t^Y\Big( S(\pi) \otimes S(\pi) \Big) D_t^Y
    \ket{\psi_t}
    =0.
\]
Also, 
\[
    \bra{\psi_t}
    D_t^Y
    \ket{\psi_t}
    = 
    N^t 
    \cdot 
    \prod_{i=1}^t
    \left(
        1
        -
        \frac{i}{N}
    \right) 
    = 
    N^t 
    \cdot 
    \left(
        1
        -
        O
        \left(
            \frac{t^2}{N}
        \right)
    \right).
\]
\end{lemma}

\begin{proof}[Proof sketch]
The proof is extremely similar to that of \cref{lem:permute-distinct}, again proceeding by induction on $t$. The case $t=1$ is again trivial. For the inductive step, we use the same decomposition 
\[ \ket{\psi_t} = \mathsf{App}_{\reg A \reg R_t} \cdot (A_t)_{\reg A \reg B} \cdot \ket{\psi}_{t-1} = \mathsf{App}_{\reg A \reg R_t} \ket{\psi_t^{\mathrm{pre}}}
\]
and equation
\[ D_t^Y \ket{\psi_t} = D_t^Y \cdot \mathsf{App}_{\reg A \reg R_t} \cdot D_{t-1}^Y \ket{\psi_t^{\mathrm{pre}}}.
\]

Next, we write $D_{t-1}^Y \ket{\psi_t^{\mathrm{pre}}}$ with $\reg R_Y$ in the \emph{Fourier} basis:
\[ D_{t-1}^Y \ket{\psi_t^{\mathrm{pre}}} = \sum_{\substack{x_1, \hdots, x_t \\ \text{distinct } y_1, \hdots, y_{t-1}\neq 0}} \ket{x_t}_{\reg A} \ket{\phi_{x_{1:t}, y_{1:t-1}}}_{\reg B} \ket{x_1,\hdots x_{t-1}, \widehat y_1, \hdots, \widehat y_{t-1}}_{\reg R_{<t}}.
\]
On this state, the effect of $D_t^Y \cdot \mathsf{App}_{\reg A \reg R_t}$ is appending with respect to an (unnormalized) \emph{Fourier basis} EPR state 
\[ \sum_{y_t \notin \{0,\, y_1,\, \hdots,\, y_{t-1}\}} \ket{\widehat{-y_t}}\ket{\widehat{y_t}}
\]
over $y_t \notin \{0,\, y_1,\, \hdots,\, y_{t-1}\}$ (controlled on $\{y_1, \hdots, y_{t-1}\}$), which tells us that 
\[D_t^Y \ket{\psi_t} = \sum_{\substack{x_1, \hdots, x_t \\ \text{distinct } y_1, \hdots, y_{t} \ne 0}} \ket{\widehat{-y_t}}_{\reg A} \ket{\phi_{x_{1:t}, y_{1:t-1}}}_{\reg B} \ket{x_1,\hdots x_{t}, \widehat y_1, \hdots, \widehat y_{t}}_{\reg R_{\leq t}}.
\]
This again completes the inductive norm calculation. In addition, an identical argument now shows that $\bra{\psi_t} D_t^Y \Big(S(\pi) \otimes S(\pi) \Big) D_t^Y \ket{\psi_t} = 0$ whenever $\pi(t)\neq t$. 

On the other hand, if $\pi(t) = t$, then an entirely analogous argument shows that

\[
\begin{aligned}
    \bra{\psi_t} D_t^Y \Big( S(\pi) \otimes S(\pi)\Big) D_t^Y \ket{\psi_t}
    &=
     (N-t)
    \bra{\psi_{t-1}}
        D_{t-1}^Y  \Big( S(\pi) \otimes S(\pi)\Big)  D_{t-1}^Y
    \ket{\psi_{t-1}}
\end{aligned}
\]
(the only difference is that the factor $(N-t)$ replaces $(N-t+1)$, from ruling out $0$ in addition to $y_1,\, \hdots,\, y_{t-1}$), completing the induction. 
\end{proof}

\begin{lemma}
    \label{lem:DNP-Y-Unitary}
    Let $\Pi_{\mathsf{DNP}_{N, t}}$ be the orthogonal projector onto the distinct nonplussed subspace (\Cref{def:dist-nonplussed-subspace}). 
    Then
    \begin{align}
        \Big| \Big|
            \Pi_{\mathsf{DNP}}^{Y}
            \ket{\mathsf{Adv}_t^{V_{U(N)}}}\Big|\Big|^2
        \ge
        1
        -
        O
        \left(
            \frac{t^2}{N}
        \right)
    \end{align}
\end{lemma}
\begin{proof}
    By \cref{corollary:squared_projector_overlap,lem:friedrich_angle_small}, we have that
    \[ \Big| \Big|
            \Pi_{\mathsf{DNP}}^{Y}
            \ket{\mathsf{Adv}_t^{V_{U(N)}}}\Big|\Big|^2
        \ge
        \Big| \Big|
            \Pi_{\mathsf{Dist}}^{Y} \cdot \Pi_{\NoPlus}^Y
            \ket{\mathsf{Adv}_t^{V_{U(N)}}}\Big|\Big|^2
        -
        O
        \left(
            \frac{t^2}{N}
        \right)
    \]
    Thus, by the quantum union bound (\cref{lemma:gao}), it suffices to prove the lemma statement for $\Pi_{\mathsf{Dist}}^Y$ and $\Pi_{\NoPlus}^Y$ rather than $\Pi_{\DNP}^Y$. 

    For $\Pi_{\mathsf{Dist}}^Y$, we begin by writing
    \begin{align} 
        \Pi_{\mathsf{Dist}}^Y 
        \ket{\mathsf{Adv}_t^{V_{U(N)}}} 
        &
        = 
        \Pi_{\mathsf{Dist}}^Y 
        \cdot 
        \Lambda_{S_t} 
        \cdot 
        \Omega_{S_t} 
        \cdot 
        \AdvApp 
        \ket{0}
        \\
        &
        = 
        \Lambda_{S_t} 
        \cdot 
        \Omega_{S_t} 
        \cdot 
        \Pi_{\mathsf{Dist}}^Y 
        \cdot 
        \AdvApp 
        \ket{0}
        \tag{by \Cref{lem:lambda-sym-commutes-with-pi-dnp,lem:omega-on-dnp}}
        \,.
    \end{align}
    Now, by \cref{lem:permute-distinct}, we have that
    \begin{align} 
        \Big\Vert 
            \Omega_{S_t} 
            \cdot 
            \Pi_{\mathsf{Dist}}^Y 
            \cdot 
            \AdvApp 
            \ket{0} 
        \Big\Vert^2 
        &
        = 
        \frac{1}{(t!)^2} 
        \Big\Vert 
            \sum_{\pi \in S_t} 
            (
                S(\pi) 
                \otimes 
                S(\pi)
            ) 
            \cdot 
            \Pi_{\mathsf{Dist}}^Y 
            \cdot 
            \AdvApp 
            \ket{0} 
        \Big\Vert^2 
        \tag{definition of $\Omega_{S_t}$}
        \allowdisplaybreaks
        \\
        &
        = 
        \frac{1}{t!} 
        \Big\Vert 
            \Pi_{\mathsf{Dist}}^Y 
            \cdot 
            \AdvApp 
            \ket{0} 
        \Big\Vert^2 
        \tag{by the first part of \Cref{lem:permute-distinct}}
        \allowdisplaybreaks
        \\
        &
        \geq 
        \frac{N^t}{t!}
        \left(
            1
            -
            O
            \left(
                \frac{t^2}{N}
            \right)
        \right)
        \tag{by the second part of \Cref{lem:permute-distinct}}
        \,. 
    \end{align}
    Finally, we know from \Cref{lem:mh-lambda-scalar} that 
    $\Lambda_{S_t}^2 \succeq \frac{t!}{N^t} \Big(1 - O\Big( \frac {t^2}{N} \Big)\Big) \cdot \Id$, 
    so we conclude that
    \begin{align}
        \left\lVert
            \Pi_{\mathsf{Dist}}^Y 
            \ket{\mathsf{Adv}_t^{V_{U(N)}}} 
        \right\rVert^2
        &
        =
        \left\lVert
            \Lambda_{S_t} 
            \cdot 
            \Omega_{S_t} 
            \cdot 
            \Pi_{\mathsf{Dist}}^Y 
            \cdot 
            \AdvApp 
            \ket{0}
        \right\rVert^2
        \\
        &
        \ge
        \frac{t!}{N^t} 
        \Big(
            1 
            - 
            O
            \Big( 
                \frac{t^2}{N} 
            \Big)
        \Big)
        \left\lVert
            \Omega_{S_t} 
            \cdot 
            \Pi_{\mathsf{Dist}}^Y 
            \cdot 
            \AdvApp 
            \ket{0}
        \right\rVert^2
        \\
        &
        \geq 
        1
        -
        O
        \left( 
            \frac{t^2}{N}
        \right)
        \,. 
    \end{align}
    Next, to handle $\Pi_{\NoPlus}$, we use the fact that
    \begin{align}
        \Big\Vert \Pi_{\NoPlus} \ket{\mathsf{Adv}_t^{V_{U(N)}}}\Big\Vert^2 &\geq \Big\Vert \Pi_{\NoPlus}^{\mathsf{FDist}} \ket{\mathsf{Adv}_t^{V_{U(N)}}}\Big\Vert^2.
    \end{align}
    An identical argument to the above then proves that 
    \begin{align}\Big\Vert \Pi_{\NoPlus}^{\mathsf{FDist}} \ket{\mathsf{Adv}_t^{V_{U(N)}}}\Big\Vert^2 &\geq 1-O\Big(\frac{t^2}{N}\Big).
    \end{align}
This completes the proof.
\end{proof}

\begin{lemma}
    \label{lem:dnp-on-twirled-unitary-recording}
    Let $C$ be sampled from a unitary 2-design $\calD$, and let $\Pi_{\mathsf{DNP}_{N, t}}$ be the orthogonal projector onto the distinct nonplussed subspace (\Cref{def:dist-nonplussed-subspace}). 
    Then
    \begin{align}
        \underset{C \gets \calD}
        {\mathbb{E}}
        \Tr[
            \Pi_{\mathsf{DNP}}^{X,Y}
            \Proj{\mathsf{Adv}_t^{V_{U(N)} \,\cdot\, C}}
        ]
        \ge
        1
        -
        O
        \left(
            \frac{t^2}{N}
        \right)
    \end{align}
\end{lemma}
\begin{proof}
    By a union bound, it suffices to prove the same statement for $\Pi_{\DNP}^X$ and $\Pi_{\DNP}^Y$ separately.

    For $\Pi_{\DNP}^X$ this follows directly from \cref{lem:C-commutes-with-compressed-unitary} and \cref{thm:proj_to_dist_cap_nu}.

    For $\Pi_{\DNP}^Y$, we begin with \cref{lem:C-commutes-with-compressed-unitary}, which tells us that 
    \begin{align} \underset{C}{\mathbb E} \Big| \Big| \Pi_{\DNP}^{Y} \ket{\mathsf{Adv}_t^{V_{U(N)}\cdot C}} \Big| \Big|^2 &= \underset{C}{\mathbb E} \Big| \Big| \Pi_{\DNP}^{Y} C^{\otimes t}_{\reg R_X}\ket{\mathsf{Adv}_t^{V_{U(N)}}} \Big| \Big|^2\\
    &= \underset{C}{\mathbb E} \Big| \Big|C^{\otimes t}_{\reg R_X} \Pi_{\DNP}^{Y} \ket{\mathsf{Adv}_t^{V_{U(N)}}} \Big| \Big|^2 \\
    &= \Big| \Big| \Pi_{\DNP}^{Y} \ket{\mathsf{Adv}_t^{V_{U(N)}}} \Big| \Big|^2,
    \end{align}
    as $C^{\otimes t}_{\reg R_X}$ is unitary and commutes with $\Pi_{\DNP}^Y$. The conclusion now follows from \cref{lem:DNP-Y-Unitary}. 
\end{proof}

\begin{lemma}
    \label{lem:dnp-on-twirled-pemutation-recording}
    Let $C$ be sampled from a unitary 2-design $\calD$, and let $\Pi_{\DNP}^{X,Y}$ be the distinct nonplussed subspace projector. 
    Then
    \begin{align}
        \underset{C \gets \calD}
        {\mathbb{E}}
        \Tr[
            \Pi_{\mathsf{DNP}}^{X,Y}
            \Proj{\mathsf{Adv}_t^{V_{S_N} \,\cdot\, C}}
        ]
        \ge
        1
        -
        O
        \left(
            \frac{t^2}{N}
        \right)
    \end{align}
\end{lemma}
\begin{proof}
    \begin{align}
        \underset{C}
        {\mathbb{E}}
        \left\lVert
            \Pi_{\mathsf{DNP}}^{X,Y}
            \ket{\mathsf{Adv}_t^{V_{S_N} \,\cdot\, C}}
        \right\rVert^2
        &
        =
        \underset{C}
        {\mathbb{E}}
        \left\lVert
            \Pi_{\mathsf{DNP}}^{X,Y}
            \cdot
            \Lambda_{P_t(N)}
            \cdot
            \Omega_{P_t(N)}
            \cdot
            C^{\otimes t}
            _{\reg{R_{X}}} 
            \cdot
            \mathsf{Adv}_{t}^{\App}
            \ket{0}
        \right\rVert^2
        \allowdisplaybreaks
        \\
        &
        =
        \underset{C}
        {\mathbb{E}}
        \left\lVert
            \Lambda_{P_t(N)}
            \cdot
            \Pi_{\mathsf{DNP}}^{X,Y}
            \cdot
            \Omega_{P_t(N)}
            \cdot
            C^{\otimes t}
            _{\reg{R_{X}}} 
            \cdot
            \mathsf{Adv}_{t}^{\App}
            \ket{0}
        \right\rVert^2
        \tag{\Cref{lem:lambda-partition-commutes-with-pi-dnp}}
        \allowdisplaybreaks
        \\
        &
        =
        \underset{C}
        {\mathbb{E}}
        \left\lVert
            \Lambda_{P_t(N)}
            \cdot
            \Pi_{\mathsf{DNP}}^{X,Y}
            \cdot
            \Omega_{S_t}
            \cdot
            C^{\otimes t}
            _{\reg{R_{X}}} 
            \cdot
            \mathsf{Adv}_{t}^{\App}
            \ket{0}
        \right\rVert^2
        \tag{\Cref{lem:omega-on-dnp}}
        \allowdisplaybreaks
        \\
        &
        \ge
        \underset{C}
        {\mathbb{E}}
        \left\lVert
            \Lambda_{S_t}
            \cdot
            \Pi_{\mathsf{DNP}}^{X,Y}
            \cdot
            \Omega_{S_t}
            \cdot
            C^{\otimes t}
            _{\reg{R_{X}}} 
            \cdot
            \mathsf{Adv}_{t}^{\App}
            \ket{0}
        \right\rVert^2
        \left(
            1
            -
            O
            \left(
                \frac{t^2}{N}
            \right)
        \right)
        \tag{\Cref{lem:mh-lambda-scalar,lem:lambda-approx-identity-partition}}
        \allowdisplaybreaks
        \\
        &
        =
        \underset{C}
        {\mathbb{E}}
        \left\lVert
            \Pi_{\mathsf{DNP}}^{X,Y}
            \cdot
            \Lambda_{S_t}
            \cdot
            \Omega_{S_t}
            \cdot
            C^{\otimes t}
            _{\reg{R_{X}}} 
            \cdot
            \mathsf{Adv}_{t}^{\App}
            \ket{0}
        \right\rVert^2
        \left(
            1
            -
            O
            \left(
                \frac{t^2}{N}
            \right)
        \right)
        \tag{\Cref{lem:lambda-sym-commutes-with-pi-dnp}}
        \allowdisplaybreaks
        \\
        &
        =
        \underset{C}
        {\mathbb{E}}
        \left\lVert
            \Pi_{\mathsf{DNP}}^{X,Y}
            \ket{\mathsf{Adv}_t^{V_{U(N)} \,\cdot\, C}}
        \right\rVert^2
        \left(
            1
            -
            O
            \left(
                \frac{t^2}{N}
            \right)
        \right)
        \allowdisplaybreaks
        \\
        &
        \ge
        1
        -
        O
        \left(
            \frac{t^2}{N}
        \right)
        \tag*{(\Cref{lem:dnp-on-twirled-unitary-recording}) \qedhere}
    \end{align}
\end{proof}

\begin{corollary}\label{cor:close-to-DNP}
    Using \Cref{lem:dnp-on-twirled-unitary-recording,lem:dnp-on-twirled-pemutation-recording,lem:proj_and_trace_distance} we get the following trace distance bounds:
    \begin{align}
        \left\lVert
            \Tr_{\reg{R}}^{\phantom{R}}
            \left[
                \underset{C \gets \calD}
                {\mathbb{E}}
                \Proj{\mathsf{Adv}_t^{V_{U(N)} \,\cdot\, C}}
            \right]
            -
            \Tr_{\reg{R}}^{\phantom{R}}
            \left[
                \underset{C \gets \calD}
                {\mathbb{E}}
                \Pi_{\mathsf{DNP}}^{X,Y}
                \Proj{\mathsf{Adv}_t^{V_{U(N)} \,\cdot\, C}}
                \Pi_{\mathsf{DNP}}^{X,Y}
            \right]
        \right\rVert_1
        \le
        O
        \left(
            \frac{t^2}{N}
        \right)
    \end{align}
    \begin{align}
        \left\lVert
            \Tr_{\reg{R}}^{\phantom{R}}
            \left[
                \underset{C \gets \calD}
                {\mathbb{E}}
                \Proj{\mathsf{Adv}_t^{V_{S_N} \,\cdot\, C}}
            \right]
            -
            \Tr_{\reg{R}}^{\phantom{R}}
            \left[
                \underset{C \gets \calD}
                {\mathbb{E}}
                \Pi_{\mathsf{DNP}}^{X,Y}
                \Proj{\mathsf{Adv}_t^{V_{S_N} \,\cdot\, C}}
                \Pi_{\mathsf{DNP}}^{X,Y}
            \right]
        \right\rVert_1
        \le
        O
        \left(
            \frac{t^2}{N}
        \right)
    \end{align}
\end{corollary}

\subsection{Putting it Together}
\label{subsec:putting_together}
Finally, we compare the cases $G = U(N)$ and $G = S_N$ of \Cref{eq:fully_expanded_adversary_action}, and show that the $G$-dependent steps are close in expectation whenever $C$ is sampled from a unitary $2$-design. This implies that queries to $PC$ for a random permutation matrix $P \in S_N$ are (nearly) indistinguishable from queries to $UC = U'$ for a Haar random unitary $U \in U(N)$, leading to the conclusion that $PC$ is a pseudorandom unitary with standard security%
\footnote{
    As mentioned, strong security, which allows other query types, requires a separate argument, which we do not provide here.
}
for any pseudorandom permutation $P$ and any unitary $2$-design $C$.

\begin{theorem}\label{thm:PRU-random-P}
    \begin{align}
        \left\lVert
            \underset{
                \substack{
                    P \gets S_N
                    \\
                    C \gets \calD
                }
            }
            {\mathbb{E}}
            \Big[
                \proj{\mathsf{Adv}_t^{P \,\cdot\, C}}
            \Big]
            -
            \underset{U \gets U(N)}
            {\mathbb{E}}
            \Big[
                \proj{\mathsf{Adv}_t^{U}}
            \Big]
        \right\rVert_1
        \le
        O
        \left(
            \frac{t^2}{N}
        \right)
    \end{align}
\end{theorem}
\begin{proof}
    By the correctness of the path-recording oracles, we know that for every $C$,
    \begin{align}
        \Tr_{\reg{R}}^{\phantom{R}}
        \!\!
        \left[
            \;
            \Proj{\mathsf{Adv}_t^{V_{U(N)} \,\cdot\, C}}
            \;
        \right] 
        &= 
        \underset{U \gets U(N)}
        {\mathbb{E}}
        \Proj{\mathsf{Adv}_t^{U\cdot C}}
        &= 
        \underset{U \gets U(N)}
        {\mathbb{E}}
        \Proj{\mathsf{Adv}_t^{U}}
    \end{align}
    and
    \begin{align}
        \Tr_{\reg{R}}^{\phantom{R}}
        \!\!
        \left[
            \;
            \Proj{\mathsf{Adv}_t^{V_{S_N} \,\cdot\, C}}
            \;
        \right] 
        &=
        \underset{P \gets S_N}
        {\mathbb{E}}
        \Proj{\mathsf{Adv}_t^{P\cdot C}}.
    \end{align}

    Therefore, by \cref{cor:close-to-DNP}, it suffices to prove that
        \begin{align}
        \left\lVert
            \Tr_{\reg{R}}^{\phantom{R}}
            \!\!
            \left[
                \underset{
                    \substack{
                        C \gets \calD
                    }
                }
                {\mathbb{E}}
                \Pi_{\DNP}^{X,Y}\Proj{\mathsf{Adv}_t^{V_{S_N} \,\cdot\, C}}\Pi_{\DNP}^{X,Y}
            \right]
            \!
            -
            \Tr_{\reg{R}}^{\phantom{R}}
            \!\!
            \left[
                \underset{
                    \substack{
                        C \gets \calD
                    }
                }
                {\mathbb{E}}
                \Pi_{\DNP}^{X,Y}\Proj{\mathsf{Adv}_t^{V_{U(N)} \,\cdot\, C}}\Pi_{\DNP}^{X,Y}
            \right]
        \right\rVert_1
        \le
        O
        \left(
            \frac{t^2}{N}
        \right)
    \end{align}
    In fact, we prove the following stronger statement: for every fixed $C$, we have that 

    \begin{align}
        \left\lVert
                \Pi_{\DNP}^{X,Y}
                \Proj{\mathsf{Adv}_t^{V_{S_N} \,\cdot\, C}}
                \Pi_{\DNP}^{X,Y}
            -
                \Pi_{\DNP}^{X,Y}
                \Proj{\mathsf{Adv}_t^{V_{U(N)} \,\cdot\, C}}
                \Pi_{\DNP}^{X,Y}
        \right\rVert_1
        \le
        O
        \left(
            \frac{t^2}{N}
        \right)
    \end{align}
    To see this, we make use of the standard manipulation\todo{add reference/citation, or not}

        \begin{align}
        &\left\lVert
                \Pi_{\DNP}^{X,Y}\proj{\mathsf{Adv}_t^{V_{S_N} \,\cdot\, C}}\Pi_{\DNP}^{X,Y}
            -
                \Pi_{\DNP}^{X,Y}\proj{\mathsf{Adv}_t^{V_{U(N)} \,\cdot\, C}}\Pi_{\DNP}^{X,Y}
        \right\rVert_1 \nonumber \\
        \le &
        2 \cdot \left\lVert \Pi_{\DNP}^{X,Y}\ket{\mathsf{Adv}_t^{V_{S_N} \,\cdot\, C}} - \Pi_{\DNP}^{X,Y}\ket{\mathsf{Adv}_t^{V_{U(N)} \,\cdot\, C}}
        \right\rVert
    \end{align}
    Now, given that our starting state is empty, we can write (for both $V = V_{U(N)}$ and $V = V_{S_N}$)

        \begin{equation}
        \ket{\mathsf{Adv}_t^{V \cdot C}}
        =
        \Big(
            \Lambda_{\calA_t}
        \Big)
        _{\reg{R}_{\le t}}
        \cdot
        \Big(
            \Omega_{\calA_t}
        \Big)
        _{\reg{R}_{\le t}}
        \cdot
        \Big(
            C^{\otimes t}
        \Big)
        _{\reg{R_{X}}} 
        \cdot
            \mathsf{Adv}_{t}^{\App} 
        \cdot \ket{0}.
    \end{equation}
Letting $\ket{\psi} = \Big(
            C^{\otimes t}
        \Big)
        _{\reg{R_{X}}} 
        \cdot
            \mathsf{Adv}_{t}^{\App} 
        \cdot \ket{0}$, we have
\begin{align}
    \left\lVert \Pi_{\DNP}^{X,Y}\ket{\mathsf{Adv}_t^{V_{S_N} \,\cdot\, C}} - \Pi_{\DNP}^{X,Y}\ket{\mathsf{Adv}_t^{V_{U(N)} \,\cdot\, C}}
        \right\rVert &= \left\lVert \Pi_{\DNP}^{X,Y}\Big(\Lambda_{P_t(N)} \Omega_{P_t(N)} - \Lambda_{S_t} \Omega_{S_t} \Big) \ket{\psi}
        \right\rVert  \\
        &= \left\lVert \Big( \Lambda_{P_t(N)} \Pi_{\DNP}^{X,Y}\Omega_{P_t(N)} - \Lambda_{S_t} \Pi_{\DNP}^{X,Y}\Omega_{S_t} \Big)\ket{\psi}
        \right\rVert \\
        &= \left\lVert \Big( \Lambda_{P_t(N)}- \Lambda_{S_t}\Big)  \Pi_{\DNP}^{X,Y}\Omega_{S_t}\ket{\psi}
        \right\rVert \\
        &= \left\lVert \Big( \Lambda_{P_t(N)}- \Lambda_{S_t}\Big)  \Pi_{\DNP}^{X,Y}\Lambda_{S_t}^+\ket{\mathsf{Adv}_t^{V_{U(N)} \,\cdot\, C}}
        \right\rVert \\
        &\leq \left\lVert \Big( \Lambda_{P_t(N)}- \Lambda_{S_t}\Big)  \Pi_{\DNP}^{X,Y}
        \right\rVert \cdot \left\lVert \Lambda_{S_t}^+
        \right\rVert. 
\end{align}
By \cref{lem:mh-lambda-scalar,lem:lambda-approx-identity-partition}, on the full-box/DNP blocks the squared reweighting matrices
\(\Lambda_{P_t(N)}^\dagger \Lambda_{P_t(N)}\) and \(\Lambda_{\mathbb C[S_t]}^\dagger \Lambda_{\mathbb C[S_t]}\) differ by a multiplicative
\(1+O(t^2/N)\). Taking square roots preserves this error up to constants, so it follows that
\begin{align}
    \left\lVert \Big( \Lambda_{P_t(N)}- \Lambda_{S_t}\Big)  \Pi_{\DNP}^{X,Y}
        \right\rVert &\leq \sqrt{\frac{t!}{N^t}} \cdot O\Big(\frac{t^2}{N}\Big).
\end{align}
Similarly, by \cref{lem:mh-lambda-scalar} we have that
\begin{align}
    \left\lVert \Lambda_{S_t}^+
        \right\rVert &\leq \sqrt{\frac{N^t}{t!}} \cdot \Big(1+O\Big(\frac{t^2}{N}\Big)\Big).
\end{align}
This completes the proof of the theorem. 
\end{proof}

The following corollary formalizes \cref{thm:main-PC-intro,thm:PC-design-intro} to conclude that $P_k \cdot C$ is either a pseudorandom unitary or a $t$-design, if $P_k$ is instantiated with respectively either a pseudorandom permutation or a $2t$-wise independent permutation.

\begin{corollary}\label{cor:PRU-design}
    Suppose that $P_k$ is a $(t,\epsilon)$-secure in-place post-quantum PRP, meaning that for all polynomial-time $t$-query quantum algorithms $\mathsf{Adv}$, we have that $\mathsf{Adv}^{P_k} \approx_\epsilon \mathsf{Adv}^{P}$ for a uniformly random permutation unitary $P$. Moreover, let $C \gets \mathcal D$ be a unitary sampled from an exact unitary $2$-design. 
    Then, $P_k \cdot C$ is a $(t, \epsilon + O(t^2/N))$-secure PRU. 
    Moreover, if $P_k$ is secure against inefficient $\mathsf{Adv}$, so is $P_k \cdot C$. 
\end{corollary}

\begin{proof}
    This immediately follows from \cref{thm:PRU-random-P} and a hybrid argument switching $P_k$ to a uniformly random $P$. 
\end{proof}

\ifanon
\else 
\section*{Acknowledgements}
The authors thank 
Henry Yuen, 
John Bostanci, 
Quynh Nguyen, 
Natalie Parham, 
Angelos Pelecanos,
Mark Zhandry, 
Joe Carolan,
and
Takashi Yamakawa
for helpful discussions.

This work was done in part while AL and BN were visiting the Simons Institute for the Theory of Computing, where BN was supported by NSF QLCI Grant No. 2016245.
BN acknowledges the following funding sources: AFOSR award FA9550-23-1-0363, NSF awards CCF-2530159, CCF-2144219, and CCF-2329939, and the Sloan Foundation.
AL was supported in part by NSF CAREER award CNS-2541300 and an E. Lawrence Keyes, Jr./Emerson Electric Co. Faculty Award.
J.W.\ was supported by the NSF CAREER award CCF-233971 and a fellowship from the Sloan Foundation.

\fi

\bibliographystyle{alpha}
\bibliography{wright,ref,abbrev3,crypto}

\end{document}